\documentclass[twoside,11pt]{article}

\usepackage[nohyperref]{jmlr2e}
\usepackage[english]{babel}

\usepackage[T1]{fontenc}
\usepackage{algpseudocode,algorithmicx,algorithm}
\usepackage{amsmath,amsfonts,amsbsy,mathtools}
\usepackage{array}
\usepackage{booktabs}
\usepackage{longtable}
\usepackage{multirow}
\usepackage{multicol}
\usepackage{ragged2e}
\usepackage{changepage}
\usepackage{rotating}
\usepackage{needspace}
\usepackage{microtype}
\usepackage{enumitem}
\usepackage{listings}
\usepackage[normalem]{ulem}
\usepackage{bigfoot}
\usepackage[]{matlab-prettifier}
\usepackage{prodint}
\usepackage{placeins}
\usepackage{url}
\usepackage{xcolor}
\definecolor{ABGKLink}{HTML}{2F4F6F}
\definecolor{ABGKCite}{HTML}{1F6F8B}
\definecolor{ABGKUrl}{HTML}{5A4E7A}
\usepackage[
    colorlinks=true,
    linkcolor=ABGKLink,
    citecolor=ABGKCite,
    urlcolor=ABGKUrl,
    filecolor=ABGKLink,
    pdfborder={0 0 0}
]{hyperref}
\hypersetup{
    pdftitle={From Risk Sets to Martingales: A Counting-Process Framework for Event-History Learning},
    pdfauthor={Elvis Han Cui},
    pdfsubject={Counting-process framework for event-history learning},
    pdfkeywords={counting processes, event-history analysis, censored learning, martingale methods}
}

\usepackage{abgk_theme}
\newcommand{\indep}{\perp \!\!\! \perp}

\setlist{leftmargin=1.35em,itemsep=2pt,topsep=3pt,parsep=0pt,partopsep=0pt}

\newcommand{\Prob}{\mathbb{P}}
\newcommand{\p}{\mathbf{P}}

\lstMakeShortInline"

\ShortHeadings{Counting-Process Framework}{Cui}
\firstpageno{1}

\begin{document}

\title{From Risk Sets to Martingales: A Counting-Process Framework for Event-History Learning}

\author{\name Elvis Han Cui \email                           elviscuihan@g.ucla.edu\\
       \addr Department of Biostatistics\\
       University of California, Los Angeles, CA 90095, USA \\
       Kuntuo: an IQVIA company, China \\
       }

\maketitle

\begin{abstract}
Counting-process notation separates predictable risk-set information from observed event jumps through decompositions of the form \(dN(t)=Y(t)\alpha(t)dt+dM(t)\). This article develops a unified event-history learning framework for censored, truncated, recurrent, multistate, and covariate-dependent data. Rather than cataloguing survival methods, the treatment translates each partially observed learning target into five recurring objects: risk process, jump process, compensator, estimating equation, and limiting argument. The framework connects right-censored survival curves, product-integral estimators, bivariate and interval-censored survival estimators, log-rank tests, Cox--Andersen--Gill regression, additive hazards, accelerated failure-time models, panel-count data, landmark prediction, semi-Markov models, Bayesian nonparametric transition models, and instrumental-variable methods. The original contribution is threefold. Computationally, the article turns risk-set sweeps, product-integral updates, interval-likelihood calculations, semi-Markov elapsed-time bookkeeping, Bayesian transition-hazard updating, and cross-fitted validation into reusable algorithms and simulation diagnostics. Theoretically, it gives proof templates for the recurring martingale, likelihood, product-integral, and empirical-process arguments, and proves a new out-of-fold compensator validation identity for cross-fitted censored learners. For applications, it maps biomedical, reliability, operational, economic, financial, literary, historical, and causal survival examples onto the same risk-set and compensator language. The resulting account provides a common mathematical language for deriving, checking, and comparing classical and machine-learning methods for censored, recurrent, and multistate event-history data.
\end{abstract}
\medskip

\clearpage
\begingroup
\setcounter{tocdepth}{2}
\tableofcontents
\endgroup
\clearpage

\phantomsection
\addcontentsline{toc}{section}{Introduction: Counting Processes for Event-History Learning}
\section*{Introduction: Counting Processes for Event-History Learning}

Survival and event-history data are intrinsically filtered observations rather than complete response variables. At a time just prior to \(t\), the analyst observes the accumulated history: which subjects are under follow-up, which covariates and prior states are available, and which individuals remain eligible for the event or transition under study. The jump at \(t\), by contrast, is revealed only when it occurs. Counting-process notation formalizes this temporal asymmetry through the basic multiplicative-intensity representation
\[
dN(t)=Y(t)\alpha(t)dt+dM(t),
\]
where \(Y(t)\) is the predictable risk process, \(Y(t)\alpha(t)dt\) is the compensator increment, and \(dM(t)\) is the martingale innovation. This decomposition supplies a common analytic basis for the Kaplan--Meier and Nelson--Aalen estimators, log-rank statistics, Cox--Andersen--Gill regression, Aalen--Johansen product integrals, and extensions to recurrent, multistate, interval-observed, and covariate-dependent event histories \citep{kaplan1958nonparametric,cox1972regression,aalen1978nonparametric,andersen1982cox,andersen1993statistical,fleming2011counting,aalen2008survival}.

This article uses the representation as an organizing framework for event-history learning rather than as a catalogue of named survival methods. Each partially observed learning problem is expressed through five objects: the risk process, the jump process, the compensator, the estimating equation, and the limiting argument. Such an organization is useful both for classical survival analysis and for modern censored-learning problems, since inferential validity in both settings depends on making the risk set, censoring mechanism, and filtration restrictions explicit.

\phantomsection
\addcontentsline{toc}{subsection}{Running Example and Scope}
\Needspace{6\baselineskip}
\subsection*{Running Example and Scope}
\begin{figure}[tbp]
\centering
\includegraphics[width=\textwidth]{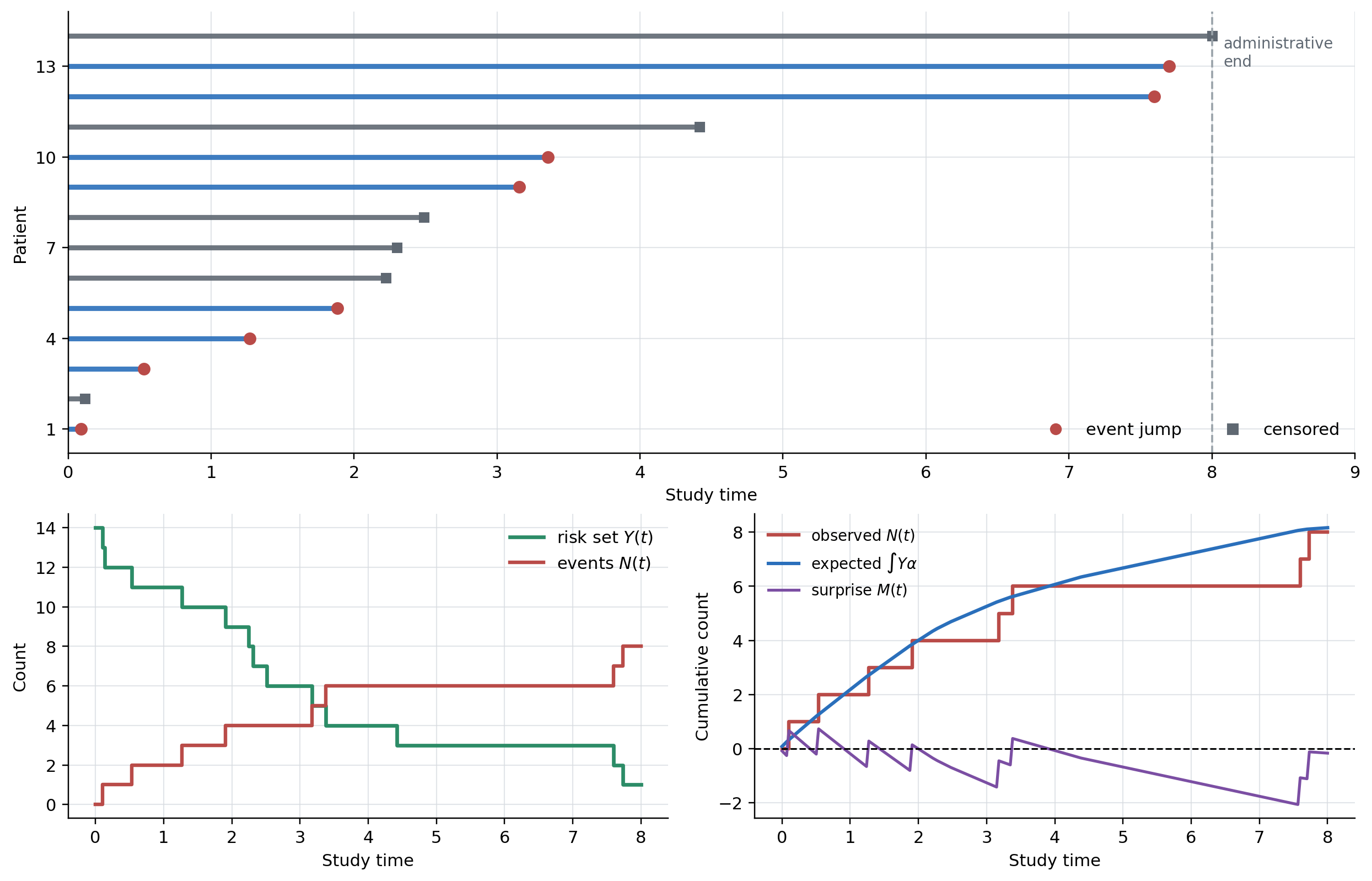}
\caption[Simulated registry in counting-process notation]{Simulated registry in counting-process notation.}
\label{fig:registry_processes}
\end{figure}

The running example is a transplant registry in which patients may relapse, die, remain under administrative follow-up, or be censored. Such records determine a predictable risk set \(Y(t)\), an event count \(N(t)\), and a compensator implied by the selected intensity model. Figure~\ref{fig:registry_processes} displays these objects together with the corresponding martingale residual. Reliability and operational examples have the same stochastic structure: units enter observation, remain at risk while operating, and may jump to degradation, repair, replacement, or failure states before censoring or administrative closure.

The scope is deliberately broad but analytically bounded. Biomedical survival and multistate data provide the main setting. Reliability, operations, economic duration records, financial point processes, spatially indexed disease registries, literary reception histories, historical office spells, and instrumental-variable survival designs are used as secondary examples only when they clarify how the admissible risk set, counted jump, compensator, or estimand changes. The cross-domain notation and data-set menus are collected in Appendix Tables~\ref{tab:example_scope}--\ref{tab:literary_history_dataset_menu}.

\phantomsection
\addcontentsline{toc}{subsection}{Contributions}
\Needspace{7\baselineskip}
\subsection*{Contributions}
The article contributes along three dimensions. First, it develops a common notation for right censoring, delayed entry, multistate transitions, recurrent events, panel counts, interval censoring, landmark prediction, semi-Markov transitions, Bayesian nonparametric transition models, and instrumental-variable survival models. Second, it presents the main estimators as proof blocks: identify the martingale, likelihood identity, or product-integral perturbation; compute the predictable variation or relevant derivative; state the censoring, positivity, and regularity conditions; and apply the appropriate limit theorem. Third, it translates recurring computations into algorithms and diagnostics, including risk-set sweeps, product-integral updates, interval-likelihood updates, semi-Markov elapsed-time bookkeeping, Bayesian transition-hazard updating, and cross-fitted validation.

The main new formal result is Theorem~\ref{thm:oof_compensator_validation}, an out-of-fold compensator validation identity for cross-fitted censored learners. Conditional on the training folds, validation-fold residual scores are martingale centered precisely when the learned intensity is calibrated in a weighted integrated-intensity sense. The result interprets cross-fitting as a filtration requirement: validation predictions may depend on the training sample and on the subject history available just before \(t\), but not on future validation jumps.

\phantomsection
\addcontentsline{toc}{subsection}{Assumptions and Organization}
\Needspace{8\baselineskip}
\subsection*{Assumptions and Organization}
Unless a result states otherwise, all limits are taken on a fixed interval \([0,\tau]\); subjects are independent observational units; covariates and at-risk indicators are predictable; counting processes have locally bounded intensities; and the relevant risk-set proportions are bounded away from zero on the region where hazards or transition probabilities are estimated. Right censoring and delayed entry are assumed conditionally independent given the observed history used by the model. Instrumental-variable results add relevance, independence, exclusion, positivity, and censoring assumptions on the event-process filtration; these are causal restrictions, not consequences of counting-process notation.

Section~1 develops right-censored survival curves and product-integral extensions. Section~2 derives one-sample and \(k\)-sample log-rank tests as martingale risk-set contrasts. Section~3 introduces covariates, panel counts, landmark prediction, Bayesian multistate learning, semi-Markov models, and modern censored-learning diagnostics. Section~4 treats causal targets with censored or competing-risk endpoints through instrumental-variable restrictions. The appendix collects the notation map, data examples, and the martingale, likelihood, product-integral, and empirical-process tools used in the proofs.

\phantomsection
\addcontentsline{toc}{subsection}{Foundational Reference Timeline}
\Needspace{7\baselineskip}
\subsection*{Foundational Reference Timeline}
\begin{figure}[tbp]
\centering
\includegraphics[width=\textwidth,height=0.70\textheight,keepaspectratio]{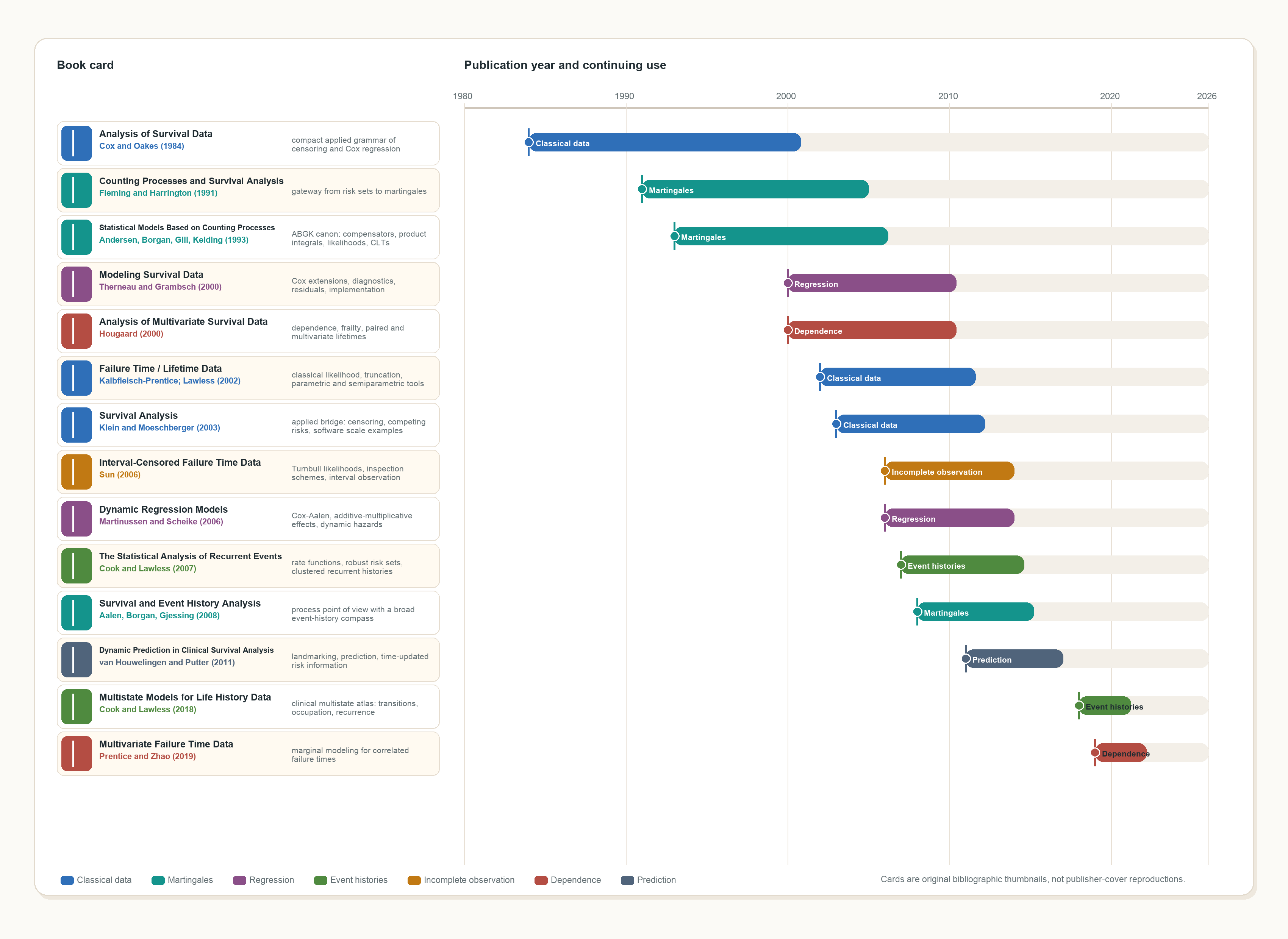}
\caption[Foundational reference timeline]{Foundational reference timeline. Book cards list the monographs used as structural guides, and the horizontal bars place each reference by publication year and continuing methodological role: classical data analysis, martingales, regression, event histories, incomplete observation, dependence, and prediction.}
\label{fig:foundational_reference_timeline}
\end{figure}

The reference frame is organized around the monographs and methodological papers that define the recurring objects used throughout the article: risk sets, compensators, martingales, product integrals, interval likelihoods, recurrent-event processes, semi-Markov elapsed time, and dynamic prediction. Figure~\ref{fig:foundational_reference_timeline} converts this reference frame into a timeline. The book cards identify the monographs that anchor the notation, while the year axis indicates how their roles extend across classical failure-time analysis, martingale calculus, regression, event histories, incomplete observation, dependence modeling, and prediction.

\phantomsection
\addcontentsline{toc}{subsection}{Formal Blocks and Simulation Companions}
\Needspace{7\baselineskip}
\subsection*{Formal Blocks and Simulation Companions}
\begin{figure}[tbp]
\centering
\includegraphics[width=\textwidth,height=0.70\textheight,keepaspectratio]{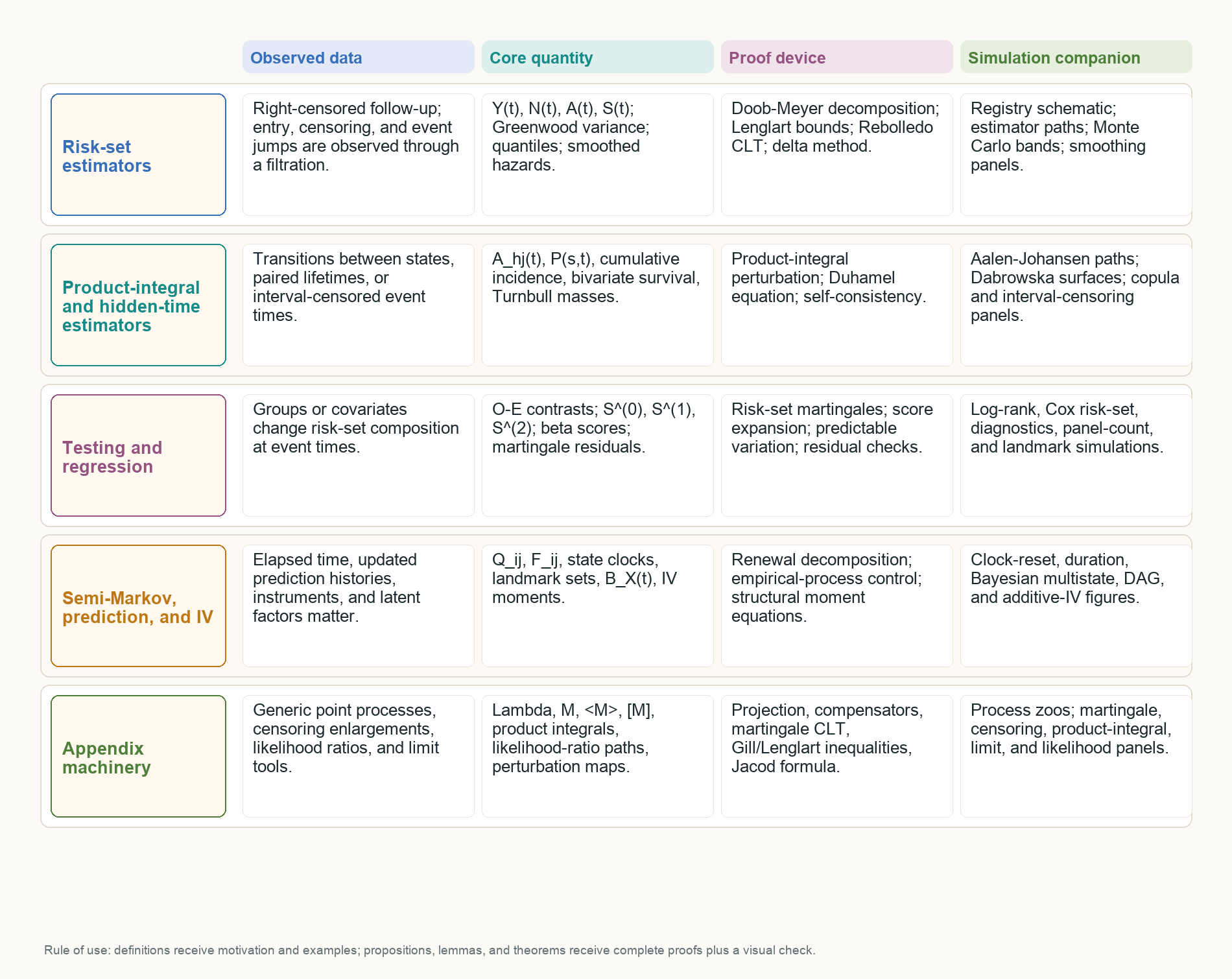}
\caption[Formal-result companion map]{Formal-result companion map. Rows connect observation schemes, stochastic quantities, proof devices, and simulation companions.}
\label{fig:formal_companion_map}
\end{figure}

The formal map in Figure~\ref{fig:formal_companion_map} identifies where the article's original computational and proof components enter the exposition. Each row connects an observation scheme with the corresponding stochastic quantities, proof device, and finite-sample simulation companion. This arrangement keeps algorithms and tables close to the estimands they support: the risk-set sweep appears with right censoring, the product-integral update appears with Aalen--Johansen estimation, the Turnbull blueprint appears with interval censoring, and the cross-fitted censored-learning wrapper appears with modern prediction diagnostics.

\begin{figure}[tbp]
\centering
\includegraphics[width=\textwidth,height=0.46\textheight,keepaspectratio]{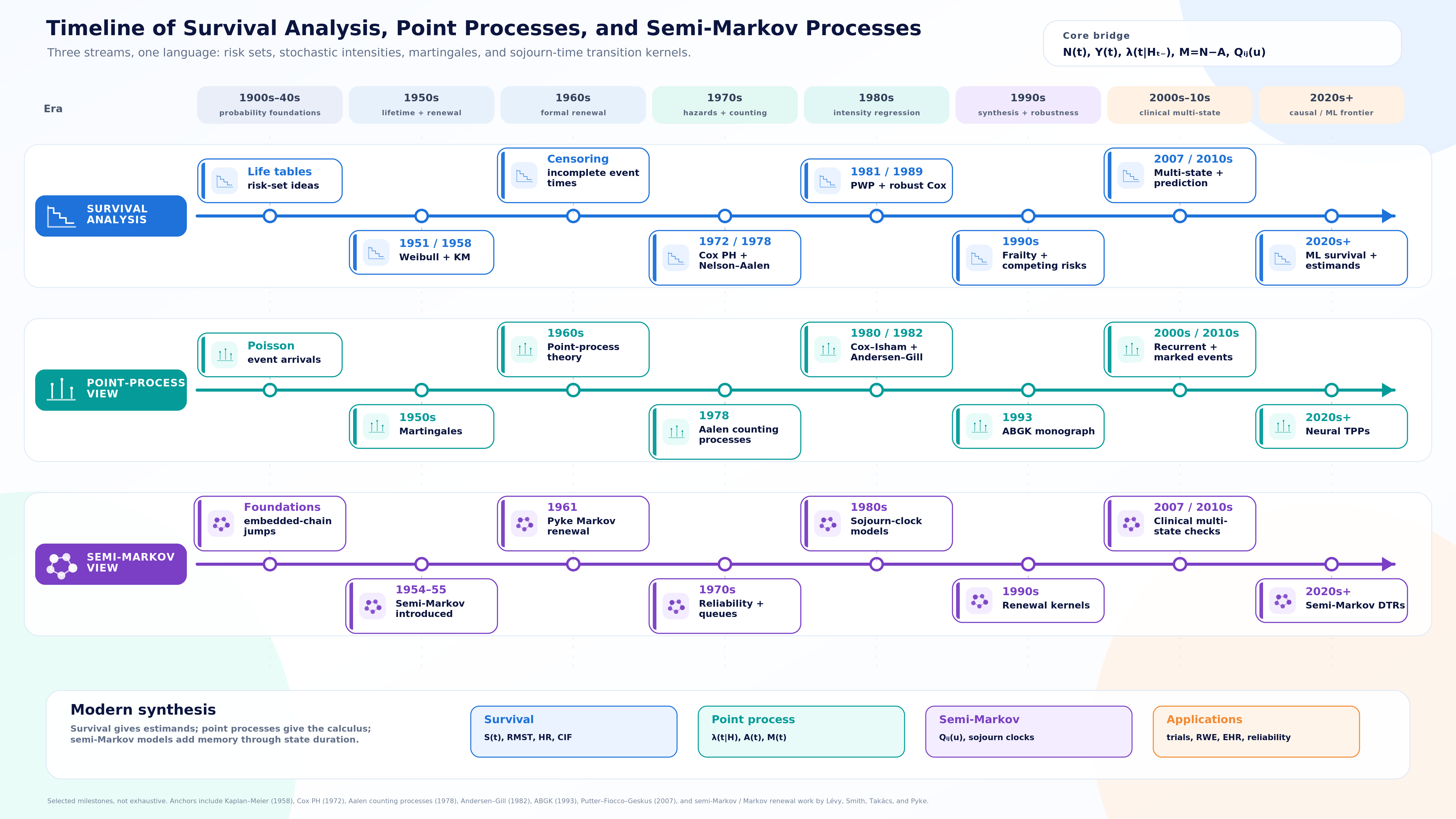}
\caption[Survival, point-process, and semi-Markov timeline]{Survival, point-process, and semi-Markov timeline. The three horizontal streams align milestones in survival analysis, point-process theory, and semi-Markov modeling, while the lower synthesis band records the shared objects used throughout the article.}
\label{fig:survival_point_semimarkov_timeline}
\end{figure}

Figure~\ref{fig:survival_point_semimarkov_timeline} places the same foundations on a broader methodological timeline. The three streams show how survival analysis, point-process calculus, and semi-Markov modeling repeatedly reuse the same bridge objects: risk sets, stochastic intensities, martingales, and sojourn-time transition kernels.

\section{From Risk Sets to Survival Curves}

The first setting is right-censored event-time data without covariates, multiple states, or causal assumptions. The available information is the risk set just before each event time and the jump observed at that time. Two quantities carry the nonparametric analysis. The cumulative hazard
\[
    A(t)=\int_0^t\alpha(s)ds
\]
records cumulative event intensity, while the survival curve
\[
    S(t)=\mathbb{P}(T>t)
\]
records the probability of remaining event-free. The Nelson--Aalen estimator estimates the first quantity by adding observed event increments divided by the number at risk; the Kaplan--Meier estimator estimates the second by multiplying the corresponding conditional survival probabilities. Once these two objects are fixed, later estimators differ mainly in the admissible risk set, the counted transition, and the estimation operation: additive hazard increment, product-integral transition update, or interval likelihood maximization.

Algorithm~\ref{alg:riskset_sweep} records the common event-time sweep behind Nelson--Aalen, Kaplan--Meier, log-rank, and Cox--Andersen--Gill calculations. The key convention is that subjects failing at time \(t\) remain in the predictable risk set \(Y(t-)\) for the increment at \(t\), and are removed only after the jump has been processed.

\begin{algorithm}[tbp]
\caption{Risk-set sweep for one-jump and regression estimators}
\label{alg:riskset_sweep}
\begin{algorithmic}[1]
\Require Records \((L_i,\widetilde T_i,\Delta_i,Z_i)\), ordered event times \(\mathcal T\), optional groups \(G_i\)
\Ensure Nelson--Aalen/Kaplan--Meier increments, log-rank contrasts, and Cox score ingredients
\State Sort all entry, event, censoring, and covariate-update times.
\State Initialize \(Y_i(t)=0\), cumulative hazards \(\widehat A_j(0)=0\), score \(U(0)=0\), and information \(I(0)=0\).
\For{each ordered event time \(t\in\mathcal T\)}
    \State Add subjects with \(L_i<t\) and remove subjects censored before \(t\); keep subjects failing at \(t\) in the risk set \(Y(t-)\).
    \State Compute \(Y(t)=\sum_iY_i(t)\), event increments \(dN_j(t)\), and covariate sums \(S^{(k)}(\beta,t)=\sum_iY_i(t)e^{\beta^TZ_i}Z_i^{\otimes k}\).
    \State Update \(d\widehat A_j(t)=dN_j(t)/Y(t)\) when the target is a cumulative hazard.
    \State Update \(\widehat S(t)=\widehat S(t-)\{1-dN(t)/Y(t)\}\) when the target is survival.
    \State Accumulate log-rank residuals \(O_g(t)-E_g(t)\) or Cox score increments \(dN_i(t)\{Z_i-\bar Z(\beta,t)\}\).
    \State Remove subjects whose censoring or terminal event occurs at \(t\) after all predictable quantities at \(t-\) have been used.
\EndFor
\State Return the requested curve, score, variance, or diagnostic residual process.
\end{algorithmic}
\end{algorithm}

\subsection{Cumulative Hazards: The Nelson--Aalen Estimator}

Suppose within a small interval $[t, t+dt)$, there is a probability $\alpha(t)dt$ of an event occurring for each subject still under observation. Thus, the expected number of events in this interval is $Y(t)\alpha(t)dt$, where $Y(t)=\sum_i \mathbb{I}(T_i \geq t)$ counts the number of individuals at risk at time $t$. The process \(Y(t)\) is predictable because it is known just before \(t\), while \(dN(t)\) is revealed only when the event time is observed. The decomposition is
\[
dN(t) = Y(t)\alpha(t)dt + dM(t),
\]
where the first term is predictable, and the second term represents the increment of a martingale. Rearranging terms gives:
\[
\frac{dN(t)}{Y(t)} - \frac{dM(t)}{Y(t)} = \alpha(t)dt.
\]
This leads to the following estimator for $A(t) = \int_0^t \alpha(s)ds$:
\[
\widehat{A}(t) = \int_{(0, t]} \frac{J(s)}{Y(s)}dN(s),
\]
where $J(s) = \mathbb{I}(Y(s) > 0)$ ensures that division is performed only when the risk set is nonempty, and $0/0 = 0$. This is the \textbf{Nelson--Aalen estimator} for the cumulative hazard function.
At each event time the estimator adds one observed jump divided by the number of subjects exposed to that jump.

As a small risk-set calculation, suppose the observed pairs $(\widetilde T_i,\Delta_i)$ are
\[
(1,1),\quad (2,0),\quad (3,1),\quad (5,1),\quad (6,0),
\]
where $\Delta_i=1$ denotes an observed event and $\Delta_i=0$ denotes censoring. The event times are $1,3,$ and $5$. Just before these times the risk-set sizes are $Y(1)=5$, $Y(3)=3$, and $Y(5)=2$. Hence
\[
\widehat A(5)=\frac{1}{5}+\frac{1}{3}+\frac{1}{2}.
\]
The Kaplan-Meier curve uses the same risk-set construction but multiplies survival factors:
\[
\widehat S(5)=\left(1-\frac{1}{5}\right)
\left(1-\frac{1}{3}\right)
\left(1-\frac{1}{2}\right).
\]
This finite-sample calculation is the scalar version of the construction used later with more states, more event types, or covariates.

\subsubsection{Variance Estimator}

The estimator is random because the jumps are random. The martingale representation identifies the part of that randomness that has accumulated by time $t$. This accumulated uncertainty is measured by predictable variation. To estimate the variance of $\widehat{A}(t)$, let $A^*(t)=\int_0^t J(s)\alpha(s)ds$ and $A(t)=\int_0^t\alpha(s)ds$. The difference $(\widehat{A}-A^*)(t)$ can be expressed as:
\[
(\widehat{A} - A^*)(t) = \int_0^t \frac{J(s)}{Y(s)} \left(dN(s) - Y(s)\alpha(s)ds\right).
\]
Since $dN(s)-Y(s)\alpha(s)ds=dM(s)$, where $M(\cdot)$ is the counting-process martingale, it follows that $\widehat{A}-A^*$ is a mean-zero martingale.
A natural variance estimator is obtained by plugging in the Nelson-Aalen estimator into the compensator representation:
\begin{equation*}
\langle\widehat{A} - A^*\rangle(t) = \int_0^t \frac{J(s)}{Y(s)}dA(s).
\end{equation*}
The corresponding variance estimator is:
\[
\widehat{\sigma}^2(t) = \int_0^t \frac{J(s)}{Y(s)^2}dN(s).
\]
Each observed event contributes roughly $1/Y(s)^2$ to the variance because the Nelson--Aalen jump itself is $1/Y(s)$. Events late in follow-up, when the risk set is small, are therefore noisier than early events with many subjects still under observation.

\subsubsection{Bias}

The only systematic bias in this basic argument comes from empty risk sets. If nobody is still observed at time $s$, the data cannot tell us about the hazard at $s$, and $J(s)$ switches the estimator off. The expected value of the Nelson--Aalen estimator can be derived as follows:
\begin{align*}
    \mathbb{E}[\widehat{A}(t)] &= \int_0^t \mathbb{E}\left(\frac{J(s)}{Y(s)}(Y(s)dA(s) + dM(s))\right) \\
    &= \int_0^t \mathbb{E}[J(s)]dA(s) \\
    &= \int_0^t \mathbb{P}(Y(s) > 0)dA(s).
\end{align*}
Thus, the bias of the Nelson-Aalen estimator is:
\[
\mathbb{E}[\widehat{A}(t)] - A(t) = -\int_0^t \mathbb{P}(Y(s) = 0)\alpha(s)ds.
\]
This sign matters: the estimator misses hazard only on stretches where the risk set has vanished, so it is biased downward relative to the full target $A(t)$.

\subsubsection{Consistency}

This first consistency theorem is the prototype for much of the article. Figure~\ref{fig:registry_processes} gives the finite-sample objects: as the sample grows, every time point with nonzero target hazard should have many subjects still at risk. The proof shows that once the risk-set denominator is large, the martingale noise in the Nelson--Aalen estimator becomes uniformly small, and the only remaining bias is the empty-risk-set correction.

\begin{theorem}[Consistency of the Nelson-Aalen Estimator]
    Suppose
    \[
    \inf_{s \in [0,t]}Y(s)\to_p \infty
    \quad\text{as } n\to\infty,
    \]
    and assume that:
    \[
    A(t) = \int_0^t \alpha(s)ds < \infty,
    \]
    then:
    \[
    \lim_{n \to \infty} \mathbb{P}\left(\sup_{s \in [0, t]} |\widehat{A}(s) - A(s)| > \epsilon\right) = 0,
    \]
    where $\epsilon > 0$ and $\widehat{A}$ is the Nelson-Aalen estimator.
\end{theorem}
\begin{proof}
Write
\[
A^*(s)=\int_0^sJ(u)dA(u),\qquad
\widehat{A}(s)-A^*(s)=\int_0^s\frac{J(u)}{Y(u)}dM(u).
\]
The second process is a locally square-integrable martingale. Its predictable variation is
\[
\left\langle\widehat{A}-A^*\right\rangle(s)
=\int_0^s\frac{J(u)}{Y(u)}dA(u).
\]
Fix $\eta>0$ and $\delta>0$. By Lenglart's inequality \citep{lenglart1977relation},
\[
\mathbb{P}\left(\sup_{s \in [0,t]} |\widehat{A}(s)-A^*(s)| > \eta\right)
\leq \frac{\delta}{\eta^2}
+\mathbb{P}\left(\int_0^t\frac{J(u)}{Y(u)}dA(u)>\delta\right).
\]
On the event $\inf_{u\leq t}Y(u)>K$, the integral on the right is bounded by
$A(t)/K$. Since $\inf_{u\leq t}Y(u)\to_p\infty$ and $A(t)<\infty$, the predictable variation converges to zero in probability. Letting first $n\to\infty$ and then $\delta\downarrow0$ gives
\[
\sup_{s\leq t}|\widehat{A}(s)-A^*(s)|\to_p0.
\]
It remains to compare $A^*$ with $A$. We have
\[
\sup_{s\leq t}|A^*(s)-A(s)|
\leq \int_0^t\{1-J(u)\}dA(u)
\leq A(t)\mathbb{I}\left(\inf_{u\leq t}Y(u)=0\right),
\]
which converges to zero in probability. The triangle inequality completes the proof.
\end{proof}

\subsubsection{Asymptotic Normality}

The asymptotic-normality theorem explains the Monte Carlo behavior later seen in Figure~\ref{fig:simulation_quantity_mc}. After multiplying by \(\sqrt n\), the accumulated martingale error no longer vanishes; it converges to a Gaussian martingale whose covariance function is the predictable variation computed from the risk sets.

\begin{theorem}[Asymptotic Normality of the Nelson-Aalen Estimator]
Let \(F\) be the distribution function of \(T_i\). Assume
\[
    F(s)<1\quad\text{for all }s\leq t,
\]
and
\[
\sup_{s \in [0, t]} \left|\frac{1}{n}Y(s) - 1 + F(s)\right| \to_p 0,
\]
then
\[
\sqrt{n}(\widehat{A} - A) \to_d U,
\]
where $U$ is a Gaussian martingale with $U(0) = 0$ and covariance structure:
\[
\text{cov}(U(s), U(t)) = \sigma(s \wedge t),
\]
where:
\[
\sigma^2(s) = \int_0^s \frac{\alpha(u)}{1 - F(u-)}du.
\]
In the case of independent censoring, where both survival time $X_i$ and censoring time $U_i$ are absolutely continuous, the variance simplifies to:
\[
\sigma^2(s) = \int_0^s \frac{p(u)}{S(u)^2 G(u)}du,
\]
where $p$ is the density of $X$, and $S$ and $G$ are the survival functions of $X$ and $U$, respectively.
\end{theorem}
\begin{proof}
Let $y(u)=1-F(u-)$ and assume, as stated, that $Y(u)/n$ converges uniformly in probability to $y(u)$ on $[0,t]$, with $\inf_{u\le t}y(u)>0$. By the consistency result just proved, the difference between $A^*$ and $A$ is asymptotically negligible, so it is enough to study
\[
M_n(s)=\sqrt{n}\{\widehat{A}(s)-A^*(s)\}
=\sqrt{n}\int_0^s\frac{J(u)}{Y(u)}dM(u).
\]
This is a martingale whose predictable variation is
\[
\langle M_n\rangle(s)
=n\int_0^s\frac{J(u)}{Y(u)^2}d\Lambda(u)
=n\int_0^s\frac{J(u)}{Y(u)}dA(u).
\]
Uniform convergence of $Y/n$ implies
\[
n\frac{J(u)}{Y(u)}\to_p\frac{1}{y(u)}
\]
uniformly away from the event of an empty risk set. Since $A$ is finite on $[0,t]$, it follows that
\[
\langle M_n\rangle(s)\to_p\int_0^s\frac{\alpha(u)}{1-F(u-)}du=\sigma^2(s)
\]
uniformly in $s\le t$. The jump sizes satisfy
\[
\sup_{s\le t}|\Delta M_n(s)|
\leq \frac{\sqrt{n}}{\inf_{u\le t}Y(u)}
\to_p0,
\]
because $\inf_{u\le t}Y(u)$ is of order $n$ in probability. Rebolledo's martingale central limit theorem \citep{rebolledo1980central} therefore gives $M_n\Rightarrow U$, where $U$ is a continuous Gaussian martingale with covariance $\sigma^2(s\wedge t)$. The asymptotically negligible term $\sqrt n(A^*-A)$ vanishes under the same empty-risk-set condition, giving the stated limit for $\sqrt n(\widehat A-A)$.

For independent censoring, the risk probability is $y(u)=S(u-)G(u-)$ and the event intensity while at risk is $\alpha(u)=p(u)/S(u)$. Substitution gives
\[
\sigma^2(s)=\int_0^s\frac{p(u)}{S(u)^2G(u)}du,
\]
with left limits suppressed in the continuous case.
\end{proof}

\subsubsection{Simultaneous Confidence Bands}

Simultaneous confidence bands are essential tools for assessing the uncertainty of the Nelson-Aalen estimator over an interval. Let \( q \) be a continuous and non-negative function on \([s, t]\), where \( 0 \leq s < t \leq \tau \). The following asymptotic result holds:
\[
\left(\frac{\sqrt{n}(\widehat{A} - A)}{1 + n\widehat{\sigma}^2}\right) q \circ \left(\frac{n\widehat{\sigma}^2}{1 + n\widehat{\sigma}^2}\right) \xrightarrow{d} \left(\frac{U}{1 + \sigma^2}\right) q \circ \left(\frac{\sigma^2}{1 + \sigma^2}\right),
\]
on \(\mathcal{D}[s, t]\), where \( U \) is a Gaussian martingale with predictable variation, and:
\[
\widehat{\sigma}^2(t) = \int_0^t \frac{J(s)}{Y(s)^2} dN(s),
\]
is the predictable variation of the martingale \(\widehat{A} - \int_0^t \alpha(s) J(s) ds\).

The processes \(\left(\frac{U}{1 + \sigma^2}\right) q \circ \left(\frac{\sigma^2}{1 + \sigma^2}\right)\) and \((qW^0) \circ \left(\frac{\sigma^2}{1 + \sigma^2}\right)\) share the same distribution, where \(W^0\) is a standard Brownian bridge. This connection enables the construction of simultaneous confidence bands.

The \(100(1-\alpha)\%\) confidence band for \(A\) on \([s, t]\) is given by:
\[
\widehat{A}(s) \pm K_{q, \alpha}(c_1, c_2) \frac{1 + n\widehat{\sigma}^2(s)}{\sqrt{n} q \left(\frac{n\widehat{\sigma}^2(s)}{1 + n\widehat{\sigma}^2(s)}\right)},
\]
where \(K_{q, \alpha}(c_1, c_2)\) represents the upper \(\alpha\)-percentile of the distribution:
\[
\sup_{x \in [c_1, c_2]} \left|q(x) W^0(x)\right|.
\]

\textbf{Common choices for \(q(x)\).}  The argument \(x=n\widehat\sigma^2/(1+n\widehat\sigma^2)\) is a transformed variance scale, not calendar time.  The choice of \(q\) therefore decides which parts of the Brownian bridge are stabilized, and because \(q\) appears in the denominator of the band, small values of \(q(x)\) produce wider bands at the corresponding part of the curve.

\begin{table}[tbp]
\centering
\caption[Brownian-bridge weights for simultaneous confidence bands]{Brownian-bridge weights for simultaneous confidence bands.}
\label{tab:q_choices}
\begingroup
\footnotesize
\setlength{\tabcolsep}{3.5pt}
\renewcommand{\arraystretch}{1.16}
\begin{tabular}{>{\RaggedRight\arraybackslash}p{0.14\textwidth}
                >{\centering\arraybackslash}p{0.14\textwidth}
                >{\RaggedRight\arraybackslash}p{0.34\textwidth}
                >{\RaggedRight\arraybackslash}p{0.29\textwidth}}
\toprule
\textbf{Choice} & \(\boldsymbol{q(x)}\) & \textbf{Effect on the band} & \textbf{Typical use} \\
\midrule
Unweighted & \(1\) & Uses \(\sup |W^0(x)|\) directly; no endpoint stabilization. & A baseline comparison on a restricted interior window.\\
HW & \(x\) & Widens the low-variance left end and is simple on the cumulative-hazard scale. & Classical Nelson--Aalen or Kaplan--Meier simultaneous bands.\\
Sqrt EP & \(\{x(1-x)\}^{1/2}\) & A compromise between unweighted and equal-precision weighting. & Useful when equal precision is too wide near the endpoints.\\
EP & \(x(1-x)\) & Stabilizes both ends of the bridge window, giving relatively balanced precision in the middle. & Central-window inference when both early and late endpoints are unstable.\\
Window & any row above with \(x\in[c_1,c_2]\) & Avoids interpreting bands where the risk set is nearly empty or the transformed variance is too close to 0 or 1. & Routine reporting choice; \(c_1,c_2\) should be stated.\\
\bottomrule
\end{tabular}
\endgroup
\end{table}

\begin{figure}[tbp]
\centering
\includegraphics[width=\textwidth]{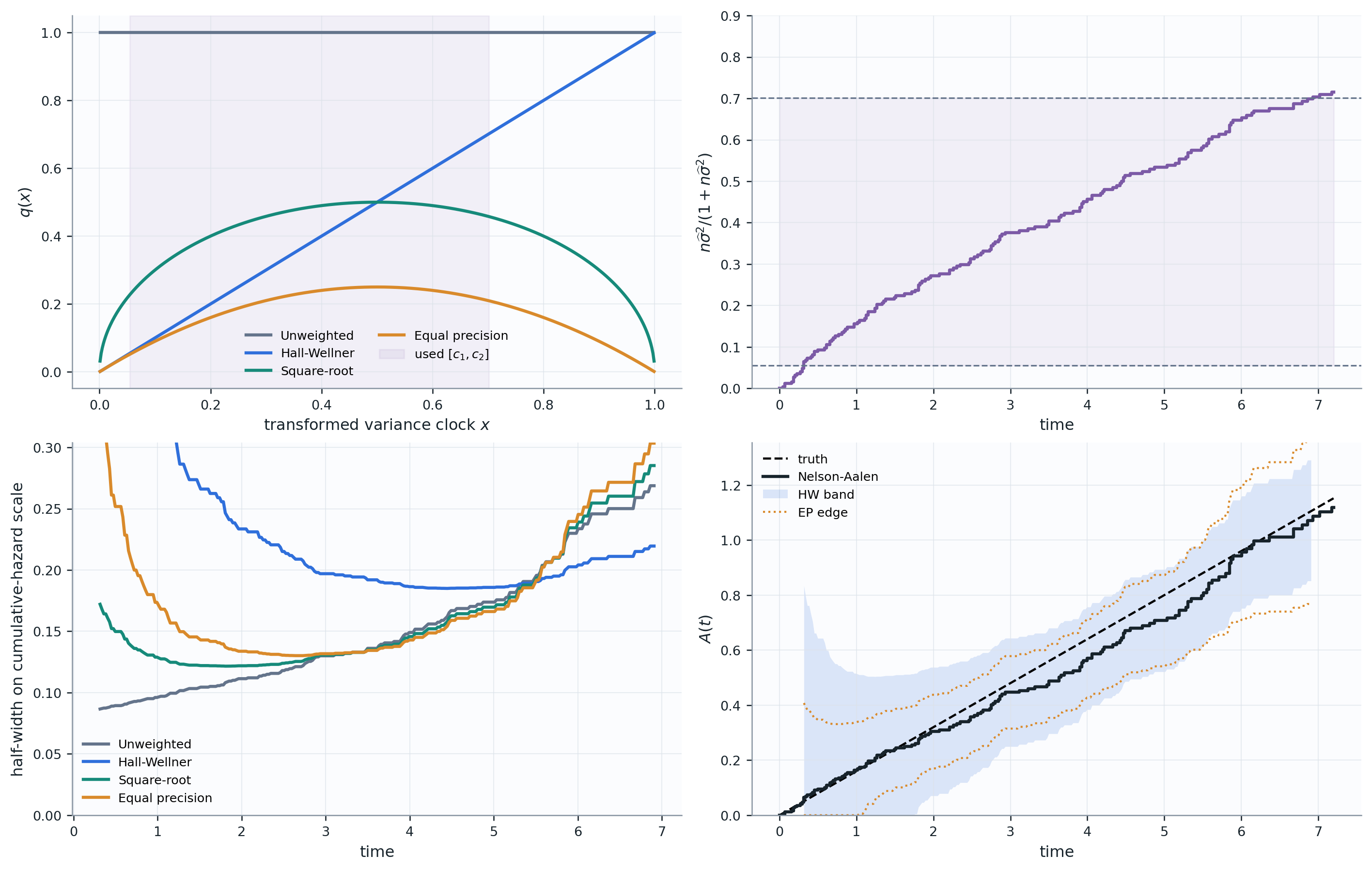}
\caption[Simulation study for Nelson--Aalen confidence bands]{Simulation study for Nelson--Aalen confidence bands.}
\label{fig:confidence_band_q_choices}
\end{figure}

Figure~\ref{fig:confidence_band_q_choices} shows the finite-sample effect of these choices in a right-censoring simulation. The first panel plots the weights themselves; the second panel shows how the Nelson--Aalen variance estimator maps calendar time into the Brownian-bridge scale; the third panel shows the resulting half-widths; and the last panel overlays Hall--Wellner and equal-precision bands on one simulated cumulative-hazard curve.

\begin{figure}[tbp]
\centering
\includegraphics[width=\textwidth]{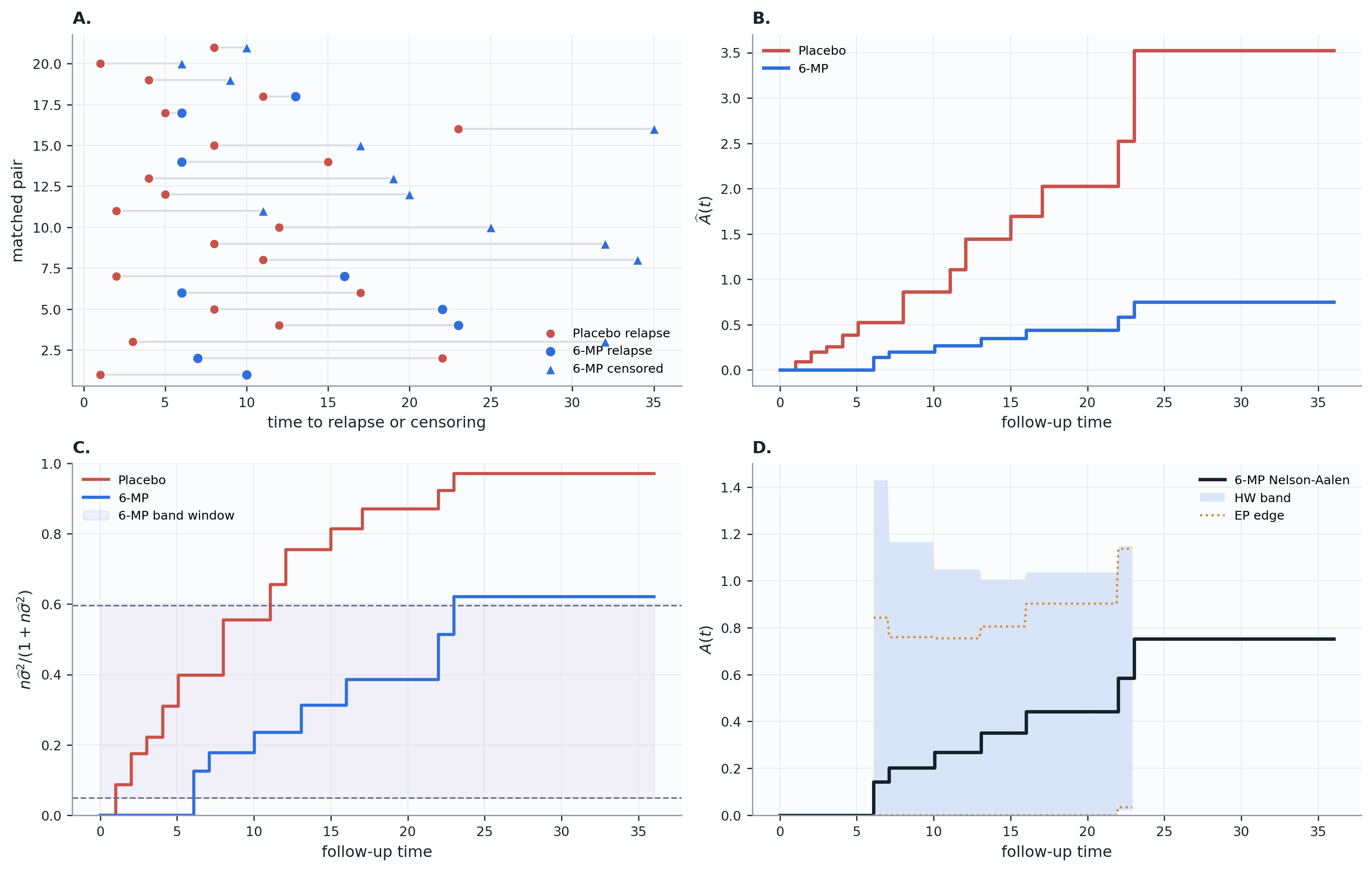}
\caption[Real-data Nelson--Aalen confidence bands]{Real-data Nelson--Aalen confidence bands for the 6-MP leukemia remission trial.}
\label{fig:real_drug6mp_confidence_bands}
\end{figure}

Figure~\ref{fig:real_drug6mp_confidence_bands} repeats the same calculation on a real medical data set: the paired 6-MP leukemia remission data of \citet{freireich1963effect}, distributed in the \texttt{drug6mp} data set from the \texttt{KMsurv} package \citep{yan2026kmsurv}. In the placebo arm all recorded times are relapse times; in the 6-MP arm censored patients remain in the risk set until their last observed remission time. The figure complements Figure~\ref{fig:confidence_band_q_choices}: the simulation displays the Brownian-bridge weights, while the real data show how the same variance clock behaves when event times are sparse and censoring is substantial.

These bands allow for simultaneous inference about the cumulative hazard function \(A(t)\) over the interval of interest, accounting for variability and uncertainty in the estimation process. In applications, the window \([c_1,c_2]\), the weight \(q\), the time unit, and the censoring convention should be reported with the figure.

\subsection{Survival Curves: The Kaplan--Meier Estimator}

The Nelson--Aalen estimator targets accumulated hazard. The corresponding survival-scale estimand is \(S(t)=\mathbb P(T>t)\), the probability of remaining event-free beyond time \(t\). At an event time \(s\), the empirical conditional survival probability is \(1-dN(s)/Y(s)\): among subjects who reached \(s\), all but the observed failures survive past that instant. Kaplan--Meier multiplies these conditional survival probabilities along the time axis.

\begin{definition}[Kaplan-Meier (KM) Estimator]
    Let $\widehat{A}$ denote the Nelson-Aalen estimator for the integrated hazard rate function. The Kaplan-Meier estimator for the survival function is defined as:
    \[
    \widehat{S}(t) = \Prodi_{s \leq t}(1 - d\widehat{A}(s)),
    \]
    which can also be expressed as:
    \[
    \widehat{S}(t) = \prod_{s \leq t} \left(1 - \frac{N(\Delta s)}{Y(s)}\right),
    \]
    where \(N(\Delta s)\) represents the number of events (e.g., deaths) at time \(s\), and \(Y(s)\) denotes the number of individuals at risk at time \(s\).
\end{definition}

\begin{figure}[tbp]
\centering
\includegraphics[width=\textwidth]{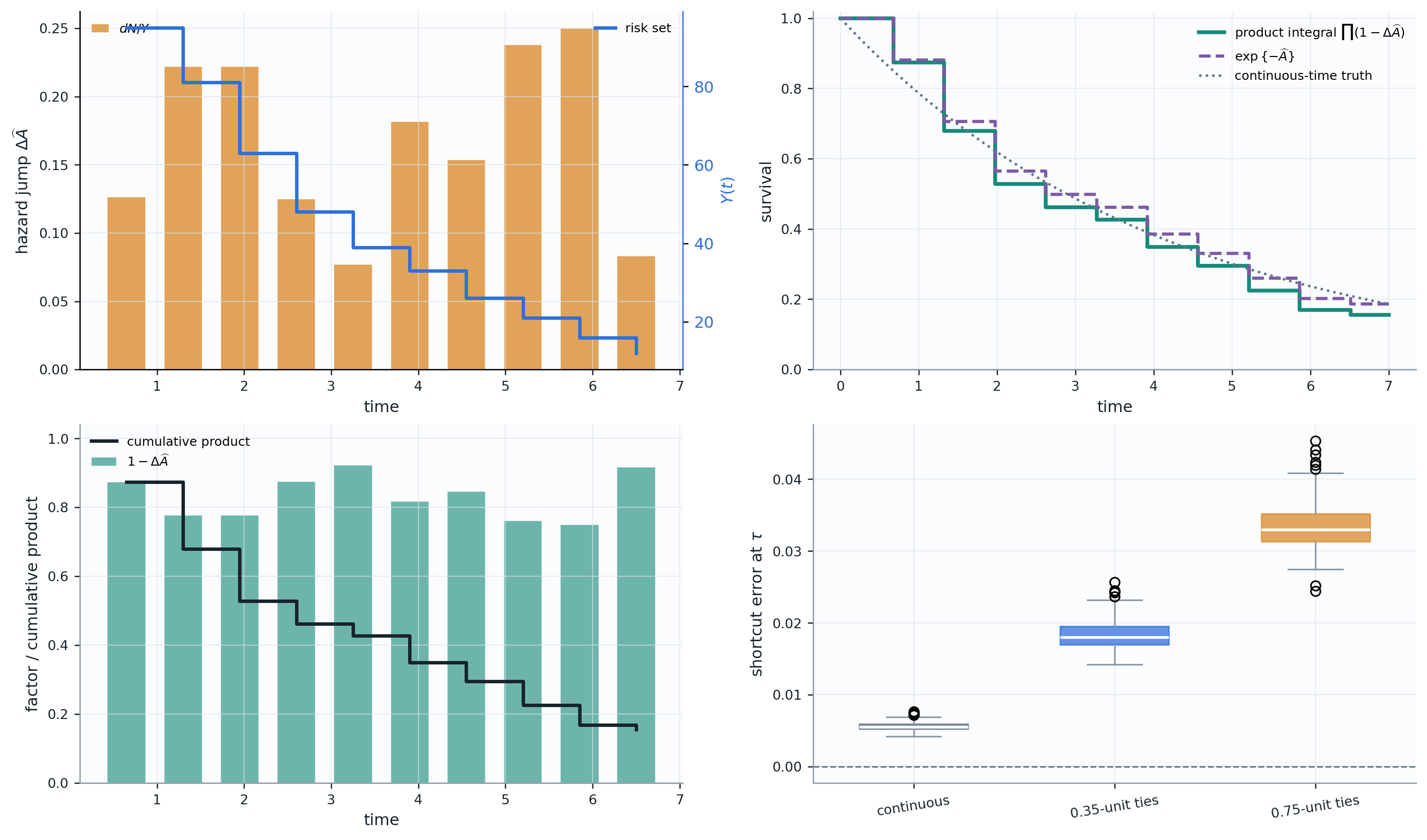}
\caption[Simulation study for the scalar product integral]{Simulation study for the scalar product integral.}
\label{fig:product_integral_scalar}
\end{figure}

Figure~\ref{fig:product_integral_scalar} illustrates the scalar product integral as a finite-sample object. When event times are grouped, the increments \(\Delta\widehat A=dN/Y\) are no longer tiny, so the product integral \(\prod(1-\Delta\widehat A)\) and the exponential approximation \(\exp\{-\widehat A\}\) visibly separate. In continuous models with very small jumps the two are close, but the product integral is the exact survival update for the observed jump process.

\subsubsection{From One Survival Curve to a Multistate Process}

A product integral can be read as a sequential update over ordered event times. In a weekly oncology cohort, suppose \(y_k\) patients are still under follow-up just before week \(k\), and \(d_k\) deaths occur during that week. The Kaplan--Meier update is not a new model assumption; it is the empirical conditional-survival update
\[
    \widehat S(k)=\widehat S(k-1)\left(1-\frac{d_k}{y_k}\right),
    \qquad
    \widehat S(t)=\prod_{k\le t}\left(1-\frac{d_k}{y_k}\right).
\]
The first factor is the empirical conditional survival probability for week \(1\). The second factor acts only on the fraction that already survived week \(1\). By week \(t\), the order of the factors is the calendar of the study. Replacing \(d_k/y_k\) by \(d\widehat A(u)\) gives the compact notation \(\Prodi_{u\le t}\{1-d\widehat A(u)\}\).

The same reading rule becomes more interesting when a patient can move through several states. In an illness--death study, state \(0\) is healthy, state \(1\) is ill, and state \(2\) is dead. At an event time \(u\), the local empirical transition matrix is
\[
\mathbf I+d\widehat{\mathbf A}(u)=
\begin{pmatrix}
1-\frac{dN_{01}(u)+dN_{02}(u)}{Y_0(u)} &
\frac{dN_{01}(u)}{Y_0(u)} &
\frac{dN_{02}(u)}{Y_0(u)}\\[3pt]
0 &
1-\frac{dN_{12}(u)}{Y_1(u)} &
\frac{dN_{12}(u)}{Y_1(u)}\\[3pt]
0 & 0 & 1
\end{pmatrix}.
\]
Multiplying these matrices in time order gives the Aalen--Johansen estimator,
\[
    \widehat{\mathbf P}(0,t)
    =
    \Prodi_{0<u\le t}\{\mathbf I+d\widehat{\mathbf A}(u)\}.
\]
Kaplan--Meier is the one-state model with an absorbing exit. Aalen--Johansen is its multistate extension: patients may stay healthy, become ill, die without illness, or die after illness, and each local matrix redistributes the probability mass that has survived to that time.

\begin{figure}[tbp]
\centering
\includegraphics[width=\textwidth,height=0.44\textheight,keepaspectratio]{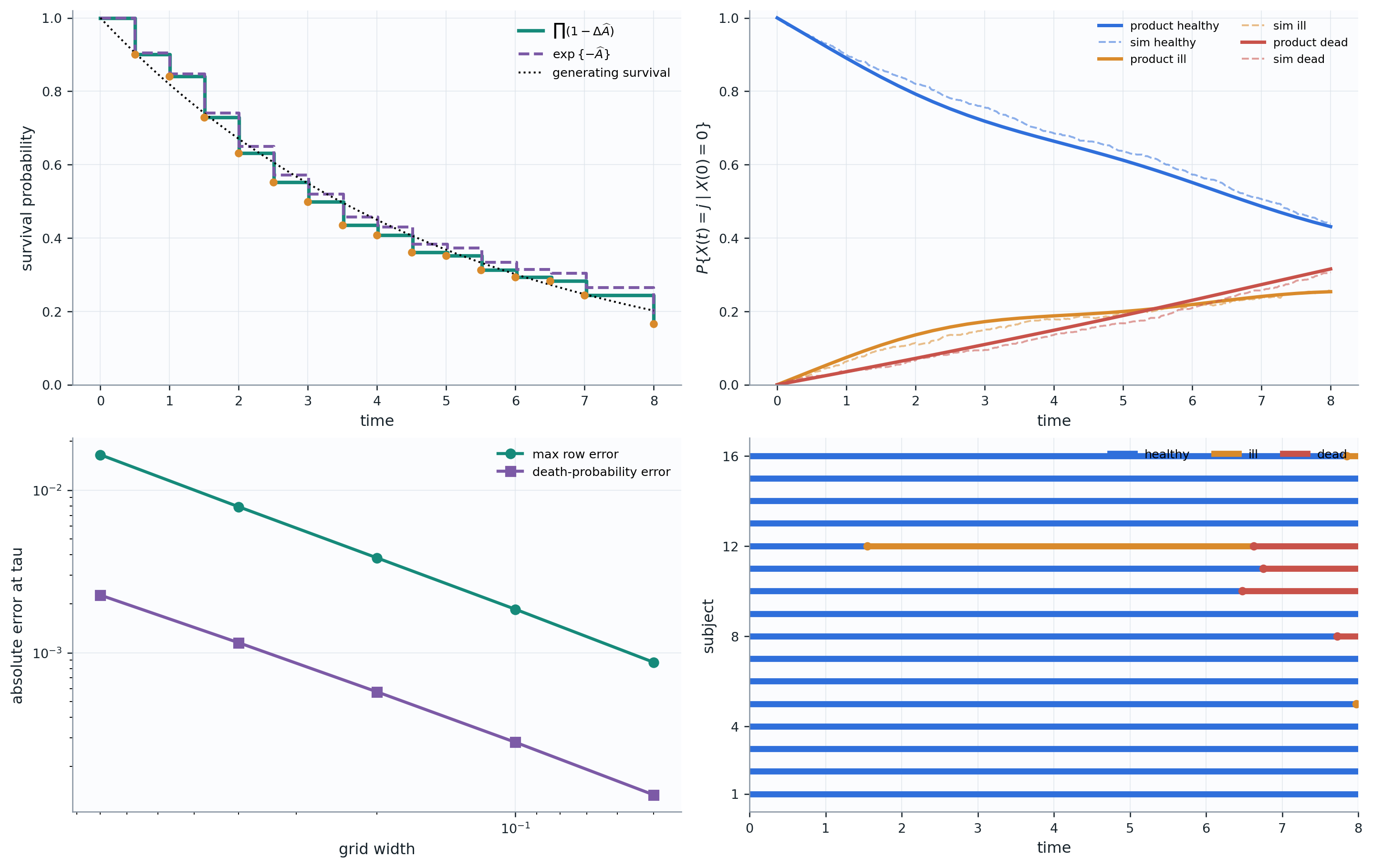}
\caption[Product integrals in survival and multistate simulation]{Product integrals in survival and multistate simulation. The upper-left panel compares the Kaplan--Meier product with its exponential approximation when deaths are grouped. The upper-right panel shows the row sums of local illness--death matrices, checking that each empirical update conserves probability. The lower panels compare product-integral transition probabilities with simulated state occupation proportions and with a plug-in exponential approximation.}
\label{fig:product_integral_survival_multistate}
\end{figure}
Figure~\ref{fig:product_integral_survival_multistate} uses the same simulated cohort in both languages. The survival panel starts with a single risk set and multiplies scalar conditional-survival factors. The multistate panels keep three state occupancies open at once and multiply stochastic matrices. Thus estimands such as ``probability alive and illness-free at year 5,'' ``probability ill but alive,'' and ``probability dead after illness'' are obtained from the Kaplan--Meier multiplication rule after replacing scalar survival factors by ordered transition matrices.

\subsubsection{Variance Estimation: Greenwood’s Formula}

Variance estimation for the Kaplan-Meier estimator is a critical step in constructing confidence intervals for the survival function; the classical plug-in form is Greenwood's formula \citep{greenwood1926natural}. Using martingale theory, we can derive the variance of the Kaplan-Meier estimator. Let \(J(t) = \mathbb{I}(Y(t) > 0)\), and define \(A^*(t) = \int_0^t J(u)dA(u)\), where \(A(u)\) may exhibit jumps. The Kaplan-Meier estimator can be expressed as:
\[
\widehat{S}(t) = \Prodi_{0 \leq s \leq t}(1 - dA^*(s)).
\]

Using the Duhamel equation and Doob-Meyer decomposition for stochastic integration, we have the following mean-zero martingale:
\[
\frac{\widehat{S}(t)}{S^*(t)} - 1 = -\int_0^t \frac{\widehat{S}(s-)}{S^*(s)}dM(s),
\]
where \(M(s)\) is a martingale and \(S^*(t)\) is the deterministic part of the survival function. The compensator of this martingale is given by:
\[
\left\langle \frac{\widehat{S}(t)}{S^*(t)} - 1 \right\rangle(t) = \int_0^t \left\{\frac{\widehat{S}(s-)}{S^*(s)}\right\}^2 \frac{J(s)}{Y(s)} (1 - \Delta A(s))dA(s).
\]

Plugging in \(S^*(t) = \widehat{S}(t)\), the variance estimator simplifies to:
\[
\widehat{\sigma}^2(t) = \int_0^t \frac{1}{Y(s)(Y(s) - \Delta N(s))}dN(s).
\]

Thus, the pointwise variance of the Kaplan-Meier estimator is:
\[
\widehat{\text{Var}}(\widehat{S}(t)) = (\widehat{S}(t))^2 \widehat{\sigma}^2(t).
\]

This result is commonly referred to as \emph{Greenwood's formula} and is widely used in applied survival analysis.

\subsubsection{Bias}

Under the assumption of random censoring, where \(X\) (survival time) and \(U\) (censoring time) are independent with distribution functions \(1 - S(t)\) and \(G(t)\), respectively, the bias of the Kaplan-Meier estimator can be bounded. Specifically, we have:
\[
0 \leq \mathbb{E}[\widehat{S}(t)] - S(t) \leq (1 - S(t))\left\{1 - S(t)(1 - G(t))\right\}^n,
\]
where \(n\) is the sample size. If \(G(t)\) is continuous, then:
\[
0 \leq \mathbb{E}[\widehat{S}(t)] - S(t) \leq (1 - S(t)) \exp\{-\mathbb{E}[Y(t)]\},
\]
where \(Y(t)\) represents the risk process. This shows that the bias decreases as the sample size increases.

\subsubsection{Consistency}

The Kaplan--Meier consistency theorem is a transformation argument rather than a new counting argument. Nelson--Aalen consistency controls the cumulative-hazard estimator; the product-integral map then sends that estimator to a survival curve. In Figure~\ref{fig:simulation_estimator_paths}, this is the reason the survival step curve follows the truth whenever the cumulative hazard estimator is stable.

\begin{theorem}[Consistency of the KM Estimator]
    Assume
    \[
        \inf_{s \in [0,t]}Y(s)\to_p\infty
        \quad\text{and}\quad
        S(t)>0 .
    \]
    Then, as \(n \to \infty\),
    \[
    \sup_{s \in [0, t]} |\widehat{S}(s) - S(s)| \to_p 0.
    \]
\end{theorem}

\begin{proof}
Let $\phi$ denote the product-integral map
\[
\phi(B)(s)=\Prodi_{u\le s}(1-dB(u)).
\]
The Kaplan-Meier estimator can be written as $\widehat S=\phi(\widehat A)$ and the target survival curve as $S=\phi(A)$, up to the same empty-risk-set correction used for the Nelson-Aalen estimator. By the preceding consistency theorem,
\[
\sup_{s\le t}|\widehat A(s)-A(s)|\to_p0.
\]
On $[0,t]$, the assumption $S(t)>0$ implies that the product integral is bounded away from zero and that $A$ has no fatal jump before $t$. The product-integral map is continuous, under the uniform norm on functions of bounded variation with jumps bounded away from one. Hence
\[
\sup_{s\le t}|\phi(\widehat A)(s)-\phi(A)(s)|\to_p0.
\]
This is exactly the desired uniform consistency of $\widehat S$.
\end{proof}

\subsubsection{Asymptotic Normality}

The next result gives the asymptotic behavior of the Kaplan-Meier estimator under a random censoring model. The event time \(X\) has survival function \(S\), the censoring time \(U\) has distribution function \(G\), and both distributions are assumed absolutely continuous.

The proof uses the same simulation logic as the previous theorem but passes it through the product-integral derivative. In Figure~\ref{fig:simulation_quantity_mc}, the Kaplan--Meier quantity \(S(6)\) behaves smoothly because small vertical errors in \(\widehat A\) become small vertical errors in \(\widehat S\); the quantile \(Q(0.5)\) is noisier because it inverts the curve.

\begin{theorem}[Asymptotic Normality of the KM Estimator]\label{thm:an_km}
    Let
    \[
        A(t)=\int_0^t\alpha(s)\,ds,
        \qquad
        \alpha(s)=-\frac{dS(s)/ds}{S(s)},
    \]
    and define
    \[
    \sigma^2(t) = \int_0^t \frac{\alpha(u)}{S(u)G_C(u)}du,
    \qquad G_C(u)=\mathbb P(U\ge u).
    \]
    Assume that \(A(t)<\infty\) and that the at-risk process satisfies
    \[
        \sup_{s \in [0, t]} \left|\frac{1}{n}Y(s) - S(s)G_C(s)\right| \to_p 0.
    \]
    Then:
    \[
    \sqrt{n}(\widehat{S} - S) \to_d -S \cdot U,
    \]
    on \(\mathcal{D}[0, t]\), where \(U\) is a Gaussian martingale with \(U(0) = 0\) and covariance:
    \[
    \text{Cov}(U(s), U(t)) = \sigma^2(s \wedge t).
    \]
    Furthermore, \(n\widehat{\sigma}^2(s)\) is uniformly weakly consistent on \([0, t]\), where \(\widehat{\sigma}^2(t)\) is Greenwood’s variance estimator.
\end{theorem}

\begin{proof}
The proof is a direct application of the functional delta method to the product-integral representation:
\[
S(t) = \Prodi_{s \in [0, t]} (1 - dA(s)).
\]
The map $\phi(A)=\Prodi(1-dA)$ is Hadamard differentiable at $A$. In the scalar case its derivative in direction $h$ is
\[
d\phi_A(h)(t)
=-S(t)\int_0^t\frac{h(du)}{1-\Delta A(u)}.
\]
When $A$ is continuous, this reduces to $d\phi_A(h)(t)=-S(t)h(t)$. The Nelson-Aalen estimator satisfies
\[
\sqrt{n}(\widehat{A} - A) \to_d U,
\]
where \(U\) is the Gaussian martingale in the previous theorem. Therefore
\[
\sqrt n(\widehat S-S)
=\sqrt n\{\phi(\widehat A)-\phi(A)\}
\Rightarrow d\phi_A(U).
\]
In the common continuous-time random-censoring model this gives $-S\cdot U$, as stated. Finally,
\[
n\widehat{\sigma}^2(t)
=n\int_0^t\frac{1}{Y(s)\{Y(s)-\Delta N(s)\}}dN(s)
\]
is obtained by substituting the empirical compensator into Greenwood's predictable variation formula. The uniform law of large numbers for the risk process and the martingale law of large numbers for $N$ imply uniform weak consistency for the limiting variance.
\end{proof}

\subsubsection{Simultaneous Confidence Bands}
The previous theorem gives pointwise intervals for $S(t)$:
\[
\widehat{S}(t)\pm c_{\alpha/2}\widehat{S}(t)\widehat{\sigma}(t),
\]
where $c_{\alpha/2}$ is the upper $\alpha/2$ quantile of $\mathcal{N}(0,1)$. The plain linear interval is easy to derive but awkward in practice because it can leave the interval $[0,1]$. Transformations repair this by doing the normal approximation on a scale better suited to survival probabilities. Two common choices are the Kalbfleisch--Prentice log--log transformation, $g\{S(t)\}=\log[-\log\{S(t)\}]$, and the Thomas--Grunkemeier square-root arcsine transformation, $g\{S(t)\}=\arcsin\{\sqrt{S(t)}\}$. The delta method gives pointwise intervals on the transformed scale, which are then mapped back to the survival scale.

For a whole curve, we need a band rather than a pointwise interval. The time change
\[
\frac{\sigma^2(t)}{1+\sigma^2(t)}
\]
maps the variance scale onto $[0,1]$, where Brownian-bridge quantiles can be tabulated. Similar to the confidence bands for the Nelson--Aalen estimator,
\[
\left(\frac{U}{1+\sigma^2}\right)q\circ \left(\frac{\sigma^2}{1+\sigma^2}\right)
=_d \left(qW^0\right)\circ\left(\frac{\sigma^2}{1+\sigma^2}\right),
\]
where $q$ is a continuous non-negative weight function on $[s,t]$ and $W^0$ is a standard Brownian bridge. Hence the $100(1-\alpha)\%$ equal-precision (EP) band is
\[
\widehat{S}(s)\pm d_\alpha(\widehat{c}_1,\widehat{c}_2)\widehat{S}(s)\widehat{\sigma}(s),
\]
where
\[
\widehat{c}_i=\frac{n\widehat{\sigma}^2(t_i)}{1+n\widehat{\sigma}^2(t_i)},\qquad i=1,2,
\]
and $d_\alpha(\widehat{c}_1,\widehat{c}_2)$ is the upper $\alpha$ quantile of
\[
\sup_{c_1\le x\le c_2}|W^0(x)\sqrt{x(1-x)}|.
\]
The Hall--Wellner (HW) band is
\[
\widehat{S}(s)\pm\frac{1+n\widehat{\sigma}^2(s)}{\sqrt{n}}e_\alpha(\widehat{c}_1,\widehat{c}_2)\widehat{S}(s),
\]
where $e_\alpha(\widehat{c}_1,\widehat{c}_2)$ is the upper $\alpha$ quantile of
\[
\sup_{c_1\le x\le c_2}|W^0(x)|.
\]
\subsubsection{Quantile Estimation: Brookmeyer-Crowley}

Quantile estimation plays a vital role in survival analysis because it translates a whole survival curve into a number that readers can use: median survival time, a \(25\)th percentile, or the time by which a target event probability has been reached. The price of that interpretability is inversion. A vertical error in \(\widehat S\) becomes a horizontal error in \(\widehat Q(p)\), and that horizontal error is large when the survival curve is nearly flat at the target probability.

Let \( Q(p) \) denote the quantile function, defined as a functional of the distribution function \( F \):
\[
Q(p) = \phi(F) = \phi(1 - S) = \inf\{x : F(x) \geq p\},
\]
where \( S \) is the survival function. Applying the functional delta method to \(\widehat{Q}(p)=\phi(1-\widehat S)\) gives
\[
\sqrt{n}(\widehat{Q}(p) - Q(p)) \xrightarrow{d} -\frac{(1 - p)U(Q(p))}{f(Q(p))},
\]
where \(f\) is the density of \(F\) and \(U\) is the Gaussian martingale appearing in the Kaplan--Meier limit theorem. Equivalently,
\[
\sqrt{n}(\widehat{Q}(p) - Q(p)) \xrightarrow{d} \mathcal{N}\left(0, \frac{(1 - p)^2 \sigma^2(Q(p))}{f(Q(p))^2}\right),
\]
where \(\sigma^2(t) = \int_0^t \alpha(u)/\{S(u)G(u)\}\,du\), as in Theorem~\ref{thm:an_km}; see \citet[pp.~275--277]{andersen1993statistical}.

The Brookmeyer--Crowley method \citep{brookmeyer1982confidence} uses this limit result by transforming the Kaplan--Meier curve and then inverting the pointwise survival interval. The confidence set is
\[
\left\{u : \frac{|g(\widehat{S}(u)) - g(1 - p)|}{|g'(\widehat{S}(u))| \widehat{S}(u) \widehat{\sigma}(u)} \leq c_{\alpha/2}\right\},
\]
where \(g\) is a variance-stabilizing transformation such as \(\log\{-\log(\cdot)\}\) or \(\arcsin\{\sqrt{\cdot}\}\), and \(c_{\alpha/2}\) is the upper \(\alpha/2\) standard-normal quantile. To estimate the variance of \(\widehat Q(p)\) directly, one must estimate \(f\{Q(p)\}\). The smoothed Nelson--Aalen estimator provides a smooth hazard, a smooth survival curve, and hence a derivative near the target quantile. Definition~\ref{def:SNA} and Figure~\ref{fig:sna_smoothing_simulation} show this smoothing step explicitly.

\subsection{State Occupation: The Aalen--Johansen Estimator}

The \textit{Aalen--Johansen estimator} extends Kaplan--Meier estimation to subjects or units that can move through several states. Instead of estimating only survival, the target may be the probability that a patient is event-free, relapsed, in remission, transplanted, or dead at time \(t\). The quantities therefore become vector- and matrix-valued. The count \(N_{hj}(t)\) records observed jumps from state \(h\) to state \(j\), the risk set \(Y_h(t)\) records how many subjects occupy state \(h\) just before \(t\), and the cumulative transition hazard \(A_{hj}(t)\) records the accumulated transition intensity from \(h\) to \(j\). The transition matrix \(\mathbf{P}(s,t)\) is the multistate analogue of a survival curve: its \((h,j)\) entry is the probability of being in state \(j\) at time \(t\) given state \(h\) at time \(s\).

\begin{definition}[\textit{Aalen-Johansen (AJ) Estimator}]
    Let \(\mathbf{A}\) denote the intensity measure of a Markov process \(X\), and let \(\mathbf{P}\) represent the transition probability matrix. Define:
    \[
    \widehat{A}_{hj}(t) = \int_0^t \frac{J_h(s)}{Y_h(s)}dN_{hj}(s),
    \]
    where \(\widehat{A}_{hj}(t)\) is the Nelson-Aalen estimator for the cumulative hazard \(A_{hj}\) for \(h \neq j\), and:
    \[
    \widehat{A}_{hh}(t) = -\sum_{j \neq h} \widehat{A}_{hj}(t).
    \]
    Then, the \textit{Aalen-Johansen estimator} for the transition probability matrix \(\mathbf{P}(s, t)\) is given by:
    \[
    \widehat{\mathbf{P}}(s, t) = \Prodi_{(s, t]} \left(\mathbf{I} + d\widehat{\mathbf{A}}(u)\right),
    \]
    where \(\widehat{\mathbf{A}} = \{\widehat{A}_{hj}\}\) is the matrix of estimated cumulative hazards, and \(\Prodi\) denotes the product integral.
\end{definition}
The product integral appears because transitions compound in order. A small transition matrix at time $u$ changes the state distribution, and the next small transition acts on the distribution already changed by previous jumps. Ordinary multiplication of conditional survival factors becomes ordered multiplication of empirical transition matrices. Figure~\ref{fig:simulation_estimator_paths} shows this idea in a competing-risks simulation, and Figure~\ref{fig:appendix_product_integral_tools} shows the same algebra as a product of local stochastic matrices.

Algorithm~\ref{alg:product_integral_blueprint} gives the computational version of the display above. It is intentionally written as a local matrix update: each event time creates a stochastic matrix \(I+D(t)\), and the estimated transition probability is the ordered product of those local matrices.

\begin{algorithm}[tbp]
\caption{Aalen--Johansen product-integral update}
\label{alg:product_integral_blueprint}
\begin{algorithmic}[1]
\Require Finite state space \(\mathcal S\), transition counts \(N_{rs}\), state-risk processes \(Y_r\), ordered transition times \(\mathcal T\)
\Ensure Estimated transition matrix \(\widehat P(0,t)\)
\State Set \(\widehat P(0,0)=I\).
\For{each \(t\in\mathcal T\)}
    \For{each ordered pair \(r\ne s\)}
        \State Compute \(d\widehat A_{rs}(t)=dN_{rs}(t)/Y_r(t)\) if \(Y_r(t)>0\), otherwise set the increment to zero.
    \EndFor
    \State Form the local matrix \(D(t)\) with off-diagonal entries \(D_{rs}(t)=d\widehat A_{rs}(t)\) and diagonal entries \(D_{rr}(t)=-\sum_{s\ne r}d\widehat A_{rs}(t)\).
    \State Update \(\widehat P(0,t)=\widehat P(0,t-)\{I+D(t)\}\).
    \State Check row sums and nonnegativity; small violations indicate empty-risk or tie-handling problems.
\EndFor
\State Return \(\widehat P(0,t)\), state-occupation probabilities, and cumulative-incidence summaries.
\end{algorithmic}
\end{algorithm}

For a clinical illness-death example, let the states be \(0=\) event-free, \(1=\) relapsed or ill, \(2=\) transplanted, \(3=\) recovered, and \(4=\) dead. Then \(Y_1(t)\) is the number of relapsed patients still under observation just before time \(t\), \(N_{12}(t)\) counts relapse-to-transplant transitions, and \(\widehat A_{12}(t)\) estimates the accumulated transition pressure from relapse to transplant. The point probability \(p_2(t)=\mathbb P\{X(t)=2\}\) estimates the fraction of comparable patients in the transplant state at time \(t\). The transition probability \(P_{13}(s,t)\) is the probability that a patient already relapsed at time \(s\) is recovered by time \(t\). The Aalen--Johansen estimator is needed because a patient can pass through several ordered disease states before the calendar reaches \(t\); a single survival curve would only say that some transition happened, not which state is occupied.

The same notation can describe a reliability process. Let \(0=\) functioning, \(1=\) degraded, \(2=\) repaired, \(3=\) replaced, and \(4=\) failed. Then \(Y_1(t)\) is the number of degraded devices still monitored at \(t-\), \(N_{12}(t)\) counts degradation-to-repair transitions, and \(P_{13}(s,t)\) is the probability that a device already degraded at time \(s\) is replaced by time \(t\). The vocabulary changes, but the statistical objects remain the eligible units, the observed movement, and the ordered transition matrices that update the state distribution.

\subsubsection{Complete Observation Setting}

Under the assumption of complete observation (i.e., no censoring), the \textit{Aalen-Johansen estimator} exhibits a simple relationship between the at-risk process and the transition probability matrix. The lemma below gives the finite-sample identity behind the estimator: if every transition is observed, multiplying the current state-count vector by the empirical one-step transition matrix exactly updates the observed state counts.

\begin{lemma}
    Suppose no censoring occurs, and the observations are complete for all \(t \in \mathcal{T}\). Let \(\mathbf{Y}(t)\) denote the row vector \((Y_1(t), \cdots, Y_k(t))\), where \(Y_h(t)\) is the number of individuals at risk in state \(h\) at time \(t\). Then:
    \[
    \mathbf{Y}(t+) = \mathbf{Y}(s+)\widehat{\mathbf{P}}(s, t),
    \]
    where \(\widehat{\mathbf{P}}(s, t)\) is the Aalen-Johansen estimator.
\end{lemma}

\begin{proof}
    It is enough to check the identity at the ordered jump times in \((s,t]\), because neither side changes between jumps. Let \(u\) be one such time and write
    \(d_{hj}(u)=\Delta N_{hj}(u)\). Complete observation implies that subjects are lost from state \(h\) only through observed transitions out of \(h\), and subjects enter state \(j\) only through observed transitions into \(j\). Thus
    \[
    Y_j(u+)=Y_j(u-)+\sum_{h\neq j}d_{hj}(u)-\sum_{k\neq j}d_{jk}(u).
    \]
    On the other hand, the \(j\)th component of
    \(\mathbf{Y}(u-)\{\mathbf I+\Delta\widehat{\mathbf A}(u)\}\) is
    \[
    Y_j(u-)+\sum_{h\neq j}Y_h(u-)\frac{d_{hj}(u)}{Y_h(u-)}
    +Y_j(u-)\left\{-\sum_{k\neq j}\frac{d_{jk}(u)}{Y_j(u-)}\right\},
    \]
    with the convention that a term with an empty risk set is zero. This is exactly the preceding display for \(Y_j(u+)\). Therefore one-step multiplication by
    \(\mathbf I+\Delta\widehat{\mathbf A}(u)\) updates the empirical state-count vector correctly at every jump time. Iterating over all jumps in \((s,t]\) gives
    \[
    \mathbf Y(t+)=\mathbf Y(s+)\prod_{s<u\le t}\{\mathbf I+\Delta\widehat{\mathbf A}(u)\}
    =\mathbf Y(s+)\widehat{\mathbf P}(s,t),
    \]
    which proves the claim.
\end{proof}

\subsubsection{Bias and Covariance Estimation}

Recall that the cumulative hazard matrix \(\mathbf{A}^* = \{A_{hj}^*\}\) is defined as:
\[
A_{hj}^*(t) = \int_0^t J_h(u)\alpha_{hj}(u)du,
\]
where \(\alpha_{hj}(u)\) is the transition intensity. Define the transition matrix:
\[
\mathbf{P}^*(s, t) = \Prodi_{(s, t]} \left(\mathbf{I} + d\mathbf{A}^*(u)\right).
\]
By the Duhamel equation and boundedness arguments, the following holds:
\[
\mathbf{M}(s, t) = \widehat{\mathbf{P}}(s, t)\mathbf{P}^*(s, t)^{-1} - \mathbf{I},
\]
where \(\mathbf{M}(s, t)\) is a \(k \times k\) martingale. If \(\mathbb{P}(Y_h(u) = 0)\) is small for \(u \in (s, t]\), then the AJ estimator is nearly unbiased.

To estimate the covariance of \(\widehat{\mathbf{P}}(s, t)\), we can use:
\[
\widehat{\text{Cov}}(\widehat{\mathbf{P}}(s, t)) = \int_s^t \widehat{\mathbf{P}}(u, t)^T \otimes \widehat{\mathbf{P}}(s, u)d[\widehat{\mathbf{A}} - \mathbf{A}^*](u)\widehat{\mathbf{P}}(u, t) \otimes \widehat{\mathbf{P}}(s, u)^T,
\]
where \([\cdot, \cdot]\) represents the quadratic variation of a local square-integrable martingale; see \citet[pp.~292--293]{andersen1993statistical}.

\subsubsection{Large Sample Properties}

The large sample properties of the \textit{Aalen-Johansen estimator} follow naturally from the properties of the Nelson-Aalen estimator. Uniform consistency is inherited from the uniform consistency of the transition-specific Nelson--Aalen estimators and the continuity of the product integral:
\[
    \sup_{u \in [s, t]} \lVert\widehat{\mathbf{P}}(s, u) - \mathbf{P}(s, u)\rVert \xrightarrow{p} 0.
\]

Asymptotic normality comes from the same chain of ideas with one additional matrix-valued derivative. Combining the Nelson--Aalen central limit theorem, Hadamard differentiability of the product integral, and the functional delta method gives
\[
    \sqrt{n} \left(\widehat{\mathbf{P}}(s, \cdot) - \mathbf{P}(s, \cdot)\right) \xrightarrow{d} \int_s^\cdot \mathbf{P}(s, u)d\mathbf{U}(u)\mathbf{P}(u, \cdot),
\]
where \(\mathbf{U} = (U_{hj})\) is a \(k \times k\) matrix-valued process. For \(h \neq j\), the \(U_{hj}\) are independent Gaussian martingales with \(U_{hj}(0) = 0\) and
\[
    \text{Cov}(U_{hj}(s), U_{hj}(t)) = \sigma^2(s \wedge t),
\]
while \(U_{hh} = -\sum_{j \neq h} U_{hj}\).

\subsection{Dependence in Paired Lifetimes: The Dabrowska Estimator}

The Dabrowska estimator extends the product-limit construction from a line to a plane. A univariate survival curve asks whether one event time exceeds one threshold. A bivariate survival curve asks whether two related event times both exceed their thresholds:
\[
S(s,t)=\mathbb P(T_1>s,T_2>t).
\]
This is needed when the scientific object is dependence: paired organs, matched family members, two recurrent gap times, or two failure modes observed on the same subject. The difficulty is that dependence has its own local behavior. We must distinguish a first-coordinate failure, a second-coordinate failure, and a joint failure inside a small rectangle of the $(s,t)$ plane.

\subsubsection{Bivariate Survival Times and Cumulative Hazards}

Let \( T = (T_1, T_2) \) represent a pair of nonnegative random variables corresponding to two survival times. The joint survival function is defined as:
\[
S(s, t) = \mathbb{P}(T_1 > s, T_2 > t),
\]
which captures the probability that both survival times exceed given thresholds \(s\) and \(t\).

\begin{definition}[Bivariate Cumulative Hazard]
The bivariate cumulative hazard function is defined as:
\[
\Lambda(s, t) = (\Lambda_{10}(s, t), \Lambda_{01}(s, t), \Lambda_{11}(s, t)),
\]
where:
\begin{align*}
    \Lambda_{11}(ds, dt) &= \frac{\mathbb{P}(T_1 \in ds, T_2 \in dt)}{\mathbb{P}(T_1 \geq s, T_2 \geq t)} = \frac{S(ds, dt)}{S(s-, t-)}, \\
    \Lambda_{10}(ds, t) &= \frac{\mathbb{P}(T_1 \in ds, T_2 > t)}{\mathbb{P}(T_1 \geq s, T_2 > t)} = -\frac{S(ds, t)}{S(s-, t)}, \\
    \Lambda_{01}(s, dt) &= \frac{\mathbb{P}(T_1 > s, T_2 \in dt)}{\mathbb{P}(T_1 > s, T_2 \geq t)} = -\frac{S(s, dt)}{S(s, t-)}.
\end{align*}
These hazard components are subject to the initial conditions:
\[
\Lambda_{10}(0, t) = \Lambda_{01}(s, 0) = \Lambda_{11}(0, 0) = 0.
\]
\end{definition}

The bivariate cumulative hazard function extends the univariate Nelson--Aalen framework by modeling hazards for both marginal survival times and their joint dependence. The terms \(\Lambda_{10}\) and \(\Lambda_{01}\) describe one-coordinate failures along the horizontal and vertical directions, while \(\Lambda_{11}\) describes mass that lands inside a two-dimensional rectangle. The interaction term \(L\) below is needed because the joint survival surface differs from the product of two marginal Kaplan--Meier curves unless the coordinates satisfy the appropriate independence condition.

\subsubsection{The Dabrowska Representation and Estimator}

The Dabrowska representation provides a product-integral decomposition of the joint survival function based on the cumulative hazard components. This result serves as the theoretical foundation for constructing nonparametric estimators. Its simulation companion is Figure~\ref{fig:dabrowska_simulation}: the theorem explains why a survival surface can be built from marginal edge hazards plus a dependence correction in the interior of the plane.

\begin{theorem}[The Dabrowska Representation]
For \((s, t)\) such that \(S(s, t) > 0\), the joint survival function can be expressed as:
\[
S(s, t)
= \Prodi_{u \leq s}\{1 - \Lambda_{10}(du, 0)\}
  \Prodi_{v \leq t}\{1 - \Lambda_{01}(0, dv)\}
  \Prodi_{u \leq s,\,v \leq t}\{1 - L(du, dv)\},
\]
where:
\[
L(du, dv)
= \frac{\Lambda_{10}(du, v-) \Lambda_{01}(u-, dv) - \Lambda_{11}(du, dv)}
        {\{1 - \Lambda_{10}(\Delta u, v-)\}\{1 - \Lambda_{01}(u-, \Delta v)\}},
\]
and the first two product-integral terms represent the marginal survival functions \(S(s, 0)\) and \(S(0, t)\), respectively. The bivariate product integral is taken as the limit when both \(\max_i |u_i - u_{i-1}|\) and \(\max_i |v_i - v_{i-1}|\) approach zero.
\end{theorem}
\begin{proof}
The proof is easiest to read first on a finite grid and then pass to the
product-integral limit.  Fix partitions
\[
0=u_0<u_1<\cdots<u_m=s,\qquad
0=v_0<v_1<\cdots<v_\ell=t,
\]
and write \(\Delta u_i=(u_{i-1},u_i]\) and
\(\Delta v_j=(v_{j-1},v_j]\).  On the axes the joint survival surface reduces
to the marginal survival curves.  Hence the ordinary one-dimensional
product-integral identity gives
\[
S(s,0)=\Prodi_{u\le s}\{1-\Lambda_{10}(du,0)\},
\qquad
S(0,t)=\Prodi_{v\le t}\{1-\Lambda_{01}(0,dv)\}.
\]
These two factors account only for the edge behavior of the surface.  What
remains is the interior dependence correction.

Consider one rectangle
\((u_{i-1},u_i]\times(v_{j-1},v_j]\), and condition on being at risk just before
the southwest corner, that is, on \(T_1\ge u_{i-1}\) and \(T_2\ge v_{j-1}\).
Inside this rectangle there are three relevant increments: a first-coordinate
failure while the second coordinate has survived past \(v_{j-1}\), a
second-coordinate failure while the first coordinate has survived past
\(u_{i-1}\), and a joint failure in the rectangle.  In the notation of the
theorem these increments are
\[
\Lambda_{10}(\Delta u_i,v_{j-1}),\qquad
\Lambda_{01}(u_{i-1},\Delta v_j),\qquad
\Lambda_{11}(\Delta u_i,\Delta v_j).
\]
If the two coordinates had no residual local dependence after the edge
increments were removed, the second-order contribution over the rectangle would
be the product
\[
    \Lambda_{10}(\Delta u_i,v_{j-1})
    \Lambda_{01}(u_{i-1},\Delta v_j).
\]
The actual second-order contribution is changed by the joint-failure mass
\(\Lambda_{11}(\Delta u_i,\Delta v_j)\).  Dividing by the survival probabilities
left after the two one-coordinate decrements gives the conditional
cross-ratio decrement
\[
1-L(\Delta u_i,\Delta v_j)
=1-
\frac{\Lambda_{10}(\Delta u_i,v_{j-1})\Lambda_{01}(u_{i-1},\Delta v_j)
      -\Lambda_{11}(\Delta u_i,\Delta v_j)}
     {\{1-\Lambda_{10}(\Delta u_i,v_{j-1})\}
      \{1-\Lambda_{01}(u_{i-1},\Delta v_j)\}}.
\]
This is the two-dimensional analogue of the univariate factor
\(1-\Lambda(\Delta u_i)\): it is the conditional multiplicative decrement
that remains after the two marginal directions have been accounted for.

Multiplying the two marginal product-limit factors and all interior
cross-ratio decrements over the grid gives a step-function approximation
\[
S_{\pi}(s,t)
= \prod_i\{1-\Lambda_{10}(\Delta u_i,0)\}
  \prod_j\{1-\Lambda_{01}(0,\Delta v_j)\}
  \prod_{i,j}\{1-L(\Delta u_i,\Delta v_j)\}.
\]
The approximation has the correct boundary values on \(t=0\) and \(s=0\), and
on every grid rectangle it satisfies the same conditional increment equation
that defines the joint survival surface through the three cumulative-hazard
components.  Letting the mesh of both partitions tend to zero turns the three
ordinary products into the two marginal product integrals and the bivariate
product integral.  The limiting function therefore satisfies the required
increment equations and boundary conditions.  Since the product-integral
equations determine the survival surface uniquely on any rectangle where
\(S\) and the denominators above stay positive, the limit is \(S(s,t)\).
\end{proof}

\subsubsection{Observed Data and Counting Processes}

Let \(X_{1i},X_{2i}\) be the paired latent event times and \(C_{1i},C_{2i}\) the corresponding censoring times. The observed data are
\(\widetilde T_{ki}=X_{ki}\wedge C_{ki}\) and
\(\delta_{ki}=\mathbb I(X_{ki}\le C_{ki})\), \(k=1,2\). Under independent coordinatewise censoring, the empirical risk sets and counting measures used by Dabrowska's estimator can be written schematically as
\begin{align*}
    N_{11}(s, t) &= \sum_{i=1}^n \mathbb{I}(\widetilde T_{1i} \leq s, \widetilde T_{2i} \leq t, \delta_{1i} = 1, \delta_{2i} = 1), \\
    Y_{11}(s, t) &= \sum_{i=1}^n \mathbb{I}(\widetilde T_{1i} \geq s, \widetilde T_{2i} \geq t), \\
    N_{10}(s | t) &= \sum_{i=1}^n \mathbb{I}(\widetilde T_{1i} \leq s, \widetilde T_{2i} \geq t, \delta_{1i} = 1), \quad
    Y_{10}(s,t) = \sum_{i=1}^n \mathbb{I}(\widetilde T_{1i} \geq s,\widetilde T_{2i}\ge t), \\
    N_{01}(t | s) &= \sum_{i=1}^n \mathbb{I}(\widetilde T_{1i} \geq s, \widetilde T_{2i} \leq t, \delta_{2i} = 1), \quad
    Y_{01}(s,t) = \sum_{i=1}^n \mathbb{I}(\widetilde T_{1i} \geq s,\widetilde T_{2i}\ge t).
\end{align*}

\subsubsection{The Dabrowska Estimator}

\begin{example}[The Dabrowska Estimator]\normalfont
Dabrowska's bivariate estimator \citep{dabrowska1988kaplan} starts from the univariate Kaplan-Meier estimators \(\widehat{S}(s, 0)\) and \(\widehat{S}(0, t)\) for the marginal survival functions. Let \(\widehat{\Lambda}_{10}(s, t)\) and \(\widehat{\Lambda}_{01}(s, t)\) be the corresponding univariate Nelson-Aalen estimators. Define:
\[
\widehat{\Lambda}(s, t) = \int_0^t \int_0^s \frac{dN_{11}(s, t)}{Y_{11}(s, t)},
\]
as the \textit{bivariate Nelson-Aalen estimator}; see \citet[p.~702]{andersen1993statistical}. Then, the \textit{Dabrowska estimator} of the joint survival function is given by:
\[
\widehat{S}(s, t) = \widehat{S}(s, 0)\widehat{S}(0, t)\Prodi_{u \leq s, v \leq t}\left(1 - \widehat{L}(du, dv)\right),
\]
where \(\widehat{L}(du, dv)\) is obtained by plugging in the Nelson-Aalen estimators for the cumulative hazard components.

\end{example}

\subsubsection{Asymptotic Properties}

The large-sample behavior of the Dabrowska estimator has the same structure as
the Kaplan--Meier estimator, but with one additional layer: the map from hazard
increments to survival is now a two-dimensional product-integral map.  The
empirical Nelson--Aalen pieces first converge to their population cumulative
hazards, and the product-integral map then transfers that convergence to the
survival surface.

\begin{theorem}[Asymptotic properties of the Dabrowska estimator]
Let \(\mathcal K\) be a compact rectangle contained in the region where
\(S(s,t)>0\), with
\[
    \mathcal K=[0,\tau_1]\times[0,\tau_2].
\]
Assume that the paired lifetimes are
i.i.d., that censoring is independent of the paired failure times, that the
marginal and joint at-risk probabilities are bounded away from zero on
\(\mathcal K\), and that all denominators appearing in the definition of
\(L\) are bounded away from zero.  Let \(\widehat S_D=\Phi(\widehat\Lambda)\)
denote the Dabrowska estimator obtained by replacing
\(\Lambda=(\Lambda_{10},\Lambda_{01},\Lambda_{11})\) by the empirical
Nelson--Aalen hazard components in the representation above.  Then:
\[
    \sup_{(s,t)\in\mathcal K}
    |\widehat S_D(s,t)-S(s,t)| \xrightarrow{p} 0 .
\]
Moreover,
\[
    \sqrt n\{\widehat S_D-S\}
    \rightsquigarrow \mathbb G_S
    \qquad\text{in } \ell^\infty(\mathcal K),
\]
where \(\mathbb G_S\) is a tight mean-zero Gaussian process.  Equivalently, for
each fixed \((s,t)\in\mathcal K\),
\[
    \sqrt n\{\widehat S_D(s,t)-S(s,t)\}
    \xrightarrow{d}N\{0,\sigma_D^2(s,t)\}.
\]
The covariance function of \(\mathbb G_S\) is obtained by applying the
Hadamard derivative of the product-integral map \(\Phi\) to the covariance
function of the Nelson--Aalen hazard limit.
\end{theorem}

\begin{proof}
Write the empirical hazard vector as
\(\widehat\Lambda=(\widehat\Lambda_{10},\widehat\Lambda_{01},
\widehat\Lambda_{11})\).  Each component has the usual Nelson--Aalen
martingale decomposition.  Schematically, for a component \(a\in\{10,01,11\}\),
\[
    \widehat\Lambda_a-\Lambda_a
    =
    \frac{1}{n}\sum_{i=1}^n
    \int \frac{1}{\widehat y_a}\,dM_{ai}
    + r_{an},
\]
where \(M_{ai}\) is the corresponding counting-process martingale,
\(\widehat y_a\) is the empirical at-risk fraction, and
\(\sup_{\mathcal K}|r_{an}|=o_p(n^{-1/2})\) under the positivity assumptions.
The uniform law of large numbers for the at-risk processes and Lenglart-type
martingale bounds give
\[
    \|\widehat\Lambda-\Lambda\|_{\mathcal K}\xrightarrow{p}0.
\]
The multivariate martingale central limit theorem gives the joint weak limit
\[
    \sqrt n(\widehat\Lambda-\Lambda)\rightsquigarrow \mathbb G_\Lambda
    \qquad\text{in the hazard-component function space,}
\]
where \(\mathbb G_\Lambda\) is mean-zero Gaussian.  Its covariance is the
predictable-variation limit of the martingales above, with cross-covariances
included when the same subject contributes to more than one component.

It remains to transfer these limits through the representation theorem.  On
\(\mathcal K\), the map
\[
    \Phi:\Lambda\mapsto
    \Prodi_{u\le s}\{1-\Lambda_{10}(du,0)\}
    \Prodi_{v\le t}\{1-\Lambda_{01}(0,dv)\}
    \Prodi_{u\le s,\,v\le t}\{1-L(du,dv)\}
\]
is continuous and Hadamard differentiable as long as the survival surface and
the denominators in \(L\) stay away from zero.  Continuity gives the uniform
consistency of \(\widehat S_D=\Phi(\widehat\Lambda)\).  The functional delta
method gives
\[
    \sqrt n\{\Phi(\widehat\Lambda)-\Phi(\Lambda)\}
    \rightsquigarrow
    D\Phi_{\Lambda}(\mathbb G_\Lambda)
    \equiv \mathbb G_S .
\]
For intuition, this derivative is obtained by differentiating the logarithm of
the three product-integral factors:
\[
\frac{\mathbb G_S(s,t)}{S(s,t)}
=
-\int_{u\le s}\frac{\mathbb G_{10}(du,0)}
                    {1-\Lambda_{10}(du,0)}
-\int_{v\le t}\frac{\mathbb G_{01}(0,dv)}
                    {1-\Lambda_{01}(0,dv)}
-\iint_{u\le s,\,v\le t}
    \frac{\mathbb G_L(du,dv)}{1-L(du,dv)},
\]
where \(\mathbb G_L\) is the linearization of the map
\((\Lambda_{10},\Lambda_{01},\Lambda_{11})\mapsto L\).  This display is the
bivariate analogue of the Kaplan--Meier influence representation.  Evaluating
the limiting process at a fixed point yields the stated pointwise normal limit,
and its variance is the covariance function of \(\mathbb G_S\) evaluated at
that point.
\end{proof}

In applications, \(\sigma_D^2(s,t)\) can be estimated either by plugging
empirical at-risk fractions and Nelson--Aalen increments into the influence
representation, or by a multiplier version of the same martingale expansion.
With i.i.d. weights \(\xi_i\) satisfying \(E(\xi_i)=0\) and
\(\operatorname{Var}(\xi_i)=1\), one forms a weighted hazard perturbation
\[
    \mathbb G_\Lambda^*
    =
    \frac{1}{\sqrt n}\sum_{i=1}^n \xi_i\,\widehat\psi_{\Lambda}(O_i),
\]
pushes it through \(D\Phi_{\widehat\Lambda}\), and uses the conditional
distribution of \(D\Phi_{\widehat\Lambda}(\mathbb G_\Lambda^*)\) to obtain
standard errors or simultaneous bands.  This is exactly the same logic used for
Kaplan--Meier bands, with the derivative now carrying the additional
two-dimensional dependence correction \citep{dabrowska1988kaplan,andersen1993statistical}.

\begin{figure}[tbp]
\centering
\includegraphics[width=0.84\textwidth,height=0.36\textheight,keepaspectratio]{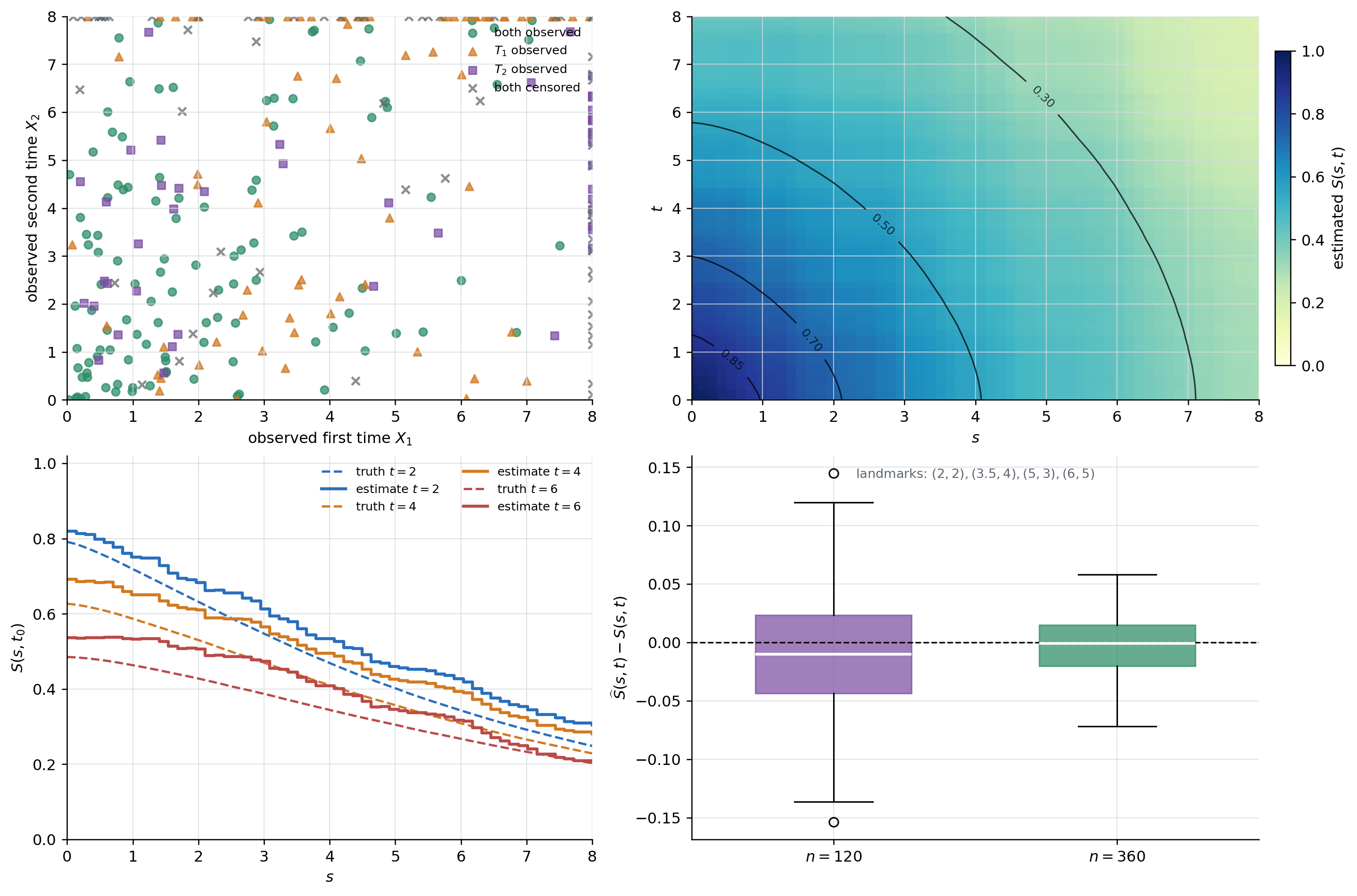}
\caption[Simulation study for Dabrowska's bivariate estimator]{Simulation study for Dabrowska's bivariate estimator.}
\label{fig:dabrowska_simulation}
\end{figure}

Figure~\ref{fig:dabrowska_simulation} gives a simulation study for this estimator. The data are generated from dependent bivariate exponential survival times with independent right censoring in the two coordinates. The plotted surface uses the same product-limit ingredients as the Dabrowska construction: marginal Kaplan--Meier estimates for \(S(s,0)\) and \(S(0,t)\), together with a two-dimensional correction for joint failures under censoring. The heat map shows the estimated joint survival surface, the black contours show the data-generating truth, and the slice plot fixes \(t\) at several values to show how the two-dimensional surface reduces to ordinary-looking survival curves along horizontal cuts. The Monte Carlo panel isolates the main difficulty: dependence must be estimated in regions where relatively few pairs remain jointly at risk.

\begin{figure}[tbp]
\centering
\includegraphics[width=0.84\textwidth,height=0.36\textheight,keepaspectratio]{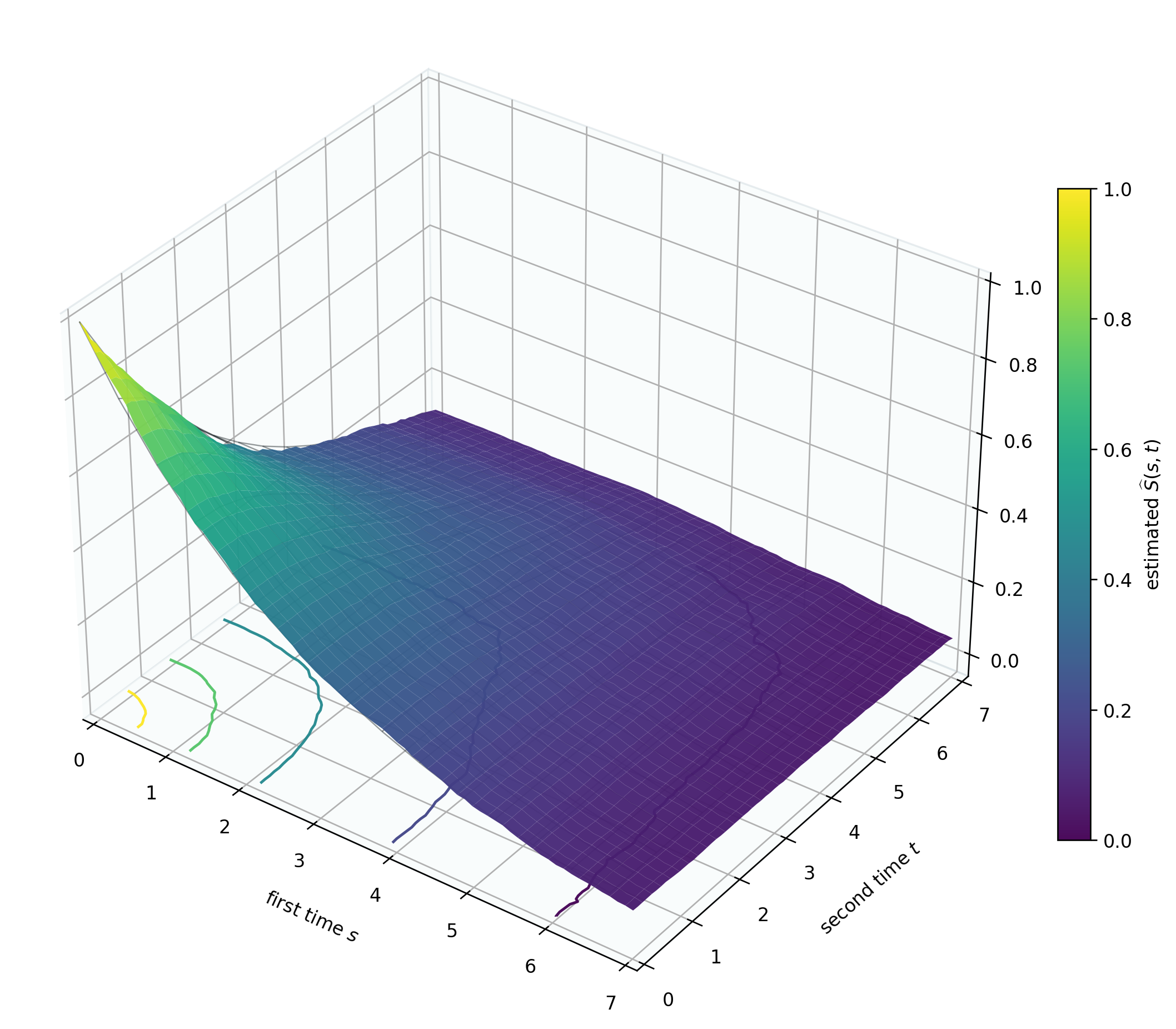}
\caption{Three-dimensional view of Dabrowska's bivariate survival estimate.}
\label{fig:dabrowska_3d_surface}
\end{figure}

Figure~\ref{fig:dabrowska_3d_surface} shows the estimator in the geometry most directly suggested by the notation: the target is a surface over the \((s,t)\) plane. The surface should decrease as either coordinate increases, while departures from a simple product of marginal curves are visible as curvature across the diagonal direction. In this simulation the colored sheet is the censoring-adjusted product-limit estimate, the black wireframe is the data-generating survival surface, and the contours on the floor show level sets of \(\widehat S(s,t)\).

\subsubsection{U-Processes, Copula Models, and Dependence Tests}

Dabrowska's estimator is a fully nonparametric surface estimator.  When the
dimension is larger than two, or when the main scientific target is a
dependence contrast rather than the entire surface, it is often more efficient
to work with pairwise functionals.  This is where \(U\)-processes enter.  Let
\[
    O_i=(\widetilde T_{i1},\ldots,\widetilde T_{id},
          \Delta_{i1},\ldots,\Delta_{id},Z_i)
\]
be the observed multivariate survival record.  For a class of symmetric kernels
\(\mathcal H=\{h_a:a\in\mathcal A\}\), define the indexed \(U\)-process
\[
    U_n(h_a)
    =
    \binom{n}{2}^{-1}\sum_{1\le i<j\le n} h_a(O_i,O_j),
    \qquad a\in\mathcal A .
\]
The index \(a\) may be a landmark time, a covariate threshold, a copula
parameter, or a weight defining a local alternative.  A censored Kendall-type
kernel for the first two coordinates has the schematic form
\[
    h_a(O_i,O_j)
    =
    W_i(a)W_j(a)
    \operatorname{sign}(\widetilde T_{i1}-\widetilde T_{j1})
    \operatorname{sign}(\widetilde T_{i2}-\widetilde T_{j2})
    \mathbb I\{(O_i,O_j)\hbox{ are comparable at }a\},
\]
where \(W_i(a)\) is usually an inverse-censoring weight built from
\(\widehat G_k(\widetilde T_{ik}-)\).  For \(d>2\), the same construction is applied to
all coordinate pairs and then averaged.  Thus a complicated multivariate
survival problem is reduced to a smooth process indexed by time, covariates, or
dependence parameters.

The key asymptotic device is the Hoeffding decomposition.  If
\(U(h)=E\{h(O_1,O_2)\}\), then
\[
\begin{aligned}
    U_n(h)-U(h)
    &=
    \frac{2}{n}\sum_{i=1}^n
    \left[E\{h(O_i,O')\mid O_i\}-U(h)\right]
    +R_n(h).
\end{aligned}
\]
where \(O'\) is an independent copy of \(O_i\).  For nondegenerate kernels, the
remainder is \(o_p(n^{-1/2})\) uniformly over well-behaved classes
\(\mathcal H\).  Consequently, the \(U\)-process behaves to first order like an
empirical process with influence function
\[
    \phi_h(O_i)=2\left[E\{h(O_i,O')\mid O_i\}-U(h)\right].
\]
This is the clean reason that rank-based dependence measures, censored
concordance statistics, and pairwise residual tests can share the same
large-sample logic as Kaplan--Meier and Nelson--Aalen estimators.

The same linearization also yields specification tests.  For a null model \(U_0(h_a)\), one can
use either a Kolmogorov--Smirnov statistic
\[
    T_{\mathrm{KS}}
    =
    \sup_{a\in\mathcal A}
    \left|
        \frac{\sqrt n\{U_n(h_a)-U_0(h_a)\}}{\widehat\sigma(a)}
    \right|
\]
or a Cramer--von Mises statistic
\[
    T_{\mathrm{CvM}}
    =
    n\int_{\mathcal A}
    \{U_n(h_a)-U_0(h_a)\}^2\,d\widehat\nu(a).
\]
The reference distribution is usually obtained by multiplier resampling of the
linear projection, for example
\[
    U_n^*(h_a)
    =
    \frac{2}{n}\sum_{i=1}^n \xi_i\widehat\phi_{h_a}(O_i),
    \qquad E(\xi_i)=0,\quad \operatorname{Var}(\xi_i)=1 .
\]
This construction yields global tests of independence, changing
concordance over follow-up, or lack of fit of a parametric dependence family.

Copula models are a complementary semiparametric way to organize multivariate
survival dependence.  A \(d\)-variate survival copula writes
\[
    P(T_1>t_1,\ldots,T_d>t_d)
    =
    \bar C_\theta\{S_1(t_1),\ldots,S_d(t_d)\},
\]
separating the marginal survival functions from the dependence parameter
\(\theta\) \citep{joe1997multivariate,nelsen2006introduction}.  The margins may
be Kaplan--Meier curves, Nelson--Aalen product integrals, Cox model fits, or
fully parametric survival curves.  For example, the Clayton survival copula is
\[
    \bar C_\theta(u_1,\ldots,u_d)
    =
    \left(\sum_{k=1}^d u_k^{-\theta}-d+1\right)^{-1/\theta},
    \qquad \theta>0,
\]
and in the bivariate case its Kendall's tau is
\(\tau=\theta/(\theta+2)\).  Thus \(\theta=0\) corresponds to independence,
while larger \(\theta\) gives stronger positive lower-tail survival dependence.

With right-censored bivariate data, a dependence pseudo-likelihood has
contributions of the schematic form
\[
\begin{aligned}
    L_i^{\mathrm{dep}}(\theta)
    \propto&
    \{\dot C_{12,\theta}(u_{i1},u_{i2})\}^{\Delta_{i1}\Delta_{i2}}
    \{\dot C_{1,\theta}(u_{i1},u_{i2})\}^{\Delta_{i1}(1-\Delta_{i2})} \\
    &\times
    \{\dot C_{2,\theta}(u_{i1},u_{i2})\}^{(1-\Delta_{i1})\Delta_{i2}}
    \{\bar C_{\theta}(u_{i1},u_{i2})\}^{(1-\Delta_{i1})(1-\Delta_{i2})},
\end{aligned}
\]
where \(u_{ik}=\widehat S_k(\widetilde T_{ik})\), and
\(\dot C_{1,\theta},\dot C_{2,\theta},\dot C_{12,\theta}\) denote the relevant
first and second partial derivatives of the survival copula.  The word
``pseudo'' matters: the marginal survival estimators are plugged in, and the
uncertainty from those nuisance estimates must be carried into the final
variance calculation.  In higher dimensions, one commonly uses pairwise
composite likelihoods or \(U\)-process estimating equations rather than the full
\(d\)-dimensional likelihood.

\begin{theorem}[Asymptotics for censored copula \(U\)-estimators]
Let \(\theta_0\) be the value satisfying
\[
\Psi(\theta_0,\eta_0)
=E\{\psi_{\theta_0,\eta_0}(O_1,O_2)\}=0,
\]
where \(\eta_0\) collects the marginal survival and censoring functions.  Suppose
\(\widehat\theta\) solves
\[
    \Psi_n(\theta,\widehat\eta)
    =
    \binom{n}{2}^{-1}\sum_{i<j}
    \psi_{\theta,\widehat\eta}(O_i,O_j)
    =
    o_p(n^{-1/2}),
\]
the kernel class is Donsker or Euclidean with an integrable envelope, the
inverse-censoring weights are bounded on the time region of interest, the
marginal estimators admit root-\(n\) linear expansions, and
\[
    A=\dot\Psi_\theta(\theta_0,\eta_0)
\]
is nonsingular.  Then
\[
    \sqrt n(\widehat\theta-\theta_0)
    =
    -A^{-1}\frac{1}{\sqrt n}\sum_{i=1}^n
    \left\{\phi_{\psi}(O_i)+\phi_{\eta}(O_i)\right\}
    +o_p(1),
\]
where \(\phi_{\psi}\) is the first Hoeffding projection of the pairwise kernel
and \(\phi_{\eta}\) is the additional contribution from estimating the margins
and censoring distribution.  Consequently,
\[
    \sqrt n(\widehat\theta-\theta_0)
    \Rightarrow
    N\left(0,\;A^{-1}\Omega A^{-T}\right),
    \qquad
    \Omega=\operatorname{Var}\{\phi_{\psi}(O)+\phi_{\eta}(O)\}.
\]
\end{theorem}

\begin{proof}
Uniform laws of large numbers for the \(U\)-process give
\(\sup_{\theta,\eta}|\Psi_n(\theta,\eta)-\Psi(\theta,\eta)|=o_p(1)\), so any
consistent approximate root lies near \(\theta_0\).  The uniform Hoeffding
decomposition gives
\[
    \sqrt n\{\Psi_n(\theta_0,\eta_0)-\Psi(\theta_0,\eta_0)\}
    =
    \frac{1}{\sqrt n}\sum_{i=1}^n \phi_{\psi}(O_i)+o_p(1).
\]
Replacing \(\eta_0\) by \(\widehat\eta\) adds the nuisance term
\(n^{-1/2}\sum_i\phi_{\eta}(O_i)+o_p(1)\) by the assumed linear expansion and
smoothness of the kernel in \(\eta\).  A Taylor expansion of
\(\Psi(\theta,\eta_0)\) around \(\theta_0\) gives
\[
    0
    =
    \Psi_n(\widehat\theta,\widehat\eta)
    =
    \Psi_n(\theta_0,\widehat\eta)
    +A(\widehat\theta-\theta_0)
    +o_p(n^{-1/2}+\|\widehat\theta-\theta_0\|).
\]
Solving this display yields the stated asymptotic linear representation.  The
central limit theorem applied to the summed influence function gives the normal
limit and the sandwich covariance matrix.
\end{proof}

\begin{figure}[tbp]
\centering
\includegraphics[width=0.88\textwidth,height=0.43\textheight,keepaspectratio]{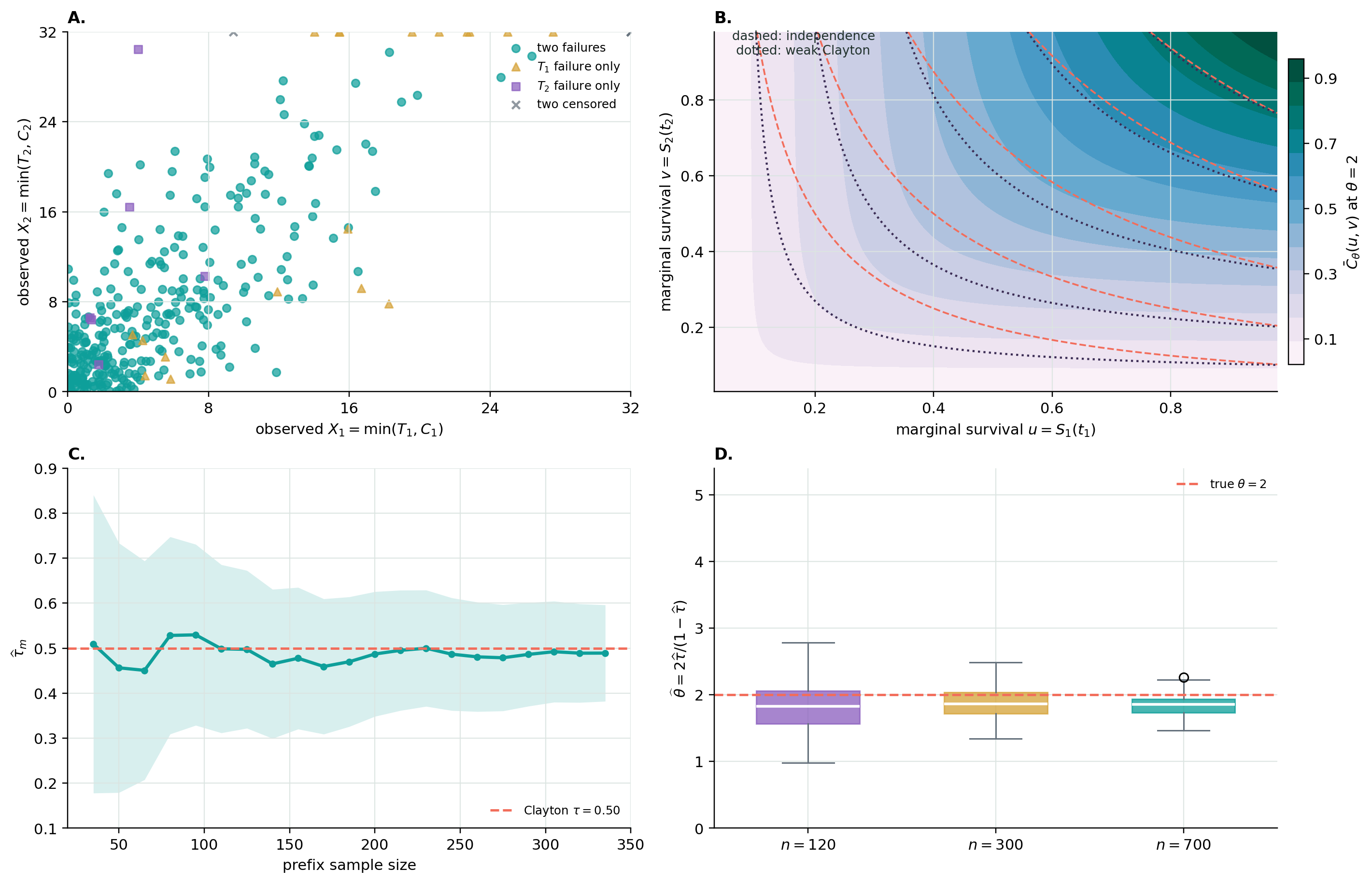}
\caption[Simulation for copula and \(U\)-process survival dependence]{Simulation for copula and \(U\)-process survival dependence.}
\label{fig:copula_u_process_simulation}
\end{figure}

Figure~\ref{fig:copula_u_process_simulation} illustrates these objects in a
small simulation.  The data are generated from a Clayton survival copula with
exponential margins and independent coordinatewise censoring.  Panel A shows
the observed bivariate survival records.  Panel B shows how the survival copula
acts on marginal survival probabilities rather than on raw time.  Panel C plots
an inverse-censoring weighted Kendall \(U\)-process along growing sample
prefixes, which makes the empirical-process stabilization visible.  Panel D
repeats the experiment and estimates \(\theta\) by the method-of-moments
identity \(\theta=2\tau/(1-\tau)\), showing the expected tightening as \(n\)
increases.

\subsubsection{References and Extensions}

Dabrowska's bivariate product-limit estimator was introduced for censored paired lifetimes by \citet{dabrowska1988kaplan}. The paper gives the product-limit representation used above and establishes large-sample properties for the joint survival estimator under independent censoring. Graphical comparison methods for bivariate survival data were then developed by \citet{dabrowska1989graphical}.

The same product-limit and martingale ideas reappear in semiparametric and multistate extensions. The Markov-renewal Cox model and bone-marrow-transplant application of \citet{dabrowska1994cox} use transition-specific risk sets and elapsed-time hazards; \citet{dabrowska1997smoothed} studies a smoothed Cox regression construction; and \citet{dabrowska2012estimation} develops transformation-model estimation for censored semi-Markov data. Broader multivariate survival references include \citet{hougaard2000analysis} and \citet{prentice2019statistical}. For copula-based dependence modeling see \citet{joe1997multivariate}, \citet{nelsen2006introduction}, \citet{genest1993statistical}, and the censored bivariate survival treatment of \citet{shih1995inferences}. The empirical-process and \(U\)-process arguments used above follow the general theory in \citet{vaart1996weak} and \citet{arcones1993limit}.

Dabrowska's one-sample conditional-survival work is the covariate-indexed analogue of the bivariate estimator: nearest-neighbor or kernel weights localize the risk set around a target covariate value, and the Kaplan--Meier product is then formed from the localized event and risk processes \citep{dabrowska1989conditionalKM}. Her modulated-renewal work with Ho extends the same elapsed-time idea to recurrent finite-state processes, where transition hazards depend on backward recurrence time and covariates rather than only on calendar time \citep{dabrowska2005modulated}.

For the present derivation, the important point is structural: the estimator combines marginal product-limit factors with a two-dimensional hazard correction. The cited papers differ in model class and covariate structure, but they retain the same counting-process ingredients: observed bivariate or transition-specific risk sets, censoring-adjusted counting processes, and martingale limits for the estimator error.

\subsection{Hidden Event Times: Turnbull's Interval-Censoring Estimator}
 
\subsubsection{Interval Censored Data}
Let $X_1,X_2,\cdots$ be a sequence of i.i.d. survival times. In practice, we only observe the interval censored data:
$$\{(L_i,R_i]:i=1,2,\cdots,n\}$$
where $0\le L_i<R_i\le\infty$ are left- and right- endpoint of an observation. That is, we only know that $X_i\in(L_i,R_i]$ instead of the exact values. 
Case I interval censoring refers to observations with either \(L_i=0\) or \(R_i=\infty\), and is also called current-status data; case II interval censoring refers to the genuinely bracketed situation \(L_i>0\) and \(R_i<\infty\).
Importantly, counting process approaches do not apply in this case because we do not have exact information to form a counting process.

A clinic-visit example makes the loss of the jump time explicit. Suppose a patient is examined at months $0,2,5,$ and $9$. If the biomarker is negative at month $2$ and positive at month $5$, the event time is not observed as a jump at month $5$; all we know is
\[
    X\in(2,5].
\]
If another patient is still negative at month $9$, the observation is $X\in(9,\infty)$. A third patient who is already positive at the first visit contributes $X\in(0,2]$. The data are therefore not a list of event times, but a list of intervals that must receive probability under the fitted distribution.
    
    Though counting process approaches do not apply, we can formulate the likelihood function of our observation:
    $$\mathcal{L}(S)=\prod_{i=1}^n\left(S(L_i)-S(R_i)\right)$$
    where $S$ is the survival function of $X$. If $L_i=R_i$ for some $i$, we replace $S(L_i)$ with its the left limit $S(L_i-)$ so that the likelihood assigns probability mass to $1-S(\Delta R_i)$. The nonparametric maximum likelihood estimation (NPMLE) is defined as
    $$\widehat{S}=\arg\max_S\log\mathcal{L}(S).$$
    
    Following \citet{sun2006statistical}, let $\{s_j\}_{j=0}^m$ be the unique ordered elements of $\{0, L_i, R_i:i=1,\cdots,n\}$ and define
    \begin{align*}
        \alpha_{ij}&=\mathbb{I}(s_j\in(L_i, R_i])\\
        p_j&=S(s_{j})-S(s_{j-1})
    \end{align*}
    for $i=1,\cdots,n$ and $j=1,\cdots,m$. If $L_i=R_i$ for some $i$, i.e., an exact observation, then we replace $\alpha_{ij}$ by $\alpha_{ij}=\mathbb{I}(s_j\in[L_i-,R_i])$ so that $\alpha_{ij}=1$ if $s_j=L_i$. Hence, the likelihood $\mathcal{L}(S)$ can be written as
    $$\mathcal{L}(S)=\prod_{i=1}^n\left(\sum_{j=1}^m\alpha_{ij}p_j\right)$$
    and NPMLE refers to the probability vector $\mathbf{p}=(p_1,\cdots,p_m)^T$.
Here \(p_j\) is the probability mass assigned to the elementary interval \((s_{j-1},s_j]\). The indicator \(\alpha_{ij}\) is the design matrix of the interval-censoring problem: it tells us whether interval \(j\) is compatible with subject \(i\)'s observation. Thus \(\sum_j\alpha_{ij}p_j\) is the probability that the fitted distribution assigns to the interval actually seen for subject \(i\).

Algorithm~\ref{alg:turnbull_blueprint} is the likelihood counterpart of the earlier risk-set algorithms. Exact event-time increments are unavailable, so each subject redistributes its probability mass over the Turnbull intervals compatible with its observed bracket.

\begin{algorithm}[tbp]
\caption{Turnbull self-consistency/EM-ICM blueprint}
\label{alg:turnbull_blueprint}
\begin{algorithmic}[1]
\Require Interval-censored observations \((L_i,R_i]\), tolerance \(\varepsilon\)
\Ensure NPMLE mass vector \(\widehat p\) on Turnbull intervals
\State Construct maximal intersections \(B_1,\ldots,B_m\) from all interval endpoints.
\State Initialize \(p_j^{(0)}>0\) with \(\sum_jp_j^{(0)}=1\).
\For{iteration \(q=0,1,2,\ldots\)}
    \State For each subject compute \(D_i^{(q)}=\sum_{j:B_j\subset(L_i,R_i]}p_j^{(q)}\).
    \State Redistribute mass by \(w_{ij}^{(q)}=p_j^{(q)}\mathbb I\{B_j\subset(L_i,R_i]\}/D_i^{(q)}\).
    \State Update \(p_j^{(q+1)}=n^{-1}\sum_iw_{ij}^{(q)}\).
    \State Optionally apply an ICM or support-reduction step to enforce the likelihood inequalities and remove inactive intervals.
    \If{\(\max_j|p_j^{(q+1)}-p_j^{(q)}|<\varepsilon\)}
        \State Stop.
    \EndIf
\EndFor
\State Return \(\widehat F(t)=\sum_{j:B_j\le t}\widehat p_j\) and the interval likelihood value.
\end{algorithmic}
\end{algorithm}
 
\subsubsection{Turnbull Intervals and the NPMLE}
    If $m$ is large, then finding $\mathbf{p}$ that maximizes $\mathcal{L}(S)$ is computationally intractable or inefficient. However, if we know in advance that some (or many) $p_j$ are $0$, then the computation will be reduced by a lot. 
\begin{lemma}[Turnbull's intervals] The $p_j$ can be nonzero only if
$s_{j-1}=L_i,\ s_j=R_k$ for some $i$ and $k$.
\end{lemma} 
\begin{proof}
The likelihood depends on the probability vector $\mathbf p$ only through the sums
\[
\sum_j\alpha_{ij}p_j=\mathbb P\{X\in(L_i,R_i]\}.
\]
Suppose an elementary interval $(s_{j-1},s_j]$ is not of the form $(L_i,R_k]$. Then it is not a maximal intersection of the observed censoring intervals. If its left endpoint is not any observed $L_i$, moving mass from $(s_{j-1},s_j]$ to the adjacent interval on the left does not remove mass from any observation interval that previously contained it. If its right endpoint is not any observed $R_k$, the same argument applies by moving mass to the adjacent interval on the right. In either case, every likelihood factor is unchanged or increased. Hence an NPMLE has a representative with zero mass on all nonmaximal elementary intervals. The only elementary intervals that can be maximal intersections have left endpoint equal to some $L_i$ and right endpoint equal to some $R_k$.
\end{proof}
    The resulting intervals $(s_{j-1},s_j]$ that can carry nonzero mass are the \emph{Turnbull intervals}, and the maximizing probability vector $\widehat{\mathbf{p}}=(\widehat{p}_1,\cdots,\widehat{p}_m)^T$ is the \emph{Turnbull estimator}.
    The Turnbull estimator is not unique in general and is usually computed by numerical algorithms such as support-reduction methods, EM, and EM-ICM \citep{turnbull1976empirical,gentleman1994maximum,anderson2017efficient}.

To determine if a candidate estimate $\widehat{\mathbf{p}}$ is the maximizer, one uses the Lagrange multiplier criterion, which is derived from graph theory and general mixture maximum likelihood theory \citep{gentleman1994maximum,hning1996interval}. Specifically, define
$$d_j(\mathbf{p})=\sum_{i=1}^n\frac{\alpha_{ij}}{\sum_{l=1}^m\alpha_{il}p_l}$$
for $j=1,\cdots,m$. Then $\widehat{\mathbf{p}}$ is the Turnbull's estimator or NPMLE if and only if $d_j(\widehat{\mathbf{p}})\le n$ for all $j$.
 
\subsubsection{Bivariate Interval Censoring}

The same observation problem becomes geometrically clearer when two event times are followed by separate visit processes. For subject \(i\), suppose the latent pair is \((T_{i1},T_{i2})\), but the data only say that
\[
    (T_{i1},T_{i2})\in
    \mathcal R_i=(L_{i1},R_{i1}]\times(L_{i2},R_{i2}],
\]
where either right endpoint may be \(\infty\). Instead of assigning probability to intervals on a line, a bivariate interval-censored NPMLE assigns probability to rectangles in the plane. On a finite grid of rectangles \(B_m\), write \(p_m=\mathbb P\{(T_1,T_2)\in B_m\}\) and let \(a_{im}\) indicate whether cell \(B_m\) is compatible with the observed rectangle \(\mathcal R_i\). The grid likelihood is
\[
    L(\mathbf p)=\prod_{i=1}^n \left(\sum_m a_{im}p_m\right),
    \qquad p_m\ge 0,\quad \sum_m p_m=1.
\]
This is the two-dimensional analogue of the Turnbull likelihood. A simple EM step redistributes each subject's probability over the cells compatible with its observed rectangle:
\[
    p_m^{\mathrm{new}}
    =\frac{1}{n}\sum_{i=1}^n
      \frac{a_{im}p_m}{\sum_\ell a_{i\ell}p_\ell}.
\]
After convergence, the estimated joint survival surface is obtained by summing fitted mass northeast of the query point,
\[
    \widehat S(s,t)=\sum_m \widehat p_m
    \mathbb I\{B_m \subset (s,\infty)\times(t,\infty)\}.
\]
The grid version is not meant to hide the computational difficulty of the exact multivariate NPMLE; rather, it makes the likelihood geometry visible. With exact paired event times one would count points. With bivariate interval censoring one sees only rectangles, and the estimator must decide how to distribute mass inside the overlapping rectangle arrangement \citep{turnbull1976empirical,sun2006statistical,gentleman1994maximum}.

\begin{figure}[tbp]
\centering
\includegraphics[width=0.84\textwidth,height=0.40\textheight,keepaspectratio]{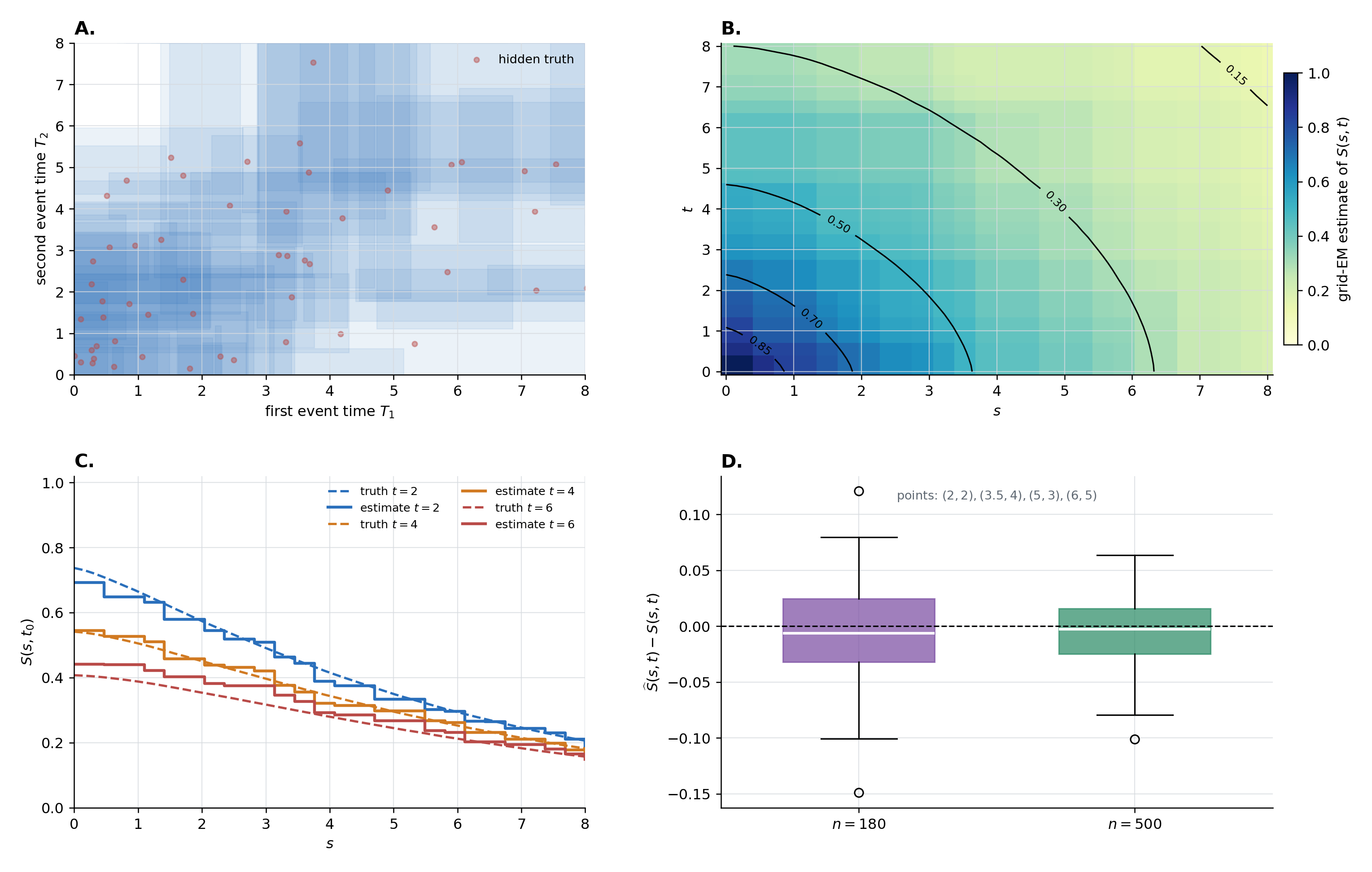}
\caption[Simulation for bivariate interval-censored paired event times]{Simulation for bivariate interval-censored paired event times.}
\label{fig:bivariate_interval_censoring_sim}
\end{figure}

\subsection{Simulation Gallery and Examples}
 
\subsubsection{Simulation Illustrations}
To connect the preceding estimators with their finite-sample behavior, we use a small simulation study. For the right-censoring examples, event times are generated from an exponential model with hazard \(0.14\), independent censoring is generated from an exponential model with hazard \(0.045\), and administrative follow-up stops at time \(12\). Thus the target cumulative hazard and survival curve are
\[
    A(t)=0.14t,\qquad S(t)=\exp(-0.14t).
\]
For the competing-risks example, the cause-specific hazards are \(0.08\) and \(0.05\), giving
\[
    F_j(t)=\frac{\alpha_j}{\alpha_1+\alpha_2}\{1-\exp[-(\alpha_1+\alpha_2)t]\},\qquad j=1,2.
\]
For the interval-censoring example, subjects are inspected at random visit times, so the observation is only the interval \((L_i,R_i]\) containing the event time.

\begin{figure}[tbp]
\centering
\includegraphics[width=0.82\textwidth,height=0.38\textheight,keepaspectratio]{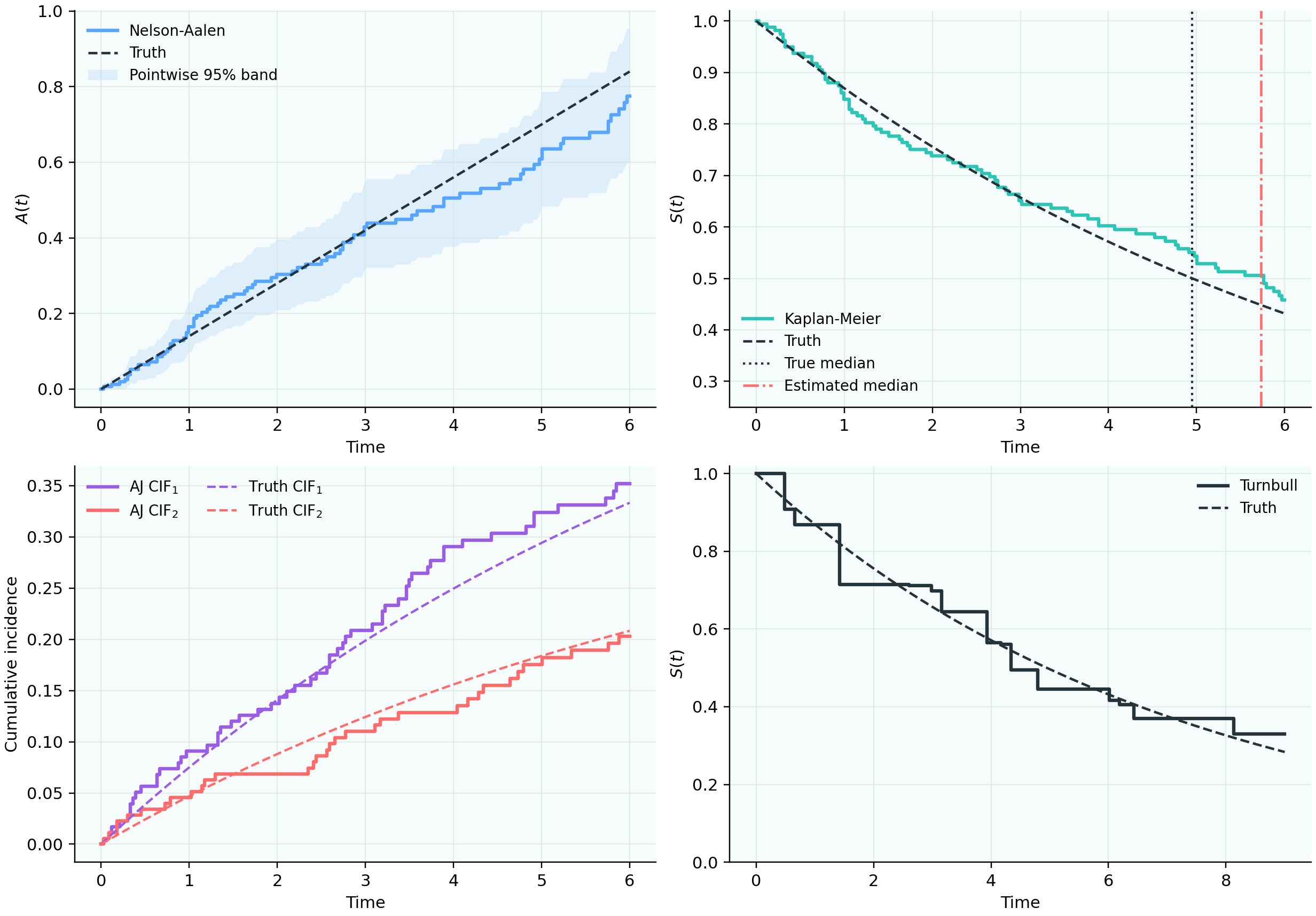}
\caption{Simulation paths for common counting-process estimators.}
\label{fig:simulation_estimator_paths}
\end{figure}

Figure~\ref{fig:simulation_estimator_paths} compares four estimators on representative simulated data sets. The Nelson-Aalen curve tracks the cumulative hazard, the Kaplan-Meier curve tracks survival and yields the median survival estimate, the Aalen-Johansen estimator recovers the cause-specific cumulative incidence functions, and the Turnbull estimator gives a step-function NPMLE under interval censoring.

\begin{figure}[tbp]
\centering
\includegraphics[width=0.78\textwidth,height=0.34\textheight,keepaspectratio]{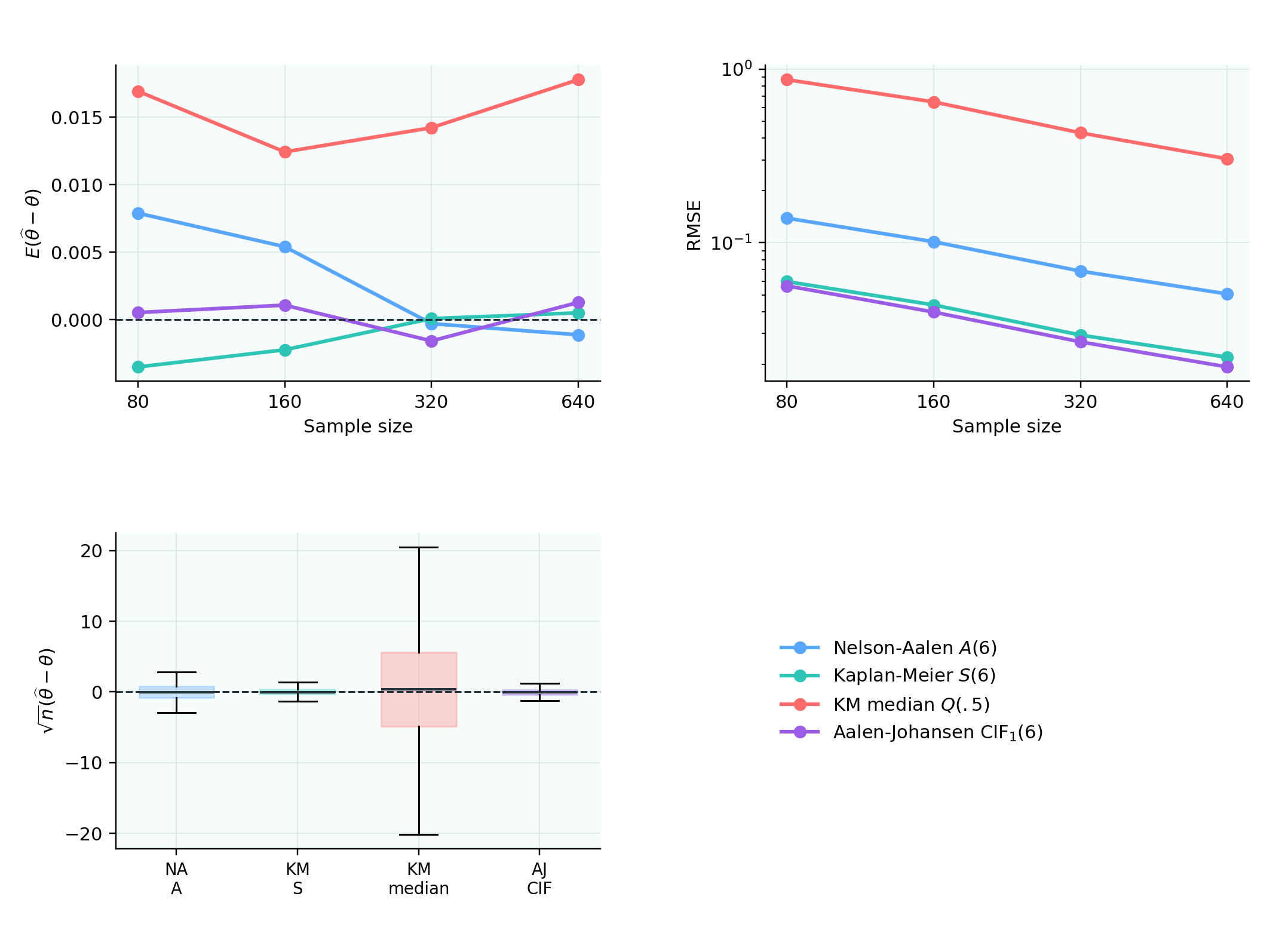}
\caption[Monte Carlo comparison of estimated survival quantities]{Monte Carlo comparison of estimated survival quantities.}
\label{fig:simulation_quantity_mc}
\end{figure}

The same data-generating mechanisms can be used to compare scalar quantities. In Figure~\ref{fig:simulation_quantity_mc}, \(700\) Monte Carlo replications are used at each sample size. The quantities are \(A(6)\), \(S(6)\), the median \(Q(0.5)\), and \(F_1(6)\), estimated respectively by Nelson-Aalen, Kaplan-Meier, the Kaplan-Meier inverse, and Aalen-Johansen. Bias is close to zero for the smooth functionals, while the median has larger variability because it depends on inverting the survival curve near \(S(t)=0.5\). The RMSE plot illustrates the expected decrease with \(n\), and the boxplots of \(\sqrt n(\widehat\theta-\theta)\) give a visual check of the central-limit-theorem scaling.

\subsubsection{Smoothed Nelson-Aalen (SNA) Estimator}
\begin{definition}[SNA]\label{def:SNA}
Let $\widehat{A}(t)$ be the NA estimator of \(A(t)=\int_0^t\alpha(s)ds\). The \emph{smoothed Nelson-Aalen (SNA) estimator} of $\alpha(t)$ is
$$\widehat{\alpha}(t)=\frac{1}{b}\int_\mathcal{T}K(\frac{t-s}{b})d\widehat{A}(s)$$
where the \emph{kernel function} $K$ is bounded, vanishes outside $[-1, 1]$, and has integral 1. The hyperparameter $b$ is referred to as the \emph{bandwidth} or \emph{window size}.

\end{definition}
The Nelson--Aalen curve is a step function because events arrive as jumps. If the scientific target is the instantaneous hazard \(\alpha(t)\), rather than its cumulative version \(A(t)\), the jumps must be smoothed. The kernel \(K\) determines how nearby event increments are averaged, and the bandwidth \(b\) determines how local the averaging is: small \(b\) gives a high-variance hazard estimate, while large \(b\) gives a smoother but more biased estimate. An SNA can be applied to any asymptotically normal cumulative-hazard estimator when the goal is to recover a hazard curve.

\begin{figure}[tbp]
\centering
\includegraphics[width=0.82\textwidth,height=0.38\textheight,keepaspectratio]{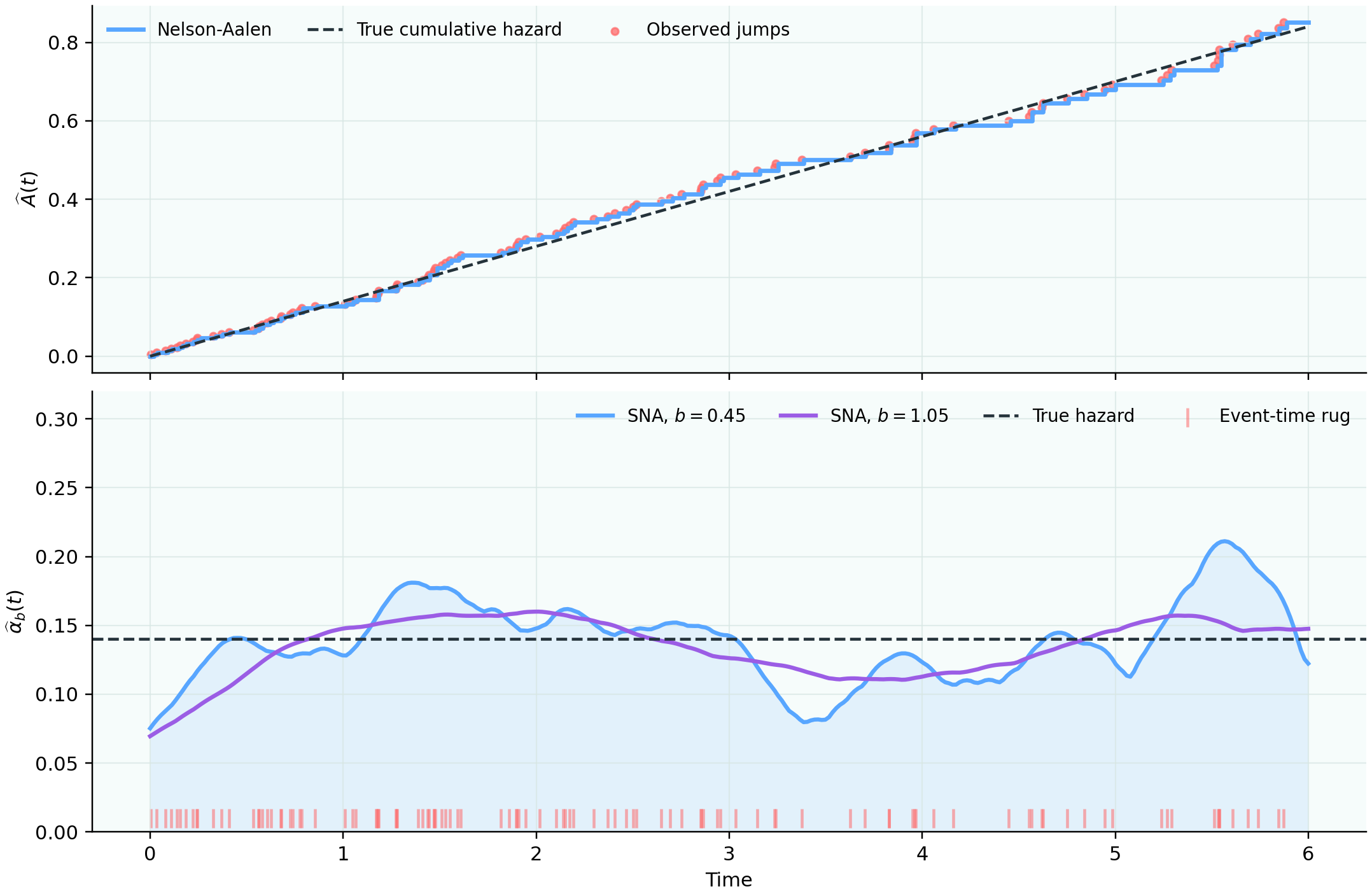}
\caption{Simulation illustration for the smoothed Nelson--Aalen estimator.}
\label{fig:sna_smoothing_simulation}
\end{figure}

Figure~\ref{fig:sna_smoothing_simulation} displays the estimator directly. The upper panel begins with the Nelson--Aalen jumps. The lower panel spreads each jump across a local neighborhood; a narrow bandwidth preserves local event-time variation, while a wider bandwidth trades detail for a more stable hazard curve. This is the same bias--variance choice that later reappears in smoothed Breslow baselines and density estimates for survival quantiles.

\subsubsection{Random Sample from a Finite-State Markov Process}
Let \(X_1(\cdot),X_2(\cdot),\ldots,X_n(\cdot)\) be \(n\) independent copies of a finite-state Markov process \(X(t)\), \(t\in\mathcal T\), with state space \(\mathbb S\), initial probabilities \(p_h=\mathbb P\{X(0)=h\}\), transition probabilities \(P_{hj}(s,t)\), and transition intensities \(\alpha_{hj}(t)\) for \(h\ne j\). For subject \(i\), define
\[
    N_{ihj}(t)=\#\{u\le t:X_i(u-)=h,\ X_i(u)=j\},\qquad h\ne j.
\]
The aggregated transition count and the state-specific risk process are
\[
    N_{hj}(t)=\sum_{i=1}^nN_{ihj}(t),\qquad
    Y_h(t)=\sum_{i=1}^n\mathbb I\{X_i(t-)=h\}.
\]
Thus \(N_{hj}(t)\) counts direct \(h\to j\) transitions up to time \(t\), and \(Y_h(t)\) is the number of sample paths in state \(h\) just before \(t\). Since
\[
    P_h(t-)=\sum_{g\in\mathbb S}p_gP_{gh}(0,t-),
\]
we have \(Y_h(t)\sim\operatorname{Binomial}\{n,P_h(t-)\}\).

By Theorem~\ref{thm:Jacobson}, \((N_{hj},h\not=j)\) is a multivariate counting process with intensity \(\lambda_{hj}(t)=\alpha_{hj}(t)Y_h(t)\), so each transition-specific cumulative intensity \(A_{hj}(t)=\int_0^t\alpha_{hj}(s)ds\) has the Nelson--Aalen estimator
\[
    \widehat{A}_{hj}(t)=\int_0^t\frac{J_h(s)}{Y_h(s)}dN_{hj}(s),\qquad
    J_h(s)=\mathbb{I}\{Y_h(s)>0\}.
\]
If \(P_h(u)>0\) for every state \(h\) and every \(u\in[s,t]\), then \(Y_h(u)\rightarrow_p\infty\), and hence
\[
    \sup_{u\in[s,t]}|\widehat{A}_{hj}(u)-A_{hj}(u)|\rightarrow_p 0
    \qquad\text{as }n\rightarrow\infty.
\]
A more detailed martingale argument, given by \citet[pp.~197--198]{andersen1993statistical}, yields
\[
    \left(\sqrt{n}(\widehat{A}_{hj}-A_{hj});h\not=j\right)\rightarrow_d(U_{hj};h\not=j)
\]
on \([s,t]\), where the \(U_{hj}\) are independent Gaussian martingales with \(U_{hj}(0)=0\) and
\[
    \operatorname{Cov}\{U_{hj}(u_1),U_{hj}(u_2)\}
    =\int_s^{u_1\wedge u_2}\frac{\alpha_{hj}(v)}{P_h(v)}dv.
\]

\noindent\textbf{Example (three-state illness-death sample).}
Let \(\mathbb S=\{0,1,2\}\), where \(0,1,2\) denote event-free, relapsed, and dead. Allow only the transitions \(0\to1\), \(0\to2\), and \(1\to2\). Four complete patient histories observed on \([0,5]\) are
\[
\begin{array}{c|l}
\text{patient} & \text{path; arrow label = transition time}\\ \hline
1 & 0\xrightarrow{2}1\xrightarrow{4}2,\\
2 & 0\xrightarrow{3}2,\\
3 & 0\xrightarrow{1}1\quad\text{and remains in }1\text{ through }5,\\
4 & 0\quad\text{and remains in }0\text{ through }5.
\end{array}
\]
The transition-specific risk sets are evaluated immediately before each jump. Hence
\[
\begin{array}{c|c|c|c}
\text{time} & \text{transition} & \text{risk set} & \text{NA increment}\\ \hline
1 & 0\to1 & Y_0(1)=4 & \Delta\widehat A_{01}(1)=1/4\\
2 & 0\to1 & Y_0(2)=3 & \Delta\widehat A_{01}(2)=1/3\\
3 & 0\to2 & Y_0(3)=2 & \Delta\widehat A_{02}(3)=1/2\\
4 & 1\to2 & Y_1(4)=2 & \Delta\widehat A_{12}(4)=1/2.
\end{array}
\]
Therefore, at \(t=5\),
\[
    \widehat A_{01}(5)=\frac14+\frac13,\qquad
    \widehat A_{02}(5)=\frac12,\qquad
    \widehat A_{12}(5)=\frac12.
\]
This finite-sample calculation shows that each transition has its own risk set and its own Nelson--Aalen increment. A transition \(0\to1\) is relapse, \(0\to2\) is death without recorded relapse, and \(1\to2\) is death after relapse. The estimator only requires that, just before each event time, we know which patients are eligible for each transition. Point probabilities describe disease-state occupation, while transition intensities describe local movement between states.

\subsubsection{Excess Mortality Models}

An excess mortality model decomposes the observed event intensity into a known background component and an unknown excess component. For subject \(i\),
\[
    \lambda_i(t)=(\gamma(t)+\mu_i(t))Y_i(t),
\]
where \(\mu_i(t)\) is a known background hazard, often obtained from an external life table, and \(\gamma(t)\) is the excess hazard common to the study cohort. The target is the cumulative excess hazard,
\[
    \Gamma(t)=\int_0^t\gamma(s)ds.
\]

The resulting decomposition is explicit. Summing the known background hazard over all items still at risk gives
\[
    Y^\mu(t)=\sum_{i=1}^n\mu_i(t)Y_i(t),
    \qquad
    Y(t)=\sum_{i=1}^nY_i(t).
\]
The observed counting process \(N(t)\) contains both ordinary background events and excess events. The Doob-Meyer decomposition subtracts the ordinary part and leaves a martingale error:
\[
    M(t)=N(t)-\int_0^tY^\mu(s)ds-\int_0^t\gamma(s)Y(s)ds.
\]
Solving this identity for the unknown cumulative excess hazard gives the Nelson-Aalen type estimator
\[
    \widehat{\Gamma}(t)
    =\int_0^t\frac{J(s)}{Y(s)}dN(s)
    -\int_0^t\frac{Y^\mu(s)}{Y(s)}ds.
\]
The first term is the empirical cumulative hazard observed in the cohort. The second term is what the known background process would have produced among exactly the same risk sets. Their difference is the excess curve.

The product-integral version writes this correction on the survival scale:
\[
    \Prodi_{0}^{t}(1-d\widehat{\Gamma}(s))
    =
    \frac{\widehat{S}(t)}
    {\exp\left(-\int_0^t\frac{Y^\mu(s)}{Y(s)}ds\right)}.
\]
Here \(\widehat{S}(t)\) is the Kaplan-Meier estimator for the observed event process, while the denominator is the survival that would be expected from background hazards alone. The left-hand side is therefore a corrected survival function: survival after factoring out the known ordinary risk.

\subsubsection{Competing Risks Model}

Competing risks arise when the first observed event can be one of several mutually exclusive types. Once a subject has experienced a first event of type \(j\), the other first-event types can no longer occur for that subject. The cause-specific incidence function therefore depends both on the event-free survival probability and on the cause-specific hazard.

Let \(0<T_1<T_2<\cdots<T_m\le t\) be the observed event times, and let \(0\) denote the initial state. If there are \(k\) possible absorbing causes, the Aalen-Johansen estimator of the transition probability matrix \(\p(0,t)=(P_{hj}(0,t))\) is \citep{aalen1978empirical,andersen1993statistical}
\[
    \widehat{\p}(0,t)=\prod_{i=1}^m\left(\mathbf{I}+\Delta\mathbf{A}(T_i)\right),
\]
where
\[
\Delta\mathbf{A}(T_i)=
\left[\begin{matrix}
    -\sum_{j=1}^k\frac{\Delta N_{0j}(T_i)}{Y_{0}(T_i)}
    &\frac{\Delta N_{01}(T_i)}{Y_{0}(T_i)}
    &\frac{\Delta N_{02}(T_i)}{Y_{0}(T_i)}
    &\cdots
    &\frac{\Delta N_{0k}(T_i)}{Y_{0}(T_i)}\\
    0&0&0&\cdots&0\\
    \vdots&\vdots&\vdots&\vdots&\vdots\\
    0&0&0&\cdots&0
\end{matrix}\right].
\]
The diagonal entry for state \(0\) subtracts all exits from the initial state, and each off-diagonal entry allocates the observed exit to one specific cause. Thus
\[
    \widehat{P}_{00}(0,t)
    =
    \prod_{i=1}^m
    \left(1-\frac{\sum_{j=1}^k\Delta N_{0j}(T_i)}{Y_{0}(T_i)}\right)
\]
is the usual Kaplan-Meier estimator for remaining event-free, regardless of cause.

For cause \(j\), induction on the event times gives
\begin{align*}
    \widehat{P}_{0j}(0,t)
    &=
    \sum_{i=1}^m
    \left\{
    \prod_{h<i}
    \left(1-\frac{\sum_{\ell=1}^k\Delta N_{0\ell}(T_h)}{Y_0(T_h)}\right)
    \right\}
    \frac{\Delta N_{0j}(T_i)}{Y_0(T_i)}\\
    &=
    \int_0^t\widehat{P}_{00}(0,u-)d\widehat{A}_{0j}(u).
\end{align*}
This is the cumulative incidence function used in standard competing-risk analysis \citep{tsiatis1975nonidentifiability,kalbfleisch2002statistical,gray1988class,fine1999proportional}:
\[
    CIF_j(t)=F_j(t)=\int_0^tS(u-)\lambda_j(u)du,
\]
where \(S(u-)\) is the probability of still being event-free just before \(u\), and \(\lambda_j(u)\) is the cause-specific hazard. The multiplicative form reflects that cause \(j\) can occur at time \(u\) only among items that have not already exited through another cause.

Regression models for competing risks must therefore state their estimand. A cause-specific Cox model writes
\[
    \lambda_j(t\mid Z)=\lambda_{j0}(t)\exp\{\beta_j^\top Z\},
\]
where the risk set contains only subjects who are still event-free just before \(t\). The coefficient \(\beta_j\) describes how \(Z\) changes the instantaneous transition intensity to cause \(j\) among subjects who have not yet failed from any cause. It is the natural model for transition mechanisms, etiologic questions, and multistate simulation. To obtain the cumulative incidence \(F_j(t\mid Z)\), however, the analyst must also model or estimate the hazards of the competing causes, because those causes remove subjects from the event-free state before cause \(j\) can occur.

The Fine--Gray model instead targets the marginal cumulative incidence for a chosen cause. Its subdistribution hazard is
\[
    \widetilde\lambda_j(t\mid Z)
    =
    \lim_{\Delta\downarrow0}
    \frac{
    \mathbb P\{t\le T<t+\Delta,\ J=j\mid T\ge t
    \ \text{or}\ (T<t,\ J\ne j),Z\}
    }{\Delta}
    =
    \frac{dF_j(t\mid Z)}{1-F_j(t-\mid Z)}.
\]
The denominator is not the ordinary event-free risk set. Subjects who have already experienced a competing event remain in the subdistribution risk set, often with censoring weights, because the target is the marginal probability \(F_j(t\mid Z)=\mathbb P(T\le t,J=j\mid Z)\). A proportional subdistribution-hazards model takes
\[
    \widetilde\lambda_j(t\mid Z)
    =
    \widetilde\lambda_{j0}(t)\exp\{\theta_j^\top Z\},
    \qquad
    F_j(t\mid Z)
    =
    1-\exp\{-\widetilde A_{j0}(t)\exp(\theta_j^\top Z)\}
\]
in the continuous-time case. Thus \(\theta_j\) is read as a direct regression effect on the cumulative-incidence curve for cause \(j\), not as a physical transition intensity among event-free subjects. Fine--Gray regression is useful when prognosis, absolute risk, or policy reporting is framed in terms of the probability of a specific first event by time \(t\). Cause-specific Cox regression is preferable when the scientific question concerns the transition mechanism itself or when all transition probabilities must be assembled through an Aalen--Johansen or multistate model. Gray's test is the corresponding nonparametric comparison for cumulative-incidence curves, while cause-specific score or log-rank tests compare cause-specific transition intensities.

In a transplant registry, the absorbing causes might be relapse without transplant, non-relapse mortality, and loss to a terminal complication. For a patient still event-free at \(u-\), the cumulative incidence of relapse is the probability that relapse occurs before the competing terminal events. A single Kaplan--Meier curve for ``any event'' marginalizes over the exit route. In a reliability study, the same formula separates direct failure, planned replacement, and removal from service while preserving the common risk set of devices still operating just before each event. Event-free survival records the total probability of avoiding all absorbing events; cumulative-incidence curves attribute the accumulated exits to specific causes.

\section{From Estimation to Comparison: Nonparametric Tests}
 \subsection{One-Sample Tests for a Specified Hazard}
 
Estimation targets the cumulative hazard. Testing a specified hazard compares the observed event process with its predictable compensator under the null model. The natural statistic is therefore an observed-minus-expected martingale transform. Consider a sequence of univariate counting processes $(N^{(n)}(t); t\in\mathcal{T})$ with intensity process $\alpha(t)Y^{(n)}(t)$. For readability, the superscript \((n)\) is omitted below. We want to test
    $$H_0:\alpha(s)=\alpha_0(s), s\in[0,t]$$
    and this is equivalent to test
    $$H_0:A(s)=A_0(s),s\in[0,t]$$
    under the assumption $A(t)<\infty$ for $t\in\mathcal{T}$. 
    
    Recall that under $H_0$,
    \begin{align*}
        \widehat{A}(t)&=\int_0^t\frac{J(s)}{Y(s)}dN(s)\\
        A_0^*(t)&=\int_0^t\alpha_0(s)J(s)ds\\
        \langle\widehat{A}-A_0^*\rangle(t)&=\int_0^t\frac{J(s)}{Y(s)}\alpha_0(s)ds
    \end{align*}
    The test statistic is based on the martingale (\ref{thm:doob_meyer_stochastic_integral})
    $$Z(t)=\int_0^tK(s)d(\widehat{A}-A_0^*)(s)$$
    where $K$ is any locally bounded predictable non-negative stochastic process satisfying $K(s)=0$ if $Y(s)=0$.
The weight \(K(s)\) lets the analyst decide which part of follow-up should matter most. Taking \(K(s)=Y(s)\) counts raw observed minus expected failures; other choices emphasize early or late departures from the null hazard.

As a clinical illustration, suppose an external registry provides a benchmark death hazard \(\alpha_0(t)\) for patients with the same diagnosis and calendar period. The process \(N(t)\) counts deaths observed by follow-up time \(t\), and \(Y(t)\) counts patients alive and still under observation just before \(t\). If the benchmark predicts
\[
    E(12)=\int_0^{12}\alpha_0(s)Y(s)ds=12
\]
deaths by month 12 but \(N(12)=18\) deaths are observed, then the one-sample log-rank statistic with \(K=Y\) is
\[
    T(12)^2=\frac{(18-12)^2}{12}=3.
\]
This calculation is a standardized accumulated difference between observed deaths and deaths predicted from the risk set under the benchmark mortality curve. A reliability version would compare observed device failures with failures predicted by a reference hazard; excess failures would appear as positive martingale residuals after accounting for how many devices were still under observation at each time.

\subsubsection{One-Sample Log-Rank Statistic}

    The test statistic is ($t$ is large enough)
    $$T(t)=\frac{Z(t)}{\sqrt{\langle Z\rangle(t)}}$$
    where $\langle Z\rangle(t)=\int_0^t \frac{K^2(s)}{Y(s)}\alpha_0(s)ds$. The special case $K(s)=Y(s)$ leads to the \emph{one-sample log-rank statistic}, i.e.,
    $$T(t)^2=\frac{(N(t)-E(t))^2}{E(t)}$$
    where $E(t)=\int_0^t\alpha_0(s)Y(s)ds=\langle Z\rangle(t)$. The ratio $\frac{N(t)}{E(t)}$ is called the \emph{standardized mortality ratio} (SMR) in epidemiology. Finally, Fleming and Harrington (1982) also suggested that $K(t)=Y(t)S_0(t)^\rho$ where $S_0(t)=\exp(-A_0(t))$.

\subsection{\texorpdfstring{\(k\)-Sample Tests}{k-Sample Tests}: Comparing Survival Distributions}
    The \(k\)-sample version tests whether several groups share the same underlying hazard after accounting for their risk-set sizes. Under the null, a group can have more or fewer failures because its risk set is larger or smaller, but not because its transition intensity is different. Consider a sequence of $k$-variate counting processes $\mathbf{N}^{(n)}=(N_1^{(n)},\cdots,N_k^{(n)})$ with intensity process $\boldsymbol{\lambda}^{(n)}=(\lambda_1^{(n)},\cdots,\lambda_k^{(n)})$, where \(\lambda_h^{(n)}(t)=\alpha_h(t)Y_h^{(n)}(t)\). For readability, the superscript \((n)\) is omitted below. We want to test
    $$H_0:\alpha_1(s)=\alpha_2(s)=\cdots=\alpha_k(s), s\in[0,t].$$
    In practice, usually $k=2$ and this reduces to a two-sample problem. To construct test statistics, similar to the one-sample case, we first define
    \begin{align*}
   \widehat{A}_h(s)&=\int_0^t\frac{J_h(s)}{Y_h(s)}dN_h(s) \\
   \widehat{A}(s)&=\int_0^t\frac{J(s)}{Y(s)}dN(s)\\
        \widetilde{A}_h(s)&=\int_0^t\frac{J_h(s)}{Y(s)}dN(s)
    \end{align*}
    for $h=1,\cdots,k$ and $Y(s), J(s)$ and $N(s)$ are aggregated processes.
Here \(\widehat A_h\) is the group-specific cumulative hazard, while \(\widetilde A_h\) is what group \(h\) would have accumulated if the pooled hazard were correct but group \(h\)'s own risk-set process were retained. Their difference is the group-specific martingale contrast.

For clinical data, group \(1\) might be a standard-care cohort and group \(2\) a treatment cohort, with \(N_h(t)\) counting deaths or relapses and \(Y_h(t)\) counting subjects in group \(h\) who are still event-free and observed just before \(t\). At each event time, the pooled null hazard assigns an expected event contribution to each group in proportion to its risk-set size. The log-rank statistic accumulates the observed-minus-expected event counts over all event times. In a reliability study, the two groups might be devices built under two manufacturing processes; \(N_h(t)\) counts failures, and the null model implies equal failure hazards after conditioning on the operating devices in the risk sets.
 
\subsubsection{Two-Sample Log-Rank Test}
    Let $K_h(t)=Y_h(t)K(t)$ be a locally bounded predictable weight process and define
    $$Z_h(t)=\int_0^tK_h(s)d(\widehat{A}_h-\widetilde{A}_h)(s)$$
    for $h=1,2,\cdots,k$; the test statistic will be based on \(Z_h(t)\). Under finite moment conditions, the martingale identities used below are
\[
\begin{aligned}
Z_h(t)&=\sum_{l=1}^k\int_0^tK(s)\left(\delta_{hl}-\frac{Y_h(s)}{Y(s)}\right)dM_l(s),\\
\langle Z_h,Z_j \rangle(t)&=\int_0^tK^2(s)\left(\delta_{hj}-\frac{Y_j(s)}{Y(s)}\right)\frac{Y_h(s)}{Y(s)}Y(s)\alpha(s)ds,\\
\operatorname{Cov}\{Z_h(t),Z_j(t)\}&=\mathbb{E}\langle Z_h,Z_j \rangle(t).
\end{aligned}
\]
The case \(k=2\) gives the usual two-sample statistic.

\begin{definition}[Two sample log rank statistic]
    Let $$\widehat{\sigma}^2_{11}(t)=\int_0^tK^2(s)\frac{Y_1(s)}{Y_1(s)+Y_2(s)}\left(1-\frac{Y_2(s)}{Y_1(s)+Y_2(s)}\right)d(N_1+N_2)(s)$$ be the estimator of $Var(Z_1(t))$ and define
    $$X^2=\frac{Z_1(t)^2}{\widehat{\sigma}^2_{11}(t)}.$$
    One choice of $K$ is $K(t)=J(t)=\mathbb{I}(Y_1(t)+Y_2(t)>0)$. The test statistic $X^2$ is called the \emph{two sample log-rank statistic} which asymptotically has a $\chi^2_1$ distribution under $H_0$.
\end{definition}
The definition is an observed-minus-expected contrast. The observed term \(O_h\) counts how many events occurred in group \(h\); the expected term \(E_h\) is the model-based count under a pooled hazard with group-specific risk-set sizes. The theorem below gives the formal martingale statement behind Figures~\ref{fig:simulation_estimator_paths} and \ref{fig:multistate_simulation}: under the null, these contrasts are centered martingales, and their predictable variation supplies the variance.
With the choice \(K(t)=J(t)\), the two-sample statistic reduces to the familiar observed-minus-expected form
\[
    Z_h(t)=O_h-E_h,\qquad h=1,2,
\]
where \(O_h=N_h(t)\) is the observed number of failures in group \(h\) and
\[
    E_h=\int_0^t\frac{Y_h(s)}{Y_1(s)+Y_2(s)}d(N_1+N_2)(s)
    =\int_0^tY_h(s)d\widehat A(s)
\]
is the expected number of failures in group \(h\), with \(\mathbb E(O_h)=\mathbb E(E_h)\) under the null. The name \emph{log-rank} was proposed by \citet{peto1972asymptotically}; see \citet[pp.~364--366]{andersen1993statistical}. In Li's notation \citep{li2021survival},
\[
\begin{aligned}
    Z_h&=\sum_{i=1}^DK(t_i)[O_{ij}-E_{ij}]\\
        &=\sum_{i=1}^DK(t_i)\left[d_{ij}-r_{ij}\frac{d_i}{r_i}\right],
\end{aligned}
\]
where \(D\), \(d_{ij}\), and \(r_{ij}\) are the numbers of death times, deaths, and subjects at risk at \(t_i\) in total and in group \(j\); \(d_i=d_{i1}+d_{i2}\) and \(r_i=r_{i1}+r_{i2}\). The corresponding variance estimate is
\[
    \widehat{\sigma}_{hh}=\sum_{i=1}^DK^2(t_i)r_{ij}\frac{d_i}{r_i}\left(1-\frac{d_i}{r_i}\right)\left(\frac{r_i-r_{ij}}{r_i-1}\right),
\]
for \(h=1,2\), which reveals the hypergeometric structure behind the risk-set comparison.

\subsubsection{Stratification}
    Suppose there are \(S\) strata, so that for each stratum \(m=1,\ldots,S\) we observe a bivariate counting process
    \[
        \mathbf N_m=(N_{1m},N_{2m})
    \]
    with intensity \(\boldsymbol\lambda_m=(\lambda_{1m},\lambda_{2m})\) of the form
    \[
        \lambda_{im}(t)=\alpha_{im}(t)Y_{im}(t),
        \qquad i=1,2.
    \]
    The stratified null hypothesis allows the baseline hazard to differ by stratum but requires equality of the two groups within each stratum:
    \[
        H_0:\alpha_{1m}(t)=\alpha_{2m}(t),
        \qquad m=1,\ldots,S.
    \]
    Using the asymptotic results in Section~\ref{sec:asym_logrank}, a stratified two-sample statistic is
    \[
        X^2=
        \frac{\left\{\sum_{m=1}^S Z_{1m}(t)\right\}^2}
        {\sum_{m=1}^S\widehat{\sigma}_{11m}^2(t)},
    \]
    which converges to a \(\chi^2_1\) distribution under \(H_0\).

Strata are often what make the comparison meaningful. In a clinical trial, treatments should not be compared across centers, disease stages, age bands, or enrollment cohorts if these factors change the baseline hazard. In a reliability study, device designs may need to be compared within manufacturing batch, installation environment, or stress-test protocol. The stratified test compares groups only within the same stratum and then adds the centered contrasts, preventing background heterogeneity from being mistaken for the effect of the group label.
 
\subsubsection{Asymptotic Null Distribution}\label{sec:asym_logrank}
    The null distribution theorem is the testing analogue of Nelson--Aalen asymptotic normality. The estimator is no longer a curve estimate but a weighted sum of centered event-time increments. Once each centered increment is written as a martingale increment, Rebolledo's theorem gives a Gaussian limit for the accumulated discrepancy, and the two-sample square becomes a chi-square statistic.
    \begin{theorem}[Asymptotic null distribution of the log-rank test] Assume that, under $H_0$, the group-specific counting processes have intensities $Y_h(t)\alpha(t)$, $h=1,\cdots,k$, and that $n^{-1}Y_h(t)\to_p y_h(t)$ uniformly on compact time intervals, where $y(t)=\sum_hy_h(t)$ is bounded away from zero on the time interval of interest. If the predictable weight $K$ is bounded, then the vector of standardized log-rank processes converges weakly:
    $$n^{-1/2}(Z_1,\cdots,Z_k)\rightarrow_d(U_1,\cdots,U_k)$$
    in $D(\mathcal{T})^k$, where the limit is a mean-zero Gaussian martingale with covariance functions
    $$\sigma_{hj}(t)=\int_0^tK^2(s)\left\{\delta_{hj}\frac{y_h(s)}{y(s)}-\frac{y_h(s)y_j(s)}{y(s)^2}\right\}y(s)\alpha(s)ds.$$
    In the two-sample case, $Z_1^2/\widehat\sigma_{11}^2$ converges to a $\chi_1^2$ distribution.
    \end{theorem}
\begin{proof}
Under $H_0$, write $dN_h(t)=Y_h(t)\alpha(t)dt+dM_h(t)$ and $Y(t)=\sum_lY_l(t)$. The pooled Nelson-Aalen increment satisfies
\[
d\widehat A(t)=\frac{J(t)}{Y(t)}\sum_l dN_l(t).
\]
Hence
\[
d\{\widehat A_h(t)-\widetilde A_h(t)\}
=\frac{J_h(t)}{Y_h(t)}dN_h(t)-\frac{J_h(t)}{Y(t)}\sum_l dN_l(t).
\]
Multiplying by $K_h(t)=Y_h(t)K(t)$ gives
\[
dZ_h(t)=K(t)J_h(t)\left\{dN_h(t)-\frac{Y_h(t)}{Y(t)}\sum_l dN_l(t)\right\}.
\]
The predictable drift cancels under $H_0$:
\[
Y_h\alpha dt-\frac{Y_h}{Y}\sum_lY_l\alpha dt=0.
\]
Thus $Z_h$ is a martingale transform of the vector $(M_1,\cdots,M_k)$:
\[
Z_h(t)=\sum_l\int_0^tK(s)J_h(s)\left\{\delta_{hl}-\frac{Y_h(s)}{Y(s)}\right\}dM_l(s).
\]
The predictable covariation of $n^{-1/2}Z_h$ and $n^{-1/2}Z_j$ is therefore
\[
\frac{1}{n}\int_0^tK^2(s)
\left\{\delta_{hj}\frac{Y_h(s)}{Y(s)}-\frac{Y_h(s)Y_j(s)}{Y(s)^2}\right\}Y(s)\alpha(s)ds,
\]
which converges uniformly in probability to $\sigma_{hj}(t)$ by the assumed risk-set convergence. Boundedness of $K$ and $Y(t)$ of order $n$ imply that the jumps of $n^{-1/2}Z_h$ vanish uniformly in probability. Rebolledo's martingale central limit theorem \citep{rebolledo1980central} gives the Gaussian martingale limit. Replacing the limiting variance by its empirical quadratic-variation estimator is justified by the same convergence argument, and the continuous mapping theorem gives the $\chi_1^2$ limit for the two-sample statistic.
\end{proof}
 
\subsection{Example: Multistate Risk-Set Contrasts}
 
\subsubsection{Testing Non-Differential Mortality}
    Let \(0=\) event-free after transplant, \(1=\) relapsed but alive, and \(2=\) dead, with transition intensities \(\alpha_{01},\alpha_{02}\), and \(\alpha_{12}\), respectively. Then by Kolmogorov's equation, the transition probabilities are
    \begin{align*}
        P_{00}(s,t)&=\exp\left(-\int_s^t(\alpha_{01}(u)+\alpha_{02}(u))du\right),\\
        P_{11}(s,t)&=\exp\left(-\int_s^t\alpha_{12}(u)du\right),\\
        P_{01}(s,t)&=\int_s^tP_{00}(s,u)\alpha_{01}(u)P_{11}(u,t)du.
    \end{align*}
    We are interested in testing \emph{non-differential mortality}, i.e., \(\alpha_{02}(t)=\alpha_{12}(t)\) for all \(t\), based on \(n\) independent copies, all in state \(0\) at \(t=0\), of the Markov process over \([0,\tau]\). The null permits relapse to change the path a patient has taken, but requires the instantaneous death hazard to be the same after conditioning on the current risk set.
    Under \(H_0\), the transition probability into the relapsed-but-alive state satisfies
    \[
        P_{01}(0,t)=\exp\left(-\int_0^t\alpha_{02}(u)du\right)\left(1-\exp\left(-\int_0^t\alpha_{01}(u)du\right)\right).
    \]
    As a finite-sample calculation, consider a transplant-registry summary with 44,561 patient-months at risk in state \(0\), event-free, with 451 deaths, and 5,024 patient-months at risk in state \(1\), relapsed but alive, with 267 deaths. With \(K(s)=Y_0(s)+Y_1(s)\), the Wilcoxon-type test statistics are 28.5 and 25.6 in two clinical strata. With \(K(s)=\mathbb{I}\{Y_0(s)+Y_1(s)>0\}\), the log-rank test statistics are 29.4 and 27.3. Relative to a standard normal reference distribution, both versions would give strong evidence against non-differential mortality: the death hazard after relapse is not the same as the death hazard before relapse after conditioning on the current risk set.

\begin{figure}[tbp]
\centering
\includegraphics[width=0.98\textwidth,height=0.52\textheight,keepaspectratio]{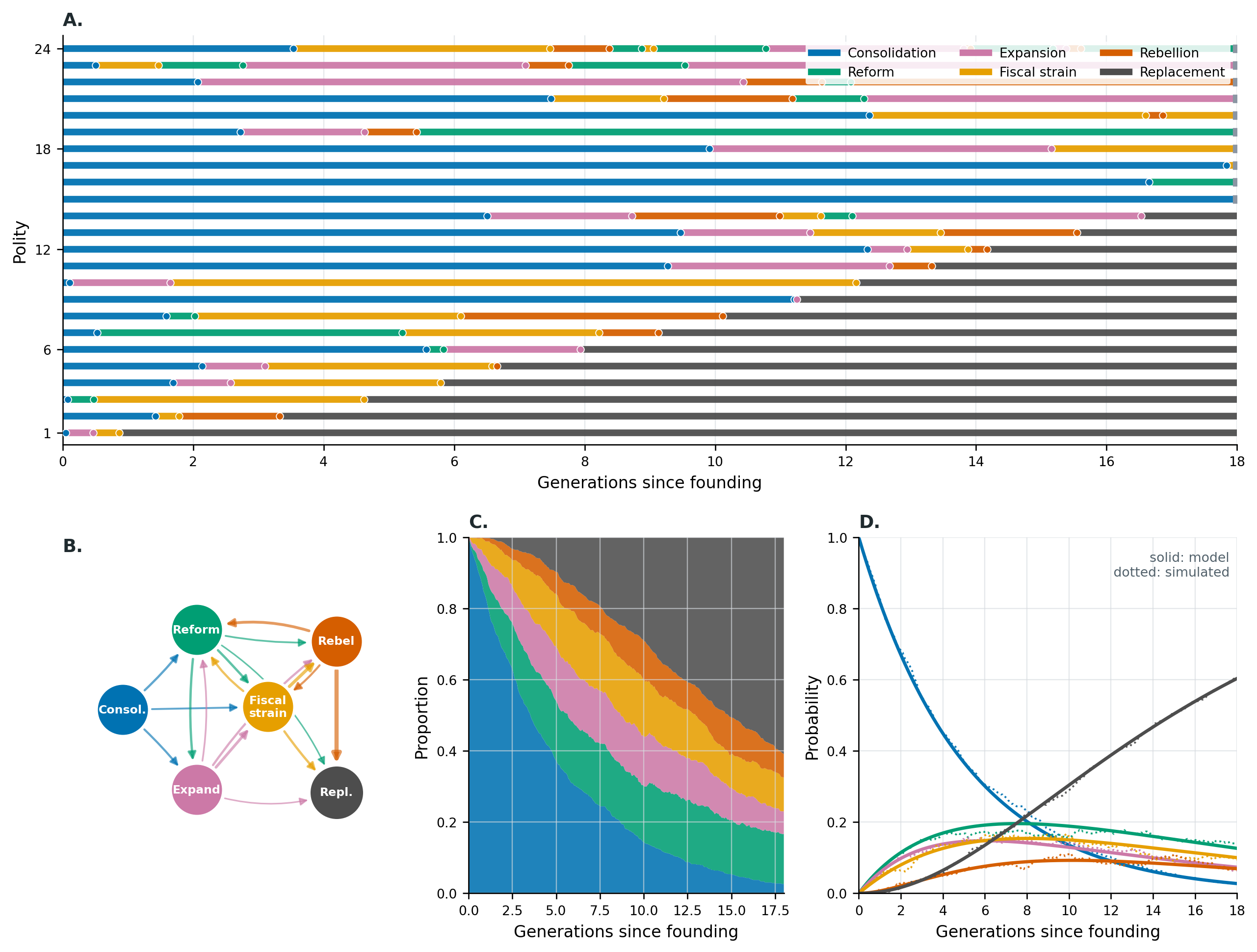}
\caption[Stylized six-state historical Markov simulation]{Stylized six-state historical Markov simulation. The dynastic-cycle example is a toy event-history model, not an empirical reconstruction; it is designed to show how transition-specific risk sets and product-integral state probabilities behave when paths can branch, recover, deteriorate, or enter an absorbing replacement state.}
\label{fig:multistate_simulation}
\end{figure}

Figure~\ref{fig:multistate_simulation} gives a richer historical analogy for the same Markov machinery. The simulated process is a stylized dynastic cycle, not a reconstruction of any particular historical data set: paths begin in consolidation and may move through reform, expansion, fiscal strain, rebellion, and eventual replacement. The panels show individual paths, the allowed transition graph, empirical state occupation, and the product-integral transition-probability target. In a large sample, the observed occupation proportions track the model probabilities. This illustrates why the Aalen-Johansen estimator is the multi-state analogue of Kaplan-Meier: it multiplies empirical transition matrices along the event-time axis, even when the state graph has several reversible and absorbing routes.

\Needspace{16\baselineskip}
\subsection{Parametric, Semiparametric, and Adaptive Tests}

The log-rank statistic is the canonical two-sample test because it targets a
well-defined conditional null: given the risk set, the predictable group
allocation of failures is determined by the common hazard.  Many tests in this
section are martingale transforms of the resulting allocation error.  They
differ in the structure imposed on the hazard, the weighting of event times,
and the class of departures emphasized by the statistic.  The testing
procedures therefore form a spectrum from fully specified parametric
likelihoods to semiparametric score tests and adaptive combinations of weighted
risk-set contrasts.

\subsubsection{Parametric Likelihood Tests}

Suppose the intensity of subject \(i\) is modeled as
\[
    \lambda_i(t;\theta)=Y_i(t)\alpha(t\mid Z_i;\theta),
    \qquad \theta\in\Theta\subset\mathbb{R}^p .
\]
For right-censored data the counting-process log likelihood, up to terms not
depending on \(\theta\), is
\[
    \ell(\theta)=
    \sum_{i=1}^n\int_0^\tau \log\{\alpha(t\mid Z_i;\theta)\}\,dN_i(t)
    -\sum_{i=1}^n\int_0^\tau Y_i(t)\alpha(t\mid Z_i;\theta)\,dt .
\]
This single expression generates the likelihood ratio, Wald, and score tests.
For a null hypothesis \(H_0:R\theta=r\), the likelihood ratio statistic is
\[
    T_{\mathrm{LR}}=
    2\{\ell(\widehat\theta)-\ell(\widehat\theta_0)\}
    \;\dot\sim\; \chi^2_{\operatorname{rank}(R)},
\]
where \(\widehat\theta\) is the unrestricted maximum likelihood estimator and
\(\widehat\theta_0\) is the constrained estimator.  The Wald statistic replaces
the likelihood comparison by the quadratic form
\[
    T_{\mathrm{W}}=
    (R\widehat\theta-r)^T
    \{R\widehat{\mathcal I}(\widehat\theta)^{-1}R^T\}^{-1}
    (R\widehat\theta-r),
\]
and the score statistic evaluates the derivative of \(\ell\) at the null
fit.  Parametric tests are powerful when the chosen hazard family is close to
the data-generating mechanism, but they can be sharply misleading when the
shape restriction is wrong \citep{kalbfleisch2002statistical,lawless2002statistical}.

Two parametric submodels illustrate the role of clock specification.  In a Weibull proportional-hazards
model,
\[
    \alpha(t\mid Z)=\rho \eta t^{\rho-1}\exp(\beta^T Z),
\]
the null \(H_0:\rho=1\) tests the exponential submodel, i.e. a constant
baseline hazard.  In a Gompertz model,
\[
    \alpha(t\mid Z)=\eta\exp(\kappa t+\beta^T Z),
\]
the null \(H_0:\kappa=0\) again tests whether the baseline hazard is flat.
These tests assess the time scale as well as a coefficient restriction.  A
rejected exponential clock indicates that the instantaneous risk changes with
age, follow-up time, exposure duration, or device wear even after the recorded
covariates have been accounted for.

\subsubsection{Cox Score Tests as Semiparametric Log-Rank Tests}

The Cox model keeps the relative-risk part parametric while leaving the
baseline hazard unspecified:
\[
    \lambda_i(t)=Y_i(t)\alpha_0(t)\exp(\beta^T Z_i).
\]
The partial-likelihood score at a candidate value \(\beta\) is
\[
    U(\beta)=
    \sum_{i=1}^n\int_0^\tau
    \{Z_i-\overline Z(t;\beta)\}\,dN_i(t),
    \qquad
    \overline Z(t;\beta)=
    \frac{\sum_jY_j(t)Z_j\exp(\beta^T Z_j)}
         {\sum_jY_j(t)\exp(\beta^T Z_j)} .
\]
Testing \(H_0:\beta=0\) with
\[
    T_{\mathrm{score}}=U(0)^T\widehat I(0)^{-1}U(0)
\]
is a semiparametric score test \citep{cox1972regression,andersen1982cox}.
When \(Z_i\) is a two-sample group indicator, this score test is the log-rank
test.  When \(Z_i\) is an ordered dose, a biomarker, or a scalar device stress
level, the same formula becomes a trend test across risk sets.  The baseline
hazard never has to be estimated in order to test the regression effect,
because it cancels inside the event-time conditional likelihood.

The same construction gives tests for time-varying effects.  Let
\[
    \lambda_i(t)=Y_i(t)\alpha_0(t)
    \exp\{\beta^TZ_i+\eta^T Z_i g(t)\},
\]
where \(g(t)\) is a chosen function such as \(t\), \(\log(t)\), or a step
function defining early and late follow-up.  The null \(H_0:\eta=0\) is a test
of proportional hazards.  Its score has the same risk-set-centering form,
\[
    U_\eta(\widehat\beta,0)=
    \sum_i\int_0^\tau
    g(t)\{Z_i-\overline Z(t;\widehat\beta)\}\,dN_i(t).
\]
Thus the test asks whether the covariate values attached to observed failures
drift systematically over event time after the fitted proportional-hazards
effect has been removed.

\subsubsection{Residual-Process and Supremum Tests}

Residual-process tests are designed for alternatives that are not captured by a single
pre-specified time interaction.  If \(t_j\) is an event time and \(i(j)\) is the
subject failing at that time, the Schoenfeld residual is
\[
    R_j=Z_{i(j)}-\overline Z(t_j;\widehat\beta).
\]
Under a correctly specified proportional-hazards model these residuals should
oscillate around zero with no systematic time pattern.  A cumulative residual
process can be written schematically as
\[
    C(t)=
    \widehat I(\widehat\beta)^{-1}
    \sum_{t_j\le t}R_j .
\]
Large values of
\[
    T_\infty=\sup_{0\le t\le\tau}\|C(t)\|,
    \qquad
    T_2=\int_0^\tau C(t)^T\,d\widehat V(t)^{-1}\,C(t),
\]
correspond to supremum and Cramer--von Mises style departures from the fitted
model \citep{lin1993checking,therneau2000modeling}.  The first statistic is
sensitive to a localized failure of the model, while the second accumulates
moderate departures over a longer time window.  The reference
distribution is often obtained by a martingale-resampling or multiplier
scheme, replacing the empirical residual increments by randomly signed or
Gaussian-weighted increments and recomputing the process many times.

\subsubsection{Crossing-Hazard, MaxCombo, and Restricted-Window Tests}

The ordinary log-rank test is most powerful against alternatives with roughly
proportional hazards.  If survival curves cross, or if treatment effects are
delayed, diluted, or concentrated early, a single weight can miss the effect.
Weighted log-rank tests use
\[
    K_{\rho,\gamma}(t)=
    \widehat S(t-)^\rho\{1-\widehat S(t-)\}^{\gamma}
\]
inside the same risk-set contrast.  The pair \((\rho,\gamma)=(1,0)\) emphasizes
early failures, \((0,1)\) emphasizes late failures, and \((0,0)\) recovers the
ordinary log-rank test \citep{fleming2011counting}.  A MaxCombo statistic takes
several standardized weighted log-rank statistics and rejects for
\[
    T_{\max}=\max_{(\rho,\gamma)\in\mathcal W} Z_{\rho,\gamma},
    \qquad
    \mathcal W=\{(0,0),(1,0),(0,1),(1,1)\}.
\]
Because the components are correlated, the null distribution is not a plain
maximum of independent standard normals; it is evaluated using the estimated
covariance matrix of the weighted score vector.  The gain is robustness: the
test can notice early, late, and more diffuse alternatives without committing
to one pattern in advance.

Another alternative is to test a clinically or scientifically interpretable
time window.  The restricted mean survival time is
\[
    \mu(\tau)=\int_0^\tau S(t)\,dt ,
\]
and a two-sample test can be built from
\[
    \widehat\Delta_\mu(\tau)=
    \int_0^\tau\{\widehat S_1(t)-\widehat S_0(t)\}\,dt ,
\]
with variance obtained from the Kaplan--Meier influence function.  A related
restricted-window log-rank test uses \(K(t)=\mathbb{I}\{a<t\le b\}\).  These
tests are attractive when only a finite horizon is meaningful: ninety-day
mortality, one-year device reliability, two-year remission, or tenure through a
fixed administrative period.  They trade some omnibus power for an estimand
that can be read directly on the time scale.

\subsubsection{Competing-Risk and Recurrent-Event Tests}

Event-history data often contain more than one kind of jump.  In a competing
risk setting, a cause-specific Cox score test asks whether a covariate changes
the instantaneous hazard of cause \(k\) among subjects still event-free.  Gray's
test instead compares cumulative incidence functions, keeping subjects who
failed from other causes in the weighted risk construction appropriate for the
subdistribution target \citep{gray1988class}.  Fine--Gray regression turns this
idea into a semiparametric score test for covariate effects on the
subdistribution hazard \citep{fine1999proportional}.  The distinction matters:
a covariate may raise the cause-specific relapse hazard but have little effect
on relapse incidence if it also raises the death hazard.

For recurrent events, the Andersen--Gill model yields a robust score test for
rate effects when each subject can jump multiple times.  Conditional-risk-set
tests such as Prentice--Williams--Peterson reset or stratify by event number,
while marginal models compare mean cumulative functions across groups
\citep{prentice1981regression,wei1989regression,lin2000recurrent}.  The null is
therefore not only ``no difference.''  It must specify whether no difference
means equal first-event hazards, equal gap-time hazards after the previous
event, equal total event rates, or equal expected cumulative burden.

\begin{table}[tbp]
\centering
\caption[Model-based and adaptive tests for event-history data]{A compact menu of parametric, semiparametric, and adaptive tests for event-history data.}
\label{tab:model_based_adaptive_tests}
\begingroup
\footnotesize
\setlength{\tabcolsep}{3.5pt}
\renewcommand{\arraystretch}{1.16}
\begin{tabular}{>{\RaggedRight\arraybackslash}p{0.18\textwidth}
                >{\RaggedRight\arraybackslash}p{0.24\textwidth}
                >{\RaggedRight\arraybackslash}p{0.26\textwidth}
                >{\RaggedRight\arraybackslash}p{0.24\textwidth}}
\toprule
\textbf{Test family} & \textbf{Null target} & \textbf{Main use case} & \textbf{Main vulnerability} \\
\midrule
Parametric LR/Wald/score & Fully specified hazard family and coefficient restriction. & The scientific clock has a plausible shape, such as exponential, Weibull, or Gompertz. & Shape misspecification can dominate the nominal test size or power.\\
Cox score/log-rank & No covariate effect under an unspecified baseline hazard. & Risk-set comparisons are trusted, but the baseline hazard is nuisance. & Low power under non-proportional or crossing hazards.\\
Time-varying-effect score & A proportional-hazards coefficient does not drift with \(g(t)\). & There is a pre-specified early, late, linear, or log-time departure. & The conclusion depends on the chosen time basis.\\
Residual-process tests & No systematic pattern remains in Schoenfeld or martingale residuals. & The alternative is exploratory or visually process-like. & Requires resampling or process approximations for calibration.\\
MaxCombo weighted tests & No group effect across several event-time weighting schemes. & Effects may be early, late, delayed, or crossing. & Multiplicity and correlation must be handled in the reference distribution.\\
RMST/window tests & Equal survival area or equal hazards over a restricted time interval. & A finite horizon is scientifically primary. & Power outside the chosen window is intentionally discarded.\\
Competing-risk tests & Equal cause-specific hazards or equal cumulative incidence, depending on target. & Multiple terminal event types change interpretation. & Cause-specific and subdistribution nulls identify different estimands.\\
Recurrent-event tests & Equal rate, gap-time hazard, event-number-specific hazard, or mean burden. & Subjects can experience repeated events. & The risk set changes with the recurrence scale chosen.\\
\bottomrule
\end{tabular}
\endgroup
\end{table}

\section{From Curves to Covariates: Regression Models}
\subsection{Multiplicative Hazards: Cox--Andersen--Gill Models}

\subsubsection{Andersen--Gill Model}
    Assume the compensator of $\mathbf{N}=(N_{hi},h=1,\cdots,k;i=1,\cdots,n)$ is
    $$\Lambda_{hi}^\theta(t)=\int_0^t\lambda_{hi}^\theta(u)du$$
    where $\theta^T=(\gamma,\beta^T)$ and
    $$\lambda_{hi}^\theta(t)=Y_{hi}(t)\alpha_{h0}(t,\gamma)r(\mathbf{\beta}^TZ_{hi}(t))$$
    and $Y_{hi}(t)$ is predictable (risk process), $\alpha_{h0}(t,\gamma)$ is the baseline hazard, $Z_{hi}$ is a type-specific covariate vector (also predictable) and $r(\cdot)$ is called the \emph{relative risk function}. The most popular choice corresponds to $r(x)=\exp(x)$, which is known as the \emph{Cox model}. We assume that $\alpha_{h0}(\cdot)$ are non-negative and integrable over $t\in\mathcal{T}$.

\begin{figure}[tbp]
\centering
\includegraphics[width=0.84\textwidth,height=0.40\textheight,keepaspectratio]{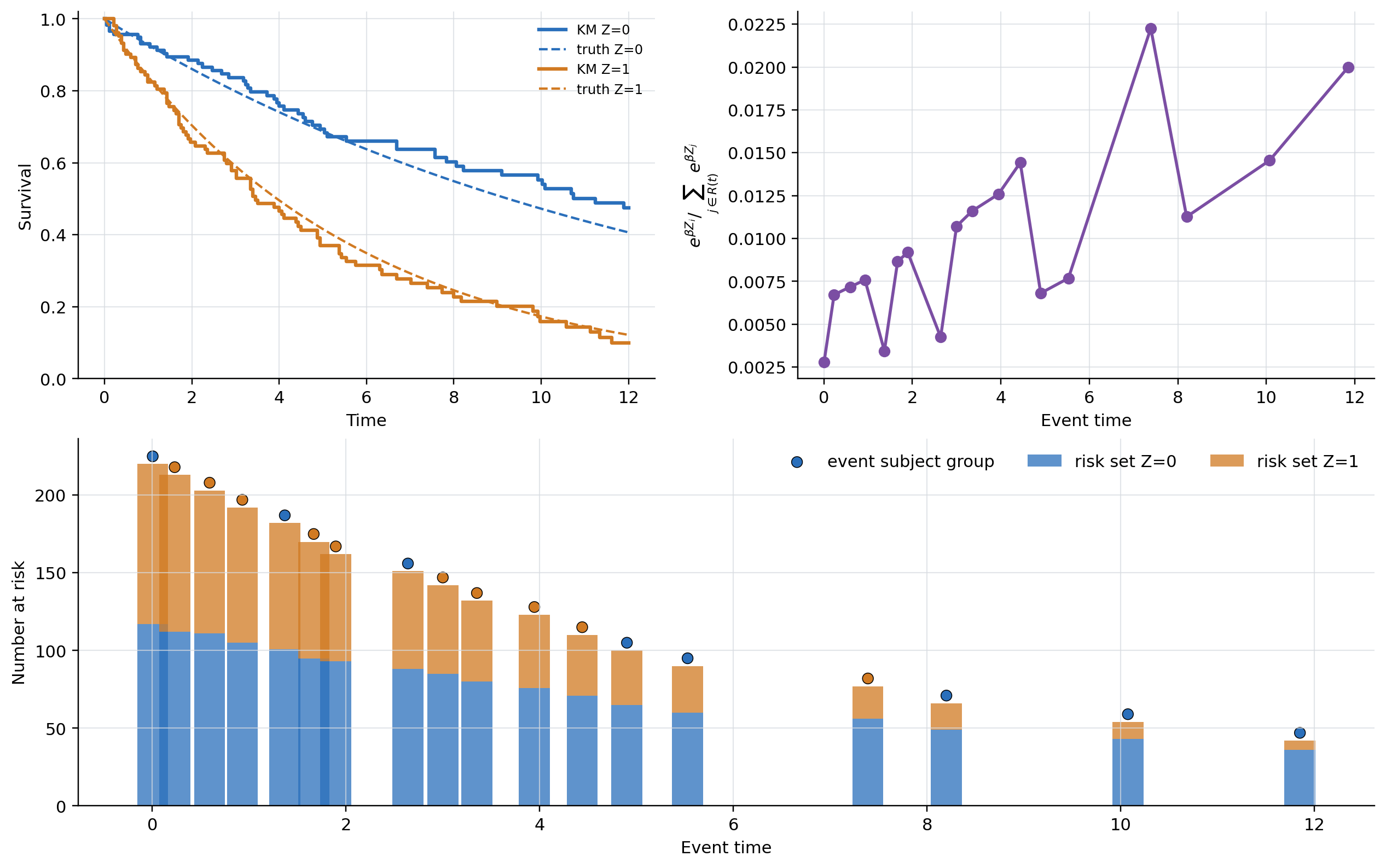}
\caption[Illustrative simulation for the Cox model]{Illustrative simulation for the Cox model.}
\label{fig:cox_riskset_simulation}
\end{figure}

The Cox partial likelihood is based on risk-set comparisons. At each observed failure time, the subject who failed is compared only with subjects who were still at risk just before that time. The baseline hazard controls how many failures occur over calendar time, but it cancels from the conditional probability of the failing subject within the current risk set. Figure~\ref{fig:cox_riskset_simulation} shows this mechanism in a simulated two-group study. The survival curves show the marginal consequence of the hazard ratio, while the stacked bars show the changing composition of the risk set at selected event times.

In a clinical Cox model, the event could be relapse and the risk set at time \(t\) would contain patients who are relapse-free, still followed, and not administratively censored just before \(t\). Covariates might include treatment, baseline severity, age, biomarker status, center, and time-dependent infection or transplant indicators. A coefficient \(\beta_{\text{treat}}<0\) corresponds to a lower instantaneous relapse hazard for treated patients among patients still at risk at the same follow-up time. In a reliability Cox model, the event could be direct failure, with covariates such as component batch, operating temperature, load, maintenance history, and time-dependent degradation status. The Cox model compares only units that are contemporaneously at risk; it does not compare a patient who has already relapsed with one who has not yet entered follow-up, or a failed device with a device not yet installed.

\subsubsection{Estimation and Partial Likelihood}
The Cox partial likelihood estimates \(\beta\) without specifying the baseline cumulative hazards.  For a fixed value of \(\beta\), define
\[
    S_h^{(0)}(\beta,t)
    =\sum_{i=1}^n r\{\beta^TZ_{hi}(t)\}Y_{hi}(t),
    \qquad
    A_{h0}(t)=\int_0^t\alpha_{h0}(u)\,du .
\]
Jacod's likelihood formula for multivariate counting processes gives, up to factors not depending on \(\beta\) and \(A_{h0}\),
\[
    L(\beta,A_0)
    \propto
    \prod_{t\in\mathcal{T}}\prod_{h=1}^k\prod_{i=1}^n
    \{dA_{h0}(t)r(\beta^TZ_{hi}(t))\}^{\Delta N_{hi}(t)}
    \exp\!\left\{-\sum_{h=1}^k\int_0^\tau
    S_h^{(0)}(\beta,u)\,dA_{h0}(u)\right\}.
\]
At an observed transition time \(t\), maximizing this expression over the jump
\(\Delta A_{h0}(t)\), with \(\beta\) fixed, yields
\[
    \Delta\widehat A_{h0}(t,\beta)
    =\frac{\Delta N_h(t)}{S_h^{(0)}(\beta,t)},
    \qquad
    \Delta N_h(t)=\sum_{i=1}^n \Delta N_{hi}(t).
\]
Substitution gives the Cox partial likelihood
\[
    L_p(\beta)=
    \prod_{t\in\mathcal{T}}\prod_{h=1}^k\prod_{i=1}^n
    \left\{
    \frac{r(\beta^TZ_{hi}(t))}{S_h^{(0)}(\beta,t)}
    \right\}^{\Delta N_{hi}(t)} ,
\]
and \(\widehat\beta\) maximizes the log partial likelihood
\[
    \ell_\tau(\beta)
    =\sum_{h=1}^k\sum_{i=1}^n
    \int_0^\tau
    \left[
    \log r\{\beta^TZ_{hi}(t)\}
    -\log S_h^{(0)}(\beta,t)
    \right]\,dN_{hi}(t).
\]
The corresponding baseline cumulative-hazard estimator is the Breslow estimator
\[
    \widehat A_{h0}(t,\widehat\beta)
    =\int_0^t\frac{dN_h(u)}{S_h^{(0)}(\widehat\beta,u)}
\]
\citep{breslow1974covariance}.  If a smooth hazard curve is desired, one may apply a kernel smoother to this cumulative-hazard estimator:
\[
    \widehat{\alpha}_{h0}(t)=
    \frac{1}{b_n}\int_0^\tau
    K_h\left(\frac{t-u}{b_n}\right)
    d\widehat{A}_{h0}(u,\widehat{\beta}).
\]

\subsubsection{Large Sample Properties}
    We assume $r(\cdot)=\exp(\cdot)$. Define the following quantities:
    \[
    \begin{aligned}
        S_h^{(0)}(\beta,t)&=\sum_{i=1}^n\exp(\beta^TZ_{hi}(t))Y_{hi}(t),\\
        \mathbf{S}_h^{(1)}(\beta,t)&=\sum_{i=1}^nZ_{hi}(t)\exp(\beta^TZ_{hi}(t))Y_{hi}(t),\\
        \mathbf{S}^{(2)}_h(\beta,t)&=\sum_{i=1}^nZ_{hi}^{\otimes 2}(t)\exp(\beta^TZ_{hi}(t))Y_{hi}(t),\\
        \mathbf{E}_h(\beta,t)&=\frac{\mathbf{S}_h^{(1)}(\beta,t)}{S_h^{(0)}(\beta,t)},\\
        \mathbf{V}_h(\beta,t)&=\frac{\mathbf{S}_h^{(2)}(\beta,t)}{S_h^{(0)}(\beta,t)}-\mathbf{E}_h(\beta,t)^{\otimes 2},
    \end{aligned}
    \]
    where $h=1,\cdots,k$ are states (components of the multivariate counting process). Further, we define the score process ($N_h=\sum_{i=1}^nN_{hi}$)
    $$\mathbf{U}_\tau(\beta)=\sum_{h=1}^k\left[\sum_{i=1}^n\int_0^\tau Z_{hi}(t)dN_{hi}(t)-\int_0^\tau \mathbf{E}_{h}(\beta,t)dN_h(t)\right].$$
The superscript notation has a simple risk-set meaning. \(S_h^{(0)}\) is the total relative-risk weight in the risk set, \(\mathbf S_h^{(1)}\) is the weighted covariate total, and \(\mathbf S_h^{(2)}\) is the weighted second moment. Thus \(\mathbf E_h\) is the covariate average among subjects currently at risk after weighting by the model, and \(\mathbf V_h\) is the corresponding risk-set covariance. The score \(\mathbf U_\tau(\beta)\) compares the covariate vector of the subject who failed with this model-weighted risk-set average. Under the model, these differences have zero predictable drift and martingale variation. Figure~\ref{fig:cox_riskset_simulation} is the finite-sample companion for the three theorems below: it shows the risk-set comparisons that become the score, information, and Breslow baseline processes in the proofs.

    \begin{theorem}[Consistency of $\widehat{\beta}$]
Assume that the parameter space is compact and contains the true value \(\beta_0\) in its interior; the predictable covariates are bounded; \(n^{-1}\mathbf S_h^{(m)}(\beta,t)\) converges uniformly in \((\beta,t)\) to \(s_h^{(m)}(\beta,t)\) for \(m=0,1,2\); \(s_h^{(0)}(\beta,t)\) is bounded away from zero on the relevant risk sets; and the limiting normalized partial log likelihood has the unique maximizer \(\beta_0\). Then the estimating equation \(\mathbf{U}_\tau(\beta)=0\) has a unique solution \(\widehat{\beta}\) with probability tending to one and \(\widehat{\beta}\rightarrow_p\beta_0\) as \(n\rightarrow\infty\).
    \end{theorem}

\begin{proof}
The score at the true parameter can be written as a martingale:
\[
\mathbf U_\tau(\beta_0)
=\sum_{h=1}^k\sum_{i=1}^n\int_0^\tau
\{Z_{hi}(t)-\mathbf E_h(\beta_0,t)\}dM_{hi}(t).
\]
Thus \(n^{-1}\mathbf U_\tau(\beta_0)\to_p0\). The normalized log partial likelihood
\(n^{-1}C_\tau(\beta)\) is concave in \(\beta\). Its predictable compensator converges uniformly on compact subsets of the parameter space to a deterministic concave function
\begin{align*}
c(\beta)
&=\sum_{h=1}^k\int_0^\tau
\left[
\beta^T s_h^{(1)}(\beta_0,t)
-s_h^{(0)}(\beta_0,t)\log s_h^{(0)}(\beta,t)
\right]\alpha_{h0}(t)\,dt,
\end{align*}
up to constants not depending on \(\beta\). Identifiability is expressed by the condition that \(c(\beta)\) has a unique maximizer at \(\beta_0\), equivalently that the limiting information matrix is positive definite. Uniform convergence of concave functions then implies that any maximizer of \(C_\tau(\beta)\) converges in probability to the unique maximizer \(\beta_0\). The same argument gives existence and local uniqueness of the root of the score equation with probability tending to one.
\end{proof}
 
\begin{theorem}[Asymptotic normality of $\widehat{\beta}$]
Under the conditions of the consistency theorem, assume in addition that the limiting information matrix \(\Sigma_\tau\) is nonsingular and that the score martingale satisfies the usual Lindeberg condition, for example through bounded covariates. Then, as \(n\rightarrow\infty\),
$$\sqrt{n}(\widehat{\beta}-\beta_0)\rightarrow_d\mathcal{N}(0,\Sigma_\tau^{-1})$$
where
\begin{align*}
    \Sigma_\tau&=\sum_{h=1}^k\int_0^\tau v_h(\beta_0,t)s_h^{(0)}(\beta_0,t)\alpha_{h0}(t)dt\\
    v_h&=\frac{s_h^{(2)}}{s_h^{(0)}}-e_h^{\otimes 2},\ e_h=\frac{s_h^{(1)}}{s_h^{(0)}}
\end{align*}
and $s^{(m)}_h$ and $v_h$ are probability limits of $\mathbf{S}_h^{(m)}$ and $\mathbf{V}_h$ defined earlier.
    
\end{theorem}
\begin{proof}
By the martingale representation of the score,
\[
n^{-1/2}\mathbf U_\tau(\beta_0)
=n^{-1/2}\sum_{h,i}\int_0^\tau
\{Z_{hi}(t)-\mathbf E_h(\beta_0,t)\}dM_{hi}(t).
\]
Its predictable variation converges in probability to
\[
\Sigma_\tau
=\sum_{h=1}^k\int_0^\tau
v_h(\beta_0,t)s_h^{(0)}(\beta_0,t)\alpha_{h0}(t)dt.
\]
The jumps are of order $n^{-1/2}$ under the usual bounded-covariate or Lindeberg condition, so Rebolledo's martingale central limit theorem \citep{rebolledo1980central} yields
\[
n^{-1/2}\mathbf U_\tau(\beta_0)\Rightarrow \mathcal N(0,\Sigma_\tau).
\]
Taylor expansion of the score around $\beta_0$ gives
\[
0=\mathbf U_\tau(\widehat\beta)
=\mathbf U_\tau(\beta_0)
-n\Sigma_\tau(\widehat\beta-\beta_0)+o_p(\sqrt n),
\]
where the derivative of the normalized score converges to $-\Sigma_\tau$ uniformly in a neighborhood of $\beta_0$. Solving the display gives
\[
\sqrt n(\widehat\beta-\beta_0)
=\Sigma_\tau^{-1}n^{-1/2}\mathbf U_\tau(\beta_0)+o_p(1),
\]
and the claimed normal limit follows.
\end{proof}
    In the special case that $k=1$ and for $i=1,\cdots,n$, $X_i$ are independent survival times given $Z_i$ and $Z_i$ are i.i.d. $p$-dimensional covariates, we have
    $$\Sigma_\tau=\int_0^\tau v(\beta_0,t)s^{(0)}(\beta_0,t)\alpha_0(t)dt.$$
 
\begin{theorem}[Joint asymptotic distribution]
Under the preceding assumptions, and assuming the risk-set limits are uniformly differentiable in a neighborhood of \(\beta_0\), $\sqrt{n}(\widehat{\beta}-\beta_0)$ and the processes
$$W_h(\cdot)=\sqrt{n}\left(\widehat{A}_{h0}(\cdot,\widehat{\beta})-A_{h0}(\cdot)\right)+\sqrt{n}(\widehat{\beta}-\beta_0)^T\int_0^\cdot e_h(\beta_0,u)\alpha_{h0}(u)du$$
$h=1,\cdots,k$, are asymptotically independent. The limiting distribution of $W_h$ is that of a mean zero Gaussian martingale with variance function
$$\omega_h^2(t)=\int_0^t\frac{\alpha_{h0}(u)}{s_h^{(0)}(\beta_0,u)}du.$$
    
\end{theorem}
\begin{proof}
Start from the Breslow estimator
\[
\widehat A_{h0}(t,\widehat\beta)
=\int_0^t\frac{dN_h(u)}{S_h^{(0)}(\widehat\beta,u)}.
\]
Insert $dN_h(u)=S_h^{(0)}(\beta_0,u)dA_{h0}(u)+dM_h(u)$ and expand
$\{S_h^{(0)}(\widehat\beta,u)\}^{-1}$ around $\beta_0$. Uniformly on compact time intervals,
\[
\sqrt n\{\widehat A_{h0}(t,\widehat\beta)-A_{h0}(t)\}
=\sqrt n\int_0^t\frac{dM_h(u)}{S_h^{(0)}(\beta_0,u)}
-\sqrt n(\widehat\beta-\beta_0)^T
\int_0^t e_h(\beta_0,u)dA_{h0}(u)+o_p(1).
\]
After moving the second term to the left, we obtain the displayed process $W_h$. The first term on the right is a martingale with predictable variation converging to
\[
\omega_h^2(t)=\int_0^t\frac{\alpha_{h0}(u)}{s_h^{(0)}(\beta_0,u)}du.
\]
Its jumps vanish by the same risk-set and boundedness assumptions used for the score. Rebolledo's theorem \citep{rebolledo1980central} gives the Gaussian martingale limit. Orthogonality of the score martingale and the Breslow martingale follows from the defining property of $\mathbf E_h(\beta_0,t)$ as the risk-set average of the covariates, which makes their limiting predictable covariation equal to zero. Joint Gaussianity then implies asymptotic independence.
\end{proof}
 
\subsubsection{Goodness-of-Fit and Model Diagnostics}
This subsection draws mainly on Chapter 11 of \citet{klein2003survival}, Section 7.3 of \citet{andersen1993statistical}, and the diagnostic treatments of \citet{fleming2011counting} and \citet{therneau2000modeling}. The common device is to convert a fitted intensity model into an observed-minus-predictable comparison. Proportional-hazards checks look for cumulative hazards that remain proportional over time; calibration checks ask whether the fitted compensator predicts the observed number of jumps; residual plots ask whether the remaining martingale part has a systematic pattern.

\begin{figure}[tbp]
\centering
\includegraphics[width=\textwidth,height=0.58\textheight,keepaspectratio]{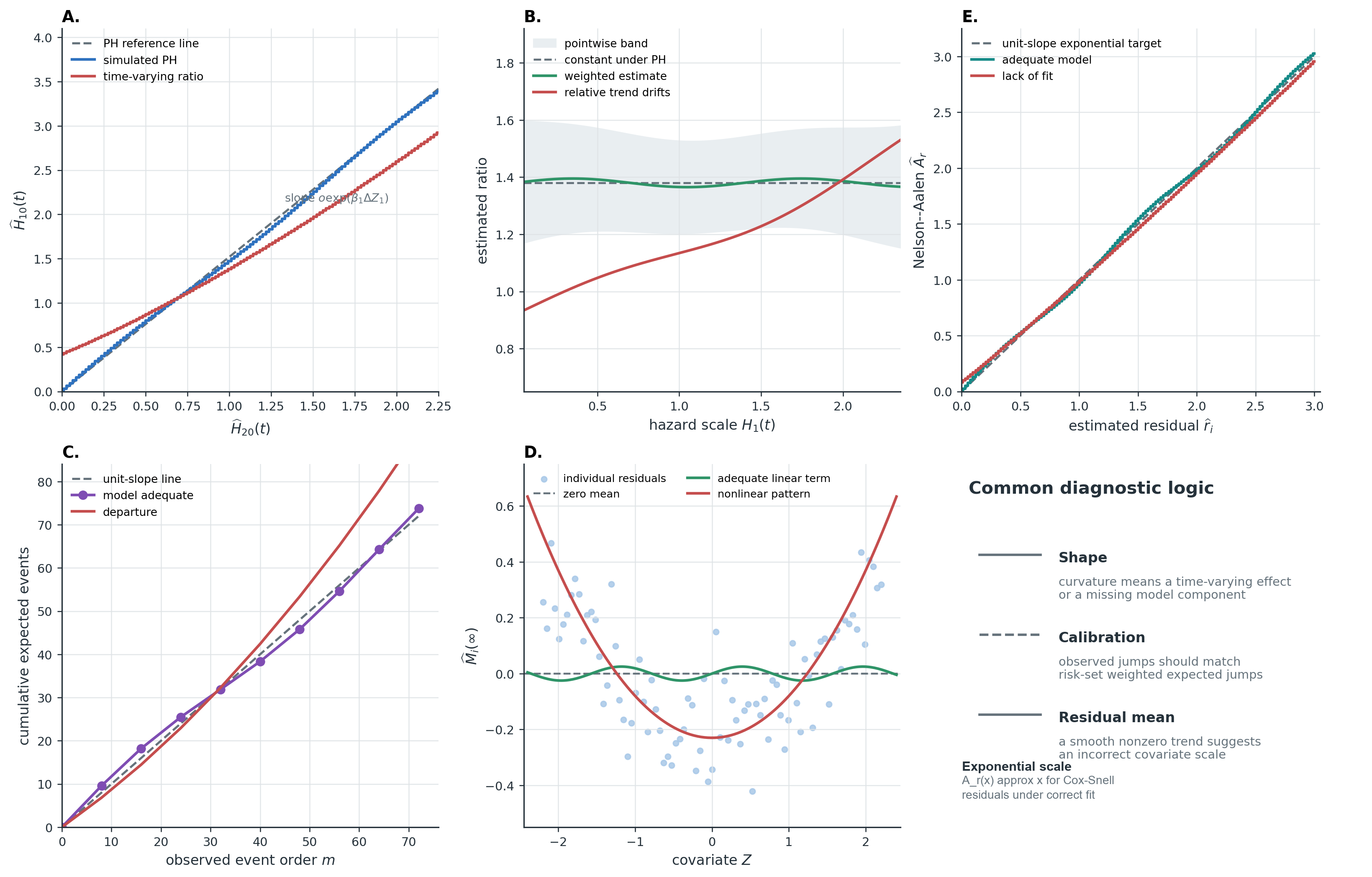}
\caption[Simulation companion for Cox model diagnostics]{Simulation companion for Cox model diagnostics. Panel A shows Andersen's cumulative-hazard plot for a two-stratum proportional-hazards check. Panel B shows the Dabrowska--Doksum--Song relative-trend idea. Panel C shows an Arjas or total-time-on-test calibration plot. Panel D shows martingale residuals against a covariate for checking functional form. Panel E shows the Cox--Snell residual plot, where a correctly specified model should produce an approximately unit-slope Nelson--Aalen curve.}
\label{fig:cox_diagnostics_simulation}
\end{figure}

Figure~\ref{fig:cox_diagnostics_simulation} is the simulation companion for the examples below. The blue, green, purple, and teal curves illustrate behavior consistent with the fitted Cox model, while the red curves show typical diagnostic departures. The point of the figure is not to provide a formal rejection rule; it records what each plot is designed to make visible before a test statistic or asymptotic approximation is used.

\begin{example}[Graphical methods: the Andersen plot]\normalfont
Let the covariate vector be \(\mathbf Z=(Z_1,\mathbf Z_2^T)^T\), where \(Z_1\) is the covariate whose proportional-hazards effect is being checked. Stratify the data by the observed values, or by a small number of categories, of \(Z_1\), fit a Cox model adjusted for \(\mathbf Z_2\) with separate baseline cumulative hazards by stratum, and write the fitted cumulative hazards as \(\widehat H_{10}\) and \(\widehat H_{20}\). If the effect of \(Z_1\) is proportional, then for two representative stratum values \(Z_{11}\) and \(Z_{12}\),
\[
    H_{10}(t)=\exp(\beta_1\Delta Z_1)H_{20}(t),
    \qquad \Delta Z_1=Z_{11}-Z_{12}.
\]
Thus a plot of \(\widehat H_{10}(t)\) against \(\widehat H_{20}(t)\) should be approximately a straight line through the origin, with slope \(\exp(\beta_1\Delta Z_1)\). Equivalently, the log-ratio \(\log\widehat H_{10}(t)-\log\widehat H_{20}(t)\) should be approximately constant over time. Panel A of Figure~\ref{fig:cox_diagnostics_simulation} shows both readings: the blue curve follows the reference line, while the red curve bends, indicating that the hazard ratio changes over follow-up.
\end{example}

\begin{example}[Dabrowska--Doksum--Song graphical test]\normalfont
\citet{dabrowska1989graphical} studied two-sample graphical procedures for comparing cumulative hazards. One summary statistic is
\[
    \widehat\theta(\tau)
    =
    \frac{\int_0^\tau L(s)\,d\widehat H_1(s)}
         {\int_0^\tau L(s)\,d\widehat H_2(s)},
\]
where \(L(s)\) is a predictable weight process. The choice \(L(s)=1\) gives an unweighted cumulative-hazard ratio, while other predictable weights emphasize selected parts of follow-up. Under a proportional-hazards relation between the two populations, the relative trend function \(H_2\circ H_1^{-1}\) is linear and the relative change function \((H_2-H_1)/H_1\) is constant. Panel B of Figure~\ref{fig:cox_diagnostics_simulation} shows the diagnostic target: a weighted estimate that stays inside a constant band is consistent with proportionality, while a smooth drift indicates a time-varying relative hazard.
\end{example}

\begin{example}[Martingale residuals: the Arjas plot]\normalfont
The Arjas plot checks whether a fitted Cox model allocates the observed event
process correctly across strata \citep{arjas1988graphical}. Define
\[
    p_i(u,\beta)
    =
    \frac{Y_i(u)\exp(\beta^T Z_i)}
         {S^{(0)}(\beta,u)},
    \qquad
    S^{(0)}(\beta,u)=\sum_{j=1}^nY_j(u)\exp(\beta^TZ_j).
\]
For a grouping variable \(h(i)\in\{1,\ldots,k\}\), let \(N_h(t)=\sum_{h(i)=h}N_i(t)\) and \(N_\cdot(t)=\sum_{i=1}^nN_i(t)\). Under the Cox model with true coefficient \(\beta\), the process
\[
    \operatorname{Arjas}(h,t)
    =
    N_h(t)-\int_0^t\sum_{h(i)=h}p_i(u,\beta)\,dN_\cdot(u)
\]
has zero predictable drift. Indeed,
\[
\begin{aligned}
\mathbf E\{dN_h(t)\mid\mathcal F_{t-}\}
&=
\alpha_0(t)\sum_{h(i)=h}Y_i(t)\exp(\beta^TZ_i)\,dt,\\
\mathbf E\{dN_\cdot(t)\mid\mathcal F_{t-}\}
&=
\alpha_0(t)S^{(0)}(\beta,t)\,dt.
\end{aligned}
\]
and therefore
\[
\begin{aligned}
\mathbf E\{d\operatorname{Arjas}(h,t)\mid\mathcal F_{t-}\}
&=
\alpha_0(t)\sum_{h(i)=h}Y_i(t)\exp(\beta^TZ_i)\,dt\\
&\quad -
\sum_{h(i)=h}\frac{Y_i(t)\exp(\beta^TZ_i)}{S^{(0)}(\beta,t)}
        \alpha_0(t)S^{(0)}(\beta,t)\,dt\\
&=0.
\end{aligned}
\]
In applications the unknown coefficient is replaced by the fitted coefficient \(\widehat\beta\).
If \(X_m^h\) is the \(m\)th ordered event time in stratum \(h\), the total-time-on-test version plots
\[
    \int_0^{X_m^h}\sum_{h(i)=h}p_i(u,\widehat\beta)\,dN_\cdot(u),
    \qquad m=1,\ldots,N_h(\tau),
\]
against \(m\). Panel C of Figure~\ref{fig:cox_diagnostics_simulation} shows the interpretation: a unit-slope line means the model's cumulative expected events agree with the observed event order in the stratum; curvature means systematic under- or over-prediction.
\end{example}

\begin{example}[Martingale residuals: functional form]\normalfont
For right-censored data fitted by a Cox model, define the estimated martingale residual
\[
    \widehat M_i(t)
    =
    N_i(t)-\int_0^tY_i(u)\exp(\widehat\beta^TZ_i)\,d\widehat A_0(u),
\]
where \(\widehat\beta\) is the partial-likelihood estimator and \(\widehat A_0\) is the Breslow estimator. These residuals estimate the part of the event process that remains after subtracting the fitted compensator. If the fitted linear predictor uses \(Z\) but the true hazard depends on a transformation \(f(Z)\), then a smooth plot of \(\widehat M_i(\infty)\) against \(Z_i\) tends to reveal the missing shape \citep[pp.~165--168]{fleming2011counting}. Panel D of Figure~\ref{fig:cox_diagnostics_simulation} displays the diagnostic criterion: a conditional mean near zero supports the fitted functional form, while a curved trend suggests using a transformation, spline, or categorical representation of the covariate.
\end{example}

\begin{example}[The Cox--Snell plot]\normalfont
Let \((T_i,D_i,Z_i)\), \(i=1,\ldots,n\), be right-censored observations and define the fitted Cox--Snell residual \citep{cox1968general}
\[
    \widehat r_i
    =
    \widehat A(T_i\mid Z_i)
    =
    \widehat A_0(T_i)\exp(\widehat\beta^TZ_i).
\]
The Cox--Snell diagnostic applies the Nelson--Aalen estimator to the pseudo-observations \((\widehat r_i,D_i)\). Under a correctly specified model, ignoring estimation error for the moment,
\begin{equation*}
    r_i=A(T_i\mid Z_i)=-\log S(T_i\mid Z_i)\sim \mathcal E(1),
\end{equation*}
because
\begin{equation*}
    \mathbb P\{A(T_i\mid Z_i)>x\mid Z_i\}
    =
    \mathbb P\{T_i>A^{-1}(x\mid Z_i)\mid Z_i\}
    =
    e^{-x}.
\end{equation*}
Hence the Nelson--Aalen curve computed from the residual times should be close to the unit-slope line through the origin. Panel E of Figure~\ref{fig:cox_diagnostics_simulation} illustrates the target curve and a typical lack-of-fit pattern.
\end{example}

\subsection{Additive Hazards: Aalen's Linear Model}

\subsubsection{Generalized Nelson--Aalen Estimator}
The additive hazard formulation follows Aalen's linear model for lifetimes \citep{aalen1989linear}.
    \begin{definition}[Aalen's additive hazard model]
        Let $\mathbf{N}(t)=(N_i(t):i=1,\cdots,n)$ be a multivariate counting process with intensity $\lambda_i(t)=\alpha_{i}(t;Z_i(t))Y_i(t)$. The \emph{additive hazard model} assumes
        $$\alpha_{i}(t;Z_i(t))=\beta_0+\beta_1(t)Z_{i1}(t)+\cdots+\beta_p(t)Z_{ip}$$
        where $\beta_j(t)$ are all locally integrable for $t\in\mathcal{T}$. More compactly, we have
        $$\mathbf{N}(t)=\int_0^t\mathbf{Y}(s)d\mathbf{B}(s)+\mathbf{M}(t)$$
        where $B_j(s)=\int_0^s\beta_{j}(s)ds, j=0,\cdots,p$ are the parameters of primary interest and $\mathbf{Y}(s)$ is an $n\times(p+1)$ matrix with rows $$Y_i(t)(1,Z_{i1}(t),Z_{i2}(t),\cdots,Z_{ip}(t))$$
        and $\mathbf{M}(t)$ is a vector of counting process martingales.
    \end{definition}

    The additive model changes the scientific scale. Cox regression represents covariates through multiplicative hazard ratios; Aalen's model represents covariates through additive hazard increments. The matrix \(\mathbf{Y}(s)\) is therefore a time-specific design matrix for the subjects still at risk, and \(d\mathbf B(s)\) is the vector of regression increments that best explains the observed jumps at time \(s\). The simulation companion is Figure~\ref{fig:iv_additive_simulation}, where the same cumulative coefficient \(B_X(t)\) becomes the causal exposure-effect curve in the IV setting.

    In a clinical additive-hazard model, an additive coefficient is an absolute hazard contribution. If \(Z_1(t)\) indicates treatment exposure, then \(dB_1(t)\) is the extra relapse-hazard increment attached to treated subjects at time \(t\), measured on the same scale as the baseline hazard. If \(Z_2(t)\) marks an acute infection, \(dB_2(t)\) is the extra short-term hazard contribution among subjects still under observation at that time. This scale targets absolute event-rate differences rather than hazard ratios. Heuristically, we have
    $$d\mathbf{N}(t)=\mathbf{Y}(t)d\mathbf{B}(t)+\mathbf{M}(dt)$$
    and this suggests the \emph{GNA} estimator:
    $$\widehat{\mathbf{B}}(t)=\int_0^tJ(s)\mathbf{Y}(s)^-d\mathbf{N}(s)$$
    where $J(s)=\mathbb{I}\{\operatorname{rank}(\mathbf{Y}(s))=p+1\}$ and $\mathbf{Y}(s)^-$ is any predictable generalized inverse of $\mathbf{Y}(s)$. The quadratic-variation estimator is
    $$\widehat{\boldsymbol{\Sigma}}(t)=\int_0^tJ(s)\mathbf{Y}(s)^-\left(\operatorname{diag}\{d\mathbf{N}(s)\}\right)(\mathbf{Y}(s)^-)^T,$$
    which estimates $[\widehat{\mathbf{B}}-\mathbf{B}^*](t)$, where $[\cdot,\cdot]$ is quadratic variation and $\mathbf{B}^*(t)=\int_0^tJ(u)d\mathbf{B}(u)$.
 
    In the special case that $\alpha_{i}(t;Z_i(t))=\beta_0$, we have
    $$\widehat{B}_0(t)=\int_0^tJ(s)\frac{\sum_{i=1}^nY_i(s)dN_i(s)}{\sum_{i=1}^nY_i(s)^2},$$
    which reduces to the classical NA estimator if $Y_i(s)$ are indicators. Further, we apply smoothed NA estimator technique to estimate $\beta_j(t)$ and $\alpha_{i}$:
    \begin{align*}
        &\widehat{\beta}_j=\frac{1}{b_j}\int_\mathcal{T}K\left(\frac{t-s}{b_j}\right)d\widehat{B}_j(s),\\
        &\widehat{\alpha}_i(t,Z_i(t))=\widehat{\beta}_0+\sum_{j=1}^p\widehat{\beta}_j(t)Z_{ij}(t).
    \end{align*}
    For large-sample properties, see \citet[pp.~575--578]{andersen1993statistical}. The additive hazards model reappears below as the baseline for the instrumental-variable construction.

\subsection{Transformation and Accelerated Failure Time Models}

\subsubsection{Transformation Models}
The transformation-model idea begins with the probability integral transform. If \(A\) is the true cumulative hazard of a continuous failure time \(X\), then
\[
    \mathbb P\{A(X)>x\}=\mathbb P\{X>A^{-1}(x)\}=\exp(-x),
\]
so \(A(X)\sim\mathcal E(1)\). In a Cox model,
\[
    S(t|Z)=\exp\{-A_0(t)\exp(\beta^TZ)\},
\]
and therefore
\[
    \log A_0(X)=-\beta^TZ+\epsilon,\qquad
    \mathbb P(\epsilon>u)=\exp\{-e^u\}.
\]
Thus the Cox model can be represented as a linear model after an unknown monotone transformation of time. This observation links proportional hazards to general transformation models \citep{kalbfleisch2002statistical,lawless2002statistical}.

In clinical and reliability studies, transformation models describe covariates that stretch or compress the event-time scale rather than only multiply an instantaneous hazard. A treatment may extend time to relapse, frailty may shorten survival time, and a harsh operating environment may compress time to device failure. The unknown transformation \(g\) identifies the time scale on which these effects are closest to linear.

\begin{example}[Transformation models]\normalfont\leavevmode\par\noindent
Let \(g\) be an unknown increasing function. A transformation model has the form
\[
    g(X)=\beta^TZ+\epsilon,
\]
where \(X\) is a possibly left-, right-, or interval-censored time-to-event outcome, and the error distribution of \(\epsilon\) is specified. The accelerated failure time (AFT) model is the special case \(g(x)=\log x\), so the covariates act additively on log survival time rather than multiplicatively on hazard \citep{buckley1979linear,lin1998accelerated}.
\end{example}
 
\subsubsection{Buckley-James Estimator}

The Buckley--James estimator is an estimating-equation method for the censored accelerated failure-time model. For uncensored observations the observed log failure time enters directly; for censored observations it is replaced by the conditional tail mean of the latent log failure time under a residual distribution. The resulting synthetic response yields an estimating equation that reduces to the ordinary least-squares normal equation when no censoring is present \citep{buckley1979linear,lin1998accelerated,jin2003rank}.

Let the latent log failure time satisfy
\[
    V_i=\log T_i=\beta^TZ_i+\epsilon_i,
\]
and let \(U_i=\log C_i\), \(V_i^o=V_i\wedge U_i\), and \(D_i=\mathbb I(V_i\le U_i)\). For a parametric error distribution with density \(f\) and survival function \(S\), the observed likelihood contribution is
\[
    \mathcal L(\beta)=\prod_{i=1}^n
    f\{r_i(\beta)\}^{D_i}S\{r_i(\beta)\}^{1-D_i},
    \qquad r_i(\beta)=V_i^o-\beta^TZ_i .
\]
The score equation can be written as
\[
    \sum_{i=1}^n Z_i
    \left[
    D_i\left\{-\frac{f'(r_i)}{f(r_i)}\right\}
    +(1-D_i)\frac{f(r_i)}{S(r_i)}
    \right]=\mathbf 0 .
\]
If \(f\) is known, this equation can be solved numerically. If \(f\) is not specified, one replaces the parametric score \(-f'/f\) by a chosen score function \(a\) and uses the conditional tail mean
\[
    m_{a,F}(r)=\mathbb E_F\{a(\epsilon)\mid \epsilon>r\}
    =\frac{\int_r^\infty a(u)dF(u)}{S_F(r)} .
\]
The resulting estimating equation is
\[
    \Psi(\beta,a,F)=
    \sum_{i=1}^nZ_i\{D_i a(r_i(\beta))+(1-D_i)m_{a,F}(r_i(\beta))\}
    =\mathbf 0 .
\]
When \(a(u)=u\) and no censoring occurs, this reduces to the ordinary least-squares normal equation for the AFT model.

\begin{lemma}[Buckley-James response]\normalfont\label{lemma:buckley_james}
Taking \(a(u)=u\) gives the Buckley--James estimating equation \citep{buckley1979linear}. Let
\[
    m_F(r)=\mathbb E_F(\epsilon\mid \epsilon>r)
    =\frac{\int_r^\infty u\,dF(u)}{S_F(r)}
\]
and define
\[
    V_i^*(\beta,F)
    =D_iV_i^o+(1-D_i)\{\beta^TZ_i+m_F(r_i(\beta))\}.
\]
Then
\[
    \Psi(\beta,u,F)=\sum_{i=1}^nZ_i\{V_i^*(\beta,F)-\beta^TZ_i\}.
\]
If \(F\) is the true residual distribution and \(U_i\) is independent of \(V_i\) conditional on \(Z_i\), then at the true value \(\beta_0\),
\[
    \mathbb E\{V_i^*(\beta_0,F)\mid Z_i\}
    =\mathbb E(V_i\mid Z_i).
\]
\end{lemma}
\begin{proof}
The algebraic identity follows by subtracting \(\beta^TZ_i\) from the synthetic response:
\[
V_i^*(\beta,F)-\beta^TZ_i
=D_i r_i(\beta)+(1-D_i)m_F(r_i(\beta)),
\]
which is exactly the summand in \(\Psi(\beta,u,F)\).

For the conditional-mean property, set \(r_i(\beta_0)=V_i^o-\beta_0^TZ_i\). On the event \(D_i=0\), \(V_i^o=U_i\), so
\[
\beta_0^TZ_i+m_F\{r_i(\beta_0)\}
=\mathbb E(V_i\mid V_i>U_i,Z_i).
\]
Conditioning first on \((U_i,Z_i)\) and using conditional independent censoring gives
\begin{align*}
\mathbb E\{V_i^*(\beta_0,F)\mid U_i,Z_i\}
&=\mathbb E\{V_i\mathbb I(V_i\le U_i)\mid U_i,Z_i\}\\
&\quad+\mathbb P(V_i>U_i\mid U_i,Z_i)
\mathbb E(V_i\mid V_i>U_i,Z_i)\\
&=\mathbb E\{V_i\mathbb I(V_i\le U_i)+V_i\mathbb I(V_i>U_i)\mid U_i,Z_i\}\\
&=\mathbb E(V_i\mid U_i,Z_i)\\
&=\mathbb E(V_i\mid Z_i).
\end{align*}
Taking expectation over \(U_i\) conditional on \(Z_i\) proves the result.
\end{proof}

In applications \(F\) is unknown and is usually estimated from the residuals, for example by a Kaplan--Meier estimator, and the estimating equation is solved iteratively. This gives the classical Buckley--James estimator and connects naturally to later rank-based AFT procedures \citep{lin1998accelerated,jin2003rank}.
 
\subsubsection{Li--Lu PBIV and Instrumental-Variable Analysis}

Instrumental variables are used when an exposure is associated with unmeasured causes of the survival outcome. A directed acyclic graph encodes the required restrictions: \(G\) is the instrument, \(X\) is the exposure, \(Z\) records measured covariates, and \(Y\) is the transformed or latent failure-time outcome. The target is \(\beta_1\), the causal change in the outcome scale induced by changing \(X\), rather than the association between \(X\) and \(Y\).
\begin{figure}[tbp]
\centering
\includegraphics[width=0.94\textwidth]{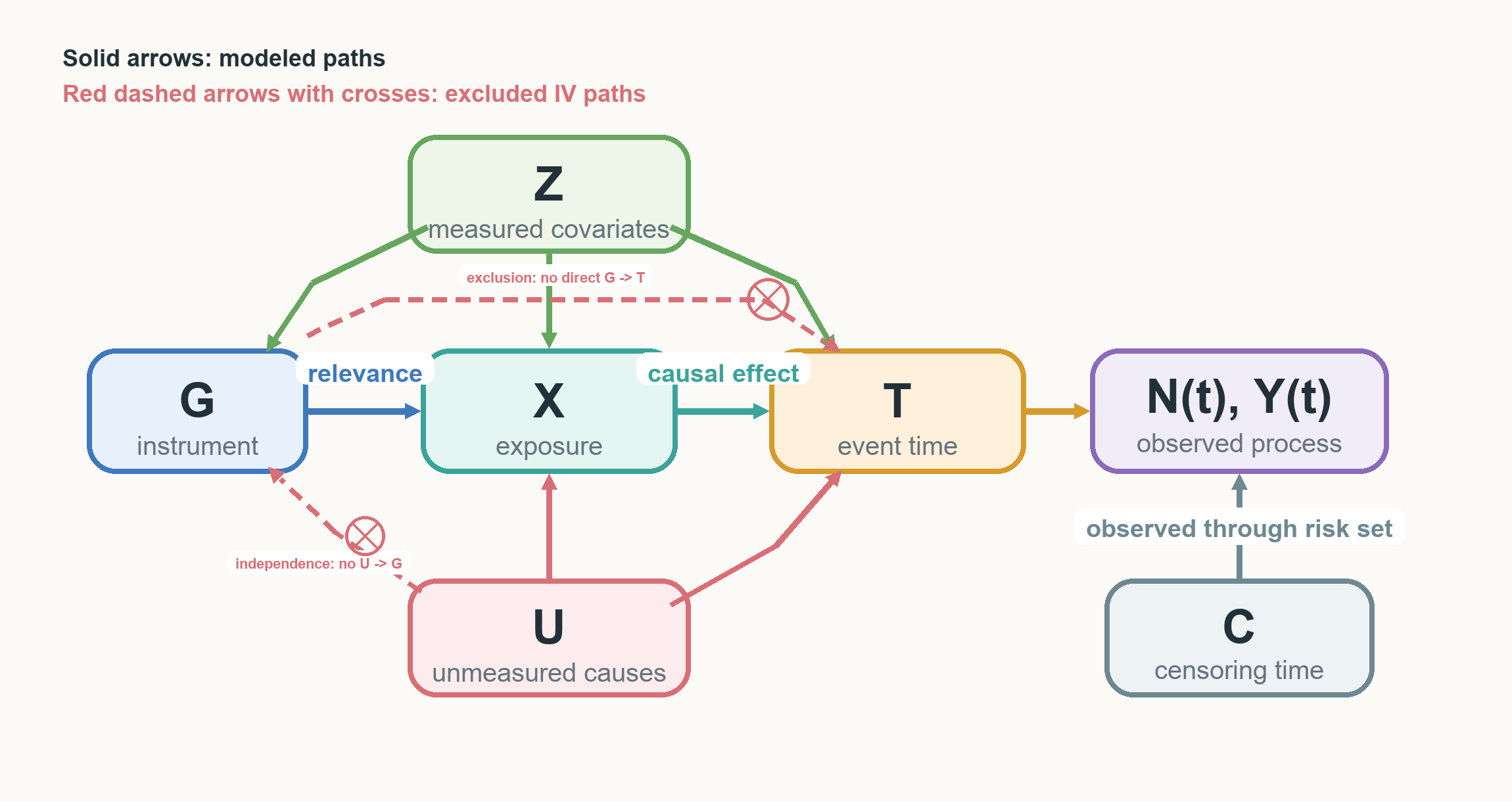}
\caption{Instrumental-variable DAG for a time-to-event outcome.}
\label{fig:IV_structure}
\end{figure}
This formulation follows Lu's dissertation \citep{lu2014dissertation}. For other graphical representations, see \citet{vanderweele2014methodological}.
 
When \(G\) refers to genetic variants, the same IV idea is called \emph{Mendelian randomization} (MR).
\begin{figure}[tbp]
\centering
\includegraphics[width=0.94\textwidth]{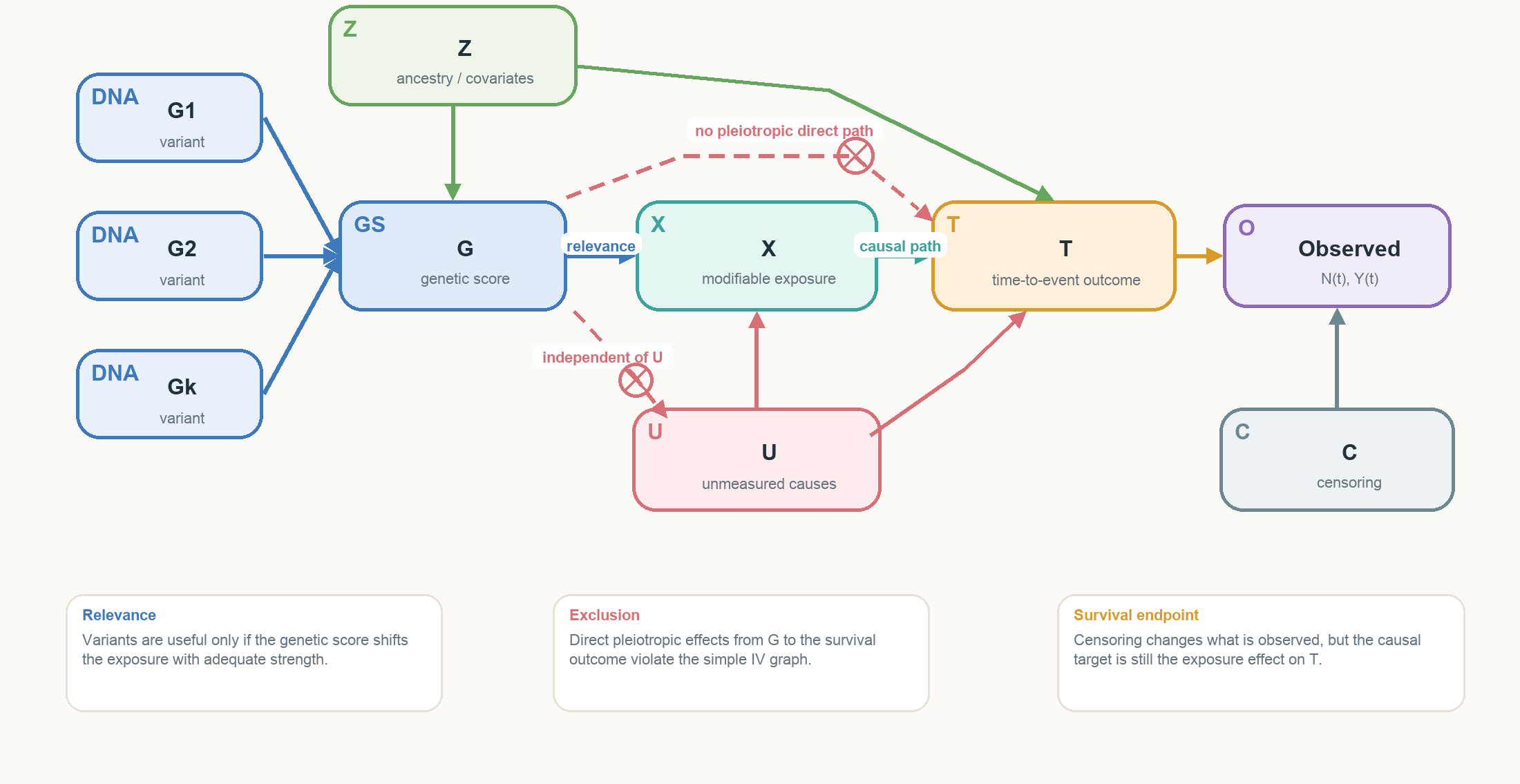}
\caption{Mendelian-randomization specialization of the survival IV graph.}
\label{fig:dag_ukb}
\end{figure}
 
The model follows \citet{li2015bayesian}:
\begin{align*}
    X &= \alpha_0+\alpha_1^TG + \alpha_2^TZ+\xi_1\\
    Y &= \beta_0+\beta_1X+\beta_2^TZ+\xi_2
    \end{align*}
The key contribution of the Li--Lu construction is to make endogeneity and censoring part of one likelihood rather than treating censoring as an afterthought to a two-stage regression. The first equation isolates the part of \(X\) predicted by the instrument and measured covariates; the second equation links the latent survival outcome to exposure and covariates. Correlation between \(\xi_1\) and \(\xi_2\) represents unmeasured common causes of exposure and outcome. Posterior inference for \(\beta_1\) therefore uses instrument-induced exposure variation while still integrating over censored or transformed event-time information.
We put parametric priors on the unknown parameters and use the instrument to separate exposure variation that is plausibly causal from exposure variation that is confounded.
\begin{align*}
    \left(\begin{matrix}\xi_1\\\xi_2\end{matrix}\right) &\sim \mathcal{N} \left(\left(\begin{matrix}\mu_1\\\mu_2\end{matrix}\right), \left(\begin{matrix}\sigma_1^2&\rho\sigma_1\sigma_2\\\rho\sigma_1\sigma_2&\sigma_2^2\end{matrix}\right)\right)\\
    \sigma_1^2&\sim InvGamma(\gamma_1,\gamma_2)\\
     \sigma_2^2&\sim InvGamma(\gamma_1,\gamma_2)\\
     \rho&\sim Uniform(-1, 1)\\
    \alpha_i&\sim\mathcal{N}(0,\zeta_i^2),\ \beta_i\sim\mathcal{N}(0,\psi_i^2)
\end{align*}
For inference, posterior samples of $\beta_1$ are drawn from the model. The \textbf{R} package is available at \url{https://github.com/ElvisCuiHan/PBIV}. Instrumental-variable analysis for time-to-event outcomes is developed further in Section~\ref{sec:iv_survival}.

\subsection{Semi-Markov Models with Duration Dependence}
 
\subsubsection{General Semi-Markov Processes}
    The treatment of semi-Markov processes follows the Markov-renewal and multistate counting-process framework of ABGK \citep{andersen1993statistical}. Related work on Markov renewal and semi-Markov survival models includes \citet{sun1992markov}, \citet{dabrowska1994cox}, and \citet{dabrowska2020stochastic}; additional references include \citet{dabrowska2012estimation}, \citet{cook2007statistical}, and \citet{cook2018multistate}.

    \begin{definition}[Semi-Markov process]
        The stochastic process
        \[
            (X,T)=\{X_n,T_n:n\in\mathbb{N}\}
        \]
        is a \emph{Markov renewal process} if the following conditions hold:
        \begin{enumerate}
        \item \(T_0=0\) a.s.;
        \item \((X_0,X_1,\cdots)\) is a discrete-time Markov chain;
        \item with \(W_n=T_n-T_{n-1}\),
        \begin{equation*}
            \mathbb{P}(W_1\le t_1,\cdots,W_n\le t_n\mid X_j,j\ge 0)
            =\prod_{k=1}^n\mathbb{P}(W_k\le t_k\mid X_{k-1},X_k).
        \end{equation*}
        \end{enumerate}
        If \(N(t)=\sum_{n\ge 1}\mathbb{I}(T_n\le t)\), then the continuous-time process
        \begin{equation*}
            X(t)=X_{N(t)}=\sum_{n=0}^\infty X_n\mathbb{I}(T_n\le t<T_{n+1})
        \end{equation*}
        is called a \emph{semi-Markov process}.
    \end{definition}

\begin{figure}[tbp]
\centering
\includegraphics[width=0.84\textwidth,height=0.40\textheight,keepaspectratio]{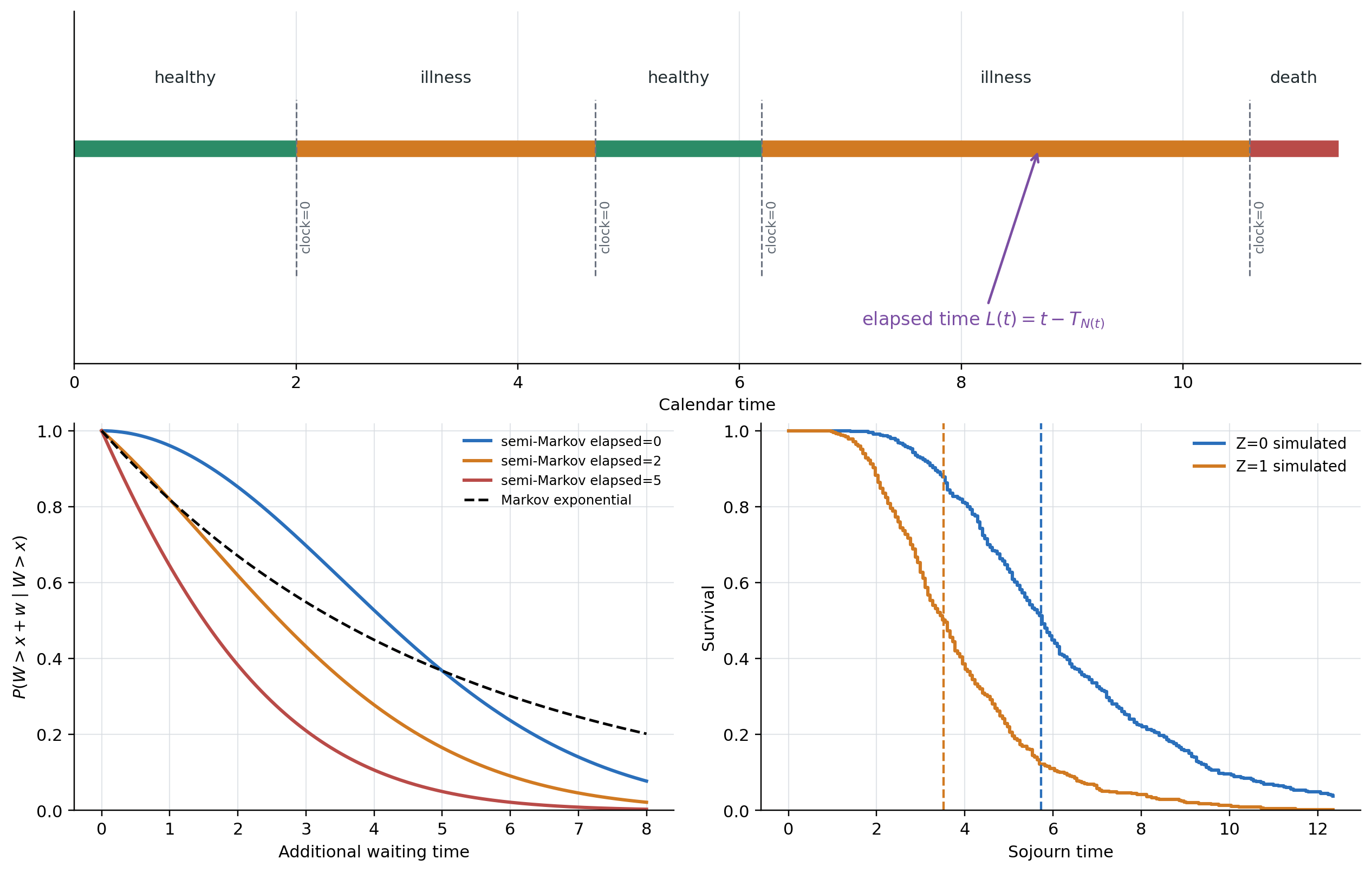}
\caption[Illustrative simulation for semi-Markov modeling]{Illustrative simulation for semi-Markov modeling.}
\label{fig:semimarkov_elapsed_time}
\end{figure}

The important difference from an ordinary continuous-time Markov chain is the elapsed sojourn time. In a Markov chain, once the current state is known, the future transition law does not depend on how long the process has already stayed there. In a semi-Markov model, the elapsed sojourn time is part of the conditioning information. For example, a patient who has just entered an illness state and a patient who has already spent five months in that state may have different residual waiting-time distributions, even though their current state label is the same. Figure~\ref{fig:semimarkov_elapsed_time} illustrates this elapsed-time reset and shows how a transformation model can shift the transition-specific sojourn distribution across covariate groups.

This duration dependence is common in clinical progression, reliability data, and operational monitoring. A patient who has been in an illness state for one month and a patient who has been in the same state for five years may have different residual waiting-time distributions. A device that has been degraded for a short interval and one that has operated in a degraded state for a long interval may also have different failure risks. A service unit that has been in an outage state for ten minutes and one that has been down for several hours may have different recovery or replacement probabilities even if both currently occupy the same state label. Semi-Markov notation keeps the state label and the elapsed sojourn time separate.

\begin{figure}[tbp]
\centering
\includegraphics[width=0.92\textwidth,height=0.48\textheight,keepaspectratio]{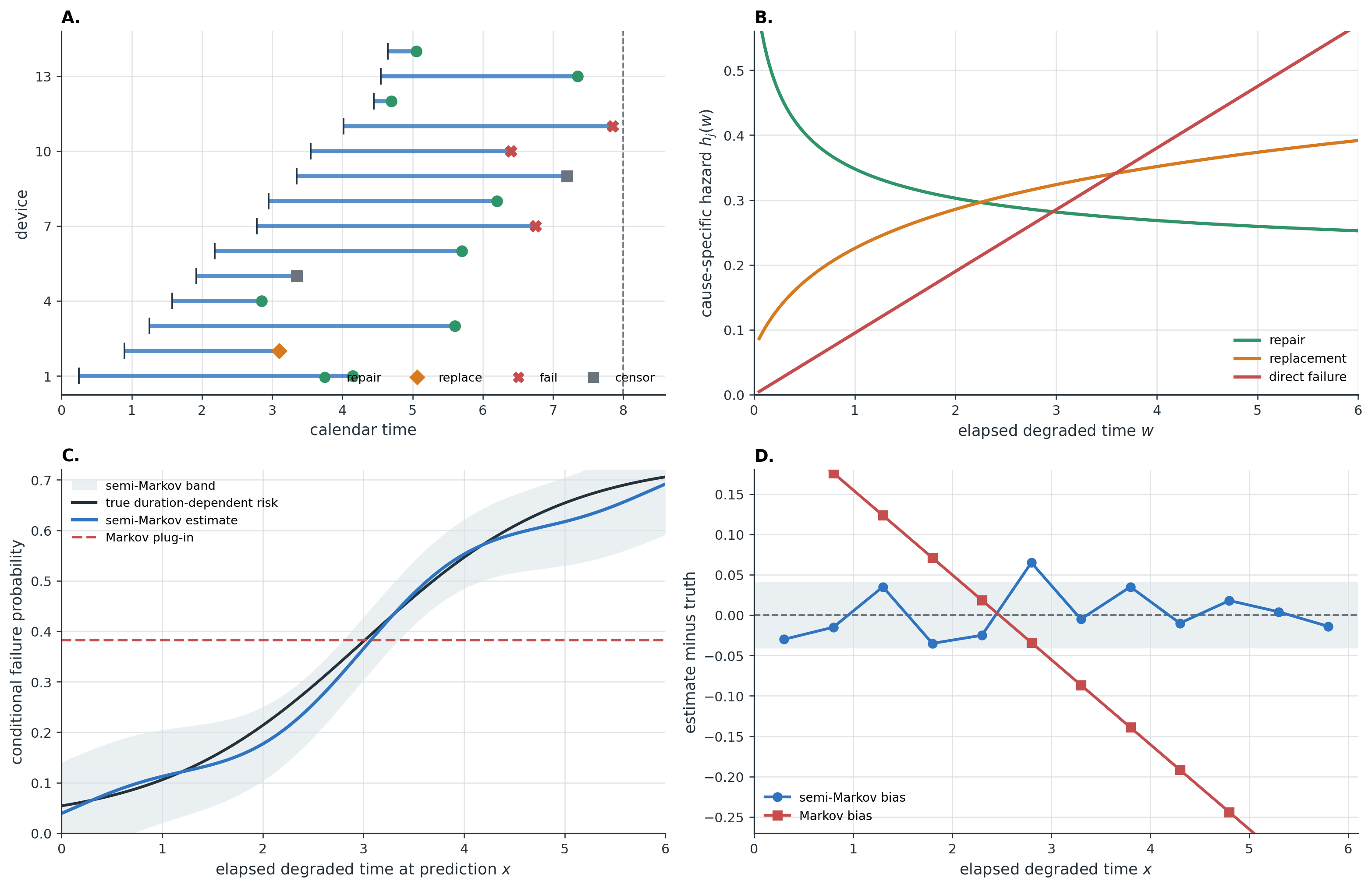}
\caption[Reliability semi-Markov simulation]{Reliability semi-Markov simulation. The figure uses degraded-device spells to show why elapsed sojourn time is part of the state description. A Markov model treats all degraded devices as exchangeable; a semi-Markov model distinguishes newly degraded units from units that have already survived a long degraded interval.}
\label{fig:semimarkov_reliability_duration_simulation}
\end{figure}

Figure~\ref{fig:semimarkov_reliability_duration_simulation} gives a second simulation, this time built around a reliability study. A device enters the degraded state at elapsed time \(0\), then exits by repair, replacement, direct failure, or administrative censoring. The data-generating mechanism is cause-specific Weibull:
\[
    h_j(w)=\frac{k_j}{a_j}\left(\frac{w}{a_j}\right)^{k_j-1},
    \qquad
    j\in\{\text{repair},\text{replacement},\text{direct failure}\},
\]
so that the exit clock is a function of elapsed degraded time \(w\), not merely of the current state label. The observed exit is
\[
    T=\min_j T_j,\qquad \Delta=\arg\min_j T_j,
\]
with administrative censoring at the end of the monitoring window. This is a standard semi-Markov problem: two devices may both be degraded, but the unit that degraded a few minutes ago and the unit that has been degraded for several hours can occupy different risk positions.

The panels show what a Markov approximation misses. Panel A displays individual degraded-state spells and the reset of the elapsed-time clock. Panel B shows that the cause-specific hazards are functions of elapsed degraded time. Panel C compares the true conditional probability of direct failure in the next two time units with an elapsed-time estimate and with a Markov plug-in estimate that is forced to be flat in \(x\). Panel D repeats the comparison over Monte Carlo samples: the Markov clock is tolerable near the average elapsed time, but it systematically misses early and late duration dependence, while the semi-Markov estimator pays variance to recover the shape.

\subsubsection{Semi-Markov Kernels and Renewal Theory}
    \begin{definition}[Semi-Markov kernel] Let $X_n$ take values in $\{1,\cdots,r\}$. The basic parameter in a semi-Markov (or Markov renewal) process is the semi-Markov kernel
    $$\mathbf{Q}(w)=\left[Q_{ij}(w)\right]$$
    where $p_{ij}=\mathbb{P}(X_{n+1}=j|X_n=i), F_{ij}(w)=\mathbb{P}(W_{n+1}\le w|X_{n+1}=j,X_n=i),$
    \begin{align*}
        Q_{ij}(w)&=\mathbb{P}(W_{n+1}\le w,X_{n+1}=j|X_n=i)=p_{ij}F_{ij}(w).
    \end{align*}
    Set
    $$\mathbf{G}(x,y)=\text{Diag}\left[\frac{S_i(y)}{S_i(x)}\right]$$
    where $S_i(w)=\sum_{j=1,j\not=i}^rp_{ij}(1-F_{ij}(w))$ so that on the diagonal, we have
        $$\mathbb{P}(W_{n+1}>y|W_{n+1}>x, X_n=i).$$
    \end{definition}
The kernel \(Q_{ij}(w)\) records the joint distribution of the next state and the waiting time: starting in state \(i\), it gives the probability that the next state is \(j\) and that the jump occurs within elapsed time \(w\). The transition probability \(p_{ij}\) controls the next-state distribution, while \(F_{ij}\) controls the transition-specific waiting-time distribution. The matrix \(\mathbf G(x,y)\) is the residual-sojourn probability, needed because prediction at calendar time \(t\) must condition on the time already spent in the current state. Figure~\ref{fig:semimarkov_elapsed_time} is the simulation companion for these quantities: the plotted elapsed-time reset is the distinction between \(Q\), \(F\), and the residual survival matrix \(\mathbf G\).

For a clinical Markov-renewal example, let \(i=\) relapsed and let \(j\) be remission, second-line treatment, severe infection, or death. Then \(p_{ij}\) specifies the next-state distribution, while \(F_{ij}(w)\) specifies how long that particular route takes. If \(R\) denotes relapse and \(D\) death, a high \(p_{RD}\) with a slow \(F_{RD}\) describes a transition that is common but usually delayed; a smaller \(p_{RD}\) with a steep \(F_{RD}\) describes a rarer transition that tends to occur soon when it occurs. In a reliability example, \(i=\) degraded and \(j\) could be repair, replacement, catastrophic failure, or administrative removal; \(Q_{ij}(w)\) separates the destination from the waiting time to that destination.
\noindent\textbf{Example (Forward recurrence time).}
Let $\mathcal{F}_{t}=\mathcal{F}_0\vee\sigma( X(s):s\le t)$ and $\gamma_t=T_{N(t)+1}-t$ be the forward recurrence time of the process given $\mathcal{F}_{t-}$. Then
    $$\mathbb{P}(\gamma_t>w|\mathcal{F}_{t-})=\sum_{i=1}^r\mathbb{I}(X(t)=i)G_{ii}(\delta_t,\delta_t+w)$$
    where $\delta_t=t-T_{N(t)}$ so that it is fully specified by $\mathbf{G}(x,y)$ defined earlier. Indeed, let $\beta_t:=T_{N(t)+1}-T_{N(t)}=W_{N(t)+1}$, then we have
    \begin{align*}
        \mathbb{P}(\gamma_t>w|\mathcal{F}_{t-})&=\mathbb{P}(T_{N(t)+1}-t>w|\mathcal{F}_{t-})\\
        &=\mathbb{P}(\beta_t>\delta_t+w|\delta_t,X_{N(t)})\\
        &=\sum_{i=1}^r\mathbb{I}(X_{N(t)}=i)G_{ii}(\delta_t,\delta_t+w).
    \end{align*}

    \textbf{Remark}. By comparison, in the continuous time Markov chain (CTMC) case,
    $$\mathbb{P}(\gamma_t>w|\mathcal{F}_{t-})=\sum_{i=1}^r\mathbb{I}(X_{N(t)}=i)\exp(-A_i(t+w)+A_i(t))$$
    where $A_i(t)=\sum_{j=1,j\not=i}^rA_{ij}(t)$. In the homogeneous case, RHS only depends on $w$ but not $t$, hence the probability depends only on the state occupied at time $t$ but not the duration time (time spent in this state).
\noindent\textbf{Example (events after time \(t\)).}
Similarly, the conditional probabilities of the first event after time \(t\) are
    $$\mathbb{P}(\gamma_t\le w,X_{N(t)+1}=j|\mathcal{F}_{t-})=\sum_{i=1}^r\mathbb{I}(X(t)=i)F_{ij}(\delta_t,\delta_t+w)$$
    where
    \begin{align*}
        F_{ij}(x,y)&=\frac{1}{S_i(x)}\int_{(x,y]}Q_{ij}(du)\\
        &=\mathbb{P}(W_{n+1}\le y, X_{n+1}=j|W_{n+1}>x, X_n=i).
    \end{align*}
    Indeed, conditioning on the state occupied at time $t$ and on the elapsed sojourn time $\delta_t$ gives
    \begin{align*}
    &\mathbb{P}\{\gamma_t\le w,X_{N(t)+1}=j\mid\mathcal{F}_{t-}\}\\
    &\quad=\sum_{i=1}^r \mathbb{I}\{X_{N(t)}=i\}
    \mathbb{P}\{W_{n+1}\le \delta_t+w,X_{n+1}=j
    \mid W_{n+1}>\delta_t,X_n=i\}\\
    &\quad=\sum_{i=1}^r \mathbb{I}\{X_{N(t)}=i\}F_{ij}(\delta_t,\delta_t+w),
    \end{align*}
    and the last conditional probability is exactly $F_{ij}(\delta_t,\delta_t+w)$.
   The transition-probability theorem below is the renewal analogue of the Aalen--Johansen product-integral representation. A path either stays in its current sojourn, or it makes a first post-\(t\) jump and then renews from the next state. The proof partitions the future according to those two possibilities.
   \begin{theorem}[Transition probabilities] Define  the matrix $\mathbf{R}$ as
   $$\mathbf{R}(y)=\mathbf{G}(0,y)+\int_0^y\mathbf{M}(du)\mathbf{G}(0,y-u)$$
   where $\mathbf{G}$ is defined earlier and $\mathbf{M}(y)=\sum_{r\ge 1}\mathbf{Q}^{(r)}(y)$ is the renewal matrix where $\mathbf{Q}^{(1)}=\mathbf{Q}$ and $\mathbf{Q}^{(r)}=\mathbf{Q}^{(r-1)}*\mathbf{Q}=\mathbf{Q}*\mathbf{Q}^{(r-1)}$. Then transition probabilities of a semi-Markov process are given by
   $$\mathbb{P}(X(t+s)=j|\mathcal{F}_{t-})=\sum_{i=1}^r\mathbb{I}(X_{N(t)}=i)\mathbb{P}_{ij}(\delta_t,\delta_t+s)$$
   where
   $$\mathbb{P}(x,y)=\delta_{ij}G_{ii}(x,y)+\sum_{k\not=i}\int_{(x,y]}F_{ik}(x,du)R_{kj}(y-u)$$
   and $R_{kj}$ is the $(kj)^{th}$ element  of the matrix $\mathbf{R}$.
   \end{theorem}
\begin{proof}
Suppose the process is in state $i$ at time $t$ and has already spent duration $x=\delta_t$ in that state. To be in state $j$ at elapsed duration $y=x+s$, there are two mutually exclusive possibilities. First, if $j=i$, the process may make no transition before elapsed time $y$, which has probability $G_{ii}(x,y)$. This gives the term $\delta_{ij}G_{ii}(x,y)$. Second, the next transition may occur at elapsed time $u\in(x,y]$ from $i$ to some $k\neq i$. The conditional probability of this first post-$t$ transition is $F_{ik}(x,du)$. After that jump, the process restarts from state $k$ with remaining time $y-u$, and the probability of being in state $j$ after that remaining time is $R_{kj}(y-u)$. Integrating over $u$ and summing over $k$ gives
\[
\delta_{ij}G_{ii}(x,y)+\sum_{k\neq i}\int_{(x,y]}F_{ik}(x,du)R_{kj}(y-u).
\]
Finally, conditioning on the random current state $X_{N(t)}$ gives the displayed formula for
$\mathbb P(X(t+s)=j|\mathcal F_{t-})$.
\end{proof}
 
\subsubsection{Cox Regression in a Markov Renewal Model}
    Define
    \begin{align*}
    \widetilde{N}_{ij}(t)&=\sum_{n\ge 1}\mathbb{I}(T_n\le t,X_n=j,X_{n-1}=i)\\
    \widetilde{N}(t)&=\sum_{i,j}\widetilde{N}_{ij}(t)\\
    L(t)&=t-T_{\widetilde{N}(t-)}\text{ (backwards recurrence time)}.
    \end{align*}
Here \(\widetilde N_{ij}\) counts observed transitions from \(i\) to \(j\), while \(L(t)\) is the elapsed time since the most recent transition. This is the quantity that makes the model semi-Markov rather than Markov: risk depends not only on the current state but also on how long the process has been there. We assume the intensity of $\widetilde{N}_{ij}(t)$ with respect to $\mathcal{F}_t=\mathcal{F}_0\vee \mathcal{N}_t$ is of form
\begin{align*}
    \Lambda_{ij}(dt)&=Y_i(t)\alpha_{ij}(L(t);Z)\\
    &=Y_i(t)\alpha_{0,ij}(L(t))e^{\beta_0^TZ_{ij}}dt
\end{align*}
where $Y_i(t)=\mathbb{I}(X(t-)=i)$ is the state-specific risk indicator and $Z$ is a vector of external covariates.
 
    We further define $W_{n}=T_n-T_{n-1}$, $T_0=0$ and
    \begin{align*}
        N_{ij}(x)&=\sum_{n\ge 0}\mathbb{I}(W_{n+1}\le x, X_n=i, X_{n+1}=j)\\
        Y_i(x)&=\sum_{n\ge 0}\mathbb{I}(W_{n+1}\ge x,X_n=i).
    \end{align*}
    Then the likelihood of observing $\{N_{ij}:i,j,\in\{1,\cdots,r\}\}$ is proportional to
    $$\Prodi_{u}\prod_{i}\left(\prod_{j\not=i}dA_{ij}(u)^{\Delta N_{ij}(u)}(1-dA_{i}(u))^{Y_i(u)-\Delta N_{ij}(u)}\right)$$
    where $A_{ij}=\int\alpha_{ij}$ and $A_i=\sum_{j\not=i}A_{ij}$ \citep[p.~681]{andersen1993statistical}. This likelihood is a product of multinomial transition probabilities.
 
Given $m$ iid realizations of the process $\{N_{ij}:i,j\le r\}$, the partial likelihood is
    \begin{align*}
    \mathcal{L}=\prod_{k=1}^m\left\{\prod_{i=1}^{r}\prod_{j\not=i}\prod_{n=1}^{n_{ijm}}\alpha_{ijk}(W_{nk})\times\exp\left(-\int_0^{\tau_k}\sum_{j\not=i}\alpha_{ijk}(L(t))dt\right)\right\}.  
    \end{align*}
where $n_{ijm}$ is the total number of transitions from $i$ to $j$ of individual $m$. Hence, a natural NA-type estimator of $A_{ij}$ is
    $$\widehat{A}_{ij}(y,\beta)=\sum_{k=1}^m\int_0^y\frac{dN_{ijk}(u)}{S_{ij}^{(0)}(u,\beta)}$$
    where $S_{ij}^{(0)}=\sum_{k=1}^mY_{ik}(x)e^{\beta^TZ_{ijk}(x)}$ is analogous to the Cox model risk-set denominator. 

Plugging $\widehat{A}_{ij}$ into $\mathcal{L}$ gives the profile likelihood for $\beta$:

    $$\mathcal{L}(\beta)=\prod_{k=1}^m\prod_{i=1}^r\prod_{j\not=i}\prod_{n=1}^{n_{ijk}}\left(\frac{\alpha_{ijk}(W_{nk})}{\sum_{k=1}^mY_{ik}(W_{nk})\exp(\beta^TZ_{ijk}(W_{nk}))}\right)^{\Delta N_{ijk}(W_{nk})}$$
    and the maximum partial likelihood is indeed
    $$\widehat{\beta}=\arg\max_\beta\mathcal{L}(\beta).$$
Finally, define
    $$H_{ij}(x)=N_{ij}(x)-\int_0^x Y_i(u)\alpha_{ij}(u)du,$$
and in general there is no filtration making \(H_{ij}\) a martingale. This is the technical obstruction that forces the semi-Markov argument to use renewal structure rather than the standard multiplicative-intensity proof alone.

    Let $\mathbf{A}(x)=[A_{ij}(x)]$ and for $x<y$, $\mathbf{G}(x,y)$ be a diagonal matrix with entries $G_{ii}(x,y)=\exp(-\sum_j(A_{ij}(y)-A_{ij}(x)))$. Let
    \begin{align*}
        \mathbf{F}(x,y)&=\int_x^y\mathbf{G}(x,u-)\mathbf{A}(du),\\
        \mathbf{H}(y)&=\mathbf{G}(0,y),\\
        \mathbf{Q}(y)&=\mathbf{F}(0,y).
    \end{align*}
    Similar to the theory part, define
    \begin{align*}
        \mathbf{F}^{(p)}&=\int_x^y\mathbf{F}(x,du)\mathbf{Q}^{(p-1)(y-u)},\\
        \mathbf{G}^{(p)}&=\int_x^y\mathbf{F}(x,du)\left[\mathbf{Q}^{(p-1)}*\mathbf{H}\right](y-u),\\
        \mathbf{G}^{(0)}&=\mathbf{G},\ \mathbf{F}^{(0)}=\mathbf{Q}^{(0)}=\mathbf{I}.
    \end{align*}

    \begin{lemma}[Predicted probability]
        Given $\mathcal{F}_{t-}$, the conditional probability that the process is in state $j$ at time $t+v$ is
        $$\mathbb{P}(X(t+v)=j|\mathcal{F}_{t-})=\begin{cases}
            \sum_{p\ge 0}Q_{ij}^{(p)}*H_{jj}(v)&\text{ if for some }n, t=T_n.\\
            \sum_{p\ge 0}G_{ij}^{(p)}(t-T_n,v+t-T_n)&\text{ if else}.
        \end{cases}$$
        The probability is denoted as $P_{ij}(t-T_n,v+t-T_n)$.
    \end{lemma}
\begin{proof}
Condition on the state occupied just before \(t\) and on the elapsed sojourn time. If \(t=T_n\), the process has just entered state \(i=X_n\), so the elapsed time is zero. To be in state \(j\) at time \(t+v\), the path may make \(p\ge0\) complete Markov-renewal transitions before \(v\), and after the last transition it must remain in state \(j\) until \(v\). The distribution of \(p\) complete transitions is the \(p\)-fold convolution \(Q_{ij}^{(p)}\), and the probability of remaining in \(j\) for the residual time is \(H_{jj}\). Summing over \(p\) gives
\[
    \sum_{p\ge0} Q_{ij}^{(p)}*H_{jj}(v).
\]

If \(t\) lies inside a sojourn interval, let \(x=t-T_n\) be the already elapsed time. The first post-\(t\) transition is governed by the conditional residual kernel \(\mathbf F(x,\cdot)\), not by the original kernel \(\mathbf Q\), because the current sojourn has already survived to age \(x\). The term \(\mathbf G(x,x+v)\) is the probability of no post-\(t\) transition before \(t+v\). If the first post-\(t\) transition occurs, the process renews, and subsequent transitions are described by the ordinary convolution powers of \(\mathbf Q\) followed by the terminal survival factor \(\mathbf H\). This is exactly the recursively defined quantity
\[
    \sum_{p\ge0}G_{ij}^{(p)}(x,x+v).
\]
Substituting \(x=t-T_n\) proves the displayed conditional probability.
\end{proof}
    \textbf{Remark}: In practice, we replace $\mathbf{A}$ by $\widehat{\mathbf{A}}$.

\subsubsection{Semi-Markov Transformation Model}
The Cox-type Markov renewal model above is built from transition-specific proportional hazards in the elapsed time $x$ since entry into the current state. A semi-Markov transformation model keeps the same renewal structure but replaces the proportional-hazards restriction by a transformation of the sojourn time. For a transition out of state $i$ into state $j$, let $W_{n+1}=T_{n+1}-T_n$ be the sojourn time and let $Z_n$ be covariates measured at, or before, entry into state $i$. A convenient formulation is
\[
    H_{ij}(W_{n+1})+\beta_{ij}^TZ_n=\varepsilon_{ij},
    \qquad X_n=i,\ X_{n+1}=j,
\]
where $H_{ij}$ is an unknown increasing transformation and the distribution of $\varepsilon_{ij}$ is specified up to no unknown parameters. Equivalently,
\[
F_{ij}(w|z)=\mathbb{P}(W_{n+1}\le w|X_n=i,X_{n+1}=j,Z_n=z)
=F_{\varepsilon,ij}\{H_{ij}(w)+\beta_{ij}^Tz\}.
\]
The semi-Markov kernel with covariates is then
\[
Q_{ij}(w|z)=p_{ij}(z)F_{\varepsilon,ij}\{H_{ij}(w)+\beta_{ij}^Tz\},
\]
where $p_{ij}(z)=\mathbb{P}(X_{n+1}=j|X_n=i,Z_n=z)$ models the next visited state. Different choices of $F_{\varepsilon,ij}$ recover familiar models. An extreme-value error yields a proportional-hazards model in elapsed time; a logistic error yields a proportional-odds model; a normal error gives an accelerated-failure-time style transformation model.

\begin{lemma}[Kernel induced by the transformation model]
For fixed current state $i$ and covariate value $z$, the transformation specification determines the conditional law of the next state and waiting time through
\[
\mathbb{P}(W_{n+1}\le w,X_{n+1}=j|X_n=i,Z_n=z)
=Q_{ij}(w|z),
\]
and the conditional survival function for remaining in state $i$ beyond elapsed time $w$ is
\[
S_i(w|z)=\sum_{j\neq i}p_{ij}(z)\overline F_{\varepsilon,ij}\{H_{ij}(w)+\beta_{ij}^Tz\}.
\]
\end{lemma}
\begin{proof}
By conditioning on the next state,
\begin{align*}
&\mathbb{P}(W_{n+1}\le w,X_{n+1}=j|X_n=i,Z_n=z)\\
&\quad=\mathbb{P}(X_{n+1}=j|X_n=i,Z_n=z)
\mathbb{P}(W_{n+1}\le w|X_n=i,X_{n+1}=j,Z_n=z),
\end{align*}
which is exactly $p_{ij}(z)F_{\varepsilon,ij}\{H_{ij}(w)+\beta_{ij}^Tz\}$. The survival probability of the sojourn time is the sum, over all possible destination states, of the probabilities that the next transition is to that state and that the waiting time exceeds $w$.
\end{proof}

For an observed transition $i\to j$ with waiting time $w$ and covariate vector $z$, the likelihood contribution is
\[
p_{ij}(z)f_{\varepsilon,ij}\{H_{ij}(w)+\beta_{ij}^Tz\}\,dH_{ij}(w).
\]
For a right-censored sojourn in state $i$ censored at elapsed time $c$, the contribution is
\[
\sum_{j\neq i}p_{ij}(z)\overline F_{\varepsilon,ij}\{H_{ij}(c)+\beta_{ij}^Tz\}.
\]
Thus estimation separates conceptually into two parts: a multinomial model for the next state and a transition-specific transformation model for the elapsed waiting time. In practice, $H_{ij}$ is treated as a monotone step function with jumps at observed transition times, and $(\beta_{ij},H_{ij})$ are estimated by profile likelihood or estimating equations. The resulting estimator has the same flavor as the Cox partial likelihood, but the risk-set weights involve the error hazard
\[
\alpha_{\varepsilon,ij}(u)=\frac{f_{\varepsilon,ij}(u)}{\overline F_{\varepsilon,ij}(u)}
\]
evaluated at $u=H_{ij}(x)+\beta_{ij}^Tz$ rather than the simple exponential weight $\exp(\beta^Tz)$.
 
\subsubsection{Semiparametric Modulated Renewal Process}

The modulated renewal model of \citet{dabrowska2005modulated} is a useful bridge between the Markov-renewal construction above and ordinary Cox regression. The process still moves through a finite state space, but the local transition risk depends on the \emph{backward recurrence time}, the elapsed time since entry into the current state. Let \(X(t)\) be the current state, let
\[
    B(t)=t-\sup\{T_m:T_m<t\}
\]
be the backward recurrence time, and let \(Z(t)\) be a predictable covariate process. For a transition \(i\to j\), a Cox-like version of the model writes
\[
    \lambda_{ij}(t|\mathcal F_{t-})
    =
    Y_i(t)\alpha_{ij}\{B(t)\}\exp\{\beta_{ij}^TZ(t)\},
\]
where \(Y_i(t)=\mathbb I\{X(t-)=i\}\). The baseline hazard \(\alpha_{ij}\) is a function of elapsed sojourn time \(B(t)\), not a common calendar-time function. This is why the model is not just a standard Andersen--Gill Cox model with a different covariate vector: after each transition the risk clock resets, and the same calendar time can correspond to different elapsed ages for different subjects.

The associated cumulative baseline is
\[
    A_{ij}(x)=\int_0^x \alpha_{ij}(u)\,du,
\]
and a kernel-smoothed Breslow-type estimator uses event increments whose elapsed ages are near \(x\). One convenient schematic form is
\[
    \widehat\alpha_{ij}(x)
    =
    \frac{\sum_k\int K_b\{x-B_k(s)\}\,dN_{ijk}(s)}
         {\sum_k\int K_b\{x-B_k(s)\}Y_{ik}(s)\exp\{\widehat\beta_{ij}^TZ_k(s)\}\,ds},
    \qquad
    \widehat A_{ij}(x)=\int_0^x \widehat\alpha_{ij}(u)\,du.
\]
The exact implementation depends on whether one smooths the baseline hazard or estimates its cumulative version by local increments, but the statistical object is the same: transition counts are aligned by elapsed time since state entry. Dabrowska and Ho use this structure to estimate regression coefficients and baseline cumulative hazards and derive consistency and asymptotic normality for censored finite-state modulated-renewal observations.

In applications, the distinction is substantive. A bone-marrow-transplant patient who has just relapsed and a patient who has remained relapsed for many months occupy the same state but may have very different transition risks. A machine that has just entered degraded operation and one that has been degraded for a long duration may also have different failure or repair rates. The modulated-renewal formulation keeps the current state, elapsed state age, and covariates visible at the same time.

\subsubsection{Asymptotics}
    The central-limit theorem for the semi-Markov estimator has a different proof structure from the earlier martingale limits. Duration dependence prevents a reduction to the same calendar-time martingale array, so the proof aggregates independent subject-level contributions instead. In Figure~\ref{fig:semimarkov_elapsed_time}, this corresponds to averaging transition-specific sojourn histories rather than summing event residuals on one common calendar-time scale.
    \begin{theorem}[Semi-Markov CLT]
    For the semi-Markov estimator studied by \citet{dabrowska1994cox} and \citet{sun1992markov}, fix a transition \(i\to j\) and an elapsed-time interval \([0,\tau]\). Suppose the Markov renewal observations are independent across subjects, covariates are bounded, risk-set limits are nondegenerate, and transition-specific cumulative hazards have finite variation on \([0,\tau]\). Assume that the estimator admits the expansion
    \[
        \sqrt m\{\widehat A_{ij}(x)-A_{ij}(x)\}
        =
        \frac{1}{\sqrt m}\sum_{k=1}^m H_{ijk}(x,\beta_0)+o_p(1)
    \]
    uniformly in \(x\in[0,\tau]\), where \(H_{ijk}(\cdot,\beta_0)\) is a measurable, centered subject-level influence process in \(\ell^\infty[0,\tau]\). Assume also that the class \(\{H_{ij1}(x,\beta_0):x\in[0,\tau]\}\) is Donsker. Then
    \[
        \widehat{W}(\cdot,\beta_0)
        =
        \frac{1}{\sqrt{m}}\sum_{k=1}^mH_{ijk}(\cdot,\beta_0)
        \rightsquigarrow W(\cdot,\beta_0)
    \]
    in \(\ell^\infty[0,\tau]\), where \(W\) is a mean-zero Gaussian process with covariance
    \[
        \mathrm{Cov}\{W(x_1,\beta_0),W(x_2,\beta_0)\}
        =
        \mathbb E\{H_{ij1}(x_1,\beta_0)H_{ij1}(x_2,\beta_0)\}.
    \]
    \end{theorem}
\begin{proof}
The displayed expansion reduces the estimator to an empirical process indexed by elapsed time. Centering of the influence process gives
\[
\mathbb E\{H_{ij1}(x,\beta_0)\}=0,\qquad x\in[0,\tau].
\]
For any finite collection \(x_1,\ldots,x_r\), the vectors
\[
\{H_{ijk}(x_1,\beta_0),\ldots,H_{ijk}(x_r,\beta_0)\},\qquad k=1,\ldots,m,
\]
are independent and identically distributed with finite second moments, so the multivariate central limit theorem gives finite-dimensional Gaussian convergence. The Donsker assumption supplies tightness in \(\ell^\infty[0,\tau]\). Therefore the empirical process \(m^{-1/2}\sum_{k=1}^mH_{ijk}(\cdot,\beta_0)\) converges weakly to a tight mean-zero Gaussian process with the covariance displayed above. The uniform \(o_p(1)\) remainder transfers the limit to \(\sqrt m\{\widehat A_{ij}-A_{ij}\}\). If \(\beta\) is estimated, the same argument applies after the usual Taylor correction for the derivative of the estimating equation with respect to \(\beta\).
\end{proof}

\subsection{Other Regression Models and Extensions}

The remaining regression models are variations on the same counting-process construction. Parametric models specify the whole intensity curve, Beran's estimator uses local neighborhoods in covariate space, and smoothed Cox or Cox--Aalen models let selected parts of the hazard vary more flexibly. In each case, the modeling decision is which part of the event intensity is specified globally and which part is allowed to adapt locally over time or covariates.

\subsubsection{Panel Count Data: Wellner--Zhang Mean Estimation}

The panel-count setting of \citet{wellner2000two} is close to recurrent-event survival analysis, but the exact jump times are not observed. For subject \(i\), the latent recurrent-event process \(N_i(t)\) is observed only at inspection times
\[
    0<T_{i1}<\cdots<T_{iK_i}\le \tau,
\]
so the data are cumulative counts \(N_i(T_{ij})\), not the individual recurrence times. The target in the basic nonparametric problem is the mean function
\[
    \Lambda_0(t)=\mathbb E\{N_i(t)\},
\]
which is nondecreasing with \(\Lambda_0(0)=0\). If we write \(T_{i0}=0\) and
\[
    \Delta N_{ij}=N_i(T_{ij})-N_i(T_{i,j-1}),\qquad
    \Delta\Lambda_{ij}=\Lambda(T_{ij})-\Lambda(T_{i,j-1}),
\]
then a convenient working likelihood is the nonhomogeneous Poisson likelihood
\[
    L(\Lambda)=
    \prod_i\prod_{j=1}^{K_i}
    \frac{\{\Delta\Lambda_{ij}\}^{\Delta N_{ij}}\exp(-\Delta\Lambda_{ij})}
    {(\Delta N_{ij})!},
\]
maximized over nondecreasing step functions \(\Lambda\). The resulting estimator is an NPMLE for the mean count curve. The paper also studies a monotone moment-type estimator, which can be read as an isotonic version of the empirical relationship between inspection time and observed cumulative count. The important lesson is not that recurrent events must be Poisson. Rather, the Poisson likelihood is a device for recovering a monotone mean function from interval counts; the scientific estimand is the mean cumulative recurrence burden.

\begin{figure}[tbp]
\centering
\includegraphics[width=0.84\textwidth,height=0.40\textheight,keepaspectratio]{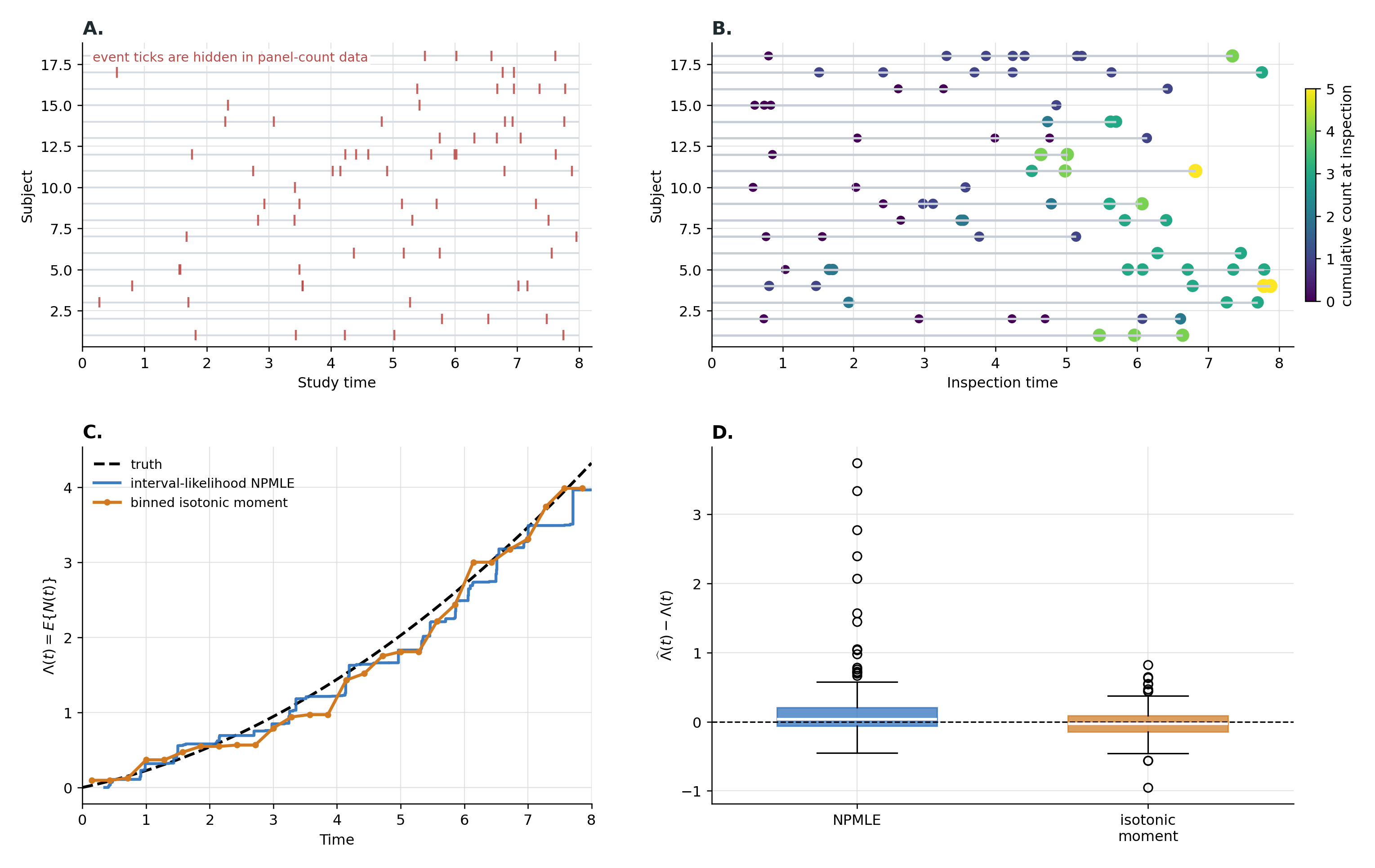}
\caption[Simulation study for the panel-count problem]{Simulation study for the panel-count problem.}
\label{fig:wellner_zhang_panel_sim}
\end{figure}

This is the conceptual bridge from exact recurrent-event data to panel-count data. With exact event times, the Nelson--Aalen increment \(dN/Y\) is attached to a known jump time. With panel counts, the jump is only known to have occurred somewhere between two inspections, so inference shifts from predictable event-time increments to interval increments \(\Delta\Lambda_{ij}\). Figure~\ref{fig:wellner_zhang_panel_sim} shows this distinction in a small simulation. Panel A shows latent recurrent-event ticks that are unobserved by design. Panel B shows the actual data: cumulative counts only at inspection visits. Panel C compares an interval-likelihood NPMLE with a binned isotonic moment estimator, and Panel D shows the finite-sample error at several time points.

\subsubsection{Landmark Analysis and Dynamic Prediction}

Landmark analysis targets future survival conditional on being event-free at a fixed prediction time \(s\). The method was introduced in response-based survival comparisons by \citet{anderson1983analysis} and later developed into a general dynamic-prediction framework by \citet{vanhouwelingen2011dynamic}; \citet{dafni2011landmark} gives an accessible methodological overview. The estimand is conditional survival after taking the landmark as the new time origin,
\[
    S_s(u\mid \mathcal H_i(s))
    =\mathbb P\{T_i>s+u\mid T_i>s,\mathcal H_i(s)\},
    \qquad u\ge 0,
\]
where \(\mathcal H_i(s)\) is the covariate history available up to \(s\). For an eligible subject with \(T_i\wedge C_i>s\), define the landmark counting and risk processes
\[
    N_i^{(s)}(u)=\mathbb I(T_i-s\le u,\Delta_i=1),
    \qquad
    Y_i^{(s)}(u)=\mathbb I(T_i\wedge C_i\ge s+u).
\]
Then a nonparametric landmark Kaplan--Meier estimator for a group \(G_i(s)=g\) is
\[
    \widehat S_s(u\mid g)
    =
    \Prodi_{0<v\le u}
    \left\{
    1-
    \frac{\sum_i \mathbb I\{G_i(s)=g,T_i\wedge C_i>s\}\,dN_i^{(s)}(v)}
         {\sum_i \mathbb I\{G_i(s)=g,T_i\wedge C_i>s\}\,Y_i^{(s)}(v)}
    \right\}.
\]
A semiparametric version fits, within the landmark risk set, a Cox model on the prediction horizon,
\[
    \lambda_s(u\mid Z_i(s))=\lambda_{0s}(u)\exp\{\beta_s^\top Z_i(s)\}.
\]
The key discipline is that the risk set is conditioned on survival to \(s\), the covariates are frozen or summarized using information available by \(s\), and time zero for prediction becomes \(s\). Starting the survival curve at baseline after grouping subjects by a future response status creates immortal-time bias: membership in the group already certifies survival up to the landmark.

In the clinical prediction analogue, patients still event-free at a post-treatment landmark are followed over a future horizon using only biomarker, treatment, and complication information observed by that landmark. The landmark \(s\) is the time at which prediction is made: after \(s\), follow-up is measured from the new origin; before \(s\), the observed history becomes covariate information rather than future leakage.

\begin{figure}[tbp]
\centering
\includegraphics[width=0.84\textwidth,height=0.40\textheight,keepaspectratio]{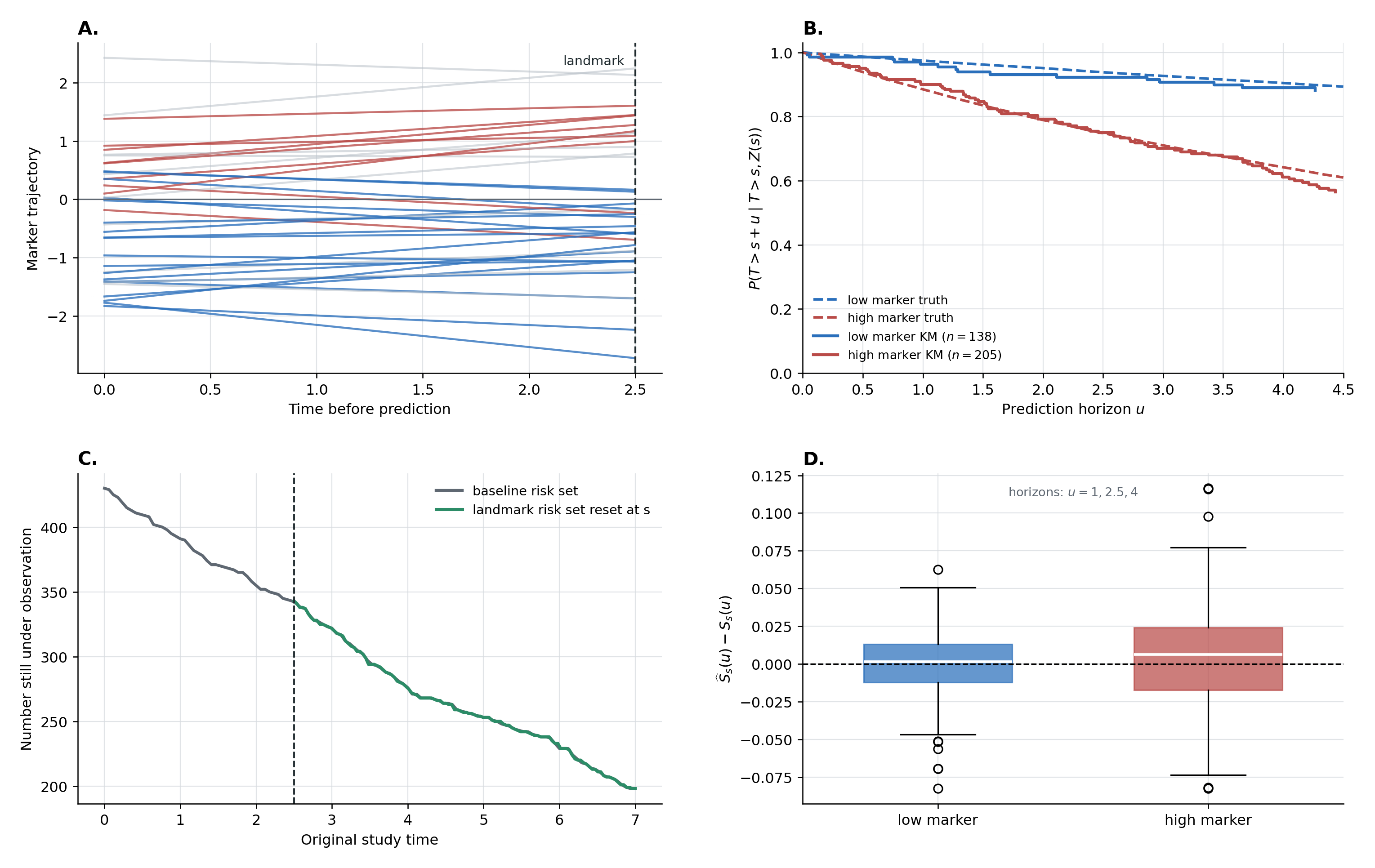}
\caption[Simulation for landmark analysis and dynamic prediction]{Simulation for landmark analysis and dynamic prediction.}
\label{fig:landmark_analysis_sim}
\end{figure}

Figure~\ref{fig:landmark_analysis_sim} gives a small dynamic-prediction simulation. A continuous marker evolves before the landmark; subjects who remain under observation at \(s=2.5\) are classified by their marker value at that time. The landmark Kaplan--Meier curves then estimate future conditional survival for the low- and high-marker groups. The risk-set panel emphasizes the reset: after \(s\), inference is based only on subjects who survived and were observed up to the landmark, not on the original baseline cohort.

\subsubsection{Bayesian Nonparametric Survival and Multistate Examples}

Bayesian nonparametric survival analysis changes the uncertainty object. Nelson--Aalen, Kaplan--Meier, and Aalen--Johansen give plug-in curves plus large-sample variance formulas. A Bayesian nonparametric model instead puts a stochastic-process prior on a survival distribution, cumulative hazard, or transition-specific hazard measure, then propagates posterior draws through the survival functional. For a one-transition survival problem, a beta-process or gamma-process style prior can be placed on \(A(t)\), and posterior draws of \(A\) are transformed into survival draws by
\[
    S(t)=\Prodi_{s\le t}\{1-dA(s)\}.
\]
This is the survival-scale version of the same product-integral operation used earlier for Kaplan--Meier.

For a multistate process, the natural BNP target is the collection of transition-specific cumulative hazards
\[
    \mathbf A(t)=\{A_{hj}(t):h\neq j\}.
\]
One may put independent process priors on the \(A_{hj}\)'s, or use a hierarchical/dependent prior when transitions or groups should share information. Posterior transition-probability draws are then obtained by the product integral
\[
    \mathbf P(0,t)=\Prodi_{0<s\le t}\{I+d\mathbf A(s)\}.
\]
This produces posterior bands for state occupation probabilities, cumulative incidence functions, and terminal-event survival probabilities. In an illness-death model, for example, a posterior draw of \((A_{01},A_{02},A_{12})\) immediately induces a posterior draw of the probability of being healthy, ill, or dead at time \(t\). Dependent Dirichlet-process survival regression \citep{deiorio2009bayesian}, semi-competing-risk BNP models \citep{xu2020bayesian}, and competing-risk priors based on completely random measures \citep{delsole2026principled} instantiate this principle.

\begin{figure}[tbp]
\centering
\includegraphics[width=0.84\textwidth,height=0.40\textheight,keepaspectratio]{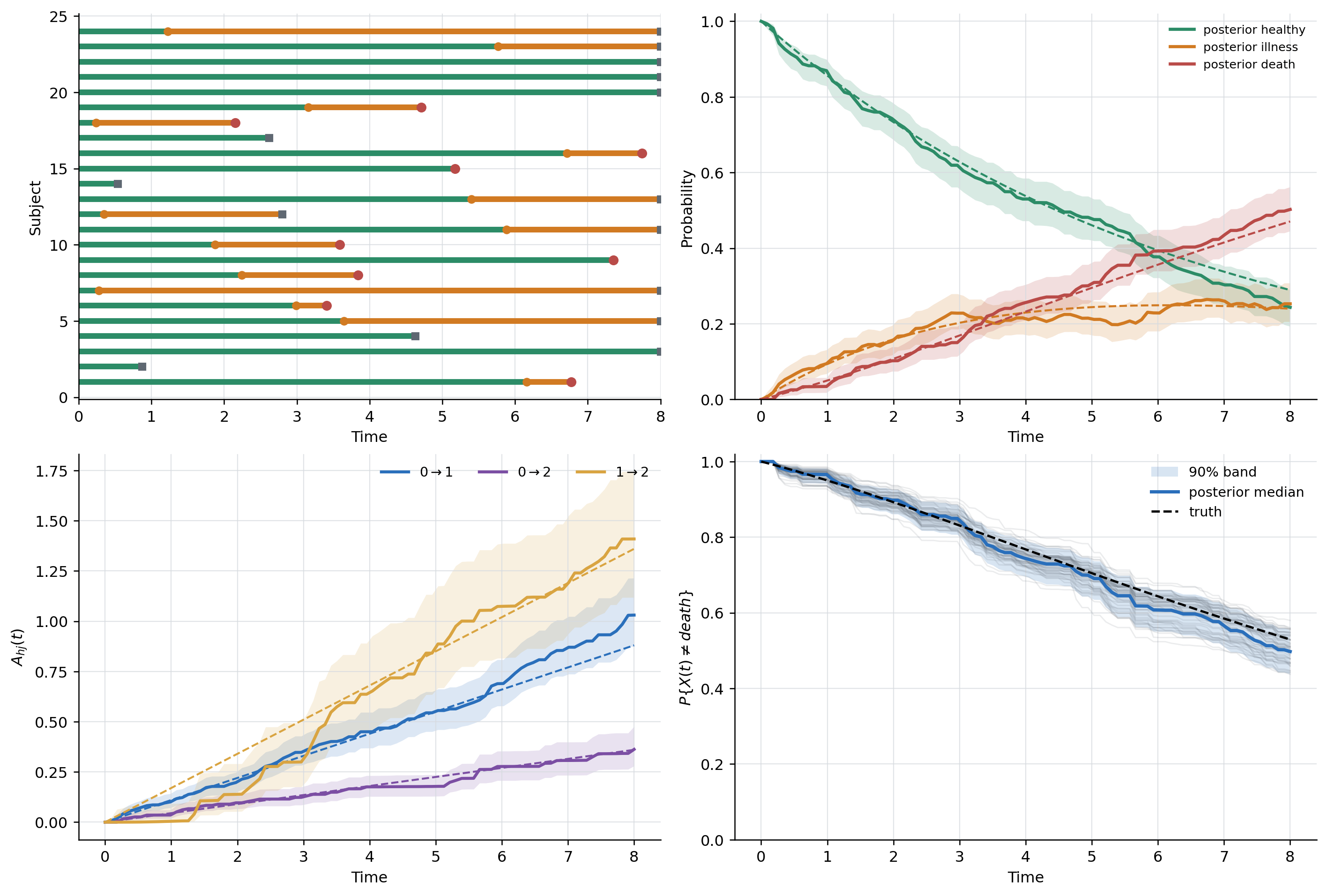}
\caption{Bayesian nonparametric multistate simulation.}
\label{fig:bnp_multistate_sim}
\end{figure}

Figure~\ref{fig:bnp_multistate_sim} gives a finite-sample version. An illness-death cohort is simulated under constant transition hazards. The analysis uses transition-specific gamma-Poisson updates on a fine time grid as a computational analogue of a gamma-process prior: each interval rate is updated by its observed transition count and accumulated risk time. Posterior rate draws are cumulated into transition-specific hazards and passed through the product integral to obtain posterior state probabilities and terminal-survival bands. The example is a simplified illustration of beta-process, beta-Stacy, and dependent Dirichlet-process survival priors \citep{hjort1990nonparametric,walker1997beta,kim1999nonparametric}.

\subsubsection{Modern Censored Learning and Out-of-Fold Validation}
\label{sec:modern_censored_learning}

The current frontier is not a single new estimator. It is a negotiation between flexible prediction and event-history discipline. Random survival forests and neural Cox models replace linear risk scores by richer functions, DeepHit and deep survival machines learn competing-risk or parametric time distributions, transformer-style models use high-dimensional records to predict multiple event types, and survival foundation models try to transfer prior knowledge across tabular time-to-event tasks \citep{ishwaran2008random,katzman2018deepsurv,lee2018deephit,kvamme2019time,nagpal2021deep,wang2022survtrace,bohm2026survpfn}. Targeted learning and doubly robust survival estimators ask for valid causal contrasts when nuisance hazards, treatment mechanisms, and censoring models are learned adaptively, with recent work targeting smooth heterogeneous treatment-effect curves under censoring and truncation \citep{cai2018onestep,rytgaard2021tmle,luo2023doubly,pryce2026targeted}. Conformal survival methods ask a different question: how to wrap a survival learner with prediction intervals or lower survival-time bounds that retain coverage under censoring, covariate shift, or time-updated histories \citep{candes2021conformal,sesia2024doubly,holmes2024twosided,farina2025doubly,wang2026history}. Multistate work is also moving toward temporally consistent prediction of hitting-time distributions from transition estimates \citep{escobarbach2026multistate}. Neural temporal point processes and neuro-symbolic event models extend Hawkes-type ideas to complex event streams, but the same compensator residuals and time-rescaling checks remain the grammar for deciding whether an intensity has actually learned the event history \citep{ogata1981lewis,bacry2015hawkes,yang2024neurosymbolic}.

Algorithm~\ref{alg:modern_censored_learning} places these learners inside the same bookkeeping used above. A flexible learner may estimate a hazard, transition probability, censoring survival, treatment mechanism, or nonconformity score; the validation step still has to respect the predictable risk set, the censoring mechanism, the target estimand, and the residual process.

\begin{algorithm}[tbp]
\caption{Cross-fitted frontier wrapper for censored learning}
\label{alg:modern_censored_learning}
\begin{algorithmic}[1]
\Require Censored event-history data \(O_i\), target \(\psi\), learner library \(\mathcal L\), folds \(1,\ldots,K\)
\Ensure Calibrated prediction, causal contrast, or conformal survival interval
\State Split subjects into \(K\) folds, preserving clusters or centers when dependence requires it.
\For{each validation fold \(k\)}
    \State Fit outcome, hazard, transition, censoring, and treatment models on the other folds using classical, forest, boosting, neural, or transformer learners from \(\mathcal L\).
    \State Predict fold-\(k\) hazards, survival curves, transition probabilities, censoring survival, and influence-function ingredients without using future information.
    \State If \(\psi\) is causal, update the initial fit along a targeting submodel or augmented estimating equation until the empirical efficient influence function is approximately zero.
    \State If \(\psi\) is predictive, compute time-dependent calibration, Brier/log losses, dynamic concordance, and compensator or martingale-residual diagnostics.
    \State If uncertainty is conformal, form censoring-aware nonconformity scores and calibrate the required lower or two-sided survival-time bound.
\EndFor
\State Aggregate fold-specific estimates; report the target estimate, uncertainty band or interval, and diagnostics for censoring, positivity, calibration, and residual time-rescaling.
\end{algorithmic}
\end{algorithm}

\begin{theorem}[Out-of-fold compensator validation]\normalfont\label{thm:oof_compensator_validation}
Let \(i=1,\ldots,n\) index independent subjects observed on \([0,\tau]\). For subject \(i\), let \(\mathcal F_i(t)\) be the observed history through time \(t\). Let \(N_i(t)\) be a one-jump or transition-specific counting process, and let \(Y_i(t)\) be predictable. Suppose that, under the data-generating law,
\[
  N_i(t)-\int_0^t Y_i(s)\lambda_{0i}(s)\,ds
\]
is an \(\mathcal F_i(t)\)-martingale, where \(\lambda_{0i}(s)\) is the true conditional event intensity.  Split the sample into fixed or independently randomized folds.  For each validation fold \(V_k\), let \(\mathcal G_{-k}\) be the sigma-field generated by the training folds and let \(\widehat\lambda_{-k,i}(t)\) be any nonnegative bounded learner fitted without using outcomes from \(V_k\).  Assume that \(Y_i(t)\widehat\lambda_{-k,i}(t)\) is predictable with respect to \(\mathcal G_{-k}\vee\mathcal F_i(t-)\).  Define the validation-fold compensator residual
\[
  \widehat R_i(t)
  =
  N_i(t)-\int_0^tY_i(s)\widehat\lambda_{-k,i}(s)\,ds,
  \qquad i\in V_k,
\]
and the integrated calibration error
\[
  B_i(t)
  =
  \int_0^tY_i(s)\{\lambda_{0i}(s)-\widehat\lambda_{-k,i}(s)\}\,ds .
\]
Then, conditional on \(\mathcal G_{-k}\),
\[
  \widehat R_i(t)-B_i(t)
\]
is a martingale with predictable variation \(\int_0^tY_i(s)\lambda_{0i}(s)\,ds\).  Consequently, for any bounded predictable validation weight \(W_i(t)\),
\[
\mathbb E\!\left[
  \sum_{i\in V_k}\int_0^\tau W_i(t)\,d\widehat R_i(t)
  \,\middle|\,\mathcal G_{-k}
\right]
=
\mathbb E\!\left[
  \sum_{i\in V_k}\int_0^\tau W_i(t)Y_i(t)
  \{\lambda_{0i}(t)-\widehat\lambda_{-k,i}(t)\}\,dt
  \,\middle|\,\mathcal G_{-k}
\right].
\]
Thus out-of-fold residual scores are centered exactly when the learner is calibrated in the corresponding weighted integrated-intensity direction.  If the weighted calibration error is \(o_p(|V_k|^{-1/2})\) and the conditional predictable variation has a nondegenerate limit, the standardized residual score has the same Gaussian martingale limit as a correctly specified counting-process score.
\end{theorem}

\begin{proof}
Fix a fold \(k\) and condition on \(\mathcal G_{-k}\).  The learner is then fixed except for its dependence on the validation subject's predictable history.  By the assumed Doob-Meyer decomposition,
\[
M_i(t)=N_i(t)-\int_0^tY_i(s)\lambda_{0i}(s)\,ds
\]
is a martingale in the enlarged filtration \(\mathcal G_{-k}\vee\mathcal F_i(t)\), because \(\mathcal G_{-k}\) is generated by subjects outside the validation fold and the subjects are independent.  Since \(Y_i(t)\widehat\lambda_{-k,i}(t)\) is predictable,
\[
\widehat R_i(t)-B_i(t)
=
N_i(t)-\int_0^tY_i(s)\lambda_{0i}(s)\,ds
=M_i(t),
\]
which proves the martingale claim and gives the predictable variation.  For bounded predictable \(W_i\), stochastic integration preserves the martingale property, so the conditional mean of \(\sum_{i\in V_k}\int W_i\,dM_i\) is zero.  Replacing \(dM_i\) by \(d\widehat R_i-Y_i(\lambda_{0i}-\widehat\lambda_{-k,i})dt\) gives the displayed identity.  The final assertion follows from the martingale central-limit theorem after the deterministic drift term induced by the weighted calibration error is \(o_p(|V_k|^{-1/2})\).
\end{proof}
 
\subsubsection{Parametric Regression Models}

\begin{example}[Parametric regression]\normalfont
Suppose the intensity process \(\lambda=(\lambda_1,\ldots,\lambda_k)\) of a multivariate counting process
\[
    \mathbf N=(N_1,\ldots,N_k)
\]
is specified by a \(q\)-dimensional parameter \(\theta=(\theta_1,\ldots,\theta_q)\in\Theta\). Write
\[
    \lambda_h(t)=\lambda_h(t;\theta).
\]
By Jacod's formula, the partial likelihood is
\[
    L_\tau(\theta)=
    \left\{\prod_{t\in\mathcal T}\prod_{h=1}^k
    \lambda_h(t;\theta)^{\Delta N_h(t)}\right\}
    \exp\left\{-\int_0^\tau\sum_{h=1}^k\lambda_h(t;\theta)\,dt\right\}.
\]
Estimation and inference are based on \(L_\tau(\theta)\). Common examples include Weibull, exponential, and Gumbel type-I distributions \citep{li2021survival}.
\end{example}
 
\subsubsection{Beran's Nonparametric Regression}

    \begin{example}[Beran's conditional NA estimator]
        Beran in 1981 proposed a method to estimate the conditional survival function $S(t|z)$ nonparametrically \citep{beran1981nonparametric}. By the product-integration notation, we have
        $$S(t|z)=\Prodi_{s\le t}\left(1-A(ds|z)\right)$$
        where $A(s|z)$ is the conditional cumulative hazard.  The \emph{Beran's conditional NA estimator} is
        $$\widehat{A}(t|z)=\int_0^t\frac{\sum_{i=1}^nW_i(z)N_i(ds)}{\sum_{i=1}^nW_i(z)Y_i(s)}$$
        where $Y_i(s)$ is the risk process for individual $i$ and
        $W_i(z)=\frac{1}{b_n}K\left(\frac{z-Z_i}{b_n}\right)$
        where $K$ is a density.
    \end{example}
    \textbf{Remark}: tests for the hypothesis $\alpha_{t,z}=\alpha_t$ are discussed in \citet[p.~580]{andersen1993statistical}.

\subsubsection{Conditional Kaplan--Meier Estimates}

Beran's estimator gives the construction; Dabrowska's conditional Kaplan--Meier theory gives the uniform large-sample control for common local weighting schemes \citep{dabrowska1989conditionalKM}. Let \(Z\) be a covariate and let \(S(t|z)=\mathbb P(T>t|Z=z)\). For a target value \(z\), define nonnegative local weights \(W_{ni}(z)\) either by nearest neighbors around \(z\) or by a kernel,
\[
    W_{ni}(z)=
    \frac{K\{(z-Z_i)/b_n\}}
         {\sum_{\ell=1}^nK\{(z-Z_\ell)/b_n\}},
\]
with the nearest-neighbor version obtained by replacing the kernel window by the \(k_n\) closest observations. The localized risk and event processes are
\[
    Y_z(t)=\sum_{i=1}^n W_{ni}(z)Y_i(t),
    \qquad
    N_z(t)=\sum_{i=1}^n W_{ni}(z)N_i(t).
\]
The conditional Nelson--Aalen and Kaplan--Meier estimators are then
\[
    \widehat A(t|z)=\int_0^t \frac{dN_z(s)}{Y_z(s)},
    \qquad
    \widehat S(t|z)=\Prodi_{s\le t}\{1-d\widehat A(s|z)\}.
\]
If \(Z\) is a treatment group or another discrete stratum, these formulas reduce to ordinary stratum-specific Kaplan--Meier curves. If \(Z\) is continuous, the estimator borrows information from nearby covariate values and the bandwidth or neighbor count controls the usual bias--variance tradeoff.

Dabrowska's contribution is important for this article because the estimator is no longer indexed only by calendar time. It is indexed by both \(t\) and \(z\), so consistency requires controlling a whole local class of weighted risk-set ratios. This is the same empirical-process issue that later appears in bivariate survival surfaces and smoothed Cox scores: the counting-process algebra is familiar, but the proof must be uniform over covariate neighborhoods.

\subsubsection{Dabrowska's Smoothed Cox Regression}
 \citet{dabrowska1997smoothed} studied the so-called \emph{smoothed Cox regression model}, where the intensity is of the form
 $$\Lambda(dt)=\mathbb{E}(N(dt)|\mathcal{F}_{t-})=Y(t)\alpha(t,X)\exp(Z^T\beta)dt,$$
 where $Y(t)$ is a 0-1 predictable process and $X$ and $Z$ are covariates. Based on i.i.d. observations, Dabrowska defined the estimator as the solution to the smoothed partial likelihood score equation \citep{martinussen2006dynamic}
 $$\sum_{i=1}^n\int_0^\tau\left\{ Z_i-\frac{S_1(X_i,s,\beta)}{S_0(X_i,s,\beta)} \right\}dN_i(s)=0,$$
 where
 $$S_k(x,s,\beta)=\sum_{i=1}^n K_b(x-X_i)Y_i(t)\exp(\beta^TZ_i(t))Z_i^{\otimes k}$$
 and $K_b(x-X_i)=\frac{1}{b}K(\frac{x-X_i}{b})$ is the kernel weight used in Beran's nonparametric regression. Following the weighted risk-set construction in \citet{andersen1993statistical}, substitution of Beran's estimator gives
 $$\widehat{A}(x,t)=\sum_{i=1}^n\int_0^t\frac{K_b(x-X_i)}{S_0(x,s,\widehat{\beta})}dN_i(s).$$
 In the absence of covariates this reduces to Beran's estimator.

\subsubsection{Cox-Aalen Hazards Model}
 As a special and more applicable case of Dabrowska's smoothed Cox regression, \citet{scheike2002additive} introduced the so-called \emph{Cox-Aalen} model by assuming a parametric form of the baseline hazard $\alpha(t,X)$:
 $$\Lambda(dt)=\mathbb{E}(N(dt)|\mathcal{F}_{t-})=Y(t)\left[X(t)^T\alpha_{t}\right]\exp(Z^T\beta)dt,$$
    which allows time-dependent covariates for the baseline hazard (though Dabrowska's formulation also allows it).


\section{From Association to Causal Questions: IV Analysis with Time-to-Event Outcomes}
\label{sec:iv_survival}
\begin{figure}[tbp]
\centering
\includegraphics[width=\textwidth,height=0.36\textheight,keepaspectratio]{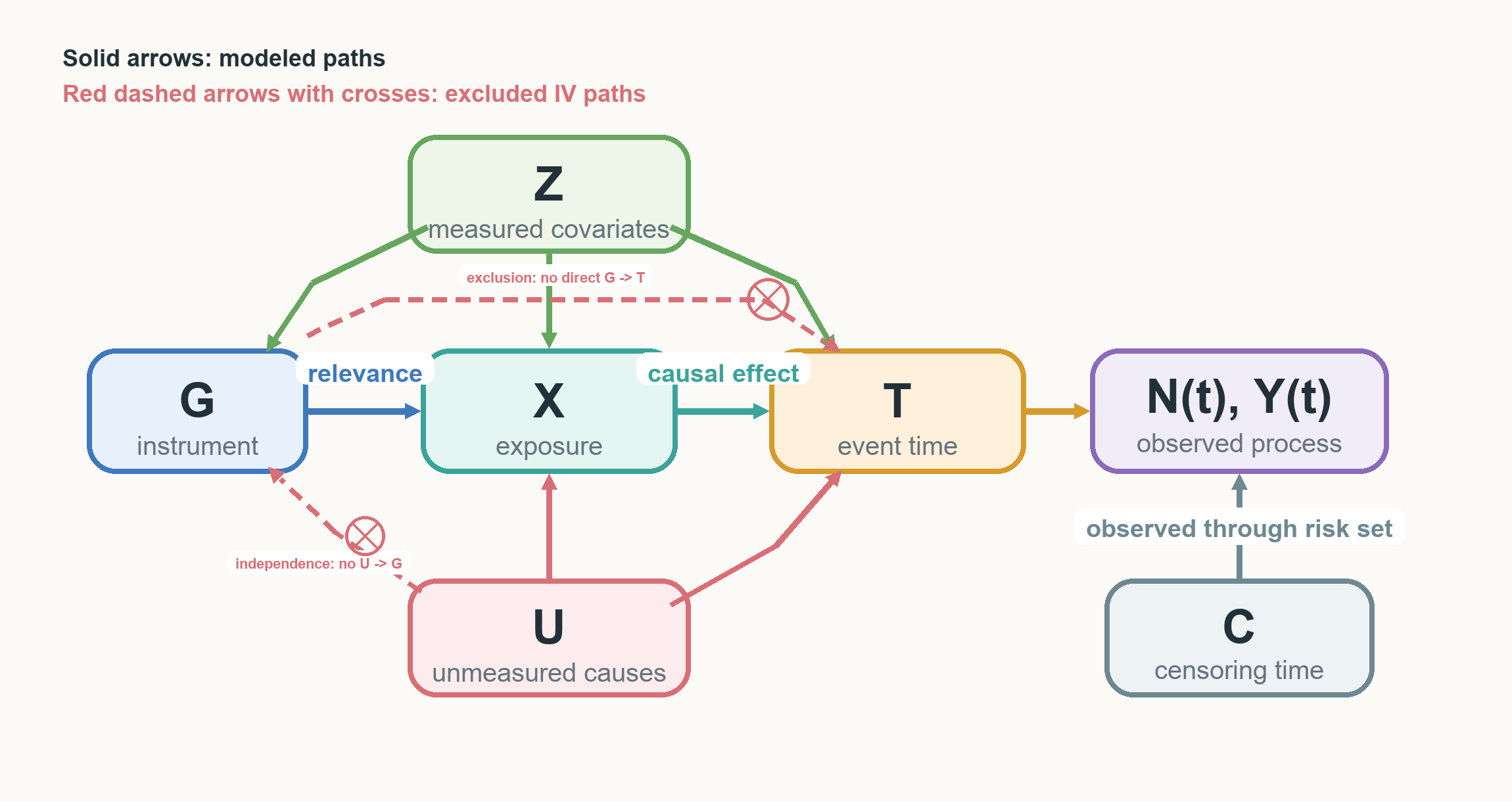}
\caption{Instrumental-variable DAG for survival outcomes.}
\label{fig:iv_survival_structure}
\end{figure}

The preceding models identify associational contrasts: among subjects still at risk, they describe how event rates vary with covariates. Causal contrasts require additional identification assumptions. If an exposure \(X\) is affected by an unmeasured factor \(U\), then a fitted exposure coefficient can combine the effect of \(X\) with the effect of \(U\). Instrumental-variable (IV) analysis introduces a variable \(G\) that predicts \(X\), is independent of unmeasured outcome determinants after conditioning on measured covariates \(Z\), and affects the survival outcome only through \(X\). Following \citet{lu2014dissertation}, a typical DAG for this problem is shown in Figure~\ref{fig:iv_survival_structure}.

If \(Y\) were fully observed and uncensored, the naive least-squares estimating equation would begin with
\[
Y=\beta_1X(U)+\beta_2^TZ+\epsilon(U).
\]
The notation makes the endogeneity explicit: both \(X\) and the residual error can depend on \(U\). After suppressing \(U\) from the notation, the problem appears as
\[
\operatorname{Cov}(X,\epsilon)\neq0,
\]
so ordinary regression no longer isolates the causal effect of \(X\). Survival data add incomplete observation, because the outcome may be right censored or interval censored, for example \(Y\in(L,R]\). The IV estimator must therefore combine a causal moment restriction with a censored-data estimating equation.

The three basic IV assumptions are as follows. First, \(G\) must be \emph{relevant}: after conditioning on \(Z\), it predicts \(X\). Second, \(G\) must be \emph{independent}: after conditioning on \(Z\), it is independent of unmeasured outcome determinants. Third, \(G\) must satisfy the \emph{exclusion restriction}: once \(X\) and \(Z\) are fixed, \(G\) has no direct path to the event time. In a Mendelian-randomization example, these assumptions correspond to genetic variants predicting an exposure such as SBP, being independent of unmeasured health determinants, and affecting time from DM to CVD only through the exposure.

A clinical IV example has the same identification requirements. Suppose \(X\) is treatment initiation and the event is relapse. A proposed instrument \(G\) might be provider preference, local prescribing tendency, or a genetic variant in a Mendelian-randomization design. Relevance requires \(G\) to change treatment exposure. Independence requires \(G\) not to be a proxy for unmeasured prognosis after conditioning on measured covariates. Exclusion requires \(G\) to affect the event time only through treatment. The counting-process component models hazards among subjects at risk, while the IV component supplies exogenous variation in \(X\).

Several models implement this idea for time-to-event outcomes, including simple two-stage least squares (TSLS), accelerated-failure-time (AFT) models, structural Cox multiplicative hazard models, Aalen additive-risk models, and competing-risk extensions built from the same ingredients \citep{martinussen2020instrumental, zheng2017instrumental}.

This section is deliberately narrower than a general theory of survival-IV identification. The displayed estimators are valid only under the stated relevance, independence, exclusion, positivity, censoring, and structural-model assumptions. When those assumptions cannot be defended, especially when the instrument is weak inside the risk sets or plausibly affects competing events directly, the estimators should be treated as sensitivity analyses rather than primary causal evidence.

PBIV provides one censored-outcome route. The next two constructions keep the counting-process vocabulary visible: first an additive-risk construction, where causal effects accumulate over time, and then a structural Cox construction, where the causal effect is expressed as a hazard ratio.

\subsection{Additive Risk Models for IV Analysis}
The additive hazard scale is well suited to IV survival analysis because effects add before they are accumulated. The causal exposure effect is then a cumulative excess-risk curve rather than a single odds ratio or hazard ratio. For subject \(i\), let
\[
    N_i(t)=\mathbb I(T_i\le t,\Delta_i=1),\qquad
    R_i(t)=\mathbb I(T_i\wedge C_i\ge t),
\]
where \(R_i(t)\) is the predictable at-risk indicator and \(C_i\) is the censoring time. A working additive hazard model writes
\[
    \lambda(t;X,G,U)
    =\beta_0(t)+\beta_X(t)X+\beta_G(t)G+\beta_U(t)U.
\]
Here \(\beta_X(t)\) is the instantaneous causal contrast attached to the exposure. Its cumulative version
\[
    B_X(t)=\int_0^t\beta_X(s)\,ds
\]
is the accumulated excess risk up to time \(t\) caused by shifting the exposure by one unit. The IV exclusion restriction requires that, once \(X\) is included, \(G\) does not directly change the hazard, so \(\beta_G(t)=0\). The unmeasured \(U\) may still enter the hazard and may also drive \(X\), which is why ordinary regression on \(X\) can be biased.

The structural version used by \citet{martinussen2017instrumental} avoids specifying the nuisance effect of \(U\) parametrically. It assumes
\[
    \mathbb{E}\left\{dN(t)\mid \mathcal F_{t-}^N,G,X,Z,U\right\}
    =
    R(t)\{d\Omega(t,Z,U)+X\,dB_X(t)\},
\]
where \(\Omega(t,Z,U)\) is an unknown nuisance accumulation. In this model the exposure shifts the additive cumulative hazard by \(X\,dB_X(t)\), while the unmeasured history \(U\) is absorbed into \(\Omega\). Identification then rests on five pieces:
\begin{enumerate}
    \item \emph{Consistency}: the observed event process is the potential event process under the observed exposure.
    \item \emph{Relevance}: \(G\) predicts \(X\) after conditioning on measured baseline covariates \(Z\).
    \item \emph{Independence}: \(G\) is independent of the unmeasured outcome determinants after conditioning on \(Z\).
    \item \emph{Exclusion}: \(G\) affects the event process only through \(X\), not through a direct hazard path.
    \item \emph{Censoring and positivity}: censoring is conditionally independent and the instrument-generated exposure contrast remains nonzero in the risk sets.
\end{enumerate}
These conditions are the survival analogue of the usual IV identifying restrictions, with one extra complication: relevance must hold inside the moving risk set \(R(t)=1\), not only at baseline.

If the first-stage exposure model is \(X=c_0+c_GG+\Delta\) with \(\mathbb{E}(\Delta\mid G)=0\), then \(V=c_0+c_GG\) is the part of the exposure predicted by the instrument. In a simplified constant-effect additive model, the survival curve conditional on \(G\) has the working representation
\[
    \mathbb{P}(T>t\mid G)
    =
    \exp\{-\widetilde B_0(t)-B_X(t)V\},
\]
where \(\widetilde B_0(t)\) collects baseline and nuisance components. This representation motivates the generalized TSLS construction below. The more structural route is the G-estimator, which uses an instrument residual to remove the nuisance \(U\) without estimating \(U\) itself.
\subsubsection{Generalized TSLS Nelson-Aalen Estimator}
The generalized TSLS estimator is a Nelson--Aalen estimator with a two-column design. Let \(c=(c_0,c_G)\), \(\widehat V_i=\widehat c(1,G_i)^T\), and let \(\widehat{\mathbf V}(t)\) be the \(n\times2\) matrix whose \(i\)th row is
\[
\{R_i(t),\,R_i(t)\widehat V_i\},
\]
where \(R_i(t)\) indicates whether subject \(i\) is still at risk. The first column estimates the nuisance cumulative baseline \(\widetilde B_0(t)\); the second column carries the instrument-predicted exposure and identifies \(B_X(t)\). The generalized Nelson--Aalen estimator for \(\mathbf B(t)^T=(\widetilde B_0(t),B_X(t))\) is \citep{andersen1993statistical,martinussen2006dynamic}
\[
\widehat{\mathbf B}(t)=\int_0^t J(u)\widehat{\mathbf V}(u)^-\,d\mathbf N(u),
\]
where \(\widehat{\mathbf V}(u)^-\) is the generalized inverse and \(J(u)=\mathbb I\{\operatorname{rank}\widehat{\mathbf V}(u)=2\}\) turns the estimator off when the two columns cannot identify two effects. Consistency follows under the IV assumptions when the causal effect is time invariant, \(\beta_X(t)=\beta_X\).
This estimator is the counting-process analogue of two-stage least squares: each event time contributes a least-squares update among subjects still at risk. The rank indicator has a substantive role. It checks whether the current risk set still contains enough instrument-induced variation in \(X\) to separate the baseline accumulation from the causal exposure accumulation. Weak instruments or late follow-up with few subjects at risk will inflate variance even when the IV assumptions are correct.

    \subsubsection{G-estimation}

More generally, let \(L\) denote the measured baseline covariates, including the earlier covariate vector \(Z\), and suppose, following \citet{martinussen2017instrumental},
\[
\mathbb{E}\left\{dN(t)|\mathcal{F}_{t-}^N,G,X,L,U\right\}
=\left\{d\Omega(t,L,U)+dB_X(t)X \right\}R(t),
\]
where \(N(t)=\mathbb I(T\le t,\Delta=1)\) is the observed one-jump counting process \citep{andersen1993statistical}, \(\mathcal{F}_{t-}^N\) is the event history just before \(t\), \(R(t)=\mathbb I(T\wedge C\ge t)\) is the observed at-risk indicator, and \(\Omega(t,L,U)\) is an unknown non-negative nuisance accumulation. The target remains
\[
B_X(t)=\int_0^t\beta_X(s)\,ds,
\]
the cumulative causal exposure effect. The central step of G-estimation is to ``blip down'' the observed risk by \(e^{B_X(t)X}\) so that the instrument has no remaining association with the adjusted event increment. This gives the moment condition
\[
\mathbb{E}\left[(G-\mathbb{E}(G|L))e^{B_X(t)X}R(t)\left\{dN(t)-dB_X(t)X\right\}\right]=0.
\]
The estimating equation is the empirical version of this identity. Without loss of generality, assume \(\mathbb{E}(G|L)=0\); otherwise replace \(G\) by \(\widetilde G=G-\mathbb{E}(G|L)\).

\begin{lemma}[G-estimation moment identity]\label{lemma:g_estimation_moment}
Under the structural mean model above, assume \(G\perp U\mid L\), the exclusion restriction holds after conditioning on \(L\), and the censoring process is conditionally independent with
\[
    \mathbb P(C\ge t\mid T\ge t,L,U,G,X)=\pi_C(t,L,U),
\]
where \(\pi_C(t,L,U)\) is positive and does not depend on \(G\) or \(X\). If \(\mathbb E(G\mid L)=0\), then
\[
\mathbb{E}\left[G e^{B_X(t)X}R(t)\left\{dN(t)-dB_X(t)X\right\}\right]=0 .
\]
\end{lemma}
\begin{proof}
First note that the structural model implies
\[
\mathbb{E}\{R(t)\mid L,U,G,X\}
=\pi_C(t,L,U)\exp\{-\Omega(t,L,U)-B_X(t)X\}.
\]
Conditioning on the event history just before \(t\) and using the structural increment model gives
\begin{align*}
&\mathbb{E}\left[G e^{B_X(t)X}R(t)\{dN(t)-dB_X(t)X\}\right]\\
&\quad=\mathbb{E}\left[
G e^{B_X(t)X}R(t)d\Omega(t,L,U)
\right]\\
&\quad=\mathbb{E}\left[
G e^{B_X(t)X}d\Omega(t,L,U)
\mathbb{E}\{R(t)\mid L,U,G,X\}
\right]\\
&\quad=\mathbb{E}\left[
G\,\pi_C(t,L,U)d\Omega(t,L,U)\exp\{-\Omega(t,L,U)\}
\right].
\end{align*}
The final factor is a function of \(t,L,U\). Since \(G\perp U\mid L\) and \(\mathbb E(G\mid L)=0\), its expectation after multiplication by \(G\) is zero.
\end{proof}

The recursive G-estimator is
\[
\widehat{B}_X(t,\widehat{\theta})
=\int_{0}^t
\frac{\sum_iG_i^c(\widehat{\theta})e^{\widehat{B}_X(s-)X_i}\,dN_i(s)}
{\sum_iR_i(s)G_i^c(\widehat{\theta})e^{\widehat{B}_X(s-)X_i}X_i},
\]
where \(G_i^c(\theta)=G_i-\mathbb{E}(G_i|L_i;\theta)\) is the residualized instrument. This residualization removes the part of \(G\) explained by measured baseline information \(L_i\). The exponential term is the blip-down factor, and the denominator is the instrument-weighted exposure contrast among subjects still at risk.

When \(X\) is binary, the recursion can be written as a Volterra integral equation \citep{dabrowska1988kaplan}
\[
A(t)=W(t)+\int_0^tA(s-)\,dU(s),
\]
where \(A(t)=e^{B_X(t)}\) and
\[
\begin{aligned}
W(t)&=\int_0^t\frac{\sum_iG_i^c(\theta)(1-X_i)}{\sum_iR_i(s)G_i^c(\theta)X_i}\,dN_i(s),\\
U(t)&=\int_0^t\frac{\sum_iG_i^c(\theta)X_i}{\sum_iR_i(s)G_i^c(\theta)X_i}\,dN_i(s).
\end{aligned}
\]
The product integral solves the Volterra equation:
\[
A(t)=W(t)+\int_0^t\left[\Prodi_{(s,t]}\left\{1+dU(v)\right\}\right]dW(s).
\]
If \(B_X(t)\) is absolutely continuous and \(\beta_X(t)=\beta_X\), a scalar summary is
\[
\widehat{\beta}_X=\int_0^\tau w(t)\,d\widehat{B}_X(t),
\]
with \(w(t)=\widetilde{w}(t)/\int_0^\tau\widetilde{w}(s)ds\), \(\widetilde{w}(t)=R_*(t)=\sum_iR_i(t)\), and \(\tau\) the upper bound of follow-up. The weight makes the constant-effect estimate emphasize time regions where the risk set still contains information.

\subsubsection{Simulation Illustration}
\begin{figure}[tbp]
\centering
\includegraphics[width=0.84\textwidth,height=0.40\textheight,keepaspectratio]{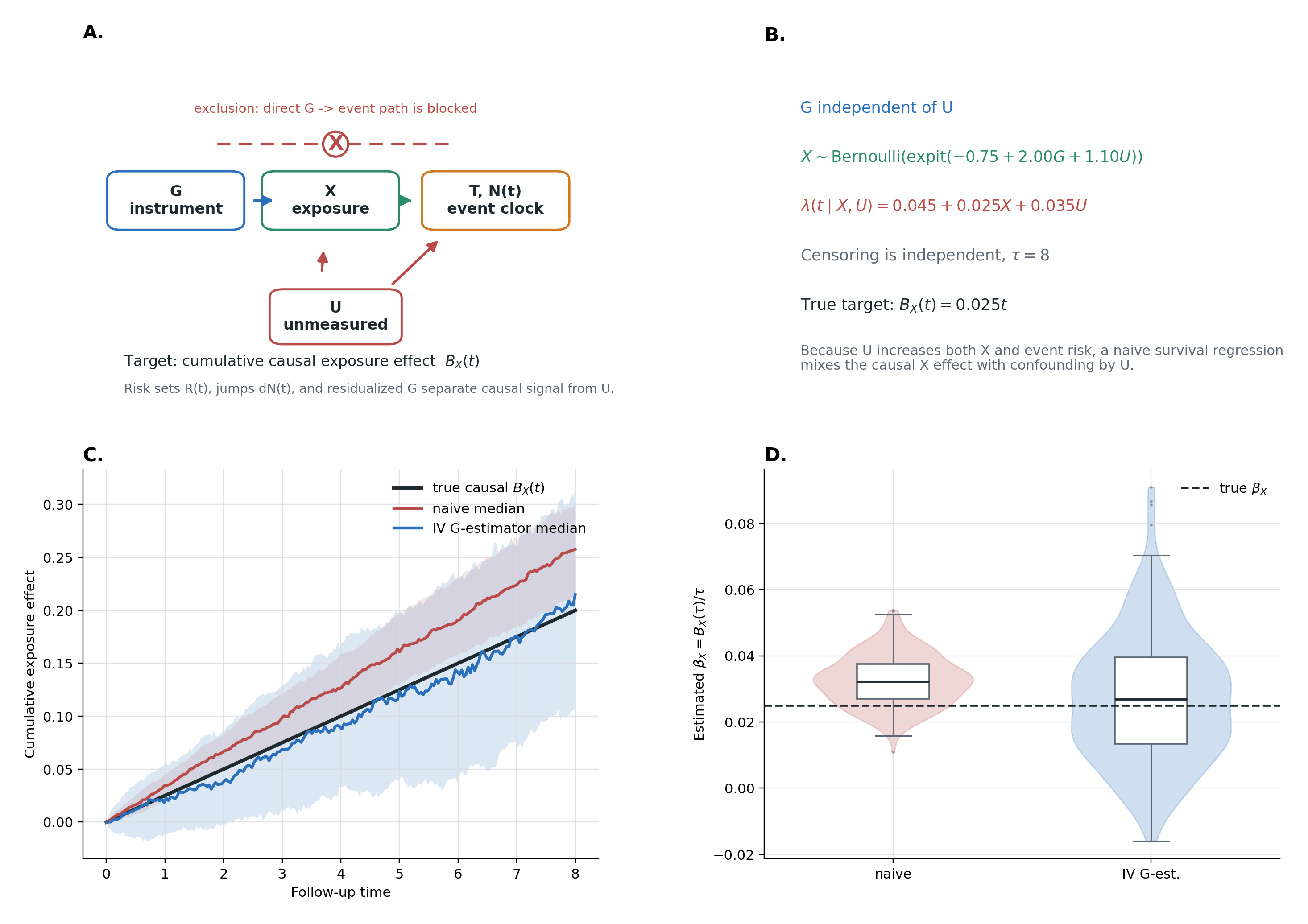}
\caption[Theory and simulation for additive-risk IV estimation]{Theory and simulation for additive-risk IV estimation.}
\label{fig:iv_additive_simulation}
\end{figure}

Figure~\ref{fig:iv_additive_simulation} gives a finite-sample experiment for the additive-risk IV idea. The simulation includes an unmeasured binary confounder \(U\) that increases both the exposure probability and the event hazard. The instrument \(G\) is independent of \(U\), shifts \(X\), and has no direct event path. Under this design, the naive additive-hazard estimator tracks the wrong curve because it attributes part of the \(U\)-effect to \(X\). The recursive IV G-estimator is centered much closer to the true cumulative effect \(B_X(t)\), but its Monte Carlo spread is wider. The simulation therefore illustrates the standard IV bias-variance tradeoff under censoring: reduced confounding bias is obtained at the cost of larger sampling variability and strong dependence on instrument relevance within the risk sets.

The following result records the consistency and influence-function argument behind the recursive estimator; it follows the structure of \citet{martinussen2017instrumental}.
    \begin{theorem}[IV G-estimator]\normalfont
Let
\[
\psi_t(O;B,\theta)=\int_0^t
G^c(\theta)e^{B(s-)X}R(s)\{dN(s)-X\,dB(s)\},
\qquad
\Psi(B,\theta;t)=\mathbb E\{\psi_t(O;B,\theta)\},
\]
where \(O=(N,R,X,G,L)\) and \(G^c(\theta)=G-\mathbb E(G\mid L;\theta)\). Let
\(\mathcal A=\dot\Psi_B(B_X,\theta_0)\) be the derivative of \(B\mapsto\Psi(B,\theta_0)\).
Suppose the following conditions hold:
\begin{enumerate}
\item consistency, relevance, conditional independence \(G\perp U\mid L\), exclusion after conditioning on \(L\), and the censoring condition in Lemma~\ref{lemma:g_estimation_moment} hold;
\item the model for $\mathbb{E}(G|L;\theta)$ is correctly specified and $\widehat\theta\to_p\theta_0$;
\item the risk-set processes satisfy a uniform law of large numbers;
\item the denominator in the recursive estimator is bounded away from zero on $[0,\tau]$;
\item the class \(\{\psi_t(\cdot;B,\theta):t\le\tau,\|B-B_X\|_\infty<\eta,\|\theta-\theta_0\|<\eta\}\) is Donsker for some \(\eta>0\);
\item \(\mathcal A\) has a bounded inverse and \(\widehat\theta\) is asymptotically linear with influence function \(\phi_\theta(O)\).
\end{enumerate}
Then, conditionally on \(L\), the G-estimator $\widehat{B}_X(t,\widehat{\theta})$ is uniformly consistent for $B_X(t)$. Furthermore,
    \[
    W_n(t)=\sqrt{n}\{\widehat{B}_X(t,\widehat{\theta})-B_X(t)\}
    \]
    converges weakly in \(\ell^\infty[0,\tau]\) to a mean-zero Gaussian process with covariance
    \[
    \Sigma(s,t)=\mathbb E\{\phi_s(O)\phi_t(O)\},
    \qquad
    \phi_\cdot(O)=-\mathcal A^{-1}
    \left\{\psi_\cdot(O;B_X,\theta_0)
    +\dot\Psi_\theta(B_X,\theta_0)\phi_\theta(O)\right\}.
    \]
    \end{theorem}
\begin{proof}
Let \(\Psi_n(B,\theta;t)=n^{-1}\sum_{i=1}^n\psi_t(O_i;B,\theta)\). By Lemma~\ref{lemma:g_estimation_moment}, \(\Psi(B_X,\theta_0;t)=0\) for all \(t\le\tau\). The stated uniform law of large numbers and stochastic equicontinuity give
\[
\sup_{t\le \tau}|\Psi_n(B_X,\widehat\theta;t)-\Psi(B_X,\theta_0;t)|\to_p0.
\]
The denominator condition and bounded inverse of \(\mathcal A\) imply local identification of the deterministic solution. Hence the empirical root \(\widehat B_X\) converges uniformly to \(B_X\).

For the limit distribution, expand the estimating equation around \((B_X,\theta_0)\) in sup norm:
\[
0=\Psi_n(\widehat B_X,\widehat\theta;t)
=\Psi_n(B_X,\theta_0;t)
+\mathcal A\{\widehat B_X-B_X\}(t)
+\dot\Psi_\theta(\widehat\theta-\theta_0)(t)+o_p(n^{-1/2}).
\]
The Donsker condition gives
\[
\sqrt n\,\Psi_n(B_X,\theta_0;\cdot)
=n^{-1/2}\sum_{i=1}^n\psi_\cdot(O_i;B_X,\theta_0)
\rightsquigarrow \mathbb G_\psi(\cdot),
\]
and asymptotic linearity of \(\widehat\theta\) gives
\[
\sqrt n(\widehat\theta-\theta_0)=n^{-1/2}\sum_{i=1}^n\phi_\theta(O_i)+o_p(1).
\]
Applying \(\mathcal A^{-1}\) to the linearized equation yields
\[
\sqrt n(\widehat B_X-B_X)(\cdot)
=n^{-1/2}\sum_{i=1}^n\phi_\cdot(O_i)+o_p(1)
\]
with \(\phi_\cdot\) as displayed in the theorem. The functional central limit theorem for the Donsker class therefore gives weak convergence in \(\ell^\infty[0,\tau]\) to a mean-zero Gaussian process with covariance \(\Sigma(s,t)=\mathbb E\{\phi_s(O)\phi_t(O)\}\).
\end{proof}
    The proof uses the same Andersen--Gill martingale expansion and empirical-process regularity conditions that underlie large-sample results for censored counting-process models \citep{andersen1982cox,shorack2009empirical}.
    To study temporal changes, we consider
    $$H_0:\beta_X(t)=\beta_X\Longleftrightarrow H_0:B_X(t)=\beta_Xt.$$

\subsection{Structural Cox Models}
The structural Cox model imposes the same IV restrictions on a multiplicative hazard scale. Let \(T^x\) be the potential survival time had the treatment been set to \(x\) by intervention. The model states \citep{martinussen2019instrumental}
\[
\frac{\lambda_{T^x}(t|X=x,G)}{\lambda_{T^0}(t|X=x,G)}=\exp(\beta x)\quad\text{for all }x.
\]
Thus \(\beta\) is the local causal log-hazard ratio. To estimate it, we use a working Cox model for the observed data,
\[
\lambda_T(t|X=x,G=g)=\lambda_0(t)\exp(\alpha_1x+\alpha_2g),
\]
where \(\lambda_0(t)\) is a nuisance baseline hazard. The IV assumptions imply
\[
T^0\indep G
\quad\text{and}\quad
S_{T^0}(t|X=x,G)=S_{T^x}(t|X=x,G)^{\exp(-\beta x)},
\]
so a candidate value of \(\beta\) transforms the observed survival curve back to the no-exposure potential survival curve. The identifying value of \(\beta\) is the one for which this transformed curve is orthogonal to the instrument. The display below records this identifying equation; a full large-sample theorem requires the same censoring, positivity, and empirical-process regularity conditions stated for the additive-risk G-estimator.
    \subsubsection{G-estimation}
     As in the additive-risk construction, identification is expressed through an orthogonality condition:
     $$\mathbb{E}(GS_{T^0}(t|X,G))=\mathbb{E}(GS_{T^0}(t))=0$$
     under the assumption that $\mathbb{E}G=0$ and
     $$S_{T^0}(t|X,G)=\exp\left(-(\int_0^t\lambda_0(t)dt)e^{\alpha_1X+\alpha_2G}e^{-\beta X}\right).$$
     The estimating equation is the empirical version of this orthogonality condition: plug in the sample average of \(G\), a Breslow estimator of \(\int_0^t\lambda_0(s)ds\), and Cox partial-likelihood estimates of \(\alpha_1\) and \(\alpha_2\). Finally, the asymptotic results are based on Andersen--Gill theory and empirical process arguments \citep{wellner2013weak}.
\section{Appendix: The Mathematical Toolkit}

The appendix collects the definitions and limit tools used in the main text. Compensators, predictable variation, optional variation, observed filtrations, product integrals, intensity measures, and likelihood ratios are process-valued objects rather than scalar parameters. The simulation studies below identify these quantities on sample paths before the formal statements begin, so that each theorem has a small computational counterpart.

\setcounter{table}{0}
\renewcommand{\thetable}{A.\arabic{table}}
\makeatletter
\@ifundefined{theHtable}{}{\renewcommand{\theHtable}{appendix.\arabic{table}}}
\makeatother

\subsection{Application Maps from the Introduction}

The cross-domain examples in the introduction all use the same bookkeeping. Table~\ref{tab:example_scope} records the risk set, counted jump, and statistical target for each application class.

\begingroup
\footnotesize
\setlength{\tabcolsep}{1.8pt}
\renewcommand{\arraystretch}{1.02}
\setlength{\LTpre}{2pt}
\setlength{\LTpost}{2pt}
\begin{longtable}{@{}>{\RaggedRight\arraybackslash}p{0.16\textwidth}
                >{\RaggedRight\arraybackslash}p{0.25\textwidth}
                >{\RaggedRight\arraybackslash}p{0.30\textwidth}
                >{\RaggedRight\arraybackslash}p{0.22\textwidth}@{}}
\caption[Examples translated into counting-process notation]{Examples translated into the same risk-set and counting-process notation.}
\label{tab:example_scope}\\[-1pt]
\hline
Application & At-risk process \(Y(t)\) & Jump process \(N_j(t)\) & Estimand or observation issue \\
\hline
\endfirsthead
\multicolumn{4}{@{}l}{\footnotesize\textit{Table~\thetable\ continued.}}\\
\hline
Application & At-risk process \(Y(t)\) & Jump process \(N_j(t)\) & Estimand or observation issue \\
\hline
\endhead
Clinical survival & Patients alive, event-free, and still under observation just before \(t\). & Death, relapse, infection, transplant, readmission, or disease-state transition. & Survival, cumulative incidence, transition probabilities, recurrent-event rates; censoring and delayed entry.\\
Illness-death histories & Patients currently in the relevant disease state and still observable at \(t-\). & Recovery, relapse, transplant, terminal event, or transition between disease states. & State-occupation probabilities, transition probabilities, and transition-specific cumulative hazards.\\
Recurrent-event follow-up & Subjects still under observation and eligible for another event at \(t-\). & Rehospitalization, infection, device alert, maintenance visit, or other repeatable event. & Mean cumulative function, rate contrasts, gap-time effects, and robust risk-set comparisons.\\
Panel-count inspection & Subjects whose cumulative event counts are observed only at scheduled or random inspections. & Increment in the latent recurrent-event count between inspection times. & Mean count function and interval likelihood when exact event times are hidden.\\
Reliability and operations & Devices, accounts, machines, or service units still eligible for a logged event at \(t-\). & Degradation, repair, replacement, direct failure, outage, recurrent alert, or inspection count. & Transition-specific hazards, mean recurrence burden, panel-count likelihoods, and landmark prediction.\\
Economic duration records & Jobs, loans, firms, subscriptions, or accounts active and observed just before \(t\). & Employment exit, prepayment, default, churn, reactivation, bankruptcy, acquisition, or liquidation. & Duration dependence, competing exits, dynamic default or churn prediction, and recurrent account events.\\
Financial point processes & Orders, cancellations, trades, failures, claims, or operational losses eligible for observation just before \(t\). & Arrival of a marked event, price jump, cancellation, claim, loss, failure, or recovery. & Conditional intensity, compensator diagnostics, excitation, clustering, and change-point monitoring.\\
Spatial disease mapping and environmental survival & Persons, facilities, communities, or spatial cells located in a study region and still observable just before \(t\). & Diagnosis, infection, death, failure, claim, or other event marked by location and time. & Spatial intensity surface, space--time compensator, spatial frailty, exposure-gradient effect, and residual maps.\\
Literary reception and book history & Works, authors, editions, or cited items present in a catalogue, archive, review corpus, or bibliography and not yet having the target reception event at \(t-\). & First review, translation, serialization installment, reprint, anthologization, canonization, censorship, archival appearance, or disappearance from print. & Reception survival curves, weighted log-rank comparisons by genre or period, competing-risk cumulative incidences, interval-censored discovery times, and panel-count reprint or citation means.\\
Historical office and institutional careers & Officials, ministers, commanders, regimes, treaties, or laws still in force, in office, or eligible for a transition just before \(t\). & Appointment, promotion, transfer, demotion, dismissal, recall, exile, death, conquest, coup, reform, or treaty termination. & Tenure survival, cause-specific exits, Cox/Aalen/AFT contrasts, Aalen--Johansen career paths, and semi-Markov sojourn effects.\\
Instrumental-variable survival & Subjects with an instrument, exposure process, endpoint process, and censoring process defined on the same filtration. & Treatment initiation, competing event, endpoint, censoring, or residualized exposure jump. & Structural cumulative survival or hazard contrasts under relevance, exclusion, independence, and positivity assumptions.\\
\hline
\end{longtable}
\endgroup

The remaining appendix tables translate the same bookkeeping into concrete data-set families. They are meant as reference menus: a reader can locate a data structure, identify the risk set and jumps, and then return to the relevant estimator, test, or model in the main text.

\begingroup
\footnotesize
\setlength{\tabcolsep}{1.8pt}
\renewcommand{\arraystretch}{1.02}
\setlength{\LTpre}{2pt}
\setlength{\LTpost}{2pt}
\begin{longtable}{@{}>{\RaggedRight\arraybackslash}p{0.19\textwidth}
                >{\RaggedRight\arraybackslash}p{0.27\textwidth}
                >{\RaggedRight\arraybackslash}p{0.29\textwidth}
                >{\RaggedRight\arraybackslash}p{0.17\textwidth}@{}}
\caption[Biomedical data-set examples and model choices]{Biomedical data-set examples and the tests or models they naturally motivate.}
\label{tab:bio_dataset_menu}\\[-1pt]
\hline
Biology or medicine data set & Observation scheme & Associated tests and models & Counting-process target \\
\hline
\endfirsthead
\multicolumn{4}{@{}l}{\footnotesize\textit{Table~\thetable\ continued.}}\\
\hline
Biology or medicine data set & Observation scheme & Associated tests and models & Counting-process target \\
\hline
\endhead
Liver disease, oncology, or transplant follow-up & Time from enrollment to death, relapse, graft failure, or transplant; labs and treatment histories may change over time. & Kaplan--Meier, Nelson--Aalen, log-rank or weighted log-rank tests, Cox model, AFT model, Schoenfeld and martingale-residual diagnostics. & Survival \(S(t)\), cumulative hazard \(A(t)\), hazard ratio, RMST, time-varying effect.\\
Randomized cancer or cardiovascular trials & Treatment arms with right censoring, delayed entry, stratification, and non-proportional hazards. & Stratified log-rank, Fleming--Harrington weights, Cox with time interaction, RMST contrast, landmark prediction. & Treatment contrast on hazard, survival, or restricted-mean scale.\\
Bone-marrow transplant and illness-death registries & Patients move among remission, relapse, transplant, adverse event, and death states. & Aalen--Johansen, transition-specific Cox, Markov or semi-Markov model, landmark state prediction, product-integral sensitivity calculation. & State occupation, transition probability, transition-specific cumulative hazard.\\
Recurrent infection, readmission, seizure, or hospitalization records & Subjects remain under observation and may experience repeated events before censoring or death. & Andersen--Gill, PWP total-time or gap-time model, frailty model, mean cumulative function, robust recurrent-event score tests. & Rate function, mean cumulative count, event-order effect, heterogeneity.\\
Competing toxicity, relapse, and death endpoints & Several terminal or semi-terminal outcomes prevent or alter interpretation of the target event. & Cause-specific Cox, Gray test, Fine--Gray model, Aalen--Johansen cumulative incidence, semi-competing-risk model. & Cumulative incidence, cause-specific hazard, subdistribution hazard.\\
Paired organs, family data, twin studies, or paired biomarkers & Two or more correlated event times are observed with censoring. & Dabrowska bivariate product-limit estimator, copula model, frailty model, Kendall \(U\)-process dependence test. & Joint survival surface, dependence parameter, marginal versus shared-risk effect.\\
Geocoded cancer, infection, mortality, or environmental-exposure registries & Subjects or spatial cells have locations; follow-up may be censored, delayed, or aggregated by region. & Spatial frailty Cox model, log-Gaussian Cox process, areal random-effects hazard, residual intensity map, space--time scan or compensator check. & Local hazard surface, spatial risk gradient, cluster detection, environmental exposure effect.\\
Screening, registry inspection, and panel-count studies & Failure or recurrence is known only between visits, or cumulative counts are read at inspections. & Turnbull NPMLE, interval-censored Cox, panel-count NPMLE, isotonic mean-count estimator, Wellner--Zhang two-sample procedures. & Event-time distribution or mean count function under hidden event times.\\
Biobank genomics and Mendelian-randomization survival endpoints & Genetic variants, exposures, event times, competing events, and censoring are recorded on the same follow-up scale. & Survival IV, structural Cox, additive-risk IV, PBIV/AFT models, weak-instrument and exclusion sensitivity checks. & Causal survival or hazard contrast under relevance, independence, exclusion, and censoring assumptions.\\
\hline
\end{longtable}
\endgroup

\begingroup
\footnotesize
\setlength{\tabcolsep}{1.8pt}
\renewcommand{\arraystretch}{1.02}
\setlength{\LTpre}{2pt}
\setlength{\LTpost}{2pt}
\begin{longtable}{@{}>{\RaggedRight\arraybackslash}p{0.19\textwidth}
                >{\RaggedRight\arraybackslash}p{0.27\textwidth}
                >{\RaggedRight\arraybackslash}p{0.29\textwidth}
                >{\RaggedRight\arraybackslash}p{0.17\textwidth}@{}}
\caption[Economic and financial data-set examples and model choices]{Economic and financial data-set examples translated into event-history notation.}
\label{tab:econ_finance_dataset_menu}\\[-1pt]
\hline
Economic or financial data set & Observation scheme & Associated tests and models & Counting-process target \\
\hline
\endfirsthead
\multicolumn{4}{@{}l}{\footnotesize\textit{Table~\thetable\ continued.}}\\
\hline
Economic or financial data set & Observation scheme & Associated tests and models & Counting-process target \\
\hline
\endhead
Unemployment, job tenure, and welfare-spell records & Entry into unemployment or employment is observed; exits may be jobs, quits, recall, retirement, or censoring. & Cox or piecewise-exponential duration model, AFT model, frailty, duration-dependence tests, left-truncation adjustment. & Exit hazard, duration dependence, covariate effect, unobserved heterogeneity.\\
Mortgage or consumer-loan performance panels & A loan can prepay, default, refinance, become delinquent, cure, or remain active; rates and borrower covariates change monthly. & Competing-risk Cox, Fine--Gray model, multi-state delinquency model, Gray test, landmark default prediction. & Prepayment or default incidence, transition probability, dynamic risk score.\\
Corporate bankruptcy and credit-default histories & Firms enter the risk set after listing or rating; default is rare and covariates are balance-sheet and market histories. & Discrete-time hazard, Cox with time-varying covariates, Aalen additive hazard, frailty, calibration and martingale-residual checks. & Default intensity, macro sensitivity, solvency transition risk.\\
Credit-card delinquency and collections data & Accounts repeatedly miss payments, cure, roll forward, enter collections, or charge off. & Andersen--Gill, PWP gap-time model, recurrent-event rate tests, multi-state transition matrix, semi-Markov cure/default model. & Mean delinquency burden, recurrence rate, state occupation, charge-off risk.\\
High-frequency trades, quotes, and limit-order books & Orders, cancellations, trades, and price jumps form marked event streams with strong history dependence. & Hawkes or marked point-process likelihood, compensator time-rescaling, likelihood-ratio test, residual autocorrelation check. & Self-excitation, cross-excitation, event clustering, market-impact intensity.\\
Venture capital, IPO, merger, or firm-life-cycle data & Firms are followed from founding, funding, listing, or rating until IPO, acquisition, liquidation, or censoring. & Aalen--Johansen, cause-specific Cox, Fine--Gray, stratified Cox, covariate-dependent transition probabilities. & Competing exit probabilities, time to liquidity, liquidation risk.\\
Bank failure, insurance claim, or operational-loss histories & Institutions or policies are at risk until failure, claim, closure, merger, or administrative censoring. & Cox process, frailty or random-effects hazard, martingale CUSUM, change-point test, excess-risk monitoring. & Failure intensity, stress-period effect, recurrent claim burden.\\
Subscription, churn, and platform-retention panels & Users subscribe, pause, renew, churn, reactivate, or receive treatment offers over calendar time. & Landmark prediction, recurrent-event model, semi-Markov sojourn model, causal survival model with time-varying treatment. & Retention survival, reactivation probability, treatment effect, elapsed-time dependence.\\
\hline
\end{longtable}
\endgroup

\begingroup
\footnotesize
\setlength{\tabcolsep}{1.8pt}
\renewcommand{\arraystretch}{1.02}
\setlength{\LTpre}{2pt}
\setlength{\LTpost}{2pt}
\begin{longtable}{@{}>{\RaggedRight\arraybackslash}p{0.19\textwidth}
                >{\RaggedRight\arraybackslash}p{0.27\textwidth}
                >{\RaggedRight\arraybackslash}p{0.29\textwidth}
                >{\RaggedRight\arraybackslash}p{0.17\textwidth}@{}}
\caption[Literary and historical data-set examples and model choices]{Literary and historical data-set examples translated into event-history notation.}
\label{tab:literary_history_dataset_menu}\\[-1pt]
\hline
Literary or historical data set & Observation scheme & Useful tests and models & Counting-process target \\
\hline
\endfirsthead
\multicolumn{4}{@{}l}{\footnotesize\textit{Table~\thetable\ continued.}}\\
\hline
Literary or historical data set & Observation scheme & Useful tests and models & Counting-process target \\
\hline
\endhead
Novel, poetry, or newspaper-review corpus & Works enter at first publication or first archival visibility; events are first review, serialized installment, translation, reprint, canonization, censorship, or loss from print. & Nelson--Aalen and Kaplan--Meier reception curves, weighted log-rank tests by genre or period, Cox/Aalen/AFT models for publisher, language, school, or era covariates. & Time to reception, canonization, or disappearance; hazard and survival contrasts.\\
Translation, reprint, and anthology histories & A work remains eligible for repeated translations, editions, anthologies, or citations while it is visible to the sampled corpus. & Andersen--Gill, PWP, WLW, recurrent-event score tests, mean cumulative function, and robust variance for within-work dependence. & Reception rate, expected cumulative editions or citations, event-order effects.\\
Competing literary fates & The first major post-publication fate is review, translation, censorship, canonization, archival loss, or out-of-print status. & Aalen--Johansen cumulative incidence, Gray-type comparisons, cause-specific Cox, Fine--Gray summaries, and transition-specific hazards. & Probability of each reception pathway and cause-specific intensity.\\
Manuscript transmission and archive discovery & Composition, copying, discovery, or catalogue entry times are bounded by dated witnesses, inventories, auctions, or library catalogues. & Turnbull interval-censoring NPMLE, interval-censored Cox, bivariate interval-censored models for two documentary clocks. & Distribution of hidden discovery or transmission times; joint survival of paired documentary events.\\
Correspondence, citation, and reception networks & Author--author, work--work, or institution--work dyads enter once both nodes exist; jumps are letters, citations, mentions, reviews, or acquisitions. & Recurrent-event rate models, multiplicative-intensity diagnostics, residual-process checks, stratified log-rank tests, and marked point-process summaries. & Communication intensity, citation burden, burstiness, cross-institution differences.\\
Ministerial, court, or parliamentary careers & Office-holders are at risk while in a post; exits are promotion, transfer, dismissal, recall, exile, death, retirement, coup, or defeat. & Tenure Kaplan--Meier curves, log-rank or weighted log-rank tests across dynasties, reigns, parties, or factions; Cox, Aalen additive, and AFT duration models. & Tenure survival, rank or faction effects, duration dependence.\\
Bureaucratic and institutional state sequences & Subjects move among examination success, local office, court office, military command, exile, retirement, reform office, and terminal exit. & Aalen--Johansen transition probabilities, transition-specific Cox models, Markov renewal summaries, and semi-Markov transformation models for elapsed time in the current state. & State occupation, transition probability, sojourn-time effect, career-path prediction.\\
Reforms, quotas, decrees, and shocks & A rule change, quota, border closure, printing ban, or patronage shock changes exposure while censoring and competing events still operate. & Landmark analysis around reform dates, structural cumulative survival or hazard models, IV sensitivity analyses, and graphical checks of exclusion, positivity, and censoring. & Dynamic effect of shocks on survival, transition, or reception probabilities.\\
\hline
\end{longtable}
\endgroup

\subsection{How to Use This Appendix}

The appendix is a technical reference map for the main text. A reader can start with the estimator of interest, locate the stochastic object it uses, and then read only the theorem needed for the corresponding proof. Table~\ref{tab:appendix_tool_map} gives that map explicitly.

\begin{table}[tbp]
\centering
\begingroup
\footnotesize
\setlength{\tabcolsep}{2.0pt}
\renewcommand{\arraystretch}{1.00}
\caption{Appendix tools and where they are used in the article.}
\label{tab:appendix_tool_map}
\begin{tabular}{@{}p{0.22\textwidth}p{0.26\textwidth}p{0.27\textwidth}p{0.17\textwidth}@{}}
\hline
Main-text target & Appendix object & What the object proves & Typical simulation companion\\
\hline
Nelson--Aalen, Kaplan--Meier, Greenwood & Compensator, predictable variation, Lenglart inequality & Uniform consistency and variance estimation for cumulative-hazard and survival processes & Martingale tools and confidence-band figures\\
Aalen--Johansen, product-limit transition probabilities & Matrix product integrals and Duhamel equation & Continuity and first-order expansion of transition-probability estimators & Product-integral and Duhamel figures\\
Log-rank tests and Cox--Andersen--Gill regression & Martingale scores, stochastic integrals, partial likelihood & Centering, information, and Gaussian score limits from risk-set comparisons & Cox risk-set and diagnostic figures\\
Dabrowska, Beran, smoothed Cox, and kernel estimators & Weighted risk sets and empirical-process control & Uniformity over covariate or bivariate-indexed classes & Bivariate survival and regression illustrations\\
Turnbull, bivariate interval censoring, Wellner--Zhang panel counts & Interval likelihoods and monotone NPMLE geometry & Self-consistency, isotonic constraints, and nonstandard proof routes & Interval-censoring and panel-count figures\\
Semi-Markov, landmark, BNP, and IV examples & Observed filtrations, residual risk sets, estimating equations & Predictable filtration components at the prediction or estimating time & Elapsed-time, landmark, BNP, and IV simulations\\
\hline
\end{tabular}
\endgroup
\end{table}

\paragraph{Standing assumptions.}
Unless stated otherwise, all processes are considered on a fixed horizon \([0,\tau]\) and are adapted to an observed filtration \(\mathbb F\). Subjects are independent observational units; covariates, treatment histories, inspection indicators, and at-risk processes are predictable; and the relevant cumulative hazards or intensity matrices have locally bounded variation. Risk-set denominators are assumed to be bounded away from zero on the time region where an estimand is reported. Right censoring, delayed entry, and inspection times are conditionally independent given the filtration specified in the corresponding section. Multistate examples exclude simultaneous transitions unless the displayed estimator explicitly handles ties. Product-integral arguments assume local jump sizes smaller than one, while likelihood and estimating-equation arguments assume the displayed denominators remain nonsingular with probability tending to one.

\paragraph{What is proved here.}
The appendix gives full derivations for the recurring tools: compensators, stochastic integration, the innovation theorem, Rebolledo's martingale central limit theorem, Gill's lemma, Lenglart's inequality, the functional delta method, Duhamel's equation, product-integral transition functions, and likelihood ratios for counting processes. For estimators whose full asymptotic theory is genuinely nonstandard, such as Turnbull's NPMLE, bivariate interval-censoring NPMLEs, Wellner--Zhang panel-count estimators, and kernel-smoothed regression estimators, the main text gives the estimating object, the proof route, the simulation companion, and the reference where the complete regularity theory is developed. This separates results handled by martingale arguments from those requiring monotone likelihood or empirical-process machinery.

\subsection{Simulation Studies for Appendix Quantities}

\begin{figure}[tbp]
\centering
\includegraphics[width=0.82\textwidth,height=0.32\textheight,keepaspectratio]{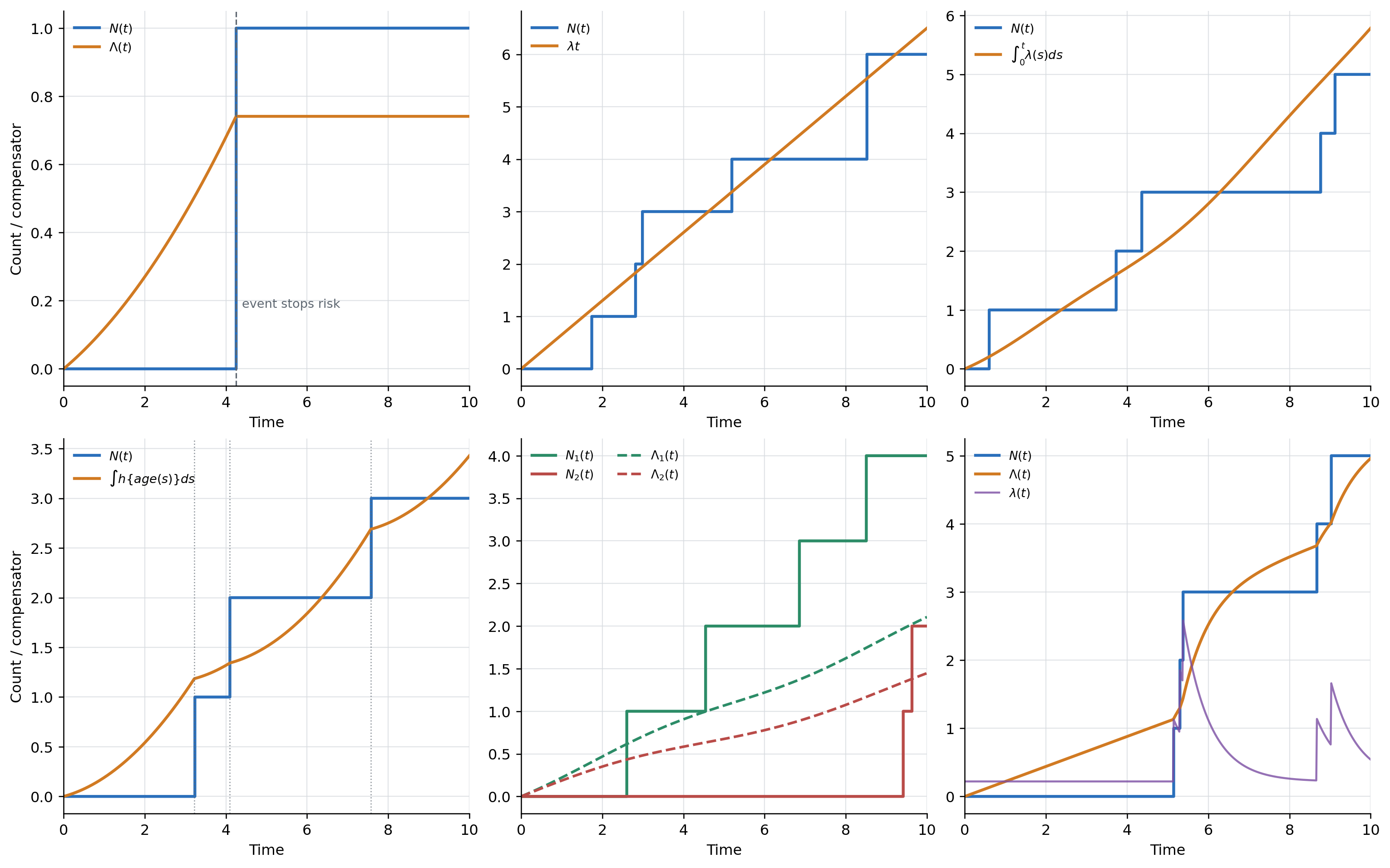}
\caption[Simulation study for common counting and point processes]{Simulation study for common counting and point processes.}
\label{fig:appendix_process_zoo}
\end{figure}

Figure~\ref{fig:appendix_process_zoo} starts with the object underlying the later martingale arguments: a point process is a random collection of event times, while a counting process is the cumulative count generated by those times. The compensator \(\Lambda\) is the predictable cumulative component of that count. For a homogeneous Poisson process it is a straight line, for a nonhomogeneous Poisson process it is a deterministic curve, for a one-jump survival process it stops accumulating after the subject leaves the risk set, for a renewal process its slope depends on elapsed time since the last jump, for a marked process it splits into component-wise compensators, and for a self-exciting process it is driven by the history of previous jumps.

\begin{figure}[tbp]
\centering
\includegraphics[width=0.82\textwidth,height=0.32\textheight,keepaspectratio]{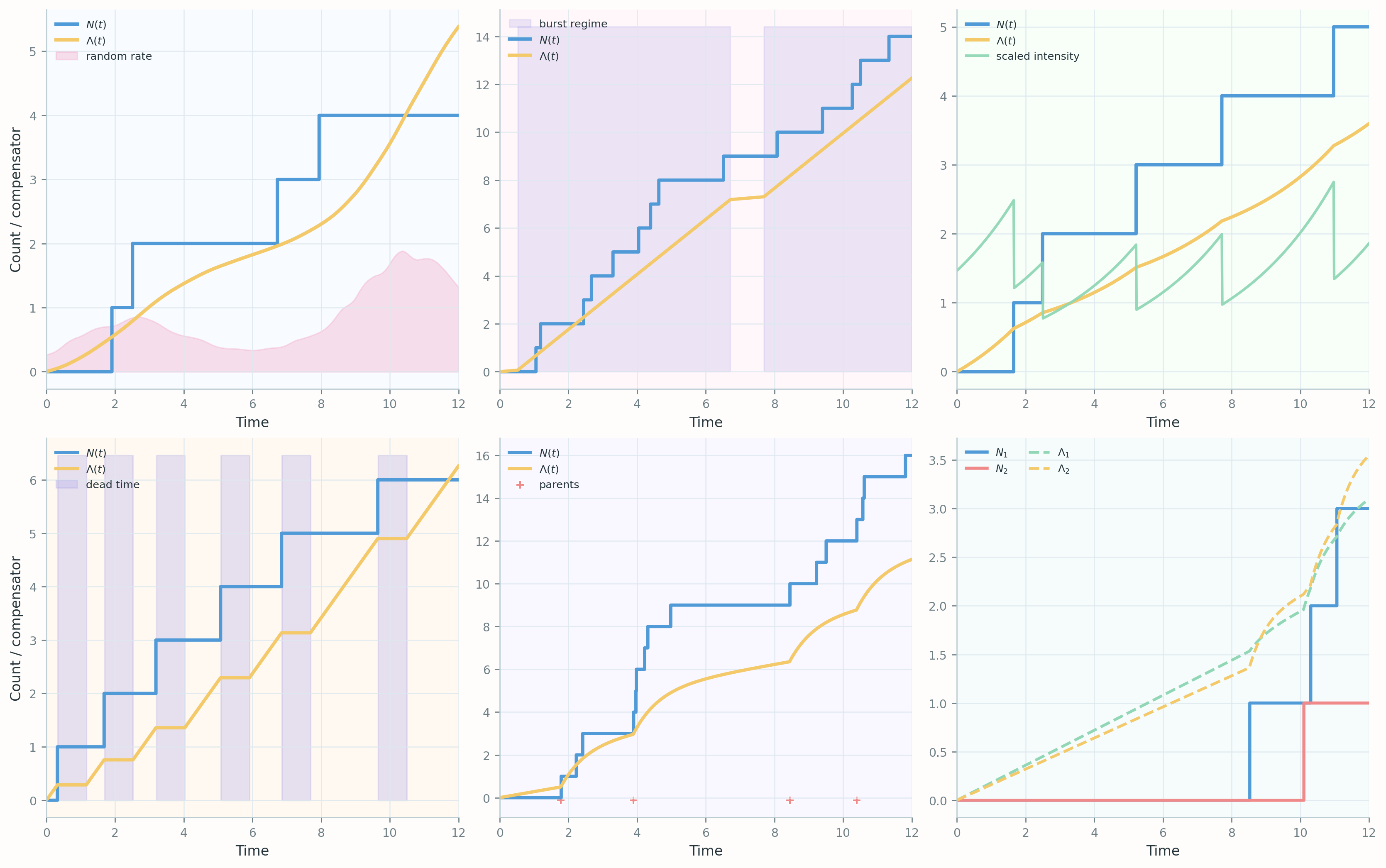}
\caption[Simulation study for additional point-process models]{Simulation study for additional point-process models.}
\label{fig:appendix_exotic_process_zoo}
\end{figure}

Figure~\ref{fig:appendix_exotic_process_zoo} extends the display to additional point-process models. A Cox process has a stochastic intensity; a Markov-modulated Poisson process changes intensity according to an unobserved or observed finite-state process; a self-correcting process increases its conditional intensity between jumps and decreases it after a jump; a refractory process excludes events during a fixed post-event interval; a parent-offspring cluster process represents delayed secondary events; and a mutually exciting marked Hawkes process lets one event type increase the future intensity of another. Spatial and space--time point processes add a location argument to the same object: events are counted over regions as well as over time, and the compensator becomes an integral over area, time, and the observed risk process \citep{moller2004statistical,baddeley2015spatial,diggle2013spatial}. These examples are not restricted to survival analysis. They show that the compensator is a predictable conditional-mean measure for a counting process, not a device specific to right-censored data.

\FloatBarrier
The zoo above is still not exhaustive. Determinantal point processes, introduced in the coincidence approach of \citet{macchi1975coincidence} and developed as fermion point fields in later accounts \citep{shirai2003random,hough2009zeros}, specify repulsion through correlation kernels:
\[
    \rho_n(x_1,\ldots,x_n)
    =
    \det\{K(x_a,x_b)\}_{a,b=1}^n .
\]

A permanental, or boson, point process replaces the determinant by a permanent,
\[
    \rho_n(x_1,\ldots,x_n)
    =
    \operatorname{per}\{K(x_a,x_b)\}_{a,b=1}^n ,
\]
and therefore represents attraction or clustering rather than inhibition \citep{shirai2003random}. These processes are most common in spatial statistics, random-matrix theory, spectral problems, and diverse subset sampling.

Their survival connection is through the first arrival and the void probability. If \(T=\inf\{t:N((0,t])>0\}\), then
\[
    S_T(t)=\Prob(T>t)=\Prob\{N((0,t])=0\}.
\]
For a determinantal process over a set \(B\), this void probability is a Fredholm determinant such as \(\det(I-K_B)\); for a permanental process it has the corresponding determinant-power form. When \(S_T\) is differentiable, the first-event hazard is \(-d\log S_T(t)/dt\), and the observed survival counting process is the stopped, censored, one-jump projection of the richer point process. Thus DPPs and permanental processes are not replacements for Kaplan--Meier or Cox regression by themselves; they become survival models only after the observation scheme, risk process, censoring mechanism, and first-event or recurrent-event target have been specified.

\FloatBarrier
\begin{table}[!t]
\centering
\caption[Point-process families and survival connections]{Point-process families and survival connections.}
\label{tab:point_process_survival_map}
\small
\begin{tabular}{p{0.23\textwidth}p{0.32\textwidth}p{0.35\textwidth}}
\toprule
Point-process family & Dependence mechanism & Survival-analysis connection \\
\midrule
Poisson and nonhomogeneous Poisson & Independent increments with deterministic mean measure & Baseline cumulative hazard gives \(S(t)=\exp\{-\Lambda(t)\}\) before censoring or covariate risk sets enter. \\
Renewal and Markov-renewal & Conditional rate depends on gap time and possibly next-state destination & Gap-time recurrent-event models and semi-Markov multistate survival models. \\
Cox, frailty, and log-Gaussian Cox & Random intensity creates overdispersion and shared latent risk & Frailty survival, spatial survival, and predictable projection when the latent field is unobserved. \\
Hawkes and cluster processes & Previous or parent events increase future event rates & Recurrent events, contagion, relapse cascades, order arrivals, and dependent terminal-event settings. \\
Self-correcting, refractory, and Gibbs processes & Prior events suppress future events or impose local interaction potentials & Minimum spacing, recovery windows, and inhibition constraints in event histories or spatial failure maps. \\
Determinantal processes & Kernel determinant makes nearby points repel & First-event survival is a void probability; useful when event locations or times are inhibited rather than clustered. \\
Permanental processes & Kernel permanent makes points attract or cluster & First-event survival is again a void probability, but the full likelihood describes clustering beyond a one-jump survival curve. \\
\bottomrule
\end{tabular}
\end{table}
\FloatBarrier

\begin{figure}[tbp]
\centering
\includegraphics[width=0.96\textwidth]{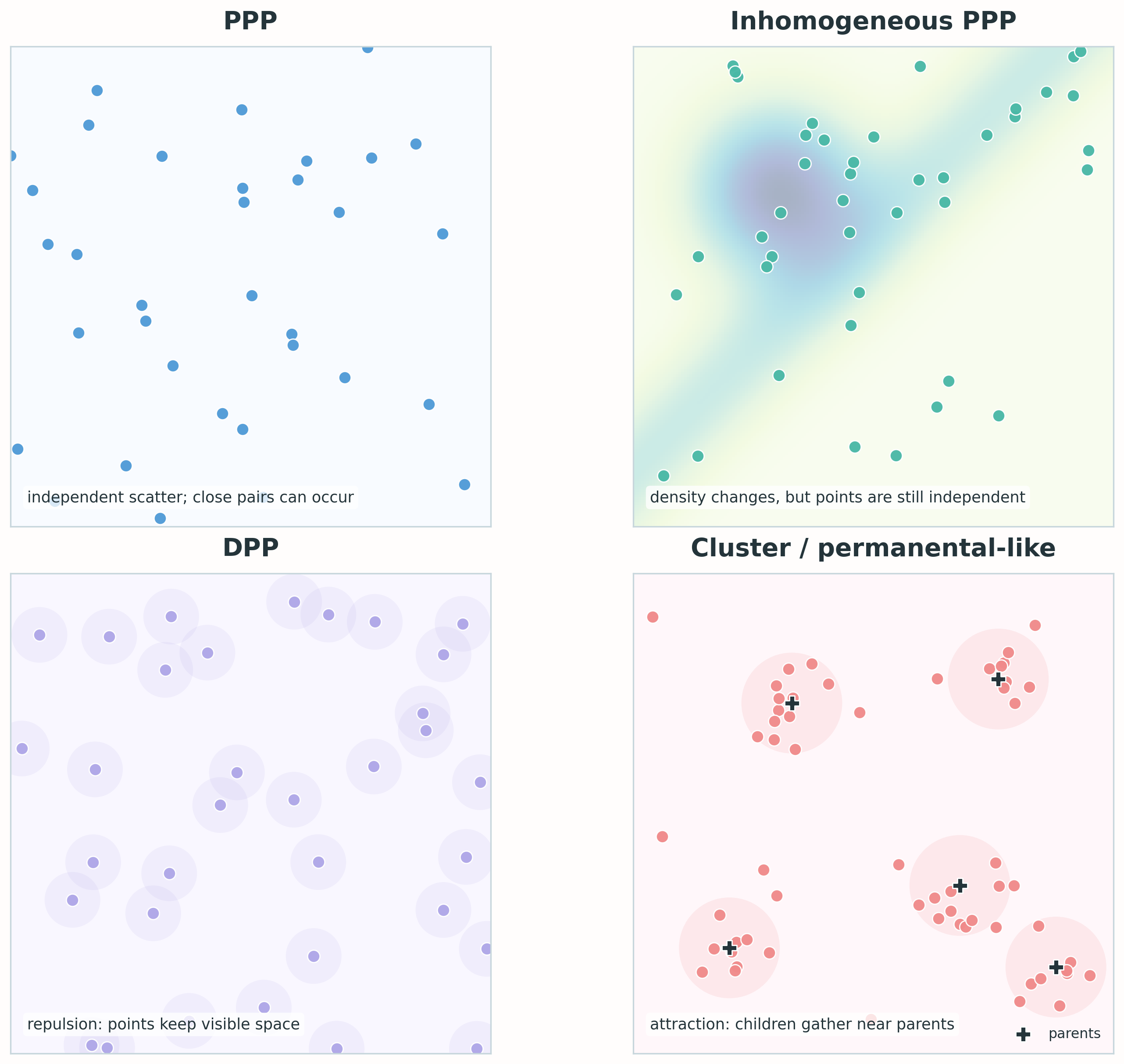}
\caption[Planar schematic for PPP, DPP, and related point processes]{Planar schematic for PPP, DPP, and related point processes.}
\label{fig:appendix_planar_point_processes}
\end{figure}

Figure~\ref{fig:appendix_planar_point_processes} gives the standard planar picture. A homogeneous PPP has no interaction among points, so close pairs can occur by chance. An inhomogeneous PPP changes density over the window but still has conditional independence given its intensity. A planar DPP is more even because nearby locations repel each other; the translucent halos are visual guides for that local spacing. A permanental-like or parent-offspring cluster pattern does the opposite: points are attracted toward latent parents, so the eye sees clumps rather than gaps. This is the cleanest way to distinguish the main interaction stories before reducing the process to first arrivals, recurrent event times, or marked survival histories.

\FloatBarrier

\begin{figure}[tbp]
\centering
\includegraphics[width=\textwidth,height=0.47\textheight,keepaspectratio]{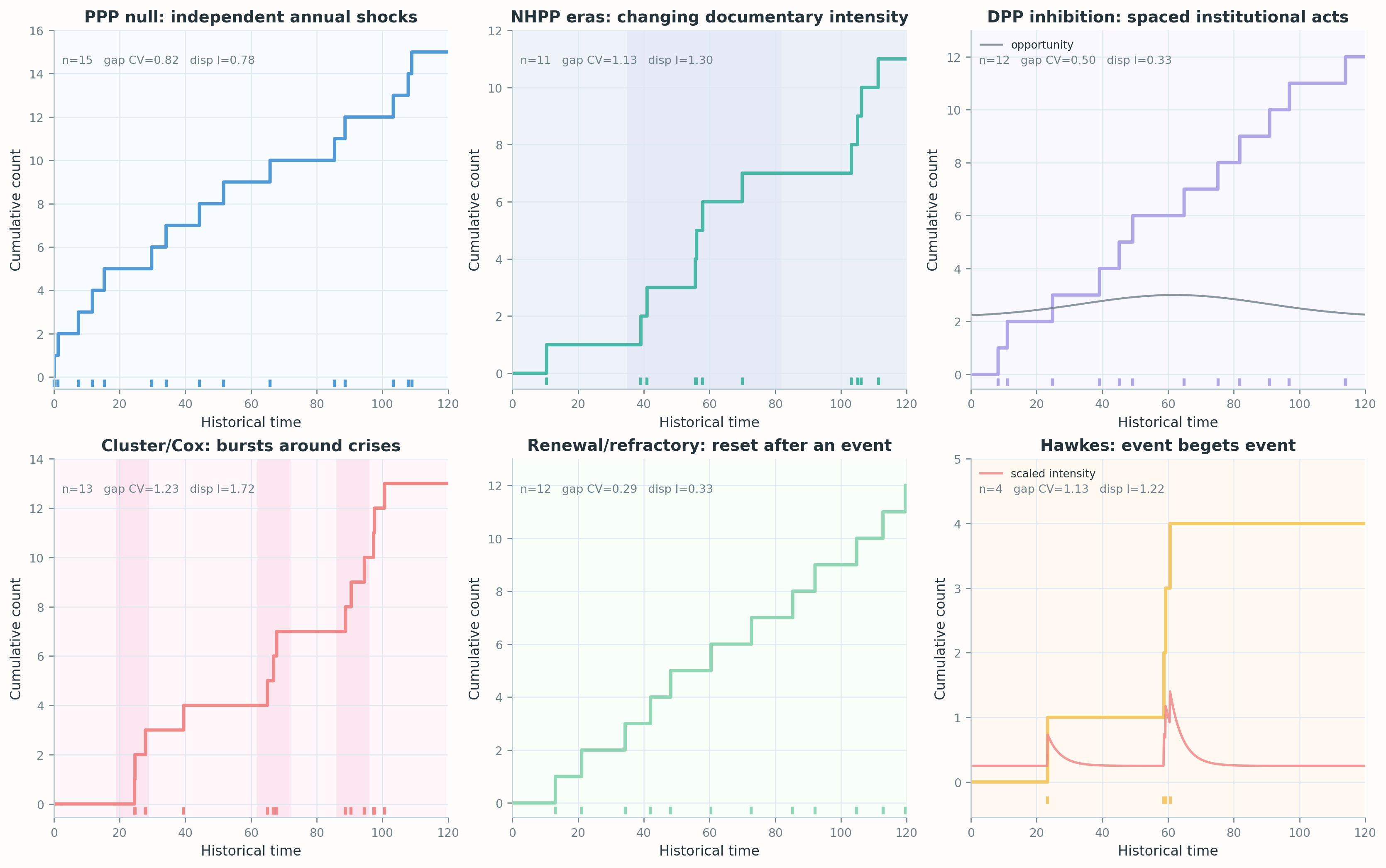}
\caption[Simulation study for DPP, PPP, and historical event-process diagnostics]{Simulation study for DPP, PPP, and historical event-process diagnostics.}
\label{fig:appendix_dpp_ppp_history_processes}
\end{figure}

Figure~\ref{fig:appendix_dpp_ppp_history_processes} then takes the same mechanisms back to one-dimensional event-history data. The horizontal axis may be read as years in an archive, and the event ticks may represent appointments, edicts, rebellions, translations, manuscript attestations, defaults, or other dated historical actions. The Poisson point process (PPP) panel is the independent-arrivals null; the nonhomogeneous Poisson process (NHPP) panel allows documentary intensity or political opportunity to change by era; the finite projection DPP panel represents inhibition, where events are possible throughout the period but tend not to occur too close together; the cluster/Cox and Hawkes panels represent bursts around crises or event-to-event contagion; and the renewal/refractory panel represents a reset after each event. The printed diagnostics are the number of events, the coefficient of variation of inter-event gaps, and the window dispersion index \(I=\operatorname{var}\{N(B_j)\}/\operatorname{E}\{N(B_j)\}\) across equal time windows. A PPP has \(I\) near one in expectation, inhibited or renewal-like histories tend to have \(I<1\), and clustered or self-exciting histories tend to have \(I>1\). In an empirical historical analysis, the workflow is therefore to define the risk intervals and marks first, fit a PPP or NHPP baseline before claiming dependence, and then use residual spacing or burst diagnostics to decide whether a DPP/Gibbs, renewal, Cox/Markov-modulated, cluster, or Hawkes-type description is more plausible. The survival-analysis reduction is obtained by asking for a first occurrence or first transition; the point-process analysis keeps the whole recurrent or marked history.

\FloatBarrier

\begin{figure}[tbp]
\centering
\includegraphics[width=0.82\textwidth,height=0.32\textheight,keepaspectratio]{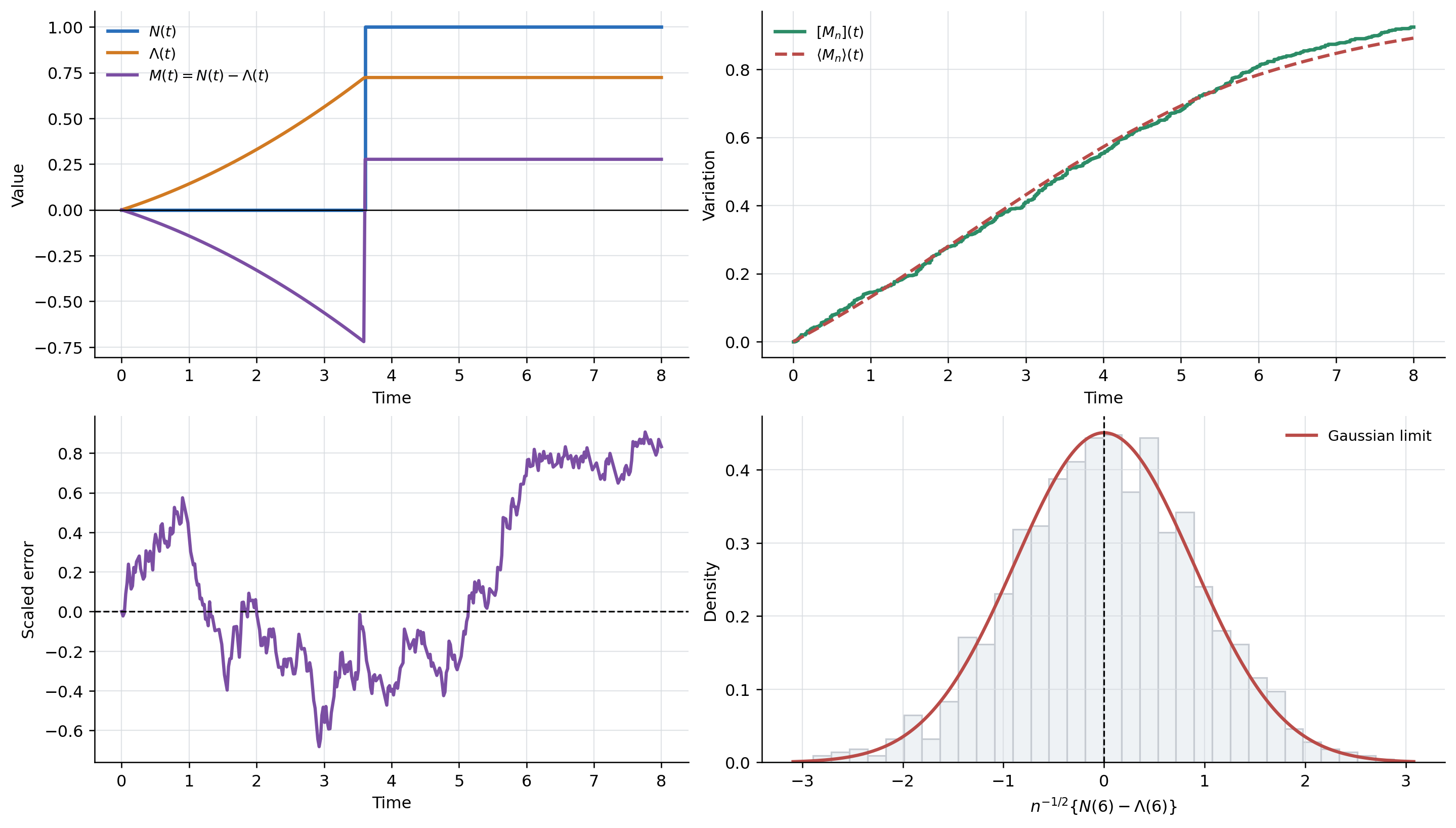}
\caption[Simulation study for appendix martingale quantities]{Simulation study for appendix martingale quantities.}
\label{fig:appendix_martingale_tools}
\end{figure}

Figure~\ref{fig:appendix_martingale_tools} illustrates the basic Doob-Meyer decomposition. In a one-jump process, \(N(t)\) records the observed jump, \(\Lambda(t)\) accumulates the predictable compensator while the subject remains at risk, and \(M(t)=N(t)-\Lambda(t)\) is the martingale residual. In an aggregated sample, the scaled error \(n^{-1/2}\{N-\Lambda\}\) fluctuates around zero, the optional variation of \(M_n\) tracks the predictable variation of \(M_n\), and the distribution at a fixed time is close to the Gaussian limit used in Rebolledo's theorem.

\begin{figure}[tbp]
\centering
\includegraphics[width=0.82\textwidth,height=0.32\textheight,keepaspectratio]{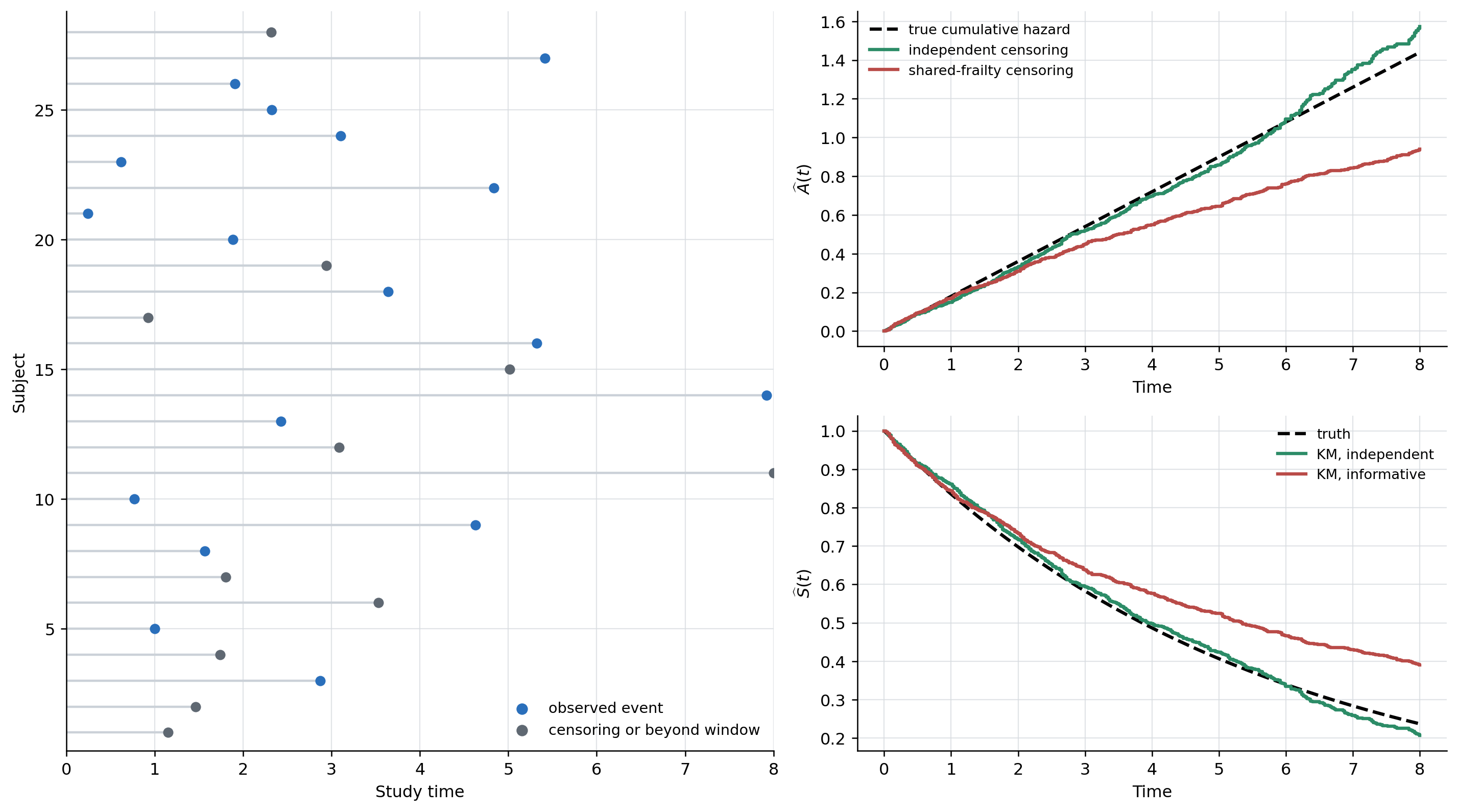}
\caption[Simulation study for independent censoring]{Simulation study for independent censoring.}
\label{fig:appendix_censoring_tools}
\end{figure}

Figure~\ref{fig:appendix_censoring_tools} focuses on the observed filtration under right censoring. The observed counting process \(N^c(t)=\int_0^tC(s)dN(s)\) has a jump only if the subject has not yet been censored. When the censoring time is independent of the event time, the Nelson-Aalen and Kaplan-Meier estimators track the true hazard and survival curves. When censoring shares unobserved frailty with the event time, the observed risk sets are distorted and the same estimators drift away from the truth. This is the data-level content of the independent-censoring definition and the innovation-theorem projection.

\begin{figure}[tbp]
\centering
\includegraphics[width=0.82\textwidth,height=0.32\textheight,keepaspectratio]{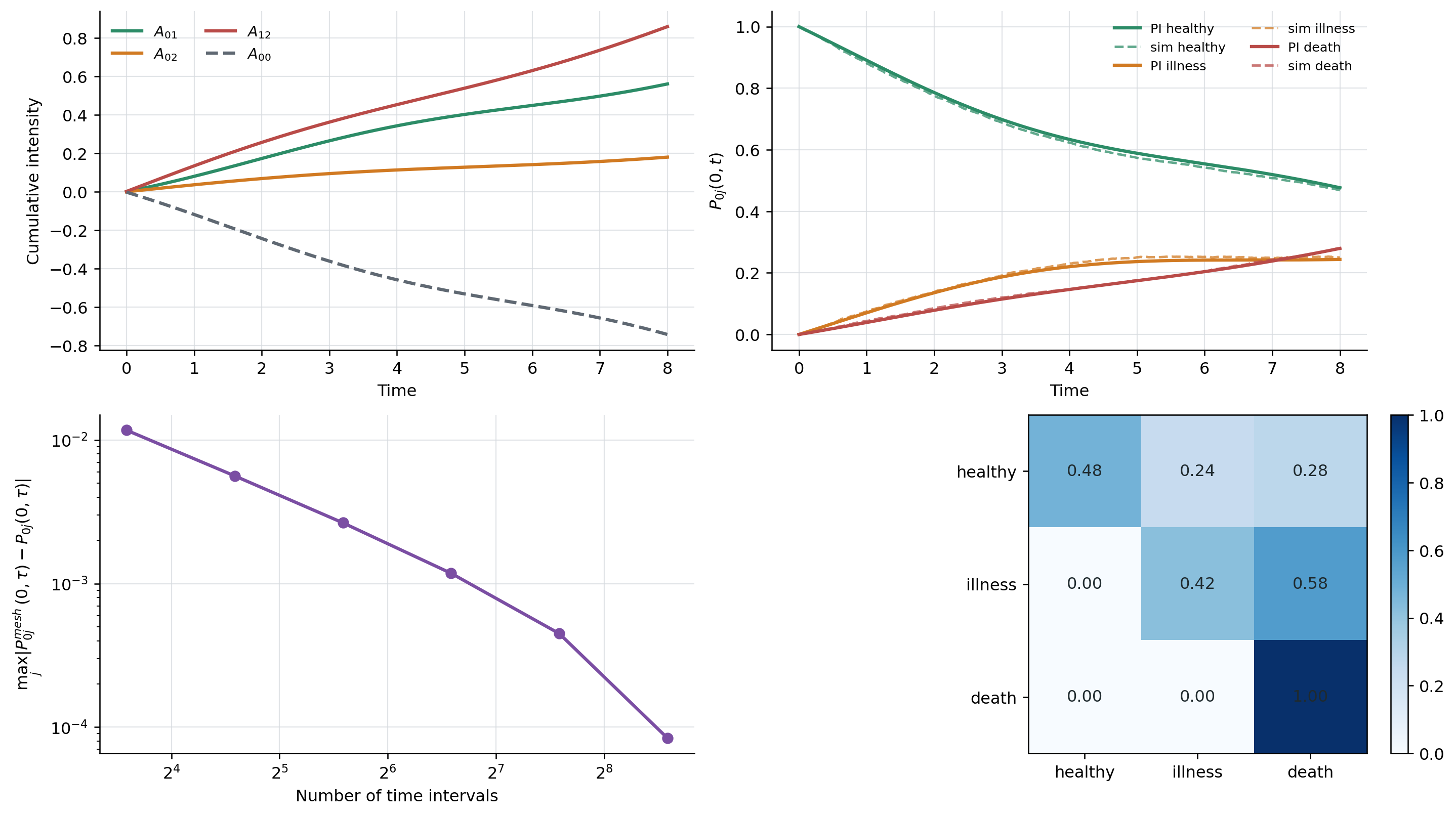}
\caption[Simulation study for product integrals]{Simulation study for product integrals.}
\label{fig:appendix_product_integral_tools}
\end{figure}

Figure~\ref{fig:appendix_product_integral_tools} gives the same treatment to product integrals and Markov intensity measures. The off-diagonal cumulative intensities \(A_{hj}\) accumulate transition-specific hazard, the diagonal entry \(A_{hh}=-\sum_{j\ne h}A_{hj}\) records the total exit hazard, and the product integral \(\Prodi(\mathbf I+d\mathbf A)\) multiplies the corresponding one-step stochastic matrices. As the mesh of the finite product is refined, the terminal transition probabilities converge to the product-integral value, and simulated state occupation probabilities follow the same transition matrix.

\begin{figure}[tbp]
\centering
\includegraphics[width=0.94\textwidth,height=0.38\textheight,keepaspectratio]{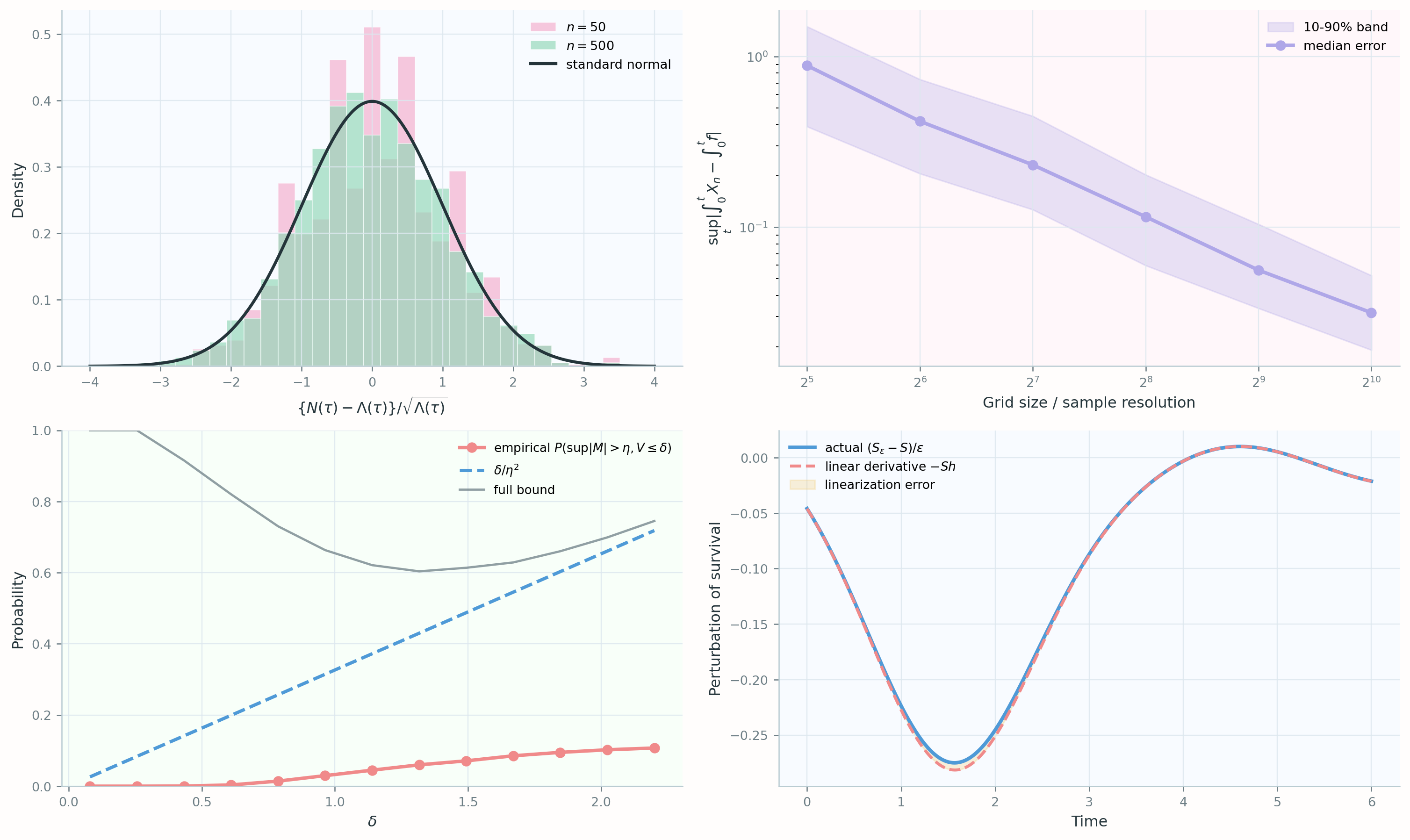}
\caption[Simulation study for appendix limit tools]{Simulation study for appendix limit tools.}
\label{fig:appendix_limit_tools}
\end{figure}

Figure~\ref{fig:appendix_limit_tools} reads four appendix limit arguments as numerical diagnostics. The Rebolledo panel compares terminal scaled martingale errors with the Gaussian limit as \(n\) grows. Gill's lemma is represented by integrated perturbations converging uniformly. Lenglart's inequality is shown as a finite-sample maximal-probability bound, and the product-integral delta-method panel compares an actual perturbation with its first-order derivative.

\begin{figure}[tbp]
\centering
\includegraphics[width=0.96\textwidth,height=0.42\textheight,keepaspectratio]{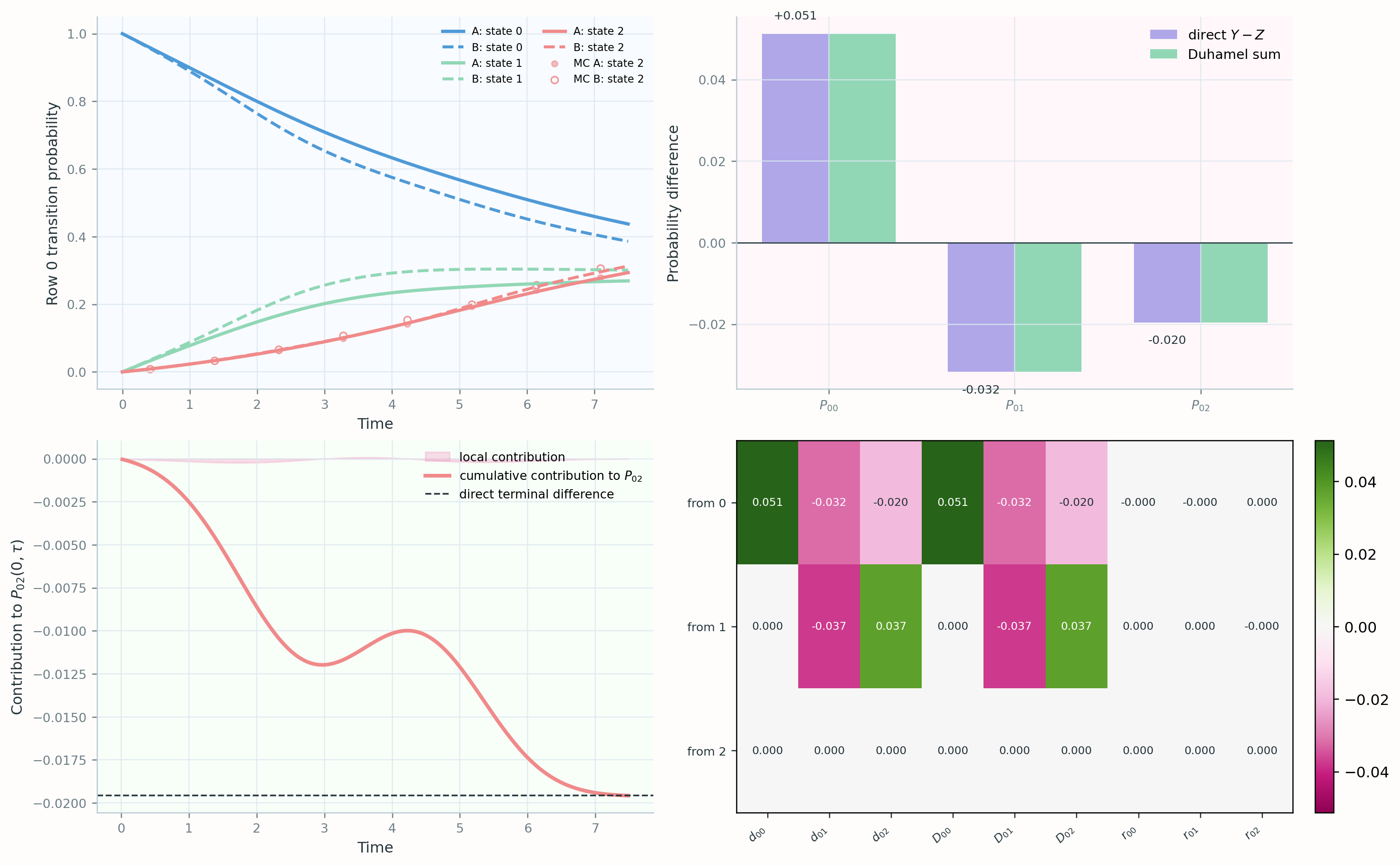}
\caption[Simulation study for Duhamel's equation]{Simulation study for Duhamel's equation.}
\label{fig:appendix_duhamel_equation}
\end{figure}

Figure~\ref{fig:appendix_duhamel_equation} turns Duhamel's equation into a finite product calculation. Two nearby three-state intensity matrices are converted into local transition factors. Their terminal product-integral difference is computed directly and again by summing local perturbation contributions, with the residual displayed as a matrix heatmap.

\begin{figure}[tbp]
\centering
\includegraphics[width=0.96\textwidth,height=0.42\textheight,keepaspectratio]{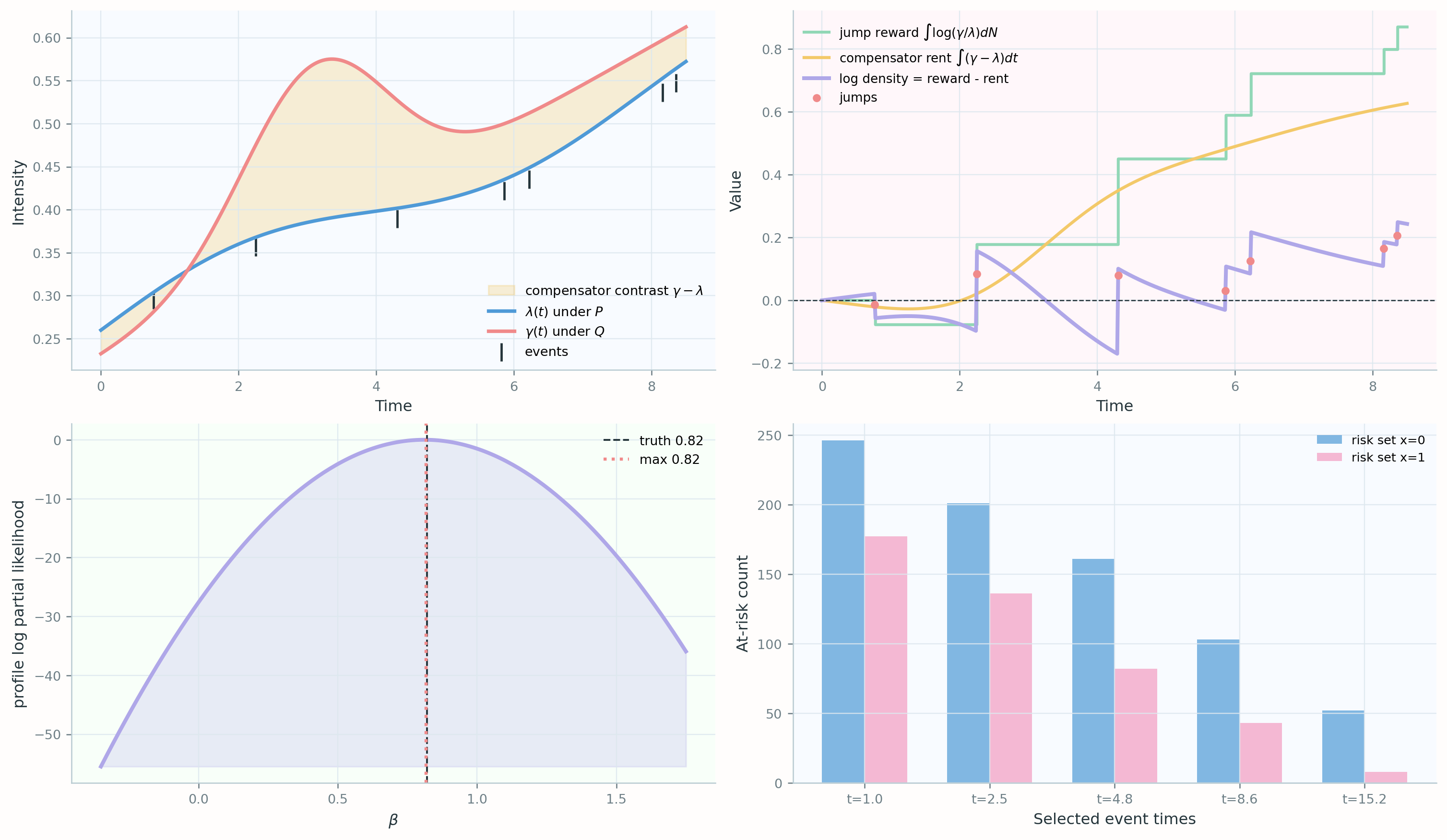}
\caption[Simulation study for likelihood ratios and partial likelihood]{Simulation study for likelihood ratios and partial likelihood.}
\label{fig:appendix_likelihood_tools}
\end{figure}

Figure~\ref{fig:appendix_likelihood_tools} gives the analogous computation for likelihood ratios and partial likelihood. A one-dimensional counting process is simulated under one intensity and reweighted toward another; the log-likelihood path jumps at event times and drifts between them. The lower panels show the Cox partial-likelihood profile and the risk-set comparisons that create it.

\subsection{Counting Processes and Martingales}
\subsubsection{Counting Processes}

The next definitions introduce the vocabulary behind Figures~\ref{fig:appendix_process_zoo}, \ref{fig:appendix_exotic_process_zoo}, \ref{fig:appendix_martingale_tools}, and \ref{fig:appendix_censoring_tools}; the later computational panels connect the same vocabulary to product integrals, Duhamel perturbations, limit arguments, and likelihood calculations. The order is important. A point process records event times. A counting process maps those event times into a right-continuous step function. The compensator is the predictable component of that step function. The martingale is the centered residual after subtracting the compensator. These definitions are not claims to prove; they are the objects that make the later Doob--Meyer and central-limit theorem possible.

\noindent\textbf{Definition (point processes and counting processes).}
A simple point process on \([0,\infty)\) is a locally finite random set of event times
\[
0<T_1<T_2<\cdots,
\]
with no two events at exactly the same time. Its associated counting process is
\[
    N(t)=\sum_{m\ge 1}\mathbb{I}(T_m\le t),\qquad t\ge 0.
\]
A marked point process records pairs \((T_m,K_m)\), where \(K_m\in\{1,\ldots,k\}\) is the event type or mark. The associated multivariate counting process is
\[
    N_h(t)=\sum_{m\ge 1}\mathbb{I}(T_m\le t, K_m=h),\qquad h=1,\ldots,k.
\]
Throughout the article we usually assume the components do not jump simultaneously, unless ties are explicitly modeled.

A spatial point process on a study region \(D\subset\mathbb R^d\) is a locally finite random set of locations \(S_m\). Its count over a measurable region \(B\subset D\) is
\[
    N(B)=\sum_m\mathbb I(S_m\in B).
\]
A space--time point process records pairs \((S_m,T_m)\) and has random measure
\[
    N(B\times(0,t])=\sum_m\mathbb I(S_m\in B,T_m\le t).
\]
If the conditional intensity is \(\lambda(s,u)\), the compensator over \(B\times(0,t]\) is
\[
    \Lambda(B\times(0,t])=\int_B\int_0^t \lambda(s,u)\,du\,ds.
\]
When events can occur only for units still observable at location \(s\) and time \(u\), the survival-analysis version replaces \(\lambda(s,u)\) by \(Y(s,u)\alpha(u\mid s,Z(s,u))\).

\medskip
\noindent\textbf{Definition (compensator).}
Let \(N\) be an adapted counting process with respect to a filtration \(\mathbb F=(\mathcal F_t)\). A predictable, nondecreasing process \(\Lambda\), with \(\Lambda(0)=0\), is the compensator of \(N\) if
\[
    M(t)=N(t)-\Lambda(t)
\]
is a local martingale. Equivalently, for every bounded predictable process \(H\),
\[
    \mathbb E\int_0^t H(s)dN(s)
    =
    \mathbb E\int_0^t H(s)d\Lambda(s).
\]
Thus \(d\Lambda(t)\) is the conditional expected jump mass at time \(t\), computed using only information available just before \(t\).

This conditional interpretation is essential. The compensator is not simply \(E\{N(t)\}\), although the two agree in some deterministic-intensity Poisson examples. It is the expected count after the analyst has conditioned on the current filtration. Changing the filtration can change the compensator: if the analyst observes a frailty variable, a censoring time, or the previous jumps of a self-exciting process, that information must enter \(d\Lambda(t)\).

The process types in Figure~\ref{fig:appendix_process_zoo} can now be read as a small list:
\begin{itemize}
\item \emph{One-jump survival process.} \(N(t)=\mathbb{I}(T\le t)\) has at most one jump. With risk process \(Y(t)=\mathbb{I}(T\ge t)\) and hazard \(\alpha(t)\), the compensator is \(\Lambda(t)=\int_0^tY(s)\alpha(s)ds\). Once the event occurs, \(Y\) becomes zero and the compensator stops.
\item \emph{Poisson and nonhomogeneous Poisson processes.} For a homogeneous Poisson process, \(\Lambda(t)=\lambda t\). For a nonhomogeneous Poisson process with deterministic rate \(\lambda(t)\), \(\Lambda(t)=\int_0^t\lambda(s)ds\). These are the cleanest cases because the future rate is fixed by calendar time.
\item \emph{Renewal and recurrent-event processes.} In a renewal process, the next waiting time is measured from the most recent jump. If \(A(t)\) is the elapsed age since the last event and \(h\) is the interarrival hazard, then \(\Lambda(t)=\int_0^t h\{A(s)\}ds\). The count is still a step function, but the predictable rate resets after each jump.
\item \emph{Marked and multivariate point processes.} Competing risks, multistate transitions, and recurrent events with event types are represented by component counts \(N_h\). Their compensators \(\Lambda_h\) track type-specific conditional mean increments, and the total count has compensator \(\sum_h\Lambda_h\).
\item \emph{Spatial and space--time point processes.} Events carry locations as well as times. A spatial Poisson process has deterministic spatial intensity \(\rho(s)\); a log-Gaussian Cox process uses \(\rho(s)=\exp\{x(s)^T\beta+w(s)\}\), where \(w\) is a spatial random field; and a space--time process uses \(\lambda(s,t)\) to describe how local risk changes over calendar time. In survival applications, the same construction becomes a spatially indexed hazard with risk process \(Y(s,t)\).
\item \emph{Self-exciting or random-intensity processes.} In a Hawkes-type process, past jumps raise the future intensity, so the compensator is history-dependent. In a Cox or frailty process, the intensity is random; the compensator depends on whether the latent environment is included in the filtration or projected out through conditional expectation.
\end{itemize}

The additional cases in Figure~\ref{fig:appendix_exotic_process_zoo} add two mechanisms. First, the environment can change the slope of the compensator, as in Cox and Markov-modulated processes. Second, previous jumps can feed back into future risk: self-correcting and refractory processes suppress the next jump, while cluster and Hawkes processes amplify it. In all six panels, the proof strategy is the same once the correct \(\Lambda\) has been identified: subtract it from \(N\), and the remaining process is the martingale residual.

\begin{example}[Spatial survival and space--time event data]\normalfont
Suppose subject \(i\) lives at location \(s_i\), has covariate history \(Z_i(t)\), and is observed until event or censoring. A spatial frailty version of the Cox model writes
\[
    \lambda_i(t\mid\mathcal F_{t-},w)
    =
    Y_i(t)\alpha_0(t)
    \exp\{\beta^TZ_i(t)+w(s_i)\},
\]
where \(w(s)\) is a spatial random field or areal random effect. Conditional on \(w\), the compensator is
\[
    \Lambda_i(t\mid w)=\int_0^tY_i(u)\alpha_0(u)
    \exp\{\beta^TZ_i(u)+w(s_i)\}\,du.
\]
If \(w\) is not observed, the compensator under the observed filtration is obtained by predictable projection. This is the survival analogue of a log-Gaussian Cox process: spatial heterogeneity enters the intensity, while censoring and delayed entry still act through the risk process. In geocoded cancer, infection, mortality, hospital-outcome, or environmental-exposure studies, residual maps and spatially smoothed martingale residuals ask whether the fitted hazard has left local clusters unexplained \citep{henderson2002modeling,banerjee2003frailty,diggle2013spatial}.
\end{example}

\begin{example}[Right-censoring model]
Let \(T\) be a nonnegative event time and let \(N(t)=\mathbb{I}(T\le t)\) be its one-jump counting process. Let \(C\) be a censoring time and write \(G(t)=\mathbb{I}(t\le C)\). The observed event process is
\[
    N^c(t)=\int_0^tG(s)dN(s)=\mathbb{I}(T\le t, T\le C),
\]
and the observed at-risk process is \(Y^c(t)=\mathbb{I}(t\le T\wedge C)\). Under independent censoring, the observed compensator is
\[
    \Lambda^c(t)=\int_0^tY^c(s)\alpha(s)ds,
\]
which is the same event hazard observed through the smaller risk set.
\end{example}

\subsubsection{Martingales and Localization}

\noindent\textbf{Definition (martingales).}
A cádlág process $\{M_t:t\ge 0\}$ defined on a stochastic basis $(\Omega,\mathcal{F},\mathbb{F},\mathbb{P})$ is called a \emph{continuous time martingale} if \(M_t\) is adapted to the filtration \(\mathcal F_t\), \(\mathbb{E}|M_t|<\infty\) for all \(t\ge0\), and \(\mathbb{E}(M_t|\mathcal{F}_s)=M_s\) a.s. for \(s<t\). The third property is the martingale property.

\medskip
\noindent\textbf{Definition (local martingales).}
A continuous-time process $\{M_t:t\ge 0\}$ is a \emph{local martingale} if there exist stopping times \(T_n\rightarrow\infty\) such that \(M^n_t=M_{t\wedge T_n}\) is a uniformly integrable martingale for each \(n\).
 
\medskip
\noindent\textbf{Definition (intensity of a counting process).}
We say that \(N_h\) has intensity process \(\lambda_h\) if \(\lambda_h\) is predictable and
\[
\Lambda_h(t)=\int_0^t\lambda_h(s)ds,\qquad t\ge0,
\]
where \(\Lambda_h\) is the compensator of \(N_h\). Equivalently, the compensator is absolutely continuous with respect to calendar time and \(\lambda_h=d\Lambda_h/dt\). If predictable atoms are present, \(\Lambda_h\) may have jumps, in which case the compensator is still well defined but cannot be represented solely by a Lebesgue density.

\begin{example}[Time-to-event outcome]
Suppose \(T\) is a nonnegative random variable. Let \(F\) be absolutely continuous with density \(f\), and write \(S=1-F\). Define \(N(t)=\mathbb{I}(T\le t)\) as the one-jump counting process. Then \(N(t)\) has compensator
\[
\Lambda(t)=\int_0^t Y(s)\alpha(s)ds,
\]
where \(Y(s)=\mathbb{I}(T\ge s)\) and \(\alpha(s)=f(s)/S(s)\). Hence \(N\) has intensity process \(\lambda(t)=Y(t)\alpha(t)\). The risk indicator is essential: after \(T\) occurs, the subject is no longer eligible for a second jump of this type, so the compensator stops accumulating.
\end{example}

\subsubsection{The Doob-Meyer Theorem and Stochastic Integration}

        \begin{theorem}[Doob-Meyer for counting processes]
        \begin{enumerate}
            \item Each component of $\mathbf{N}$ can be uniquely decomposed as
        $M_h=N_h-\Lambda_h$
        where $\Lambda_h$ is predictable and non-decreasing (called the compensator of $N_h$) and $M_h$ is a mean $0$ local martingale.
        \item Further, the compensator (predictable covariation) and the quadratic covariation of $M_h$ and $M_{h'}$ are        \begin{align*}
            \langle M_h, M_{h'}\rangle&=\delta_{hh'}\Lambda_h-\int\Delta\Lambda_hd\Lambda_{h'}\\
            [M_h]&=N_h-2\int\Delta\Lambda_hdN_h+\int\Delta\Lambda_hd\Lambda_h\\
            [M_h, M_{h'}]&=\delta_{hh'}N_h-\int\Delta\Lambda_hdN_{h'}-\int\Delta\Lambda_{h'}dN_h+\int\Delta\Lambda_h d\Lambda_{h'}
        \end{align*}
        where $\delta_{hh'}$ is the Kronecker delta function.
        \end{enumerate} 
    \end{theorem}
\begin{proof}
For each locally integrable counting process $N_h$, the process is an increasing submartingale. The Doob-Meyer decomposition for submartingales gives
\[
    N_h(t)=M_h(t)+\Lambda_h(t),
\]
where $\Lambda_h$ is predictable, increasing, and starts from zero, and $M_h$ is a local martingale. The decomposition is unique: if $N_h=M_h+\Lambda_h=M_h'+\Lambda_h'$, then $M_h-M_h'=\Lambda_h'-\Lambda_h$ is both a predictable finite-variation process and a local martingale. After localization its total variation has expectation zero, so it is indistinguishable from a constant; because both compensators start from zero, the constant is zero.

It remains to compute the covariations. Since
\[
    \Delta M_h(s)=\Delta N_h(s)-\Delta\Lambda_h(s),
\]
and no two components of $\mathbf{N}$ jump simultaneously,
\[
    \Delta N_h(s)\Delta N_{h'}(s)=\delta_{hh'}\Delta N_h(s).
\]
Therefore
\begin{align*}
[M_h,M_{h'}](t)
&=\sum_{s\le t}\Delta M_h(s)\Delta M_{h'}(s)\\
&=\delta_{hh'}N_h(t)
  -\int_{(0,t]}\Delta\Lambda_h(s)dN_{h'}(s)
  -\int_{(0,t]}\Delta\Lambda_{h'}(s)dN_h(s)\\
&\quad
  +\int_{(0,t]}\Delta\Lambda_h(s)d\Lambda_{h'}(s).
\end{align*}
Setting $h=h'$ gives the displayed formula for $[M_h]$.

The predictable covariation is the dual predictable projection of the quadratic covariation. The integrands $\Delta\Lambda_h$ and $\Delta\Lambda_{h'}$ are predictable, hence the compensators of the first three terms above are
\[
\delta_{hh'}\Lambda_h,\qquad
\int_{(0,t]}\Delta\Lambda_h(s)d\Lambda_{h'}(s),\qquad
\int_{(0,t]}\Delta\Lambda_{h'}(s)d\Lambda_h(s).
\]
The last term is already predictable. Since
\[
\int\Delta\Lambda_{h'}d\Lambda_h=\int\Delta\Lambda_hd\Lambda_{h'}
\]
both sides being the sum of $\Delta\Lambda_h(s)\Delta\Lambda_{h'}(s)$ over common predictable jump times, the compensator is
\[
\langle M_h,M_{h'}\rangle(t)
=\delta_{hh'}\Lambda_h(t)-\int_{(0,t]}\Delta\Lambda_h(s)d\Lambda_{h'}(s).
\]
Equivalently, $[M_h,M_{h'}]-\langle M_h,M_{h'}\rangle$ is a local martingale, which proves the claimed predictable covariation.
\end{proof}
 
        \begin{lemma}[Integration-by-parts]
    Suppose that $f$ is a cádlág function of bounded variation on $I=(0,t]$ and $f(0)=0$. Then
    \[
    f^2(t)=2\int_{(0,t]}f(u-)df(u)+\sum_{u\le t}\{\Delta f(u)\}^2,
    \]
    where the sum extends over the jumps of $f$.
    \end{lemma}
\begin{proof}
Decompose \(f=f^c+f^d\) into its continuous bounded-variation part and its pure-jump part. On intervals where \(f\) is continuous, the ordinary Stieltjes chain rule gives
\[
    d\{f^2(u)\}=2f(u)df(u)=2f(u-)df(u),
\]
because \(f(u)=f(u-)\) at continuity points. At a jump time \(u\),
\[
    f^2(u)-f^2(u-)
    =\{f(u-)+\Delta f(u)\}^2-f^2(u-)
    =2f(u-)\Delta f(u)+\{\Delta f(u)\}^2.
\]
Summing the continuous contribution and all jump contributions over \((0,t]\), and using \(f(0)=0\), yields the displayed identity.
\end{proof}
\begin{example}[Quadratic variation]
If $M$ is a local martingale, then we have $M^2(t)=Z_t+[M]_t$ where $Z_t=2\int_{(0,t]}M(u-)M(du)$ and $[M]_t$ is the quadratic variation process, i.e., $[M]_t=\sum_{u\le t} M(\Delta u)^2$.
\end{example}
 
        \begin{theorem}[Doob-Meyer for stochastic integration]\label{thm:doob_meyer_stochastic_integral}
    Let $M_h$ and $\Lambda_h$ be the local martingale and the compensator of $N_h$. If $H_h$'s are locally bounded predictable processes, then $\int H_hdM_h$ is a local square integrable martingale and
    \begin{align*}
        \left\langle \int H_h dM_h,\int H_{h'}dM_{h'}\right\rangle&=\int H_hH_{h'}d\langle M_h,M_{h'}\rangle\\
        \left[ \int H_h dM_h,\int H_{h'}dM_{h'}\right]&=\int H_hH_{h'}d[M_h,M_{h'}]
    \end{align*}
    where $\langle\cdot,\cdot\rangle$ and $[\cdot,\cdot]$ are predictable and optional covariation processes.
    \end{theorem}
\begin{proof}
First suppose $H_h$ and $H_{h'}$ are bounded elementary predictable processes. Then the stochastic integrals are finite sums of increments of local square integrable martingales and are therefore local square integrable martingales. For a jump time $s$,
\[
    \Delta\left(\int_0^\cdot H_h(u)dM_h(u)\right)(s)=H_h(s)\Delta M_h(s),
\]
and similarly for $h'$. Hence
\[
\left[ \int H_hdM_h,\int H_{h'}dM_{h'}\right](t)
=\sum_{s\le t}H_h(s)H_{h'}(s)\Delta M_h(s)\Delta M_{h'}(s)
=\int_{(0,t]}H_hH_{h'}d[M_h,M_{h'}].
\]
The predictable covariation is the predictable compensator of this quadratic covariation. Since $H_hH_{h'}$ is predictable and locally bounded, the compensator of the last integral is
\[
    \int_{(0,t]}H_h(s)H_{h'}(s)d\langle M_h,M_{h'}\rangle(s).
\]
For general locally bounded predictable integrands, approximate by elementary predictable processes and localize so that the integrands are bounded and the predictable variations are finite. The stochastic integrals and their covariations converge in the usual ucp sense, which preserves both displayed identities.
\end{proof}
 
\subsubsection{The Innovation Theorem}

\begin{theorem}[The innovation theorem]
Let $\mathbb{F}$ and $\mathbb{G}$ be two nested filtrations, i.e., $\mathcal{F}_t\subset \mathcal{G}_t$ for all $t$. Suppose the multivariate counting process $\mathbf{N}$ is adapted to both filtrations and has intensity process $\lambda$ with respect to $\mathbb{G}$. Then there exists an $\mathcal{F}_t$-predictable process $\gamma$ such that 
$$\gamma(t)=\mathbb{E}(\lambda(t)|\mathcal{F}_{t-})$$
If $\lambda(t)$ is also $\mathcal{F}_t$-predictable, then $\gamma(t)=\lambda(t)$.
\end{theorem}
\begin{proof}
Let $H$ be a bounded $\mathbb{F}$-predictable process. Since $\mathbb{F}\subset\mathbb{G}$, $H$ is also $\mathbb{G}$-predictable. If $\lambda$ is the $\mathbb{G}$-intensity of $N$, then
\[
    \mathbb{E}\int_0^t H(s)dN(s)
    =\mathbb{E}\int_0^t H(s)\lambda(s)ds.
\]
Taking conditional expectation with respect to $\mathcal{F}_{s-}$ inside the last integral gives
\[
    \mathbb{E}\int_0^t H(s)\lambda(s)ds
    =\mathbb{E}\int_0^t H(s)\mathbb{E}\{\lambda(s)\mid\mathcal{F}_{s-}\}ds.
\]
Thus the $\mathbb{F}$-predictable projection
\[
    \gamma(s)=\mathbb{E}\{\lambda(s)\mid\mathcal{F}_{s-}\}
\]
satisfies
\[
    \mathbb{E}\int_0^t H(s)dN(s)
    =\mathbb{E}\int_0^t H(s)\gamma(s)ds
\]
for every bounded $\mathbb{F}$-predictable $H$. This identity characterizes the compensator of $N$ in the smaller filtration, so $\int_0^t\gamma(s)ds$ is the $\mathbb{F}$-compensator. If $\lambda$ is already $\mathbb{F}$-predictable, then it is $\mathcal{F}_{s-}$-measurable, and the conditional expectation equals $\lambda(s)$.
\end{proof}
\begin{example}[Frailty model for clustered survival data]\normalfont\leavevmode\par\noindent
Suppose subjects are grouped in clusters \(c=1,\ldots,C\), such as families, hospitals, counties, paired organs, or repeated units from the same operating environment. Let \(N_{ci}(t)\) count the event for subject \(i\) in cluster \(c\), let \(Y_{ci}(t)\) be the predictable at-risk indicator, and let \(Z_{ci}(t)\) be predictable covariates. A shared-frailty Cox model introduces a positive latent variable \(W_c\), usually normalized so that \(\mathbb E(W_c)=1\), and in the enlarged filtration containing \(W_c\) sets
\[
    \lambda_{ci}^{\mathbb G}(t)
    =
    Y_{ci}(t)W_c\alpha_0(t)\exp\{\beta^\top Z_{ci}(t)\}.
\]
Conditional on \(W_c\), the compensator is
\[
    \Lambda_{ci}^{\mathbb G}(t)
    =
    \int_0^t
    Y_{ci}(u)W_c\alpha_0(u)\exp\{\beta^\top Z_{ci}(u)\}\,du .
\]
Cluster members are conditionally independent given the frailty, but they are positively associated marginally because they share the same latent multiplier. If \(W_c\) is not observed, the compensator in the observed filtration is obtained by predictable projection:
\[
    \lambda_{ci}^{\mathbb F}(t)
    =
    Y_{ci}(t)\alpha_0(t)\exp\{\beta^\top Z_{ci}(t)\}
    \mathbb E\{W_c\mid\mathcal F_{t-}\}.
\]
Previous failures in the same cluster update the conditional mean of \(W_c\), so an early cluster failure raises the projected intensity for the remaining cluster members. The frailty term therefore changes the compensator itself; it is not only a robust variance correction. A gamma frailty with variance parameter \(\theta\) gives the classical shared-frailty model, with \(\theta=0\) reducing to the ordinary Cox model. This is the counting-process version of the clustered survival example in \citet[Example~2.3.4]{martinussen2006dynamic}.
\end{example}
 
\noindent\textbf{Definition (independent censoring).}
Let \(\mathbb F=(\mathcal F_t)\) and \(\mathbb G=(\mathcal G_t)\) be nested filtrations on \([0,\tau]\), with \(\mathcal F_t\subset\mathcal G_t\). Let \(N\) be a univariate counting process adapted to both filtrations. Suppose that, in the smaller filtration,
\[
    \mathbb E\{dN(t)\mid\mathcal F_{t-}\}=\alpha_{\mathbb F}(t)dt,
\]
or equivalently that the \(\mathbb F\)-compensator is \(\int_0^t\alpha_{\mathbb F}(s)ds\). Enlarging the filtration by censoring information is called \emph{independent for \(N\)} on \([0,\tau]\) when the event compensator is unchanged:
\[
    \mathbb E\{dN(t)\mid\mathcal G_{t-}\}
    =
    \mathbb E\{dN(t)\mid\mathcal F_{t-}\}
    =
    \alpha_{\mathbb F}(t)dt.
\]
The innovation theorem explains how the smaller-filtration compensator is obtained by predictable projection from a larger filtration. The equality above is an assumption: it says that the extra censoring history in \(\mathbb G\) carries no additional predictable information about the event jump of \(N\) beyond \(\mathbb F\).

\begin{example}[Right-censored counting processes]
Let \(X\) be a nonnegative event time with hazard rate \(\lambda\). The complete event process is
\[
    N(t)=\mathbb{I}(X\le t),
\]
and the complete at-risk process is
\[
    Y(t)=\mathbb{I}(t\le X).
\]
Thus the complete event compensator is
\[
    \Lambda(t)=\int_0^tY(s)\lambda(s)ds.
\]
Let \(U\) be the right-censoring time and let
\[
    C(t)=\mathbb{I}(t\le U)
\]
be the censoring at-risk indicator. The observed event and observed risk processes are
\[
    N^c(t)=\int_0^t C(s)dN(s)
          =\mathbb I(X\le t,\ X\le U),
    \qquad
    Y^c(t)=C(t)Y(t)=\mathbb I(t\le X\wedge U).
\]
The complete event filtration, enlarged event--censoring filtration, and observed filtration are
\[
    \mathcal F^X_t=\sigma\{N(s):0\le s\le t\},\qquad
    \mathcal G_t=\sigma\{N(s),C(s):0\le s\le t\},
\]
and
\[
    \mathcal O_t=\sigma\{N^c(s),Y^c(s):0\le s\le t\}.
\]
Including \(Y^c\) is essential: the observed data include whether the subject is still at risk just before \(t\). The smaller filtration \(\sigma\{N^c(s):s\le t\}\) alone cannot distinguish a censored subject from an event-free subject who remains under observation.
\end{example}

\begin{proposition}[Observed compensator after censoring]\normalfont\leavevmode\par\noindent
In the right-censoring setup above, assume independent censoring for the event
process: the \(\mathbb G\)-compensator of \(N\) is still
\[
    \int_0^tY(s)\lambda(s)ds.
\]
Then the \(\mathbb O\)-compensator of the observed counting process \(N^c\) is
\[
    \Lambda^c(t)=\int_0^tY^c(s)\lambda(s)ds
    =\int_0^tC(s)Y(s)\lambda(s)ds.
\]
Equivalently,
\[
    M^c(t)=N^c(t)-\int_0^tY^c(s)\lambda(s)ds
\]
is an \(\mathbb O\)-martingale.
\end{proposition}
\begin{proof}
First note the predictability facts. The processes \(Y(t)=\mathbb I(t\le X)\) and \(C(t)=\mathbb I(t\le U)\) are left-continuous indicator processes, hence predictable in any filtration to which they are adapted. Therefore \(Y^c(t)=C(t)Y(t)\) is \(\mathbb O\)-predictable, because \(Y^c\) is included in the observed filtration.

We verify the compensator by its defining integral identity. Let \(H\) be any bounded \(\mathbb O\)-predictable process. Since \(\mathbb O\subset\mathbb G\), \(H\) is also \(\mathbb G\)-predictable. Since \(C\) is left-continuous and \(\mathbb G\)-adapted, \(HC\) is a bounded \(\mathbb G\)-predictable process. Using \(dN^c(t)=C(t)dN(t)\), we get
\[
    \mathbb E\int_0^tH(s)dN^c(s)
    =
    \mathbb E\int_0^tH(s)C(s)dN(s).
\]
By independent censoring, the \(\mathbb G\)-compensator of \(N\) is \(\int_0^tY(s)\lambda(s)ds\). Applying the compensator identity for \(N\) in the enlarged filtration with the predictable integrand \(H(s)C(s)\) gives
\[
    \mathbb E\int_0^tH(s)C(s)dN(s)
    =
    \mathbb E\int_0^tH(s)C(s)Y(s)\lambda(s)ds.
\]
Since \(C(s)Y(s)=Y^c(s)\), this becomes
\[
    \mathbb E\int_0^tH(s)dN^c(s)
    =
    \mathbb E\int_0^tH(s)Y^c(s)\lambda(s)ds.
\]
The last display holds for every bounded \(\mathbb O\)-predictable \(H\). By the uniqueness characterization of compensators, \(\int_0^tY^c(s)\lambda(s)ds\) is the \(\mathbb O\)-compensator of \(N^c\), and the centered process \(M^c\) is an \(\mathbb O\)-martingale.

The same argument gives the informal one-step conditional expectation formula. Because \(C(t)\) is \(\mathcal G_{t-}\)-measurable and \(Y^c(t)\) is \(\mathcal O_{t-}\)-measurable,
\begin{align*}
    \mathbb E\{dN^c(t)\mid\mathcal O_{t-}\}
    &=\mathbb E\{C(t)dN(t)\mid\mathcal O_{t-}\}\\
    &=\mathbb E\!\left[\mathbb E\{C(t)dN(t)\mid\mathcal G_{t-}\}\mid\mathcal O_{t-}\right]\\
    &=\mathbb E\{C(t)Y(t)\lambda(t)dt\mid\mathcal O_{t-}\}\\
    &=Y^c(t)\lambda(t)dt.
\end{align*}
Thus no information is being suppressed: the observed process keeps the same event hazard \(\lambda(t)\), but only while the subject is still observed and event-free.
\end{proof}

\subsection{Limit Theorems for Survival Processes}
\subsubsection{Rebolledo's Martingale Central Limit Theorem}
The martingale central limit theorem used throughout the article is the Rebolledo theorem \citep{rebolledo1980central}.

    \begin{theorem}[Rebolledo's martingale central limit theorem]
        Let $\{\mathbf{M}^{(n)}_t:t\in\mathcal{T}\}$ be a vector of $k$ local square integrable martingales for each $n=1,2,\cdots$. For $h=1,\cdots,k$, define $$M_{\epsilon h}^{(n)}(t)=\sum_{s\le t}M_h^{(n)}(\Delta s)\mathbb{I}(|M_h^{(n)}(\Delta s)>\epsilon|)$$ where $\epsilon$ is positive. Next, let $\mathbf{M}^{(\infty)}$ be a continuous Gaussian vector martingale with $\langle\mathbf{M}^{(\infty)}\rangle=[\mathbf{M}^{(\infty)}]=\mathbf{V}$, a continuous deterministic $k\times k$ positive definite matrix-valued function on $\mathcal{T}$, with positive semidefinite increments, zero at zero. 
        Assume that $\langle\mathbf{M}^{(n)}\rangle\rightarrow_P \mathbf{V}(t)$ for all \(t\in\mathcal T\) as \(n\rightarrow\infty\), and that $\langle{M}^{(n)}_{\epsilon l}\rangle\rightarrow_P0$ for all \(t\in\mathcal T\), all \(l\), and all \(\epsilon>0\).
        Then
        $$\mathbf{M}^{(n)}\rightarrow_d\mathbf{M}^{(\infty)}\text{ in }(D(\mathcal{T}))^k\text{ as }n\rightarrow\infty$$
        and $\langle\mathbf{M}^{(n)}\rangle$ and $[\mathbf{M}^{(n)}]$ converge uniformly on compact subsets of $\mathcal{T}$, in probability, to $\mathbf{V}$.
    \end{theorem}
\begin{proof}
It is enough to prove convergence on an arbitrary compact interval $[0,\tau]\subset\mathcal{T}$. By localization we may assume that the martingales are square integrable and that their predictable variations are bounded on $[0,\tau]$.

First consider finite-dimensional distributions. For any fixed vector $a\in\mathbb{R}^k$, the scalar martingale
\[
    L_n(t)=a^\top\mathbf{M}^{(n)}(t)
\]
has predictable variation $a^\top\langle\mathbf{M}^{(n)}\rangle(t)a$, which converges in probability to $a^\top\mathbf{V}(t)a$. The large-jump condition implies the Lindeberg condition for $L_n$, because a jump of $L_n$ larger than $\epsilon$ is contained in the union of the events that at least one component jump is larger than $\epsilon/(k\max_h|a_h|)$. Applying the scalar martingale central limit argument to the exponential martingale of $L_n$ gives
\[
    (L_n(t_1),\ldots,L_n(t_m))
    \Rightarrow
    (L(t_1),\ldots,L(t_m)),
\]
where $L$ is a mean-zero Gaussian martingale with covariance
\[
    \operatorname{cov}\{L(s),L(t)\}=a^\top\mathbf{V}(s\wedge t)a.
\]
The Cramer-Wold device gives convergence of the finite-dimensional distributions of $\mathbf{M}^{(n)}$ to those of the Gaussian martingale $\mathbf{M}^{(\infty)}$.

It remains to prove tightness. Rebolledo's tightness criterion states that a sequence of square integrable martingales is tight in $D[0,\tau]$ if its jumps vanish in probability and its predictable variations are tight and have asymptotically continuous increments. The convergence
\[
\langle\mathbf{M}^{(n)}\rangle\to_p\mathbf{V}
\]
and the continuity of $\mathbf{V}$ give the increment condition. The Lindeberg condition gives
\[
    \sup_{t\le\tau}|\Delta M_h^{(n)}(t)|\to_p0
\]
for each component; otherwise the compensator of the large-jump process would be bounded away from zero. Hence the vector sequence is tight. Combining tightness with the finite-dimensional convergence proves
\[
    \mathbf{M}^{(n)}\Rightarrow\mathbf{M}^{(\infty)}
\]
in $(D[0,\tau])^k$.

Finally, for each pair $(h,h')$, the process
\[
    [M_h^{(n)},M_{h'}^{(n)}]-\langle M_h^{(n)},M_{h'}^{(n)}\rangle
\]
is a local martingale. Its predictable variation is controlled by the compensators of products of jumps, and those vanish uniformly on compact intervals by the same large-jump condition and bounded predictable-variation argument. Lenglart's inequality therefore gives
\[
    \sup_{t\le\tau}\left|[M_h^{(n)},M_{h'}^{(n)}](t)-\langle M_h^{(n)},M_{h'}^{(n)}\rangle(t)\right|\to_p0.
\]
Since $\langle\mathbf{M}^{(n)}\rangle\to_p\mathbf{V}$ and $\mathbf{V}$ is continuous, the convergence of both predictable and optional covariations is uniform on compact subsets.
\end{proof}

Computationally, Rebolledo's theorem corresponds to two simulation diagnostics: the jump sizes of the scaled martingale shrink, and the predictable variation stabilizes. The first panel of Figure~\ref{fig:appendix_limit_tools} uses one-jump survival samples with sizes \(n=80\) and \(n=320\). The terminal standardized error already has the right center at \(n=80\), but the larger sample gives the noticeably cleaner Gaussian shape.
 
\subsubsection{Gill's Lemma}

    \begin{lemma}[Gill, 1983]\label{lemma:gill} Suppose $X^{(n)}(s)\rightarrow_p f(s)$ as $n\rightarrow\infty$ for all $s$ and the deterministic function $f(s)$ is integrable over $[0,\tau]$. Furthermore, for all $\delta>0$, there exists an integrable deterministic envelope $k_\delta$ such that
    $$\lim\inf_{n}\mathbb{P}(|X^{(n)}(s)|\le k_\delta(s)\text{ for all }s)\ge 1-\delta$$
    Then
    $$\sup_t\left|\int_0^t X^{(n)}(s)ds-\int_0^t f(s)ds\right|\rightarrow_p 0$$
    
    \end{lemma}
\begin{proof}
Fix \(\varepsilon>0\) and \(\eta>0\). Choose \(\delta<\eta/2\) and let
\[
A_n=\{|X^{(n)}(s)|\le k_\delta(s)\text{ for all }s\le\tau\}.
\]
By assumption, \(\liminf_n\mathbb P(A_n)\ge 1-\delta\). On \(A_n\),
\[
\sup_{t\le\tau}\left|\int_0^t\{X^{(n)}(s)-f(s)\}ds\right|
\le \int_0^\tau |X^{(n)}(s)-f(s)|ds .
\]
The integrand multiplied by \(\mathbb I(A_n)\) is bounded by
\(k_\delta(s)+|f(s)|\), an integrable deterministic function, and it converges to zero in probability for each fixed \(s\). The dominated convergence theorem in probability therefore gives
\[
\int_0^\tau |X^{(n)}(s)-f(s)|\mathbb I(A_n)ds\to_p0.
\]
Hence, for large \(n\),
\[
\mathbb P\left(
\sup_{t\le\tau}\left|\int_0^t X^{(n)}(s)ds-\int_0^t f(s)ds\right|>\varepsilon
\right)
\le \mathbb P(A_n^c)+o(1)\le \delta+o(1).
\]
Letting \(\delta\downarrow0\) proves the asserted uniform convergence in probability.
\end{proof}

\begin{example}[A computational Gill check]\normalfont
On a fixed grid, set \(X_n(t)=f(t)+\varepsilon_n(t)\), where the perturbation is bounded by a deterministic envelope and its pointwise scale decreases with \(n\). Compute the running sums
\[
    I_n(t_j)=\sum_{\ell\le j}X_n(t_\ell)\Delta t,\qquad
    I(t_j)=\sum_{\ell\le j}f(t_\ell)\Delta t .
\]
The diagnostic is \(\max_j|I_n(t_j)-I(t_j)|\). The Gill panel in Figure~\ref{fig:appendix_limit_tools} repeats this calculation and shows the error collapsing as the envelope-controlled perturbation gets smaller.
\end{example}
 
\subsubsection{Lenglart's Inequality}
The maximal inequality used to control martingale errors is Lenglart's domination inequality \citep{lenglart1977relation}.

    \begin{lemma}[Lenglart's inequality]
    Let $M$ be a local square integrable martingale with $M(0)=0$ and let $\tau$ be the endpoint of $M$. Then
    \begin{align*}
        \mathbb{P}\left(\sup_{t\le\tau}|M(t)|>\eta\right)\le\frac{\delta}{\eta^2}+\mathbb{P}(\langle M\rangle(\tau)>\delta)
    \end{align*}
    for any positive $\eta$ and $\delta$.
    \end{lemma}
\begin{proof}
By localization it is enough to consider square-integrable martingales. Let
\[
T=\inf\{t\le\tau:|M(t)|>\eta\},\qquad
S=\inf\{t\le\tau:\langle M\rangle(t)>\delta\}.
\]
On the event \(\{T\le\tau,\langle M\rangle(\tau)\le\delta\}\), the stopping time \(S\) is not reached before \(T\), and \(|M(T)|>\eta\). Therefore
\[
\eta^2\mathbb P\{T\le\tau,\langle M\rangle(\tau)\le\delta\}
\le
\mathbb E\{M^2(T\wedge S\wedge\tau)\}.
\]
Since \(M^2-\langle M\rangle\) is a martingale, optional stopping gives
\[
\mathbb E\{M^2(T\wedge S\wedge\tau)\}
=\mathbb E\{\langle M\rangle(T\wedge S\wedge\tau)\}
\le \delta.
\]
Consequently
\[
\mathbb P\left(\sup_{t\le\tau}|M(t)|>\eta\right)
\le
\mathbb P\{T\le\tau,\langle M\rangle(\tau)\le\delta\}
+\mathbb P\{\langle M\rangle(\tau)>\delta\}
\le \frac{\delta}{\eta^2}+\mathbb P\{\langle M\rangle(\tau)>\delta\}.
\]
\end{proof}

\begin{example}[Using Lenglart as a simulation bound]\normalfont\leavevmode\par\noindent
For each Monte Carlo path, record
\[
    S=\sup_{t\le\tau}|M(t)|,
    \qquad
    V=\langle M\rangle(\tau).
\]
For a grid of \(\delta\)'s, compare the empirical probability
\(\widehat{\mathbb P}(S>\eta)\) with
\[
    \delta/\eta^2+\widehat{\mathbb P}(V>\delta).
\]
The right-hand side is often conservative, but it has the right role: it turns control of the final predictable variation into control of the whole path. Figure~\ref{fig:appendix_limit_tools} displays this comparison after capping the bound at one.
\end{example}

\subsubsection{Where Empirical Process Theory Enters}

Most one-dimensional survival estimators in this article can be handled by martingale laws of large numbers, martingale central limit theorems, and continuous mapping arguments. Empirical-process theory enters when the index set is no longer just calendar time. This happens for covariate-indexed estimators such as Beran's conditional survival curve, bivariate-indexed objects such as Dabrowska's estimator, kernel-smoothed Cox scores, bootstrap diagnostic processes, and semiparametric estimating equations with nuisance parameters. In those settings the proof must control not only one process \(M(t)\), but a whole class \(\{f_\eta(O):\eta\in\mathcal H\}\) uniformly over an index \(\eta\). The references used here are \citet{vaart1996weak}, \citet{shorack2009empirical}, and the survival-specific treatments cited in the main text.

\noindent\textbf{Definition (Glivenko--Cantelli and Donsker classes).}
For i.i.d. observations \(O_1,\ldots,O_n\), write
\[
    P_nf=\frac1n\sum_{i=1}^n f(O_i),
    \qquad
    \mathbb G_nf=\sqrt n(P_n-P)f .
\]
A class \(\mathcal F\) is \emph{Glivenko--Cantelli} if
\[
    \sup_{f\in\mathcal F}|P_nf-Pf|\to_p0,
\]
and is \emph{Donsker} if \(\mathbb G_n\), viewed as a process indexed by \(f\in\mathcal F\), converges weakly in \(\ell^\infty(\mathcal F)\) to a tight Gaussian process. In this article the first property supplies uniform consistency of risk-set averages and estimating functions; the second supplies Gaussian limits for covariate-indexed scores or residual processes.

\begin{lemma}[Empirical-process transfer for estimating equations]\label{lemma:ep_transfer}\leavevmode\par\noindent
Let \(\Psi_n(\eta)=P_n\psi_\eta\) and \(\Psi(\eta)=P\psi_\eta\), where
\(\eta\) belongs to a semimetric space \(\mathcal H\). Suppose \(\eta_0\) is
the unique solution of \(\Psi(\eta)=0\), the class
\(\{\psi_\eta:\eta\in\mathcal H_0\}\) is Glivenko--Cantelli on a neighborhood
\(\mathcal H_0\) of \(\eta_0\), and \(\Psi\) is separated away from zero outside
every neighborhood of \(\eta_0\). Then any approximate root \(\widehat\eta\)
satisfying \(\|\Psi_n(\widehat\eta)\|=o_p(1)\) is consistent. If, additionally,
the local class is Donsker, \(\Psi\) is differentiable at \(\eta_0\) with
nonsingular derivative \(\dot\Psi_{\eta_0}\), and
\[
    \Psi_n(\widehat\eta)
    =
    \Psi_n(\eta_0)+\dot\Psi_{\eta_0}(\widehat\eta-\eta_0)+o_p(n^{-1/2}),
\]
then
\[
    \sqrt n(\widehat\eta-\eta_0)
    =
    -\dot\Psi_{\eta_0}^{-1}\mathbb G_n\psi_{\eta_0}+o_p(1).
\]
\end{lemma}
\begin{proof}
Uniform convergence gives
\[
    \sup_{\eta\in\mathcal H_0}\|\Psi_n(\eta)-\Psi(\eta)\|\to_p0.
\]
If \(\widehat\eta\) stayed outside a fixed neighborhood of \(\eta_0\) with nonvanishing probability, the separation condition would make \(\|\Psi(\widehat\eta)\|\) bounded away from zero on that event, contradicting \(\|\Psi_n(\widehat\eta)\|=o_p(1)\). Hence \(\widehat\eta\to_p\eta_0\). For the second claim, subtract \(\Psi(\eta_0)=0\) and use the displayed local expansion:
\[
0=\Psi_n(\widehat\eta)+o_p(n^{-1/2})
=P_n\psi_{\eta_0}+\dot\Psi_{\eta_0}(\widehat\eta-\eta_0)+o_p(n^{-1/2}).
\]
Multiplying by \(\sqrt n\) and using \(P\psi_{\eta_0}=0\) gives the stated asymptotically linear representation. The Donsker condition then identifies the weak limit of \(\mathbb G_n\psi_{\eta_0}\).
\end{proof}

For the present article, the lemma serves as an operational regularity guide rather than as an instruction to verify entropy conditions in every section. Nelson--Aalen, Kaplan--Meier, Aalen--Johansen, and the log-rank statistic usually need martingale arguments plus continuous mapping. Beran, Dabrowska, smoothed Cox, Cox--Aalen, model diagnostics indexed by time and covariates, and IV estimating equations with residualized instruments need empirical-process regularity because the estimator is selected from a class. Panel-count and interval-censored NPMLEs need still more: monotonicity, self-consistency, and likelihood geometry, not merely Donsker convergence.
 
\subsubsection{The Functional Delta Method}

\begin{lemma}[The functional $\Delta$-method]\label{lemma:fdm}
Let $T_n$ be a sequence of random elements of a Banach space $B$, $a_n\rightarrow\infty$ a real sequence such that $a_n(T_n-\theta)\rightarrow_dZ$ for some fixed $\theta\in B$ and a random element $Z\in B$. Suppose $\phi: B\rightarrow B'$ is Hadamard differentiable at $\theta$. Then
$$a_n(\phi(T_n)-\phi(\theta))\rightarrow_d d\phi(\theta)\cdot Z$$
where $d\phi(\theta)$ is the derivative of $\phi$ at $\theta$.
\end{lemma}
\begin{proof}
By the Skorohod representation theorem for separable versions of the weak convergence, we may work on a probability space on which
\[
    h_n=a_n(T_n-\theta)\to Z
    \qquad\text{almost surely in }B .
\]
Let \(t_n=a_n^{-1}\). Then \(t_n\downarrow0\) and
\[
    T_n=\theta+t_nh_n .
\]
Hadamard differentiability of \(\phi\) at \(\theta\) means that whenever \(h_n\to h\) in \(B\) and \(t_n\downarrow0\),
\[
    \frac{\phi(\theta+t_nh_n)-\phi(\theta)}{t_n}
    \to d\phi(\theta)h
    \quad\text{in }B'.
\]
Applying this deterministic implication pathwise with \(h=Z(\omega)\) gives
\[
    a_n\{\phi(T_n)-\phi(\theta)\}
    =
    \frac{\phi(\theta+t_nh_n)-\phi(\theta)}{t_n}
    \to d\phi(\theta)Z
    \quad\text{almost surely}.
\]
Almost-sure convergence on the Skorohod space implies convergence in distribution in the original space.
\end{proof}

The functional delta method is the bridge from a process limit to the statistical quantities
reported in the paper. In survival analysis the primitive weak convergence result is often
for a cumulative hazard, a vector of cumulative transition intensities, or an empirical
score process. The object finally plotted or interpreted is usually a transformation of that
primitive process: a survival curve, a transition probability, a restricted mean survival
time, a quantile, or a confidence interval on a stabilized scale. The technical task is
therefore to identify the derivative of the map that turns the primitive process into the
reported quantity. This is the same logic used throughout empirical-process treatments of
survival estimators \citep{andersen1993statistical,shorack2009empirical,wellner2013weak}.

\begin{figure}[tbp]
\centering
\includegraphics[width=0.92\textwidth,height=0.48\textheight,keepaspectratio]{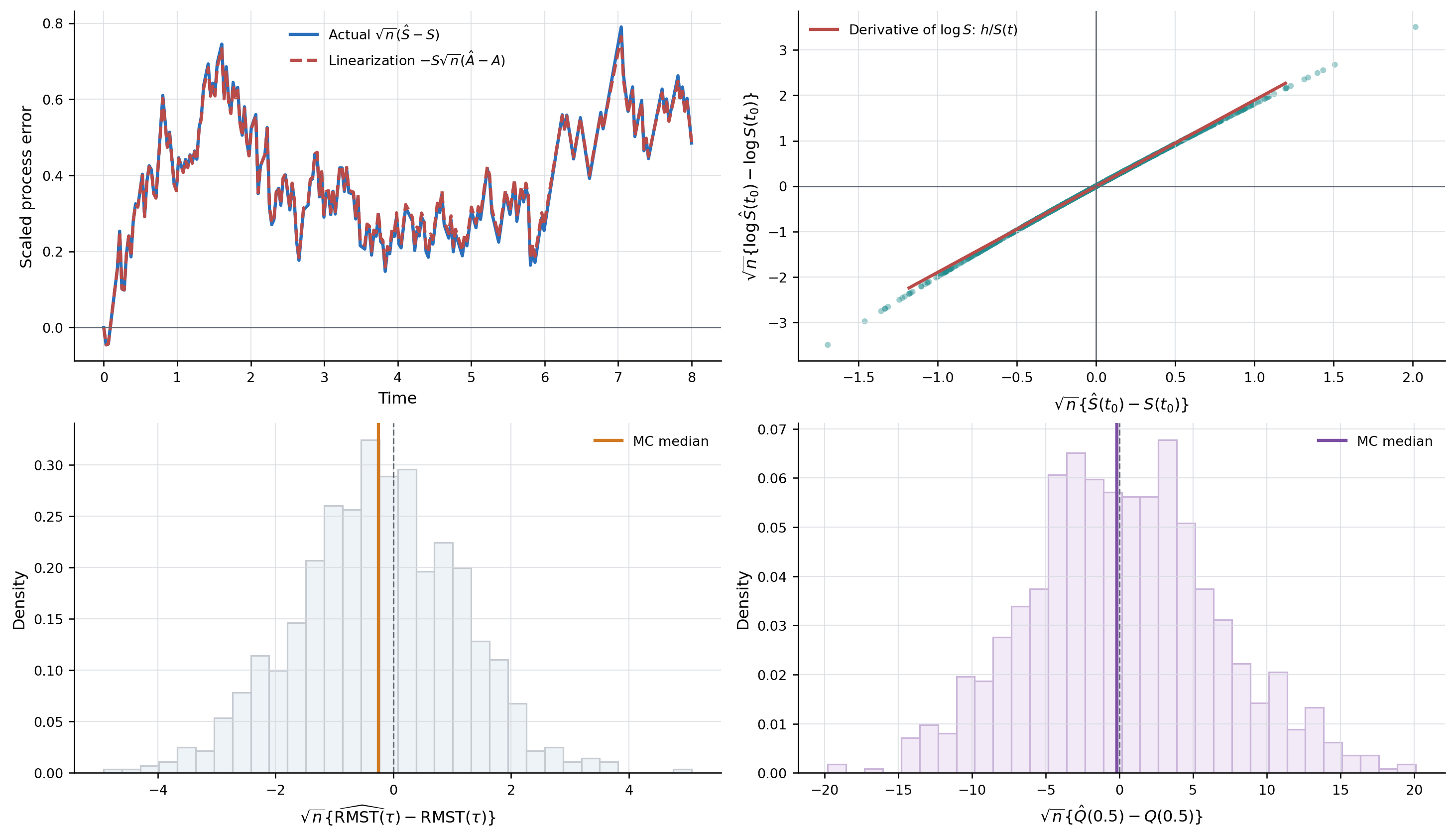}
\caption[Simulation study for the functional delta method]{Simulation study for the functional delta method. A Nelson--Aalen process error is pushed through several survival functionals: the product-integral map for Kaplan--Meier, a log transformation for pointwise survival inference, integration for restricted mean survival time, and inversion for a survival quantile.}
\label{fig:functional_delta_method_sim}
\end{figure}

Figure~\ref{fig:functional_delta_method_sim} shows the finite-sample interpretation of the lemma. The
upper-left panel compares the actual Kaplan--Meier process error with the first-order
linearization obtained by multiplying the Nelson--Aalen error by \(-S(t)\). The two curves
are nearly indistinguishable in this sample. The other panels use Monte Carlo repetitions to
show what happens to scalar functionals. Smooth maps such as \(\log S(t_0)\) and
\(\int_0^\tau S(t)dt\) produce stable, nearly linear transformations of the survival-process
error, while the median survival time is more volatile because it inverts a step function and
divides by the local density at the crossing point.

\begin{example}[Product-Integration]\label{eg:prodi_fdm} Let $\phi$ be defined by
$$\phi(\mathbf{X})=\Prodi(\mathbf{I}+d\mathbf{X})$$
and elements of $\mathbf{X}$ are cádlag and with total variation bounded by $M$. Then $\phi$ is Hadamard differentiable with derivative given by
$$(d\phi(\mathbf{X})\cdot \mathbf{H})=\int_{s\in[0,t]}\Prodi_{[0,s)}(\mathbf{I}+d\mathbf{X})\mathbf{H}(ds)\Prodi_{(s,t]}(\mathbf{I}+d\mathbf{X})$$
where the last integral is defined by application (twice) of the integration by parts formula. In practice, $\mathbf{X}$ may correspond to the Nelson-Aalen estimator and $\phi(\cdot)$ is the Kaplan-Meier estimator so that functional delta-method applies.
\end{example}

\begin{example}[Pointwise transformations]\normalfont
Let \(\widehat S(t_0)\) be a survival estimator at a fixed time \(t_0\), and suppose
\[
    \sqrt n\{\widehat S(t_0)-S(t_0)\}\rightsquigarrow Z(t_0).
\]
For any differentiable scalar function \(g\) with \(g'\{S(t_0)\}\neq0\), the ordinary delta
method is the one-dimensional version of Lemma~\ref{lemma:fdm}:
\[
    \sqrt n\{g(\widehat S(t_0))-g(S(t_0))\}
    \rightsquigarrow g'\{S(t_0)\}Z(t_0).
\]
Three common pointwise transformations have derivatives
\[
\begin{array}{ccl}
g(s)=\log s
&\Longrightarrow&
g'(s)=1/s,\\[2pt]
g(s)=\log\{-\log s\}
&\Longrightarrow&
g'(s)=1/\{s\log s\},\\[2pt]
g(s)=\arcsin(\sqrt{s})
&\Longrightarrow&
g'(s)=1/\{2\sqrt{s(1-s)}\}.
\end{array}
\]
These transformations are the formal reason transformed confidence intervals can respect
the range \(0\le S(t)\le1\) better than a plain linear interval.
\end{example}

\begin{example}[Restricted mean survival time]\normalfont
For a fixed horizon \(\tau\), define the restricted mean survival time functional
\[
    \phi(S)=\int_0^\tau S(t)dt.
\]
This map is linear and continuous on \(D[0,\tau]\) under the uniform norm, so its derivative is
itself:
\[
    d\phi(S)h=\int_0^\tau h(t)dt.
\]
Consequently, if \(\sqrt n(\widehat S-S)\rightsquigarrow Z\), then
\[
    \sqrt n\left\{\int_0^\tau \widehat S(t)dt-\int_0^\tau S(t)dt\right\}
    \rightsquigarrow
    \int_0^\tau Z(t)dt.
\]
This is the cleanest functional-delta example in the article: the limiting Gaussian process is
smoothed by integration, which is why restricted-mean summaries are often more stable than
late-tail survival probabilities.
\end{example}

\begin{example}[Survival quantiles]\normalfont
Let \(Q(p)=\inf\{t:S(t)\le 1-p\}\), and assume that \(S\) is continuously differentiable in a
neighborhood of \(Q(p)\) with density \(f\{Q(p)\}=-S'\{Q(p)\}>0\). The inverse map is
Hadamard differentiable tangentially to continuous perturbations at the crossing point, with
derivative
\[
    dQ_p(S)h=\frac{h\{Q(p)\}}{f\{Q(p)\}}.
\]
Therefore
\[
    \sqrt n\{\widehat Q(p)-Q(p)\}
    \rightsquigarrow
    \frac{Z\{Q(p)\}}{f\{Q(p)\}}.
\]
The denominator explains the instability of quantiles near flat parts of the survival curve:
if the curve crosses the level slowly, a small vertical error in \(\widehat S\) becomes a large
horizontal error in \(\widehat Q(p)\).
\end{example}

\subsection{Product Integrals and Markov Processes}
The definitions in this subsection are the formal version of Figure~\ref{fig:appendix_product_integral_tools}. The off-diagonal cumulative intensities determine the accumulated transition hazard, the diagonal entries make each local matrix conserve probability, and the product integral orders those local matrices over time.

\noindent\textbf{Definition (product integral).}
Let $\mathbf{X}(t),t\in\mathcal{T}$, be a $p\times p$ matrix of cádlág functions of locally bounded variation. Define
\[
    \mathbf{Y}=\Prodi(\mathbf{I}+d\mathbf{X})
\]
as the \emph{product-integral} of $\mathbf{X}$ over intervals of the form $[0,t]$, where
\[
    \mathbf{Y}(t)
    =\Prodi_{s\in[0,t]}(\mathbf{I}+\mathbf{X}(ds))
    =\lim_{\max|t_i-t_{i-1}|\rightarrow0}
      \prod(\mathbf{I}+\mathbf{X}_{t_i}-\mathbf{X}_{t_{i-1}}).
\]
Here $0=t_0<t_1<\cdots<t_n=t$ is a partition of $[0,t]$ and the matrix product is taken from left to right.
 
    The following product-integral representation is due to Jacod; see \citet{dabrowska2020stochastic} for the formulation used here.
    \begin{theorem}[Jacod's product-integral representation] Let $A$ be a cádlag increasing function on $[0,\tau]$ where $0<\tau\le \infty$ such that $0\le A(\Delta u)<1$ for $u<\tau$ and satisfying either
    $$A(\tau-)<\infty,\ A(\Delta\tau)=1$$
or $$A(\tau-)=\infty,\ A(\Delta \tau)=0.$$
Then $S(t)=\Prodi_{u\le t}(1-dA(u))$ is a survival function of a nonnegative random variable and $$\tau=\inf\{t:S(t)=0\}$$
is its upper support point.
    \end{theorem}
\begin{proof}
The scalar product integral has the explicit form
\[
    S(t)=\exp\{-A_c(t)\}\prod_{u\le t}\{1-\Delta A(u)\},
\]
where $A_c$ is the continuous part of $A$. Since $A$ is increasing and $0\le\Delta A(u)<1$ for $u<\tau$, every factor before $\tau$ lies in $(0,1]$. Therefore $S$ is nonincreasing, cadlag, satisfies $S(0)=1$, and is strictly positive on every compact subinterval of $[0,\tau)$.

If $A(\tau-)<\infty$ and $\Delta A(\tau)=1$, then the product integral has a final factor $1-\Delta A(\tau)=0$, so $S(\tau)=0$. If $A(\tau-)=\infty$ and $\Delta A(\tau)=0$, then
\[
    \log S(t)=-A_c(t)+\sum_{u\le t}\log\{1-\Delta A(u)\}
    \le -A_c(t)-\sum_{u\le t}\Delta A(u),
\]
because $\log(1-x)\le -x$ for $0\le x<1$. The right-hand side tends to $-\infty$ as $t\uparrow\tau$, so $S(t)\downarrow0$.

Define $F(t)=1-S(t)$ for $t<\tau$ and $F(t)=1$ for $t\ge\tau$. The preceding paragraph shows that $F$ is nondecreasing, right-continuous, starts at zero, and tends to one. Hence $F$ is a distribution function on $[0,\infty)$ and $S$ is its survival function. Since $S(t)>0$ for $t<\tau$ and $S(\tau)=0$ (or $S(t)\downarrow0$ as $t\uparrow\tau$ when $\tau=\infty$), the upper support point is exactly
\[
    \inf\{t:S(t)=0\}=\tau.
\]
\end{proof}

    \begin{example}[Cox's lemma]
    Let $T$ be a nonnegative random variable and $S(t)=\mathbb{P}(T\ge t)$ be its survival function. Let $\Lambda(t)$, $\Lambda(dt)=-S(dt)/S(t-)$, $\Lambda(0)=0$ be the associated cumulative hazard. Then we have
    $$S(t)=\Prodi_{s\le t}(1-\Lambda(ds))=\prod_{s\le t}(1-\Lambda(\Delta s))\exp(-\Lambda_c(t))$$
    where $\Lambda_c$ is the continuous part of $\Lambda$.
    \end{example}
 
    \begin{theorem}[Duhamel's Equation] Let $\mathbf{Y}=\Prodi(\mathbf{I}+d\mathbf{A})$ and $\mathbf{Z}=\Prodi(\mathbf{I}+d\mathbf{B})$. Then
    $$\mathbf{Y}(t)-\mathbf{Z}(t)=\int_{s\in[0,t]}\Prodi_{[0,s)}(\mathbf{I}+d\mathbf{A})(\mathbf{A}(ds)-\mathbf{B}(ds))\Prodi_{(s,t]}(\mathbf{I}+d\mathbf{B}).$$
    
    \end{theorem}
\begin{proof}
Take a partition $0=t_0<t_1<\cdots<t_m=t$ and write
\[
    A_i=\mathbf{I}+\mathbf{A}(t_i)-\mathbf{A}(t_{i-1}),\qquad
    B_i=\mathbf{I}+\mathbf{B}(t_i)-\mathbf{B}(t_{i-1}).
\]
For finite products,
\[
    A_1\cdots A_m-B_1\cdots B_m
    =\sum_{i=1}^m A_1\cdots A_{i-1}(A_i-B_i)B_{i+1}\cdots B_m,
\]
which follows by adding and subtracting the products in which the first $i$ factors come from $A$ and the remaining factors come from $B$.

Now refine the partition so its mesh tends to zero. By the definition of the product integral,
\[
A_1\cdots A_{i-1}\to \Prodi_{[0,s)}(\mathbf{I}+d\mathbf{A}),\qquad
B_{i+1}\cdots B_m\to \Prodi_{(s,t]}(\mathbf{I}+d\mathbf{B}),
\]
and $A_i-B_i=\mathbf{A}(dt_i)-\mathbf{B}(dt_i)$. The bounded-variation assumption on $\mathbf{A}$ and $\mathbf{B}$ gives convergence of the Riemann-Stieltjes sums to
\[
    \int_{s\in[0,t]}\Prodi_{[0,s)}(\mathbf{I}+d\mathbf{A})
    (\mathbf{A}(ds)-\mathbf{B}(ds))
    \Prodi_{(s,t]}(\mathbf{I}+d\mathbf{B}),
\]
which proves the identity.
\end{proof}

Figure~\ref{fig:appendix_duhamel_equation} shows the same identity without limiting notation. Choose a grid \(0=t_0<\cdots<t_m=t\), form local factors
\[
    A_i=\mathbf I+\Delta\mathbf A_i,\qquad B_i=\mathbf I+\Delta\mathbf B_i,
\]
and compute both terminal products \(A_1\cdots A_m\) and \(B_1\cdots B_m\). The Duhamel contribution of grid cell \(i\) is
\[
    A_1\cdots A_{i-1}(A_i-B_i)B_{i+1}\cdots B_m .
\]
Summing these cellwise contributions reproduces the direct product difference up to numerical roundoff. This is the computational role of Duhamel's equation in Aalen--Johansen expansions: a complicated terminal transition-probability error is rewritten as the accumulated effect of local cumulative-intensity errors, transported forward and backward by product integrals.

\begin{example}[Grid computation of a Duhamel perturbation]\normalfont\leavevmode\par\noindent
In a three-state illness-death model, let \(\mathbf A\) and \(\mathbf B\) differ
only by a small change in the \(0\to1\) and \(1\to2\) transition hazards. A
direct recomputation of
\[
    \Prodi(\mathbf I+d\mathbf A)-\Prodi(\mathbf I+d\mathbf B)
\]
tells us the terminal error, but not where it came from. The Duhamel sum assigns
that error to time cells and transition channels. In Figure~\ref{fig:appendix_duhamel_equation},
most of the terminal \(0\to2\) difference is explained by mid-follow-up cells
where the perturbation first moves mass into state \(1\) and then lets the
\(1\to2\) hazard carry it to state \(2\).
\end{example}
 
\subsubsection{Markov Processes}

\noindent\textbf{Definition (intensity measure of a Markov process).}
Suppose the off-diagonal elements of the $p\times p$ matrix function $\mathbf{A}$ are nondecreasing cádlág functions, zero at zero, \(A_{hh}=-\sum_{j\not=h}A_{hj}\), and \(A_{hh}(\Delta t)\ge -1\) for all \(t\). Then \(\mathbf{A}\) is a \emph{locally finite intensity measure} of a Markov process on a time interval \(\mathcal{T}\). The function \(A_{hj}\) is the integrated intensity function for transitions from state \(h\) to state \(j\), and \(A_{hh}\) is the negative integrated intensity function for transitions out of state \(h\).

\subsubsection{A Key Theorem}

\begin{theorem}[Relation among PI, MP and CP]\label{thm:Jacobson}
Let the matrix function $\mathbf{A}$ correspond to an intensity measure and define
$$\mathbf{P}(s,t)=\Prodi_{(s,t]}(\mathbf{I}+d\mathbf{A}),\ s\le t;s,t,\in\mathcal{T}.$$
Then $\mathbf{P}$ is the \emph{transition matrix} of a Markov process $\{X(t):t\in\mathcal{T}\}$ with state space $\{1,2,\cdots,p\}$ and intensity measure $\mathbf{A}$.

Given that $X(t_0)=h$, the process remains in \(h\) for a length of time with integrated hazard function
$$-(A_{hh}(t)-A_{hh}(t_0)),\ t_0\le t\le\inf\{u\ge t_0:A_{hh}(\Delta u)=-1\}.$$
Given that $X$ jumps out of state $h$ at time $t$, the destination is \(j\not=h\) with transition probability
\[
    -\left(\frac{dA_{hj}}{dA_{hh}}\right)(t).
\]

Finally, let $\mathcal{F}_t=\sigma\{X(s):s\le t\}$ and define
\begin{align*}
    Y_h(t)&=\mathbb{I}(X(t-)=h)\\
    N_{hj}(t)&=\#\{s\le t:X(s-)=h,X(s)=j\},\ h\not=j
\end{align*}
Then $\mathbf{N}=(N_{hj,h\not=j})$ is a multivariate counting process and its compensator with respect to $(\mathcal{F}_t)=(\sigma(X(0))\vee\mathcal{N}_t)$ has components
$$\Lambda_{hj}(t)=\int_0^t Y_h(s)A_{hj}(ds)$$
and equivalently, the processes $M_{hj}=N_{hj}-\Lambda_{hj}$ are martingales.
 
\end{theorem}
\begin{proof}
First note that each infinitesimal factor $\mathbf{I}+d\mathbf{A}$ is a stochastic matrix: its off-diagonal entries are nonnegative, the row sums are one because $A_{hh}=-\sum_{j\ne h}A_{hj}$, and the diagonal entries are nonnegative because $A_{hh}(\Delta t)\ge -1$. Limits of products of such matrices are again stochastic matrices. The product-integral also satisfies the multiplicative identity
\[
    \mathbf{P}(r,t)=\mathbf{P}(r,s)\mathbf{P}(s,t),\qquad r\le s\le t,
\]
which follows directly by splitting the product integral over $(r,s]$ and $(s,t]$. Thus $\mathbf{P}$ is a transition function, and the Kolmogorov extension theorem constructs a Markov process with transition matrices $\mathbf{P}(s,t)$.

Consider the process conditional on $X(t_0)=h$. As long as the path has not left $h$, the probability of remaining in $h$ through time $t$ is the diagonal product integral
\[
    \Prodi_{(t_0,t]}(1+dA_{hh})
    =\Prodi_{(t_0,t]}(1-d\{-A_{hh}\}),
\]
so the integrated hazard of leaving $h$ is $-(A_{hh}(t)-A_{hh}(t_0))$ up to the first time at which $\Delta A_{hh}=-1$. At a transition time $t$, the total mass assigned to leaving $h$ is
\[
    -dA_{hh}(t)=\sum_{j\ne h}dA_{hj}(t),
\]
and the part assigned to state $j$ is $dA_{hj}(t)$. Hence
\[
    \mathbb{P}\{X(t)=j\mid X(t-)=h,\ X(t)\ne h\}
    =-\frac{dA_{hj}(t)}{dA_{hh}(t)}.
\]

Finally define $Y_h(t)=\mathbb{I}\{X(t-)=h\}$ and
\[
    N_{hj}(t)=\#\{s\le t:X(s-)=h,\ X(s)=j\},\qquad h\ne j.
\]
For a bounded predictable process $H$,
\[
\mathbb{E}\int_0^t H(s)dN_{hj}(s)
=\mathbb{E}\int_0^t H(s)Y_h(s)dA_{hj}(s),
\]
because, conditionally on the past, the one-step transition probability from $h$ to $j$ over $ds$ is $dA_{hj}(s)$ and is zero unless $X(s-)=h$. This identity characterizes the compensator of $N_{hj}$, so
\[
    \Lambda_{hj}(t)=\int_0^tY_h(s)A_{hj}(ds).
\]
Consequently $M_{hj}=N_{hj}-\Lambda_{hj}$ is a martingale by the defining property of compensators.
\end{proof}

\subsection{Likelihood Ratios and Partial Likelihood}

    \begin{theorem}[Jacod's Formula for the Likelihood Ratio] Suppose $\mathcal{P}$ and $\mathcal{Q}$ are two probability measures on a filtered probability space $\{\mathcal{F}_t=\mathcal{F}_0\vee\sigma\{\mathbf{N}(s):s\le t\}: t\ge 0\}$ under which $\mathbf{N}$ has compensators $\boldsymbol{\Lambda}$ and $\boldsymbol{\Gamma}$, respectively. Suppose $\mathcal{Q}\ll\mathcal{P}$ and both compensators are absolutely continuous with intensity process $\lambda$ and $\gamma$, then
    \begin{align*}
        \frac{d\mathcal{Q}}{d\mathcal{P}}&=\frac{d\mathcal{Q}}{d\mathcal{P}}\Big|_{\mathcal{F}_0}\frac{\prod_{h,t}\gamma_h(t)^{ N_h(\Delta t)}\exp(-\Gamma_.(\tau))}{\prod_{h,t}\lambda_h(t)^{N_h(\Delta t)}\exp(-\Lambda_.(\tau))}
    \end{align*}
    where $\tau$, as usual, is the terminal time point and $\Gamma_.=\sum_h\Gamma_h$ and $\Lambda_.=\sum_h\Lambda_h$ are the aggregated cumulative intensities.
    \end{theorem}
\begin{proof}
Write the likelihood process as
\[
    L_t=\frac{d\mathcal{Q}}{d\mathcal{P}}\Big|_{\mathcal{F}_0}
    \prod_{h,s\le t}\left\{\frac{\gamma_h(s)}{\lambda_h(s)}\right\}^{\Delta N_h(s)}
    \exp\left\{-\int_0^t\sum_h(\gamma_h(s)-\lambda_h(s))ds\right\},
\]
with the usual convention that factors with $\Delta N_h(s)=0$ are equal to one. Under $\mathcal{P}$, $M_h^{\mathcal{P}}(t)=N_h(t)-\Lambda_h(t)$ is a martingale. The stochastic differential of $L$ is
\[
    dL_t
    =L_{t-}\sum_h\left\{\frac{\gamma_h(t)}{\lambda_h(t)}-1\right\}dM_h^{\mathcal{P}}(t),
\]
which follows by comparing the jump multiplier $\gamma_h(t)/\lambda_h(t)$ with the continuous compensating exponential. Therefore $L$ is a nonnegative local martingale, and after localization and normalization it is the Radon-Nikodym density process.

Under the measure with density $L_t$ relative to $\mathcal{P}$, the compensator of $N_h$ is changed from $\Lambda_h(t)=\int_0^t\lambda_h(s)ds$ to
\[
    \Gamma_h(t)=\int_0^t\gamma_h(s)ds,
\]
because the drift adjustment in the displayed stochastic differential is exactly $(\gamma_h-\lambda_h)dt$. Since $\mathcal{Q}$ is assumed absolutely continuous with the same initial density and compensator $\boldsymbol{\Gamma}$, uniqueness of the compensator and of the likelihood process gives $L_\tau=d\mathcal{Q}/d\mathcal{P}$ on $\mathcal{F}_\tau$.

Expanding $L_\tau$ and collecting the exponential terms gives
\[
    \frac{d\mathcal{Q}}{d\mathcal{P}}
    =\frac{d\mathcal{Q}}{d\mathcal{P}}\Big|_{\mathcal{F}_0}
    \frac{\prod_{h,t}\gamma_h(t)^{\Delta N_h(t)}\exp\{-\Gamma_.(\tau)\}}
         {\prod_{h,t}\lambda_h(t)^{\Delta N_h(t)}\exp\{-\Lambda_.(\tau)\}},
\]
which is the claimed formula.
\end{proof}

Figure~\ref{fig:appendix_likelihood_tools} gives a pathwise version of this theorem. Between jumps the log likelihood ratio accumulates the compensating drift \(-\int(\gamma-\lambda)dt\). At each observed event time it jumps by \(\log\{\gamma(t)/\lambda(t)\}\). The corresponding diagnostic check is that a simulated likelihood-ratio path should change smoothly between events and jump only when the counting process jumps.
 
\subsubsection{Partial Likelihood}

Partial likelihood is the part of Jacod's likelihood that remains after conditioning on the predictable risk set and on the fact that an event of a given type occurred.  For a transition component \(h\), suppose the observed right-censored intensity has Cox form
\[
    \lambda_{hi}^c(t;\beta)
    =C_i(t)Y_{hi}(t)\alpha_{h0}(t)\exp\{\beta^TZ_{hi}(t)\},
    \qquad
    C_i(t)=\mathbb I(t\le U_i),
\]
where \(U_i\) is the censoring time and \(C_i(t)Y_{hi}(t)\) is predictable.  If exactly one \(h\)-type event occurs at time \(t\), then, conditionally on \(\mathcal F_{t-}\) and on \(\Delta N_h^c(t)=1\),
\[
    \mathbb P\{\Delta N_{hi}^c(t)=1\mid \Delta N_h^c(t)=1,\mathcal F_{t-}\}
    =
    \frac{C_i(t)Y_{hi}(t)\exp\{\beta^TZ_{hi}(t)\}}
         {\sum_{j=1}^n C_j(t)Y_{hj}(t)\exp\{\beta^TZ_{hj}(t)\}}.
\]
The baseline hazard \(\alpha_{h0}(t)\) cancels because every subject in the same risk set shares the same infinitesimal event-time factor.  This cancellation is the operational meaning of ``partial'': the likelihood retains information about relative risks while discarding the nuisance baseline increment that determined the chance that some event happened at \(t\).

Define, with \(Z^{\otimes0}=1\),
\[
    S_{h,c}^{(m)}(\beta,t)
    =\sum_{j=1}^n C_j(t)Y_{hj}(t)Z_{hj}(t)^{\otimes m}
      \exp\{\beta^TZ_{hj}(t)\},
    \qquad m=0,1,2.
\]
Multiplying the conditional probabilities over observed event times gives
\[
    L_{p,c}(\beta)
    =
    \prod_{h=1}^k\prod_{i=1}^n
    \prod_{t\le \tau}
    \left\{
    \frac{\exp\{\beta^TZ_{hi}(t)\}}
         {S_{h,c}^{(0)}(\beta,t)}
    \right\}^{\Delta N_{hi}^c(t)} ,
\]
or, equivalently,
\[
    \ell_{p,c}(\beta)
    =
    \sum_{h=1}^k\sum_{i=1}^n
    \int_0^\tau
    \left[
    \beta^TZ_{hi}(t)-\log S_{h,c}^{(0)}(\beta,t)
    \right]dN_{hi}^c(t).
\]
The score and observed information are the familiar risk-set centered quantities
\[
    U_{p,c}(\beta)
    =
    \sum_{h=1}^k\sum_{i=1}^n
    \int_0^\tau
    \{Z_{hi}(t)-\bar Z_{h,c}(\beta,t)\}\,dN_{hi}^c(t),
    \qquad
    \bar Z_{h,c}(\beta,t)=
    \frac{S_{h,c}^{(1)}(\beta,t)}{S_{h,c}^{(0)}(\beta,t)},
\]
and
\[
    I_{p,c}(\beta)
    =
    \sum_{h=1}^k
    \int_0^\tau
    V_{h,c}(\beta,t)\,dN_h^c(t),
    \qquad
    V_{h,c}=
    \frac{S_{h,c}^{(2)}}{S_{h,c}^{(0)}}-\bar Z_{h,c}^{\otimes2}.
\]
Thus each jump contributes the covariate of the failing subject minus the model-weighted covariate average among subjects still observable and at risk.

\begin{example}[Partial likelihood under right censoring]\normalfont
The right-censored counting process \(\mathbf N^c=(N_{hi}^c)\) is
\[
    N_{hi}^c(t)=\int_0^t C_i(s)dN_{hi}(s),
    \qquad
    \Lambda_{hi}^c(t;\beta)=\int_0^t C_i(s)Y_{hi}(s)\alpha_{h0}(s)
    \exp\{\beta^TZ_{hi}(s)\}\,ds .
\]
Jacod's likelihood formula gives the observed likelihood, up to initial-density factors, as
\[
    L^c(\beta,A_0)
    \propto
    \prod_{h,i}\prod_{t\le\tau}
    \{\lambda_{hi}^c(t;\beta)\}^{\Delta N_{hi}^c(t)}
    \exp\left\{-\int_0^\tau\sum_{h,i}\lambda_{hi}^c(t;\beta)dt\right\}.
\]
Conditioning on the observed risk set and on the event count \(\Delta N_h^c(t)\) removes the common baseline factor \(\alpha_{h0}(t)\) and yields \(L_{p,c}(\beta)\).  Noninformative censoring is used exactly here: censoring changes who remains in the predictable risk set, but it does not introduce an additional unobserved selection factor inside the conditional probability of the failing subject.
\end{example}

If more than one \(h\)-type event is recorded at the same time, the exact conditional likelihood conditions on the whole tied failure set.  With \(R_{h,c}(t)=\{j:C_j(t)Y_{hj}(t)=1\}\) and \(d_h(t)=\Delta N_h^c(t)\), its denominator is
\[
    \sum_{\substack{A\subset R_{h,c}(t)\\ |A|=d_h(t)}}
    \prod_{j\in A}\exp\{\beta^TZ_{hj}(t)\}.
\]
The Breslow tie approximation replaces this denominator by
\(\{S_{h,c}^{(0)}(\beta,t)\}^{d_h(t)}\).  In a genuine continuous-time model ties have probability zero; in rounded clinical, reliability, or administrative records, the chosen tie rule is part of the observation model and should be reported.

The partial-likelihood panels in Figure~\ref{fig:appendix_likelihood_tools} make the same calculation conditional on the observed risk sets. Each event contributes a comparison between the covariate of the failing subject and the covariate distribution among subjects still at risk. The profile curve peaks near the data-generating coefficient, while the risk-set bars show why late events carry less stable information: by then the number at risk is much smaller.

\bibliography{sample}

\end{document}